\documentclass[a4paper,11pt]{article}
\pdfoutput=1 

\usepackage{jheppub} 
\usepackage{amsmath,amssymb}
\usepackage{graphicx}
\usepackage{subcaption}
\usepackage[percent]{overpic}
\usepackage[dvipsname]{xcolor}
\usepackage{multirow}
\usepackage{bbm}

\def\be{\begin{equation}}
\def\ee{\end{equation}}

\newcommand{\CD}{\mathcal{D}}
\newcommand{\CS}{\mathcal{S}}
\newcommand{\CO}{\mathcal{O}}
\newcommand{\CK}{\mathcal{K}}
\newcommand{\CE}{\mathcal{E}}
\newcommand{\CN}{\mathcal{N}}
\newcommand{\CC}{\mathcal{C}}

\newcommand{\fp}{{{\mathfrak{p}}}}

\newcommand{\IC}{\mathbb{C}}
\newcommand{\IP}{\mathbb{P}}
\newcommand{\IZ}{\mathbb{Z}}
\newcommand{\IF}{\mathbb{F}}
\newcommand{\IR}{\mathbb{R}}

\newcommand{\Ext}{\mathrm{Ext}}
\newcommand{\ch}{\mathrm{ch}}
\renewcommand{\-}{\text{-}}
\renewcommand{\(}{\left(}
\renewcommand{\)}{\right)}

\title{Exponential BPS graphs and D-brane counting on toric Calabi-Yau threefolds: Part II}

\author[a]{Sibasish Banerjee}
\author[b]{Pietro Longhi}
\author[c]{Mauricio Romo}

\affiliation[a]{Weyertal 86-90, Department of Mathematics, University of Cologne, 50679, Cologne, Germany }

\affiliation[b]{Institute for Theoretical Physics, ETH Zurich, 8093, Zurich, Switzerland}

\affiliation[c]{Yau Mathematical Sciences Center, Tsinghua University, Beijing, 100084, China}

\emailAdd{
{sbanerje@math.uni-koeln.de},
longhip@phys.ethz.ch, 
mromoj@tsinghua.edu.cn
}

\abstract{
We study BPS states of 5d $\mathcal{N}=1$ $SU(2)$ Yang-Mills theory on $S^1\times \mathbb{R}^4$. 
Geometric engineering relates these to enumerative invariants for the local Hirzebruch surface $\mathbb{F}_0$. We illustrate computations of Vafa-Witten invariants via exponential networks, verifying fiber-base symmetry of the spectrum at certain points in moduli space, and matching with mirror descriptions based on quivers and exceptional collections.
Albeit infinite, parts of the spectrum organize in families described by simple algebraic equations.
Varying the radius of the M-theory circle interpolates smoothly with the spectrum of 4d $\mathcal{N}=2$ Seiberg-Witten theory, recovering spectral networks in the limit. }

\begin{document} 

\maketitle
\flushbottom

\section{Introduction}

In \cite{Eager:2016yxd} Eager, Selmani and Walcher introduced the notion of exponential networks, a framework to study BPS states of Type IIA string theory on a toric Calabi-Yau threefold $X$.
Mirror symmetry plays a central role, since exponential networks provide a way to study special Lagrangian cycles in the mirror geometry $Y$, effectively probing the dual spectrum of $D3$ branes in Type IIB string theory on $Y$ rather than actual $D4\-D2\-D0$ boundstates on $X$.
It is well known that the mirror of a toric threefold is captured by the geometry of a Riemann surface, the mirror curve $\Sigma$. Exponential networks build on two key observations of \cite{Klemm:1996bj}. The first one, is that the spectrum of special Lagrangians of $Y$ can be studied by focusing on the simpler problem of studying calibrated Lagrangian cycles on $\Sigma$. 
The second key idea is that, viewing $\Sigma$ as an algebraic curve in $(\IC^*)^2$, one may choose a projection $\pi:\Sigma\to \IC^*$ and study \emph{saddles} on $\IC^*$ corresponding to images of the calibrated cycles on $\Sigma$ under the projection map $\pi$.
One of the main results of \cite{Eager:2016yxd} was a striking match between the spectrum of saddles on $\IC^*$ and the spectrum of stable representations of the quiver associated to $X$. 
A gap in this correspondence was the lack of a way to \emph{count} the BPS states represented by each saddle: on the quiver side one may compute the Euler characteristics (or their refinement to Poincar\'e polynomials) of quiver representations varieties, but there was no direct way to compute these from the viewpoint of exponential networks.
An attempt to fill this gap was put forth by the authors of the present paper in~\cite{Banerjee:2018syt}. 
By endowing exponential networks with additional combinatorial data attached to saddles, an algorithm to compute the BPS index of any saddle was provided.
Successful tests include the computation of boundstates of $D0$ branes in $\IC^3$, and of $D2\-D0$ boundstates in $\CO(-p)\oplus \CO(p-2)\to \IP^1$ for $p=0,1$ \cite{Banerjee:2018syt, Banerjee:2019apt}.
The BPS spectra that we found in those cases coincided with expectations coming, for instance, from computations  of Gopakumar-Vafa invariants via topological strings.

In the present paper we explore a yet richer class of examples, featuring compact four-cycles. 
BPS states in these models include boundstates of $D4$ branes, which are not directly captured by closed topological strings, and which are counted by Vafa-Witten invariants~\cite{Vafa:1994tf}.
Moreover, the simultaneous presence of compact two-cycles and compact four-cycles 
changes the behavior of the BPS spectrum dramatically compared to examples without four-cycles.
For one thing, one may empirically observe that the spectrum is generally much richer, and harder to describe in closed form.
More fundamentally, the spectrum is now sensitive to geometric moduli due to wall-crossing phenomena involving $D4$ and $D2$ branes.
We focus on the case $X=K_{\IF_0}$, and study the BPS spectrum at a special point $\mathbf{Q}_0$ in the moduli space of $X$ (a definition can be found below). 
We also derive equations characterizing generating functions of \emph{infinitely many} BPS indices at once.
For example the spectrum of $n(\overline{D2}_f\-2D0)$ boundstates is packaged by a generating function $Q(x)= \prod_{n\geq 1}(1-x^n)^{n\Omega(n)}$ obeying the algebraic equation
\be\label{eq:algebraic-eq-7}
	Q(x) = 1 + x \, Q(x)^7 \,.
\ee

Since the point $\mathbf{Q}_0$ corresponds to a symmetric choice of the moduli under fiber-base exchange, it is natural to ask whether the BPS spectrum is also symmetric under exchange of $D2_f$ and $D2_b$ ($D2$ branes wrapping fiber and base $\IP^1$s, respectively). 
By direct inspection, we find that this is indeed the case. 
In fact, exponential networks provide a natural set of `basis' saddles which makes the fiber-base symmetry of the spectrum manifest.
This is in stark contrast with quiver descriptions of BPS states, where fiber-base exchange symmetry is not manifest even at $\mathbf{Q}_0$. 

The geometric nature of exponential networks lends itself nicely to exploring the behavior of the BPS spectrum as moduli of $X$ are tuned to certain limits.
Starting from M-theory on $X\times S^1\times \IR^4$, which engineers 5d $\CN=1$ Yang-Mills theory on $S^1\times \IR^4$, we explore how the spectrum varies with the radius $R$ of the M-theory circle. 
In the limit $R\to 0$ the spectrum simplifies dramatically as some BPS states become infinitely heavy, and we recover the spectrum of $SU(2)$ Seiberg-Witten theory. In the 4d limit, exponential networks reduce to spectral networks.
Back at radius $R=1$, we also consider a limit in which some of the Kahler moduli grow infinitely large. Again, some states become heavy and decouple, while others display a more subtle behavior with a jump in the BPS index while remaining at finite mass. Eventually we recover the spectrum of the half-geometry studied previously in \cite{Banerjee:2019apt}.
Last, but not least, we use exponential networks to derive the BPS quiver for $K_{\IF_0}$ and the associated potential, matching with recent studies based on quivers and exceptional collections.

\subsection*{Brief review of BPS counting with exponential networks}

This paper is a companion to \cite{Banerjee:2019apt, Banerjee:2018syt}, to which we refer for definitions and technical details. 
To provide context for the results in this paper, and in an attempt to provide a self-cointained interpretation for them, we recollect the key ideas behind exponential networks.

The main idea behind the formulae of \cite{Banerjee:2018syt} came from switching the viewpoint from string theory to the associated quantum field theory.
Geometric engineering assigns to $X$ a gauge theory with eight supercharges, whose BPS spectrum precisely coincides with the spectrum of $D4\-D2\-D0$ 
branes that exponential networks are supposed to capture.
From the viewpoint of the field theory, the mirror curve $\Sigma$ corresponds to a Seiberg-Witten curve for the Coulomb branch geometry, corroborating the correspondence between BPS states and cycles on $\Sigma$. 
Another very useful viewpoint on the mirror curve, is that it arises as the quantum-corrected (in genus zero) moduli space of an A-brane supported on a suitable non-compact Lagrangian $L\subset X$. In many cases of interest, this coincides in fact with the Aganagic-Vafa toric brane \cite{Aganagic:2000gs, Aganagic:2001nx}. 
Translated into field theory, a $D4$ wrapping $L$ gives rise to a codimension-two defect in the gauge theory engineered by $X$, the A-brane moduli space is identified with the space of vacua of the defect theory. 

From the viewpoint of the defect theory, exponential networks coincide with solutions to the BPS equations characterizing solitons between two vacua. The counting of these solitons is well-understood in terms of the CFIV index \cite{Cecotti:1992qh}, whose computation can be conveniently approached through $tt^*$ geometry \cite{Cecotti:1991me, Cecotti:1992rm}. 
The study of BPS spectra of 2d (2,2) models and their wall-crossing properties is a subject with a long history.
On the other hand, the study of BPS spectra of \emph{defects}, such as 2d (2,2) models \emph{coupled} to a 4d $\CN=2$ gauge theory, is a much more recent endeavor.
In seminal work \cite{Gaiotto:2011tf}, Gaiotto, Moore and Neitzke explained how the BPS spectrum of such defects actually encodes the BPS spectrum of the bulk theory to which they couple.
The power of this observation lies in the fact that the defect spectrum is far easier to compute systematically, leading to the birth of spectral networks \cite{Gaiotto:2012rg}.
The approach to BPS counting with exponential networks is derived from the same principle, with the main difference stemming from the replacement of a 2d-4d system by a 3d-5d one. The latter arises naturally in M-theory on $X\times S^1\times \IR^4$ with an M5 brane wrapping $L$. A circle-uplift of $tt^*$ geometry developed by Cecotti, Gaiotto and Vafa \cite{Cecotti:2013mba} leads to an interpretation of the network in terms of BPS states of the 3d defect theory. Another fundamental role of 3d $tt^*$ geometry  is to provide a meaningful way to count these states. 
Through the bulk-defect coupling, the BPS spectrum on the defect encodes the BPS spectrum of the bulk, and the formulae  of \cite{Banerjee:2018syt} provide a way to extract this information from exponential networks.

\subsection*{Organization of this paper}
In section \ref{sec:geometry} we introduce the geometry of $K_{\IF_0}$ and its mirror. We also comment on the 4d limit to the Seiberg-Witten curve of 4d $\CN=2$ Yang-Mills theory.
Section \ref{sec:exc-coll-mirror} is devoted the identification of cycles on the mirror curve $\Sigma$ with sheaves on $X$. 
Section~\ref{sec:spectrum-conifold-point} contains the core of this paper, namely a detailed analysis of exponential networks for $X$ and an extensive description of the BPS spectrum.
In section \ref{eq:limits-of-spectrum} we consider a 4d limit obtained by shrinking the M-theory circle, and a limit to the half-geometry obtained by sending one of the Kahler moduli to infinity.
Section \ref{sec:conclusions} concludes with comments on the results obtained, puzzles that we encountered, and relation to various other approaches to the problem studied in this paper.
The Appendix contains detailed plots of BPS states, addenda to computations, and bonus technical material pertaining to the general framework of exponential networks.

\section{Geometry of local $\IF_0$ and its mirror}\label{sec:geometry}

This section introduces the problem studied in this paper. We begin with a brief review of geometric engineering of $5d$ $\CN=1$ gauge theories via M-theory on toric Calabi-Yau threefolds.
We will then review how BPS states of the 5d gauge theory arise from M2 and M5 branes wrapped on compact cycles, and how these are mapped to special Lagrangian branes in the mirror geometry, where exponential networks allow to count stable boundstates.
At the end of this section we include a detailed analysis of the mirror geometry to lay the groundwork for our later study of BPS states with exponential networks.

\subsection{Geometric engineering from M-theory}

The five-dimensional $SU(2)$ super Yang-Mills theory with trivial Chern-Simons coupling can be engineered in M-theory \cite{Seiberg:1996bd, Morrison:1996xf, Douglas:1996xp, Katz:1996fh} by considering the spacetime geometry $\IR^4\times S^1\times X$ with
\be
	X = K_{\IF_0}
	\qquad
	\text{where} 
	\qquad
	\IF_0 = \IP^1_b\times\IP^1_f  \,,
\ee
and $K_{\IF_0}$ denotes the total space of the canonical bundle of $\IF_0$.
We will compactify the theory on a circle of finite radius $\IR^4\times S^1_R\times X$, and effectively study the 4d $\CN=2$ theory of the Kaluza-Klein modes of the 5d theory.
For this reason, we switch to a more appropriate Type IIA description from now on.

The fiber $\IP^1_f$ is associated with the $U(1)$ symmetry surviving on the Coulomb branch, while the base $\IP^1_b$ is associated with the topological $U(1)$ symmetry.
BPS states arise from branes wrapping even-dimensional cycles of $X$, with charges valued in $H^{\mathrm{even}}(X)$, which is four-dimensional. 
In the case of pure $SU(2)$ gauge theory, W-bosons arise from D2-branes wrapping $\IP^1_f$, while the magnetic monopole (a string in 5d) arises from a D4-brane (M5 brane in 5d) wrapped on the entire $\IF_0$. 
Recall that the the tension of BPS branes in M-theory are 
\be 
M = \frac{{\mathrm{vol}} (\CC)}{\ell_p^3}, \quad T = \frac{{\mathrm{vol}} (E)}{\ell_p^6}
\ee 
where $\ell_p$ is the 11d Planck length. 
Consider then the compactification  on the  M-theory circle, with $R = \ell_s g_s$ as the radius and $\alpha' = \ell_s^2 = \ell_p^3/R$,
where $g_s$ is the string coupling constant and $l_s$ string lengthscale. The resulting masses of BPS particles are then (up to numerical factors)
\be
M_{\mathrm{D0}} = \frac{1}{R}, \quad M_{\mathrm{D2}} = \frac{{\mathrm{vol}} (\CC)}{R\alpha'}, \quad 
M_{\mathrm{D4}} =  \frac{{\mathrm{vol}} (E)}{R\alpha'^2}. 
\ee 
The main goal of this paper is to study the spectrum of these and other BPS states, and understand their properties in different regions of the Coulomb moduli space.

On the one hand, the geometric definition of exponential networks was originally motivated by mirror symmetry in String Theory \cite{Klemm:1996bj, Eager:2016yxd}. On the other hand the construction of a nonabelianization map, which allows to perform the counting of BPS states, was obtained via a field-theoretic reinterpretation \cite{Banerjee:2018syt}.
Either way, one is led to study the local mirror geometry on the Type IIB side. If $X$ is toric, as in our case, the mirror  has the general form  $uv = F(x,y)$ where $u,v \in \IC$ and $x,y \in \IC^*$. The mirror curve
\be 
\Sigma : \quad F(x,y) = 0
\ee
captures all the physics on the mirror side, and in particular all kinds of BPS states arising from boundstates of D0-branes, D2-branes
 wrapping either fiber or base $\IP^1$, and D4-branes, map to special Lagrangian one-cycles on  $\Sigma$, calibrated by a differential one-form.
In the case of interest to us, $\Sigma$ is a torus with four punctures. 

\paragraph{The geometry and its mirror.} 
The mirror geometry can be computed starting from the toric description of $X$. Starting from a $U(1) \times U(1)$ gauged linear sigma model with charges $(-2,1,1,0,0)$ and 
$(-2,0,0,1,1)$ under the two $U(1)$'s,  $X$ is obtained via the symplectic quotient 
\be 
K_{\IF_0} = \{
	-2\vert w_1\vert^2 + \vert w_2\vert^2 + \vert w_3\vert^2 = 2 {\mathrm Re} \log \vert Q_b\vert, 
	\  
	-2\vert w_1\vert^2 + \vert w_4\vert^2 + \vert w_5\vert^2 = 2 {\mathrm Re} \log \vert Q_f\vert 
	\}/(S^1)^2
\ee  
where $(w_1,..,w_5)$ are coordinates on $\IC^5$. 
The curve $\Sigma$ can be obtained by considering the Hori-Vafa mirror \cite{Hori:2000kt}
\be 
x_1^{-2} x_2 x_3 = Q_b^2, \quad x_1^{-2}x_4x_5 = Q_f^2, \quad x_1+x_2+x_3+x_4+x_5 = 0.
\ee   
Choosing the patch $x_1 = 1$ and $x_2= -Q_b x,\, x_4 = -Q_f y$, one obtains 
\be \label{eq:mirror-curve}
 Q_b (x+x^{-1}) + Q_f(y+y^{-1}) -1 = 0. 
\ee 

\paragraph{Exponential networks and BPS state counting.}
Having reviewed the definition of the mirror geometry, we now explain how exponential networks enter the story, following the original motivations from M-theory 
\cite{Eager:2016yxd} and the field theoretic reinterpretations in our previous work \cite{Banerjee:2018syt, Banerjee:2019apt}.
Consider a codimension-two defect engineered by wrapping a single M5-brane on $L\times S^1 \times \IR^2$, where $L$ is special Lagrangian on $X$. 
The low energy dynamics of the defect is described by a $3d$ $\CN=2$ theory $T[L]$ on $S^1 \times \IR^2$, whose field content is determined by the geometry of $L$. 
This theory couples to the bulk 5d theory engineered by M-theory on $X$, this 3d-5d coupling is vital to the counting of 5d BPS states via exponential networks.
An in-depth discussion can be found in our previous, here we just indicate the relevant BPS states that our exponential networks count. 

Let us specialize to the case in which  $L$ is a toric brane, for simplicity. Then $T[L]$ admits a description as a U(1) gauge theory with a finite number of charged chiral multiplets coupled to the bulk and background fields. 
Vevs and Wilson lines of the latter play the role of twisted masse for $T[L]$. 
In a suitable regime, the 5d degrees of freedom may be integrated away, leaving an effective 3d-5d twisted superpotential for the 3d degrees of freedom \cite{Gaiotto:2013sma}.
This encodes information about the 5d gauge theory, in fact the critical points coincide with the 5d Seiberg-Witten curve, which is also identified with the mirror curve of $(X,L)$.
The purpose of exponential networks is to probe this geometry and extract information about 5d BPS states.

For concreteness, suppose the $5d$ theory has gauge group $SU(N)$. The $3d$ theory is then described by a circle-uplift of a 2d, $\CN= (2,2)$, U(1) GLSM to s 3d, $\CN=2$
gauge theory with a charged chiral miltiplet that transforming in the fundamental representation of $SU(N)$. 
The 3d-5d coupling consists of a minimal coupling for the $3d$ chiral field to the $5d$ vectormultiplet (restricted to the defect). 
Then the quantum moduli space of the vacua can is argued to  coincide with the mirror curve for the toric brane. 

Concerning the BPS states of the 5d theory, they can be identified on the one hand with cycles on the Seiberg-Witten curve. On the other hand, going back to the String Theory engineering, they also map to special Lagrangian branes in the mirror. These A-branes are supported on compact lagrangian submanifolds, which furthermore admit a projection down to $\Sigma$. 
The calibrating differential on $\Sigma $ is  $\lambda = \log y \, d\log x$ and descends directly from the holomorphic top form on the mirror threefold \cite{Klemm:1996bj}.  The identification of the mirror curve and the Seiberg-Witten curve ties together the field-theoretic and string-theoretic viewpoints on these BPS states as one-cycles on $\Sigma$.

The study of calibrated cycles leads naturally to the geometric definition of exponential networks.
BPS states are counted by the \emph{BPS index} (\emph{a.k.a.} second helicity supertrace), which can be computed from properties of a  ``nonabelianization map'' associated with each exponential network.
This map was constructed in  \cite{Banerjee:2018syt} following inspiration from spectral networks and 2d-4d framed wall-crossing \cite{Gaiotto:2011tf, Gaiotto:2012rg}. In later sections, we will make explicit use of this approach to determine the BPS spectrum for the case in interest.

\subsection{Mirror curve}

The mirror geometry can be presented as a conic bundle degenerating over an algebraic curve \cite{Hori:2000kt}
\be
	uv = F(x,y) \qquad (u,v,x,y)\in \IC^2\times (\IC^*)^2 \,.
\ee
With a suitable choice of coordinates, the mirror curve $\Sigma$ reads
\be\label{eq:5d-curve}
	 {Q_b(x+x^{-1}) + Q_f(y+y^{-1})-1 = 0} \,,
\ee
and depends on two complex moduli $Q_{b,f}\in (\IC^*)^2$.
There is a manifest $\IZ_2\times\IZ_2$ symmetry taking $x,y$ to the respective inverse.

We will be interested in a presentation of this curve as a two-fold ramified covering 
\be
	\pi:\Sigma\to \IC^*_x\,.
\ee
The number of sheets depends on a choice of framing, we will work with the choice implied in (\ref{eq:5d-curve}).
As a consequence of the $\IZ_2$ symmetry acting on $y$, the two sheets obey the identity
\be
	y_+ y_- = 1 \,,
\ee
for all $x$. This implies
\be
	\lambda_++\lambda_- = (\log y_++\log y_-) \, d\log x \equiv 0\,.
\ee
Below we will be interested in the four-dimensional limit of this curve, where this condition maps to the tracelessness property of the ${\mathfrak{su}}(2)$ Higgs field of the Hitchin system associated to 4d $\CN=2$ super Yang-Mills. \footnote{The $\IZ_2$ symmetry $x\to x^{-1}$, is also preserved in the 4d limit.}

The branching locus of $\pi$ consists of four points
\be\label{eq:bpts-positions}
	x_{\sigma_1,\sigma_2}=  \frac{1+2 \sigma_1  {Q_f}+\sigma_2\sqrt{-4 {Q_b}^2+4 {Q_f}^2 +4 \sigma_1 {Q_f}+1}}{2 {Q_b}}	
\ee
for $\sigma_i=\pm1$.
Branch points coincide when 
\be\label{eq:bp-collisions}
	Q_b \pm Q_f = \pm \frac{1}{2}
\ee
with signs chosen independently, corresponding to four codimension-one strata in the moduli space of the curve.
Overall, there are six singular strata in the moduli space of the curve, at loci $(Q_b, Q_f)$ 
\be\label{eq:discriminant-locus}
\begin{split}
	& \CD_{b}:=\{(0,Q)\}
	\\
	& 
	\CD_{1}:=\{(Q,Q+1/2)\}
	\\
	&
	\CD_{3}:=\{(Q,Q-1/2)\}
\end{split}
\qquad
\begin{split}
	& 
	\CD_{f}:=\{(Q,0)\}
	\\
	& 
	\CD_{2}:=\{(Q,-Q+1/2)\}
	\\
	&
	\CD_4:=\{(Q,-Q-1/2)\}
\end{split}
\ee
with $Q\in \IC^\times$ a local coordinate along each divisor.

The mirror curve has four punctures, two above $x=0$ and two above $x=\infty$.
In the limit $x\to0$ the two sheets go respectively to zero and to infinity
\be
	y_\pm \sim -\(\frac{Q_f}{Q_b} x\)^{\pm 1} + \dots
\ee
By $\IZ_2$ symmetry, the same behavior occurs in the limit $x\to\infty$, therefore all four punctures are of \emph{logarithmic} type with $\lambda \sim (\log x)^2$.
We label punctures at position $(x,y)$ as follows
\be
	\fp_1 = (0,0)
	\qquad
	\fp_2 = (0,\infty)
	\qquad
	\fp_3 = (\infty,\infty)
	\qquad
	\fp_4 = (\infty,0)
\ee

\subsection{Four-dimensional limit}\label{sec:4d-limit-geometry}

To discuss the four-dimensional limit, it is necessary to introduce dependence on the radius $R$ of the M-theory circle.
One way to do this, is to notice that the curve $F(x,y)=0$, endowed with the 1-form 
\be
	\lambda = \log y\,d\log x \ \ \in\Omega^1(\IC^*)^2
\ee
pulled back to $\Sigma$, coincides with the spectral curve of a relativistic Toda system. 
If $1/R$ is identified (up to rescaling by numerical constants) with the the speed of light, the limit $R\to 0$ recovers to the non-relativistic Toda system associated to $SU(2)$ Seiberg-Witten theory. It was thus proposed in \cite{Nekrasov:1996cz} that, at finite radius, $\Sigma$ is the Seiberg-Witten curve of 5d $\CN=1$ super-Yang-Mills theory on $S^1_R\times \IR^4$  with vanishing Chern-Simons coupling.

We take both coordinates $x,y$ and moduli $Q_b, Q_f$ to depend on $R$ through the parametrization adopted in \cite{Nekrasov:1996cz}
\be\label{eq:5d4dsubs}
	x=e^{q} \,,
	\qquad
	y = e^{\sqrt{2} i R p }\,,
	\qquad
	Q_b = -\Lambda^2 R^2\,,
	\qquad
	Q_f = \frac{1}{2} R^4 U(R)^2\,.
\ee
In order to take the limit $R\to 0$, we further assume that $U(R)$ behaves as follows
\be\label{eq:Uu}
	{Q_f = \frac{1}{2} R^4 U^2 = \frac{1}{2}(1 + 2 R^2 u + \dots)} 
\ee
Expanding (\ref{eq:5d-curve}) near $R=0$
\be
	0 =- 2R^2\(
	\frac{p^2}{2}  + \Lambda^2\cosh q - u
	\)
	+O(R^3) \,,
\ee
we  recognize the spectral curve of the Toda system associated to 4d $\CN=2$  $SU(2)$ super Yang-Mills.
In this limit, the differential becomes
\be\label{eq:5d-differential}
	{\lambda = \log y\, d\log x  = i\sqrt{2} R\,  p\, dq}
\ee
By an appropriate rescaling we define
\be\label{eq:4d-differential}
	{\lambda_{4d} = \frac{1}{ i\sqrt{2} R}\lambda = p\frac{dx}{x}} \,,
\ee
in terms of which the curve becomes
\be\label{eq:4d-curve}
	{\lambda_{4d}^2 = \(-\frac{\Lambda^2}{x^3} +\frac{2u}{x^2} - \frac{\Lambda^2}{x}\) \, dx^2} \,.
\ee
This is a two-fold covering of the $x$-plane with ramification at two branch points, and at two irregular punctures located at $x=0,\infty$.
This coincides with the class $\CS$ presentation, which will be useful to make contact with spectral networks.

It is interesting to observe what happens to branch points and punctures in this limit.
Using the $R$-dependent parametrization (\ref{eq:5d4dsubs})-(\ref{eq:Uu}) into the expressions  (\ref{eq:bpts-positions}) for the positions of branch points yields 
\be
\begin{split}
	x_{-,-}  &\to 
	\frac{u + \sqrt{ \left(u^2-\Lambda ^4\right)}}{\Lambda ^2},
	\\
	x_{-,+} &\to 
	\frac{u-\sqrt{ \left(u^2-\Lambda ^4\right)}}{\Lambda ^2 },
\end{split}
\qquad
\begin{split}
	x_{+,-} & \to
	-\frac{\Lambda ^2 R^2}{2}+O\left(R^3\right),
	\\
	x_{+,+} & \to
	-\frac{2}{\Lambda ^2 R^2}+O\left(1\right)
\end{split}
\ee
Two branch points remain at finite distance, while $x_{\pm,-}$ end up respectively at $x=0,\infty$.
Those remaining at finite distance are $x_{-,\pm}$, as was to be expected. Indeed, since $y_+ y_-=1$, at branch points of the 5d curve one must have $y=\pm 1$, corresponding to $\log y$ equal to $0$ or to $i\pi$, moreover one has $y_{\pm}(x_{\sigma_1,\sigma_2}) = -\sigma_1 $. In the 4d limit we expect branch points to lie at $\lambda = 0$, because of tracelessness of the Higgs field.
The two other branch points, namely $x_{+,\pm}$ end up absorbed into punctures at $x=0,\infty$. 
When this happens, a branch point merges the two sheets of $\pi$, inducing a mutual cancellation of the logarithmic singularities that are initially present on each sheet. 
What remains after this process is a standard class $\CS$ (irregular) puncture.

\section{Exceptional collections in the mirror}\label{sec:exc-coll-mirror}

A problem that arises as soon as one considers geometries with compact four-cycles, is how to relate Lagrangian cycles, representing A-branes, on the mirror side and sheaves on the toric side.
In the context of exponential networks, we are concerned with certain compact special Lagrangian cycles in the mirror Calabi-Yau (the conic bundle), which are in one-to-one correspondence with calibrated 1-cycles on the mirror curve $\Sigma$. 
Therefore we focus on the question of establishing a map between these 1-cycles in $\Sigma$ and sheaves on $X$.

In certain cases, such as when $X$ corresponds to line bundles over $\mathbb{P}^{1}$,  it is easy to find an answer because the periods of cycles on $\Sigma$ coincide exactly with linear combinations of (complexified) Kahler volumes of compact 2-cycles on the toric side, and the $D0$ central charge ($Z_{D0} = 2\pi/R$ in our choice of normalization).
In general however this is no longer true: instanton corrections are present, and contribute nontrivially to the mirror map, making the problem significantly more challenging.

We will discuss two approaches. The first one is a systematic approach developed a long time ago. The second one is more pedestrian and tailored to the case of interest to us. 
While the material of this section is of fundamental importance to interpret results of exponential networks in any setup involving compact four-cycles, readers who are mainly interested in the results on the BPS spectrum can safely skip ahead. 

\subsection{A view from the Fukaya category}\label{sec:fuk-cat}

In this work we encounter that the counting of certain class of BPS objects on the 5d theory corresponds to calibrated cycles in $\Sigma$. The purpose of this section is to motivate this interpretation from a categorical point of view. We are dealing with toric CY 3-folds $X$, which can be written as $K_{Z}$ where $Z$ is a smooth compact Fano manifold\footnote{Several features carries over if we consider a more general situation such as $X=\oplus_{n}\mathcal{L}_{n}$ where $\mathcal{L}_{n}$ are line bundles over a Fano base.}. The BPS states in the 5d theory are mapped to bound states of B-branes on $X$, that is, to objects in $D^{b}Coh(X)$. However, we are not dealing with all the objects in $D^{b}Coh(X)$ but rather with the full subcategory $\mathcal{D}$ of objects supported on the Fano base $Z$ (the zero section of $K_{Z}$, for  a more precise definition see \cite{Bridgeland:2005fr}). When $Z$ posses a full strong exceptional collection (which is always our case), is known that $D^{b}Coh(K_{Z})$ is equivalent to $D^{b}(\mathrm{Mod-}Q)$, the derived category of a quiver $Q$ with superpotential, constructed from the exceptional collection \cite{Bridgeland:2005fr}, moreover, the full subcategory $\mathcal{D}$ is equivalent to the full triangulated subcategory $D^{b}_{0}(\mathrm{Mod-}Q)$ of $D^{b}(\mathrm{Mod-}Q)$ generated by the simple modules $T_{j}$ (i.e., the modules whose dimension vector is $1$ at a single node of $Q$ and zero for the rest). We then have
\be
\mathcal{D}\cong	D^{b}_{0}(\mathrm{Mod-}Q)
\ee
by homological mirror symmetry we have that the category $D^{b}Coh(K_{Z})$ is equivalent to the derived Fukaya-Seidel category of the Lefshetz fibration (i.e. LG model) $W:(\mathbb{C}^{*})^{n}\rightarrow \mathbb{C}$ where $W$ is the mirror potential of $Z$ and the mirror curve to $K_{Z}$ is given by $\Sigma=W^{-1}(0)$, we denote this relation as
\be
D^{b}Coh(Z)\cong	D^{b}\mathcal{F}uk(W)
\ee
indeed, $\mathcal{F}uk(W)$ is a subcategory of $\mathcal{F}uk(\Sigma)$ generated by a distinguished basis of vanishing cycles $(\gamma_{1},\ldots,\gamma_{k})$ \cite{2000math......7115S,2000math.....10032S,seidel2008fukaya}, a concept we will define below. Hence the relevant category to study the mirror of $\mathcal{D}$ is $D^{b}\mathcal{F}uk(\Sigma)$ (more comments below) and also we have a quiver theory interpretation via $D^{b}_{0}(\mathrm{Mod-}Q)$. In order to make this concepts more precise for our purposes, we need to define the distinguished basis of vanishing cycles. Given $W:(\mathbb{C}^{*})^{n}\rightarrow \mathbb{C}$ such that all is critical points $p_{k}\in \mathrm{Crit}(W)$ are isolated and nondegenerate (which is the case for example for mirrors of Fano manifolds and our present situation), we can associate a vanishing path to it, this is a map $\gamma_{k}:[0,1]\rightarrow \mathbb{C}$ satisfying
\begin{itemize}
 \item $\gamma_{k}(0)=0$.
 \item $\gamma_{k}(1)=p_{k}$.
 \item $\gamma(t)\not\in \mathrm{Crit}(W)$ for $0< t <1$.
\end{itemize}
The condition $\gamma_{k}(0)=0$ can be changed to $\gamma_{k}(0)=\lambda$ where $\lambda$ is any regular value of $W$ i.e. $W^{-1}(\lambda)$ is smooth. In our case $0$, is a regular value and is convenient to use it, since $\Sigma=W^{-1}(0)$. Then, a distingushed set of vanishing paths $\{\gamma_{k}\}_{k=1}^{s}$ where $s=|\mathrm{Crit}(W)|$ is an ordered set satisfying
\begin{itemize}
 \item $\{\gamma_{k}(0)\}_{k=1}^{s}=\mathrm{Crit}(W)$.
 \item For $k\neq k'$, the paths $\gamma_{k}$, $\gamma_{k'}$ intersect only at $0$.
 \item $\mathrm{arg}(\gamma'_{1}(0))>\cdots >\mathrm{arg}(\gamma'_{s}(0))$.
\end{itemize}
at $p_{k}$ a 1-cycle $L_{k}\subset \Sigma=W^{-1}(0)$ pinches off, we call this 1-cycle a vanishing cycle, associated with $p_{k}$. The objects of $\mathcal{F}uk(W)$ are given by the ordered set $\{L_{k}(0)\}_{k=1}^{s}$ (or the Lefshetz thimbles $D_{k}$ associated to $\gamma_{k}$, that satisfy $\partial D_{k}=L_{k}$) called a distinguished set of vanishing cycles or more precisley, by small perturbations of them that intersect transversely in $\Sigma$. Their morphisms can be then defined by
\begin{eqnarray}
Hom(L_{i},L_{j})=\begin{cases}
                   R^{|L_{i}\cap L_{j}|},  &  \text{if } i<j\\
                   R\cdot \mathrm{id}, &  \text{if } i=j\\
                   0,  &  \text{if } i>j
                  \end{cases}
\end{eqnarray}
where $R$ is the base ring (usually the Novikov ring). Note that this definition of morphisms is in agreement with the fact, found in \cite{Hori:2000ck} that the open string index between A-branes in a LG model satisfying $\Im(W(p_{k}))< \Im(W(p_{k'}))$ always vanishes. The full definition of $\mathcal{F}uk(W)$ as a directed $A_{\infty}$-category (and $D^{b}\mathcal{F}uk(W)$) was given by P. Seidel \cite{2000math......7115S, 2000math.....10032S, seidel2008fukaya}, here we collected the relevant facts for us. The category $D^{b}\mathcal{F}uk(W)$ is defined in such a way that it does not depend on the choice of distinguish basis of vanishing cycles and any two basis are related by mutations. Is important to remark though that the category $D^{b}\mathcal{F}uk(W)$ is triangulated, hence comes with an appropriate defintion of a mapping cone $\mathrm{Cone}(\phi:L\rightarrow L')$ associated a morphism  $\phi\in Hom(L,L')$ between two objects. The cone can be, roughly, defined as a surgery or connected sum between the Lagrangians $L$ and $L'$ (for more details see \cite{2000math......7115S,2000math.....10032S, seidel2008fukaya}). This then implies that it makes sense to think about the bound states betwen branes as saddles. As a final remark/warning, we point out that, for the purposes of this work, we treat the categories $D^{b}\mathcal{F}uk(\Sigma)$, $\mathcal{D}$ and $D^{b}Coh(Z)\cong	D^{b}\mathcal{F}uk(W)$ as equivalent, it seems that for BPS numbers we are computing in the present paper, the distinction is irrelevant, nevertheless is a point that deserves further clarification. In the next section we will see how to map the vanishing cycles $L_{j}$ to objects in $\mathcal{D}$.

\subsection{From sheaves to thimbles via helices: a review}

In \cite{Kontsevich:lectures, Zaslow:1994nk} it was noticed that there is a certain parallel between the sructure of the derived category of coherent sheaves on a Kahler manifold $X$ and the approach to soliton counting pioneered in works of Cecotti and Vafa.
These observations were sharpened somewhat in the work of Hori, Iqbal and Vafa \cite{Hori:2000ck}, and formalized rigorously by Seidel \cite{2000math......7115S} and in particular Auroux, Katzarkov, Orlov \cite{Auroux:2008xno} in the context of homological mirror symmetry for weighted projective planes, including $\mathbb{F}_{0}$, which is the surface relevant for us. 

We begin on the toric side, from exceptional bundles on the toric Calabi-Yau threefold~$X$. Recall that an exceptional sheaf $E$ over a $d$-dimensional manifold (which for us will be $\IF_0$) with $c_1>0$, is defined by the condition that
\be
	\Ext^0(E,E) = \IC\,,
	\qquad
	\Ext^{i>0}(E,E)=0 \,.
\ee
An exceptional collection is an ordered set of exceptional sheaves $\{E_1,\dots, E_n\}$ such that, if $i<j$ then $\Ext^k(E_i, E_j) = 0$ for $k\neq k_0$ for a certain $k_0>0$, and $\Ext^k(E_j, E_j) = 0$ for all $k\geq 0$.
One may also define a bilinear product on all sheaves 
\be	
	\chi(E,F) = \sum_{i=0}^{d} (-1)^i\, \dim_{\IC} \Ext^{i}(E,F) \,.
\ee
For an exceptional collection, the matrix $S_{ij} = \chi({E_i, E_j})$ is thus $\delta_{ij} + A_{ij}$ where $A$ is strictly upper-triangular.

Given an exceptional collection, one may generate many new ones by an operation called \emph{mutation}. We will not need the full description of what a mutation does. It suffices to recall that there are two types of mutation: a right-mutation and a left-mutation, associated to an exceptional sheaf $E_i$ and its neighbor $E_{i+1}$. 
They act as follows
\be
	L_{E_i}(E_i,E_{i+1}) = (L_{E_i}(E_{i+1}),E_i)\,,
	\qquad
	R_{E_{i+1}}(E_i,E_{i+1}) = (E_{i+1},R_{E_{i+1}}(E_i))\,,
\ee
and the Chern-characters of the new elements in the exceptional collection are related to the previous ones by
\be\label{eq:mutation-exc-col}
\begin{split}
	\pm \ch(L_{E_i}(E_{i+1})) & = \ch(E_{i+1}) - \chi(E_i, E_{i+1})\, \ch(E_i)  \\
	\pm \ch(R_{E_{i+1}}(E_{i})) & = \ch(E_{i}) - \chi(E_i, E_{i+1})\, \ch(E_{i+1})  
\end{split}
\ee
Clearly, these two relations are inverse to each other, and in fact it is known that mutations satisfy braid group relations.
Given any such exceptional collection, one may define a helix by introducing $E_{i+n} = R_{E_{i+n-1}}\circ \dots\circ  R_{E_{i+1}} (E_{i})$ and $E_{i-n} = L_{E_{i-n+1}}\circ \dots\circ  L_{E_{i-1}} (E_{i})$. This generates the whole derived category, see e.g.  \cite{Tomasiello:2000ym, Govindarajan:2000vi} for examples of applications in string theory.

In string theory, exceptional sheaves arise in the study of D-branes wrapping cycles of a Kahler manifold, where the groups $\Ext^i(E,E)$ correspond to ground states of open strings stretching from $E$ to $E'$ with fermion number $F=i$ .

On the mirror side the relevant D-branes are A-branes wrapping special Lagrangian cycles. In the case of toric Calabi-Yau threefolds the Fukaya category may be studied directly on the mirror curve $\Sigma$, where special Lagrangians correspond to calibrated one-cycles.
Unlike in Seiberg-Witten theory, the relevant one cycles are, generally speaking, not the whole $H_1(\Sigma,\IZ)$, but a  certain sublattice thereof, whose description bears striking similarities with the helices arising on the toric side.

Recall that $\Sigma$ is an algebraic curve $F(x,y)=0$ in $\IC^*\times \IC^*$. For generic choice of complex moduli, this is a smooth curve. There are however critical points $(x^*_i, y^*_i)\in \IC^*\times \IC^*$ for $F(x,y)$ lying away from $\Sigma$. 
As the complex moduli are varied, some of the critical points may end up on $\Sigma$, this happens in complex co-dimension one.
In fact, when this happens, a cycle of $\Sigma$ pinches, and the choice of moduli corresponds to one of the singular divisors in moduli space (such as the $\CD_i$ in (\ref{eq:discriminant-locus})).
The moral of this story, is that each critical point of $F$ is associated to an element of $H_1(\Sigma,\IZ)$, and the association is unique if we choose a trivialization of the moduli space. An explicit example of this map will be discussed below in subsection \ref{sec:thimbles-sheaves}.
At generic points in moduli space the curve is smooth, and avoids all critical points. In other words $F(x^*_i, y^*_i) \neq 0$ for all $i=1,\dots, n$. 
Let us denote  $w_i = F(x^*_i, y^*_i)$ the critical values of $F(x,y)$, these will be functions of the complex moduli.

Recalling that $F(x,y)$ defines a potential for the Landau-Ginzburg model describing the mirror geometry, it follows from works of Cecotti and Vafa on the classification of 2d $(2,2)$ theories \cite{Cecotti:1992rm}, that the spectrum of D-branes on the mirror admits a description in terms of Lefshetz thimbles emanating from $w_i$.  
Each thimble obtained in this way is dual to an element of an exceptional collection, with the ordering corresponding to how thimbles attach to the anchoring point, which for us will be  at $F(x,y)=0$.
Whenever a thimble crosses a critical point, this results in a change of basis described by Picard-Lefshetz theory. On the toric side, this is mirrored by a mutation of the exceptional collection.

\subsection{Mapping Lefshetz thimbles and exceptional sheaves in $\IF_0$}\label{sec:thimbles-sheaves}

There are several ways to determine a correspondence between thimbles of critical points of $F(x,y)$, and exceptional sheaves on the toric side.
A general strategy is to study suitable variations of the moduli, such as shifts in the B-field, and leverage the parallel between mutations of exceptional collections and Picard-Lefshetz jumps of a system of thimbles. See \cite{Hori:2000kt} for details and examples. 

\begin{figure}[h!]
\begin{center}
\includegraphics[width=0.4\textwidth]{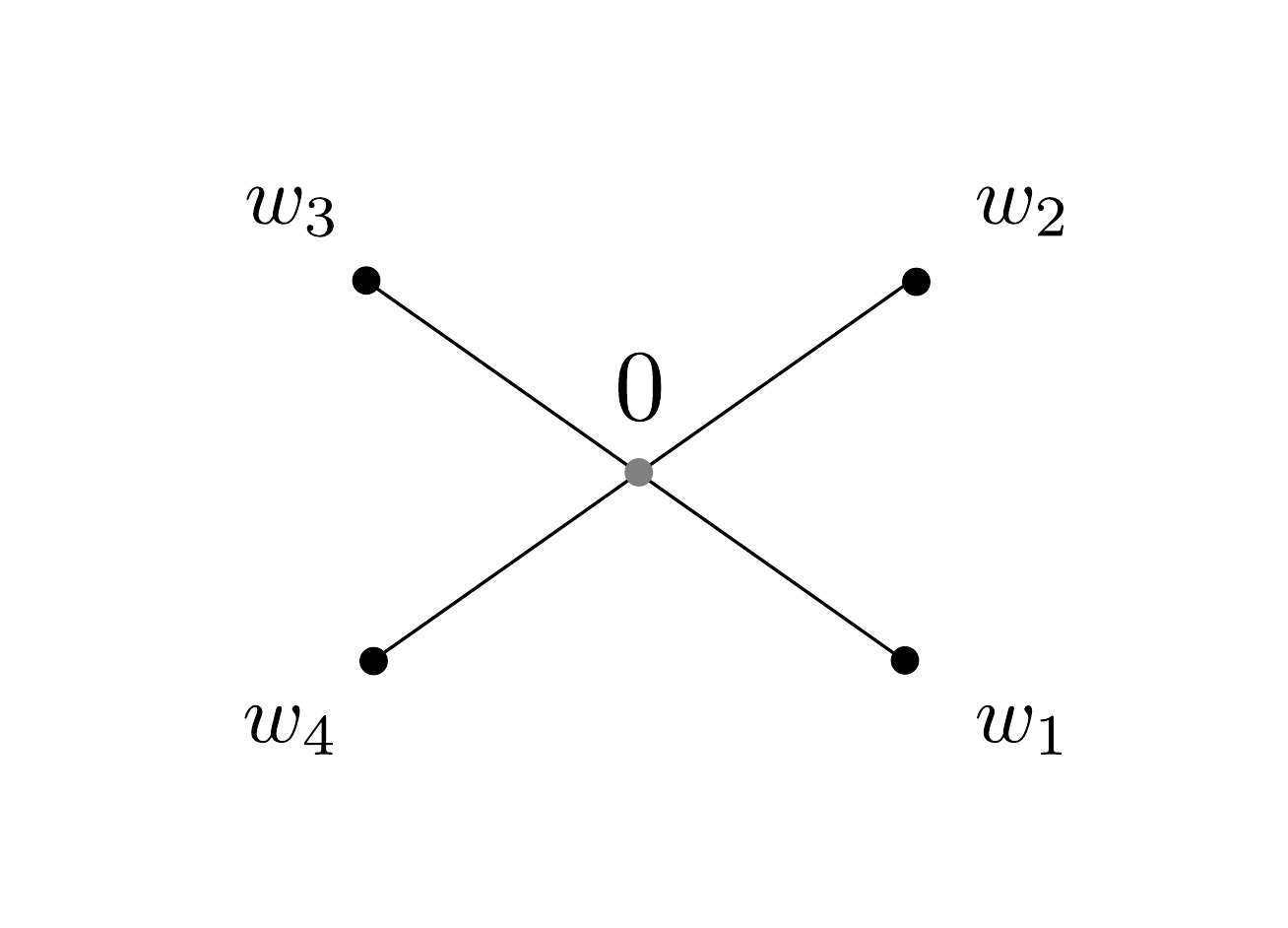}
\caption{Lefshetz thimbles for the mirror curve at $Q_b=i, Q_f=1$.}
\label{fig:Qb-eq-i-thimbles}
\end{center}
\end{figure}

In our case, however, we can take a shortcut since the identification between thimbles and exceptional sheaves has already been worked out in the literature for the case of interest to us.
For this purpose, we consider the curve  (\ref{eq:mirror-curve})  at the point $Q_b=i, Q_f=1$. The positions of the four critical values 
\be\label{eq:crit-values}
\begin{split}
		& 
		w_{1} = -1 - 2 Q_b + 2Q_f \,,
		\qquad
		w_{2} = -1 + 2 Q_b + 2Q_f \,,
		\\
		& 
		w_{3} = -1 + 2 Q_b - 2Q_f \,,
		\qquad
		w_{4} = -1 - 2 Q_b - 2Q_f \,.
		\\
\end{split}
\ee
are sketched in Figure \ref{fig:Qb-eq-i-thimbles}, which also shows our choice of thimbles.

\begin{figure}[h!]
\begin{center}
\includegraphics[width=0.23\textwidth]{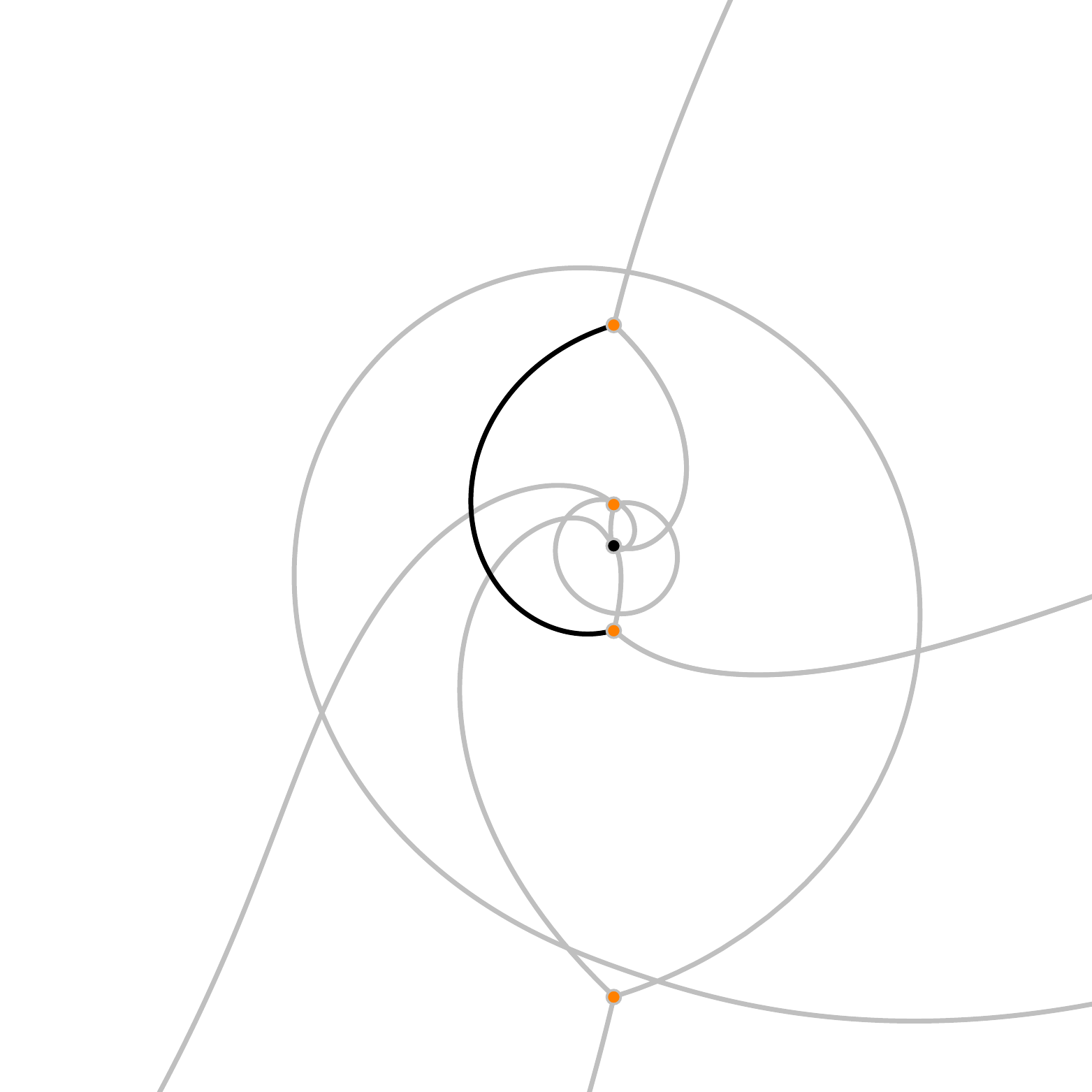}
\includegraphics[width=0.23\textwidth]{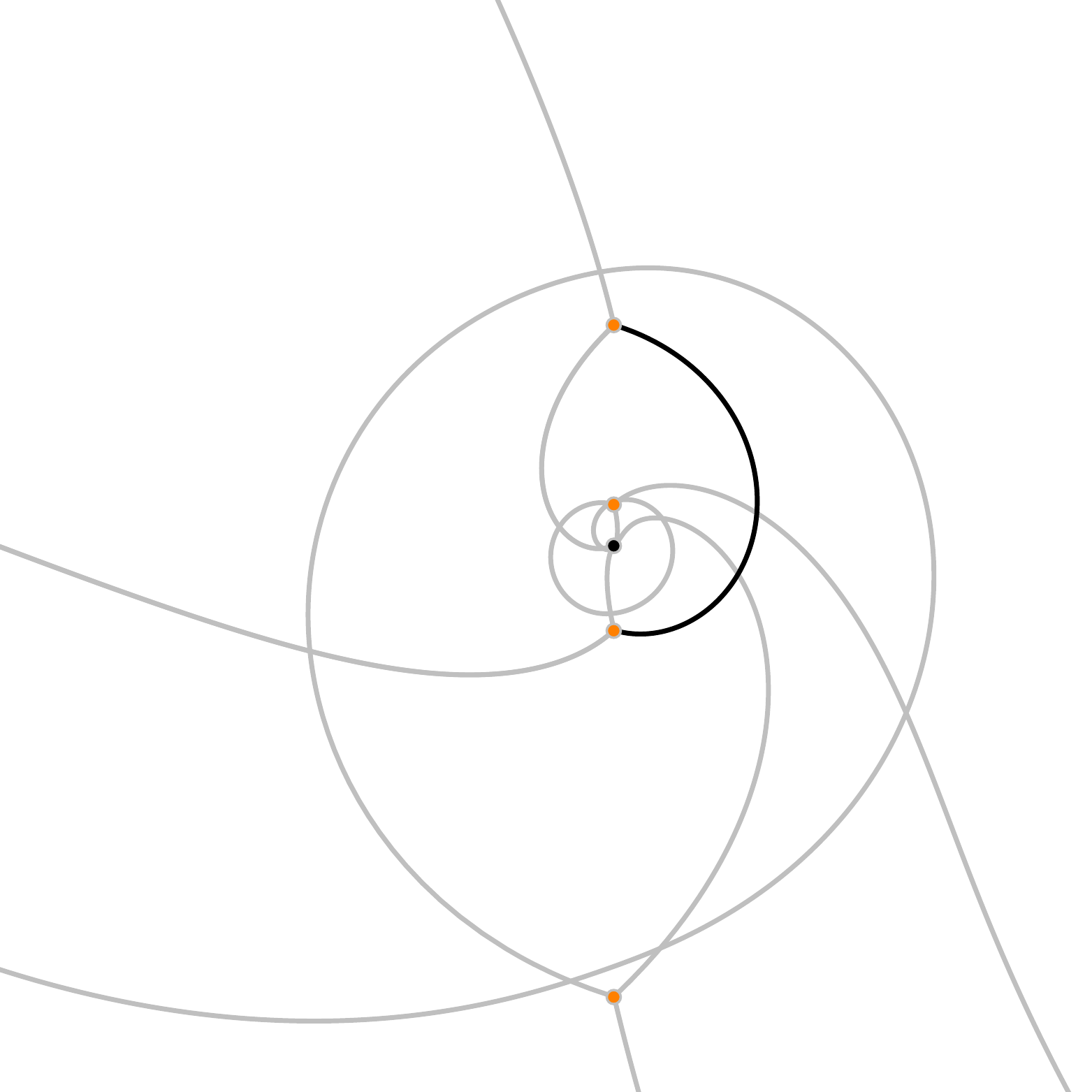}
\includegraphics[width=0.23\textwidth]{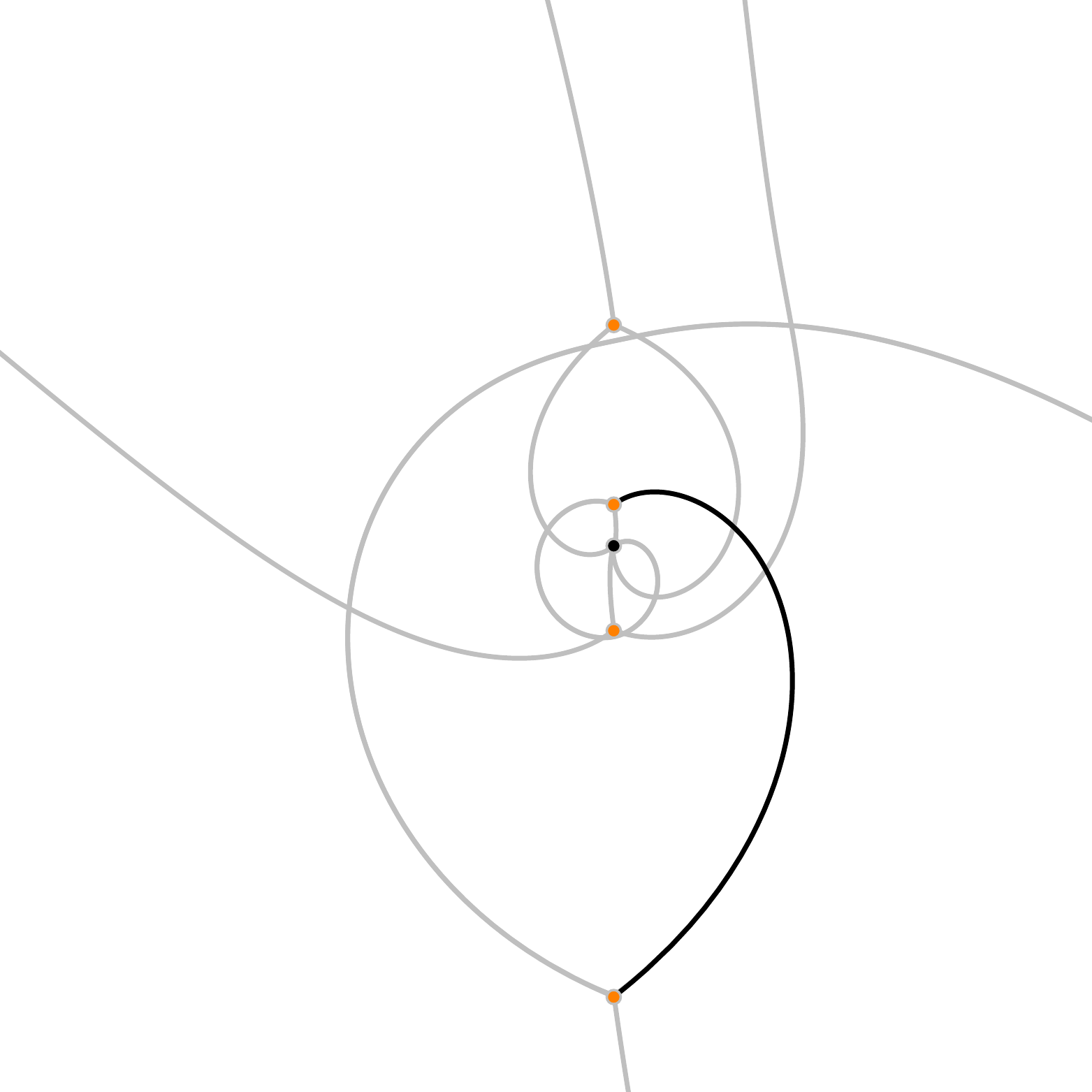}
\includegraphics[width=0.23\textwidth]{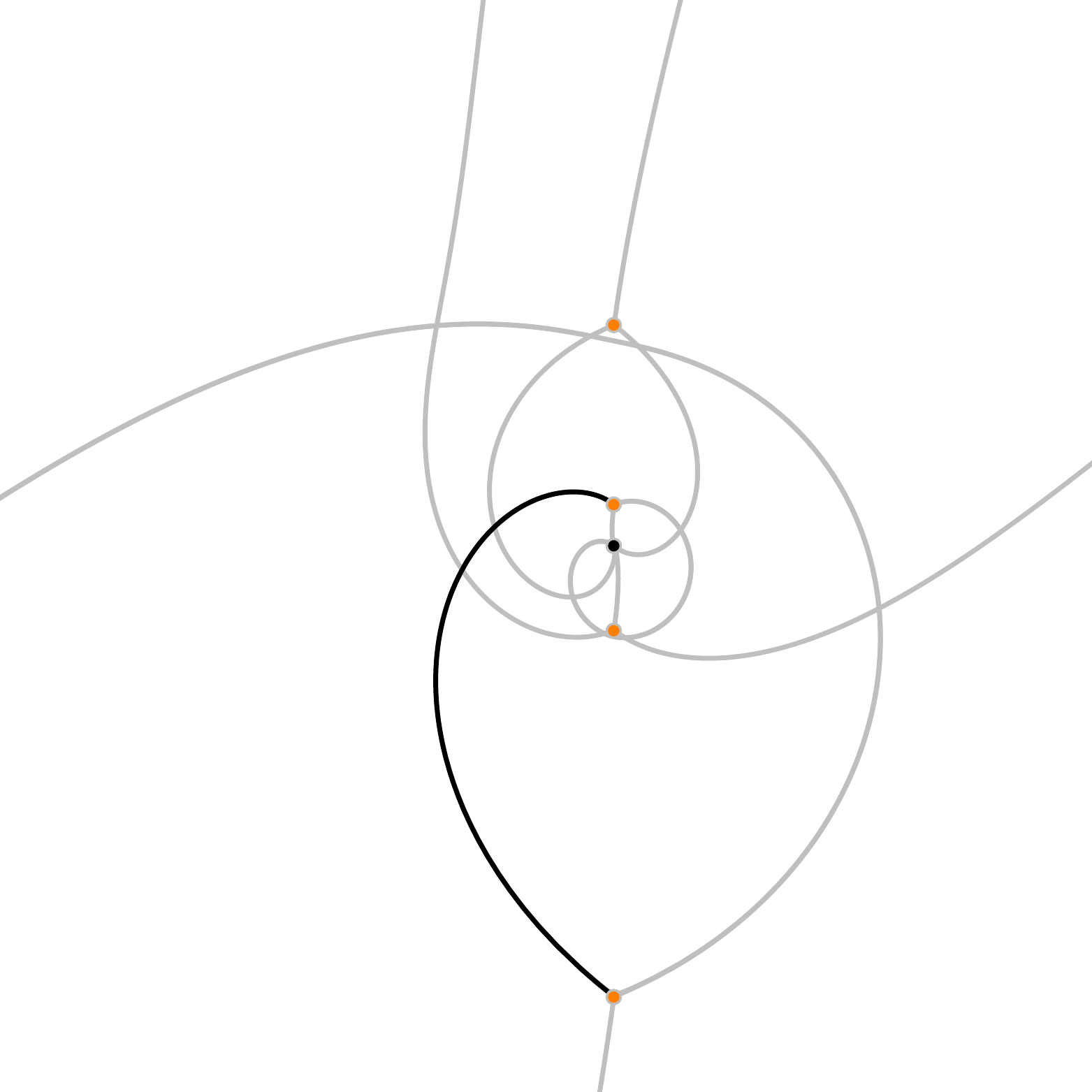}
\caption{Saddles of the four cycles pinching when $w_i=0$. From left to right: $w_1,\dots, w_4$.}
\label{fig:pinching-cycles}
\end{center}
\end{figure}

Thimbles are anchored at $W=0$ which corresponds to the mirror curve, and attach to one of the four critical points. Varying $(Q_b, Q_f)$ so that $w_i$ moves along the corresponding thimble, one finds that a cycle of $\Sigma$ pinches once $w_i$ ends up at the origin. Figure \ref{fig:pinching-cycles} shows the four saddles of the exponential network corresponding to the four pinching cycles.\footnote{Incidentally, these saddles and thimbles match with those studied in \cite{Franco:2016qxh} in the context of brick models.}

\begin{figure}[h!]
\begin{center}
\includegraphics[width=0.35\textwidth]{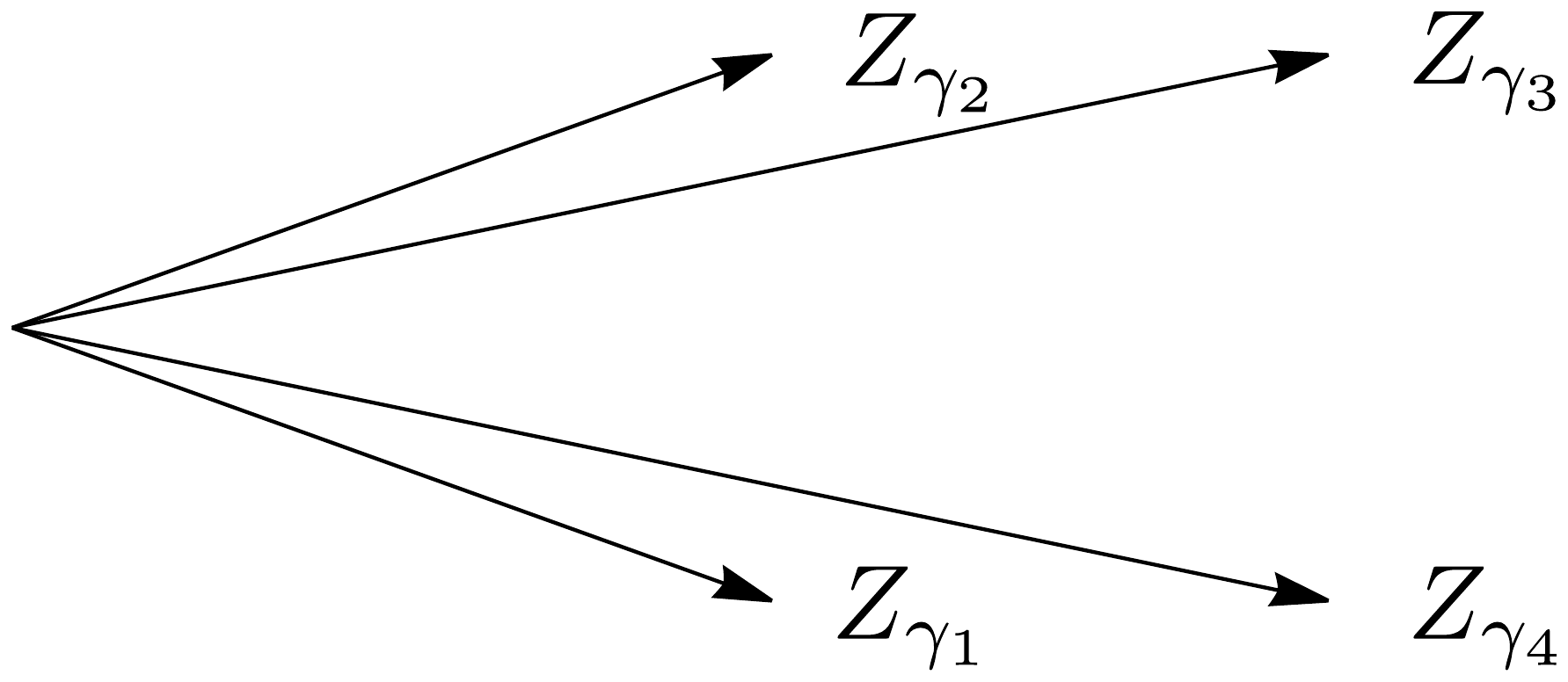}
\caption{Periods of the four vanishing cycles.}
\label{fig:pinching-cycles-Zs}
\end{center}
\end{figure}

Each saddle lifts to a cycle on $\Sigma$, we fix the orientation by requiring that central charges (periods of $\frac{1}{2\pi}\lambda $) are nearly aligned. The central charges of these saddles are shown in Figure \ref{fig:pinching-cycles-Zs}. We denote the cycles with these orientations by $\gamma_1,\dots, \gamma_4$.

The exceptional collection associated to the thimbles we chose is well-studied, and known to be 
$(\CO(0,0), \CO(1,0), \CO(1,1), \CO(2,1))$,
see for example \cite{Ueda:2006wy}. The map between cycles and exceptional sheaves is then as follows\footnote{To compare with \cite[Figure 3.1]{Ueda:2006wy}, note that 
authors studied the curve $W = x - x^{-1} + y+y^{-1}$.
Up to an inessential shift by $-1$, this is related to (\ref{eq:mirror-curve}) by rotating $x\to -i\, x$ in the latter with $Q_b = i, Q_f=1$.
We thus identify $(w_1,w_2,w_3,w_4)$ with $(c_4, c_3, c_2, c_1)$, whose corresponding sheaves are $(\CO(0,0), \CO(1,0), \CO(1,1), \CO(2,1))$}
\be\label{eq:cycles-sheaves}
	\gamma_1 : \  \CO(0,0) \,,
	\qquad
	\gamma_2 : \  \CO(1,0) \,,
	\qquad
	\gamma_3 : \  \CO(1,1) \,,
	\qquad
	\gamma_4 : \  \CO(2,1) \,.
\ee
From these sheaves, we may deduce the corresponding fractional brane charges\footnote{The relation is obtained by considering the dual collection, see for example \cite[eq. (4.22)]{Beaujard:2020sgs}}
\be\label{eq:cycles-branes}
	\gamma_1 : \  D4 \,,
	\qquad
	\gamma_2 : \  \overline{D4}\- D2_f \,,
	\qquad
	\gamma_3 : \   \overline{D4}\-\overline{D2}_f\-D2_b\-D0  \,,
	\qquad
	\gamma_4 : \   D4\-\overline{D2}_b \,.
\ee
Relations (\ref{eq:cycles-sheaves}) and (\ref{eq:cycles-branes}) are the dictionaries we needed.

In following sections we will study the mirror geometry at a slightly different point in moduli space, namely  $Q_b=-1, Q_f=1$. 
This amounts to a rotation of $Q_b \to i\, Q_b$, under which the geometry of the vanishing cycles of Figure \ref{fig:pinching-cycles} changes as shown in Figure \ref{fig:exp-net-sym-point}. By following the deformation, we can track $\gamma_i$. The dictionary is given below in (\ref{eq:basic-charges}).
Notice that, at the  point $Q_b=-1, Q_f=1$, the curve (\ref{eq:mirror-curve}) has a symmetry under exchange of fiber and base $x\leftrightarrow y$, with $Q_b\to -Q_b$ and $Q_f\to -Q_f$. This is reflected in the exchange of $w_2\leftrightarrow w_4$ in (\ref{eq:crit-values}), which correspondings to the exchange of $\gamma_2\leftrightarrow \gamma_4$ in (\ref{eq:cycles-branes}), consistently with the expected  $D2_b\leftrightarrow D2_f$.\footnote{The apparent need for a change in orientation of certain cycles, as well as the fact that $\gamma_3$ seems not to be exactly invariant may worry a scrupulous reader. We will see below in (\ref{eq:D-brane-charge-dictionary}) that a more natural, and in fact fiber-base symmetric, choice of `basis cycles' arises from networks. }

\subsection{An alternative derivation of the map between thimbles and D-branes}

For the case of interest to us, namely $X= K_{\IF_0}$, we can also take an alternative and more direct route to identifying thimbles for the mirror curve $\Sigma$ with D-branes on the toric Calabi-Yau threefold $X$.

Recall that M-theory on $X$ engineers 5d $\CN=1$ SU(2) Yang-Mills theory \cite{Seiberg:1996bd, Morrison:1996xf, Douglas:1996xp, Katz:1996fh}. In the limit where the M-theory circle shrinks to zero size, this reduces to 4d $\CN=2$ Seiberg-Witten theory \cite{Seiberg:1994rs}. In this limit the mirror curve approaches the 4d Seiberg-Witten curve, as discussed in Section \ref{sec:4d-limit-geometry}. We further show in Section \ref{sec:4d-limit-spectrum} that the singular divisor $\CD_2$ maps to the dyon singularity in the Coulomb branch. Since the dyon has charge $(e,m) = (1,-1)$, this leads to the identification of the  cycle $\gamma_2$ (which shrinks at $\CD_2$) with $D2_f\-\overline{D4}$. 
Noting that fiber-base duality exchanges $w_2$ with $w_4$ in (\ref{eq:crit-values}), one readily deduces that $\gamma_4$ is $\overline{D2}_b\-D4$.
Similarly, the monopole point is seen to coincide with the divisor where $\gamma_1$ vanishes, leading to the identification of the latter with a pure $D4$ brane. 
As an extra check, one may take a limit $Q_b\to\infty$ leading to the half-geometry $\CO(0)\oplus\CO(-2)\to \IP^1$ where the surviving 2-cycle is known to be $D2_f$, and this must correspond to the boundstate of monopole and dyon. We also perform this check in following sections.

Finally, to determine the D-brane charge mirror to $\gamma_3$, we may resort to simple numerics. Computing periods of each cycle, we already have identified $Z_{D4}, Z_{D2_{b}}, Z_{D2_f}$ from the identifications of $\gamma_{1}, \gamma_2,\gamma_4$. We also know that $Z_{D0} = 2\pi/R$ independently of moduli. By computing the numerical period $Z_{\gamma_3}$, see Appendix \ref{app:conifold-point}, we deduce that the dual D-brane charge is precisely $D0\-{D2}_b\-\overline{D2}_f\-\overline{D4}$.

These charges coincide exactly with the ones obtained via mirror symmetry in (\ref{eq:cycles-sheaves}), providing a strong independent check on the dictionary between D-branes on the toric side and homology cycles on the mirror curve.

\section{BPS spectrum at a special point}\label{sec:spectrum-conifold-point}

\begin{figure}[h!]
\begin{center}
\includegraphics[width=0.55\textwidth]{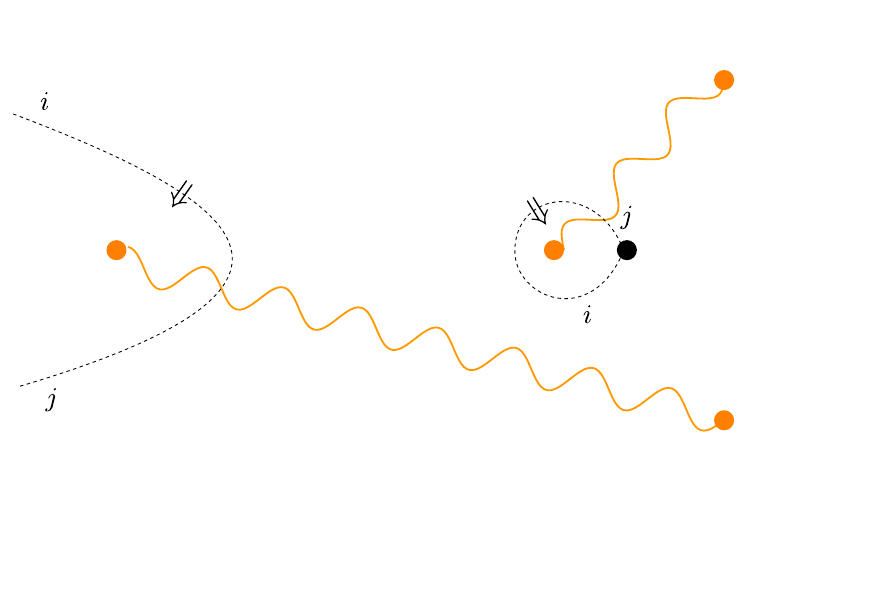}
\caption{Trivialization for mirror curve at the point $\mathbf{Q}_0$. Dashed lines are logarithmic cuts, running on $+/-$ sheets as indicated. The arrow $\Rightarrow$ indicates the direction in which crossing the cut induces a logarithmic shift $\log y\to\log y+2\pi i$.}
\label{fig:curve-trivialization}
\end{center}
\end{figure}

To begin investigating the spectrum of BPS states, we must choose a point in the moduli space parameterized by $Q_b,Q_f$.
Since this model is expected to exhibit wall-crossing, the spectrum that we will find will be sensitive to this choice.
A natural choice is to consider the \emph{fiber-base symmetric locus} determined by $Q_b=Q_f$. 
Just like the discriminant locus (\ref{eq:discriminant-locus}), this is also complex-codimension one, and the two loci meet at points.
One should also note that a change of coordinates $x\to -x$ and $y\to -y$ can be undone by changing signs to $Q_b, Q_f$. 
Since rotations of $\IC^*_{x,y}$ do not affect the physics of BPS states,  the full extent of the symmetric locus is actually
\be
	Q_b  \pm Q_f = 0 \,.
\ee
For the rest of this section we will study the BPS spectrum at $Q_b=-1, Q_f=1$. We refer to this as $\mathbf{Q}_0$. 
This does not belong to any of the singular divisors (\ref{eq:discriminant-locus}) and the geometry of the curve is non-singular.
For later reference, we fix once and for all a choice of trivialization of the covering map\footnote{In fact, we also specify a choice of trivialization for the logarithmic covering $\tilde\pi:\tilde\Sigma\to\Sigma$.} $\pi:\Sigma\to \IC^*_x$ as depicted in Figure \ref{fig:curve-trivialization}

\subsection{Basic saddles and quiver}

The spectrum of BPS states is encoded by saddles of exponential networks for the curve $\Sigma$. Recall that the network is determined by a choice of phase $\vartheta$ through the differential equation
\be
	(\log y_j - \log y_i +2\pi i n) \, \frac{d\log x}{d\tau}  \in  e^{i\vartheta} \IR^+
\ee
for a trajectory, known as $\CE$-wall, of type $(ij,n)$.
Whenever two or more $\CE$-walls intersect, new ones may be generated. The specifics depend on certain \emph{soliton data} attached to each $\CE$-wall, we refer to \cite{Banerjee:2018syt} for background.
Saddles appear at distinguished values of $\vartheta$, corresponding to $\arg Z_\gamma$ where $\gamma$ is the charge of the BPS state in question.
To capture the full spectrum, it is thus necessary to plot networks for various values of $\vartheta$ and record all saddles that occur. The BPS index $\Omega(\gamma)$ can be computed form the soliton data attached to the degenerate $\CE$-walls forming the saddle. Some saddles have a simple topology, while other are more complicated, possibly involving several degenerated $\CE$-walls joined at intersections or branch points.

\begin{figure}[h!]
\begin{center}
\includegraphics[width=0.45\textwidth]{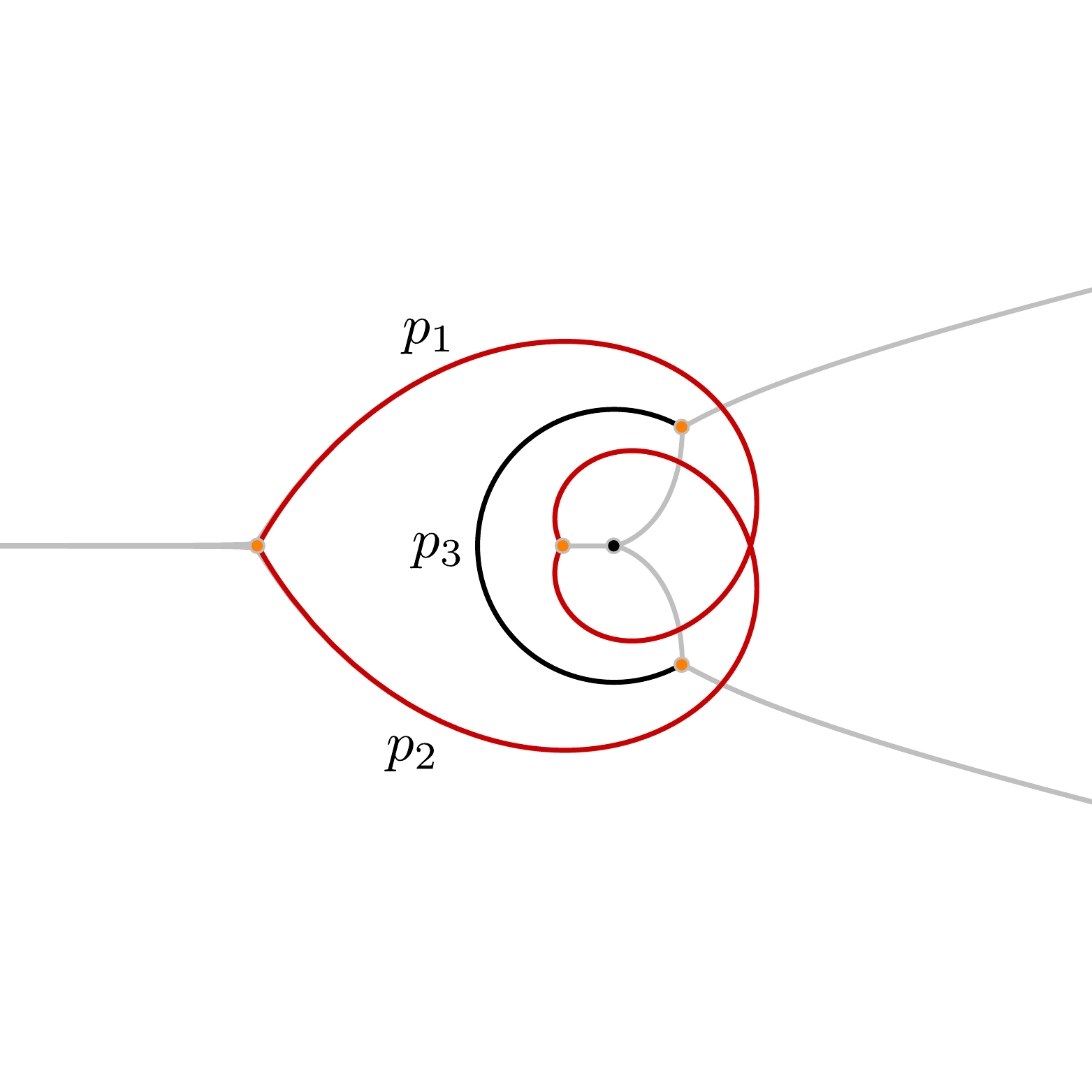}
\hspace*{.05\textwidth}
\includegraphics[width=0.45\textwidth]{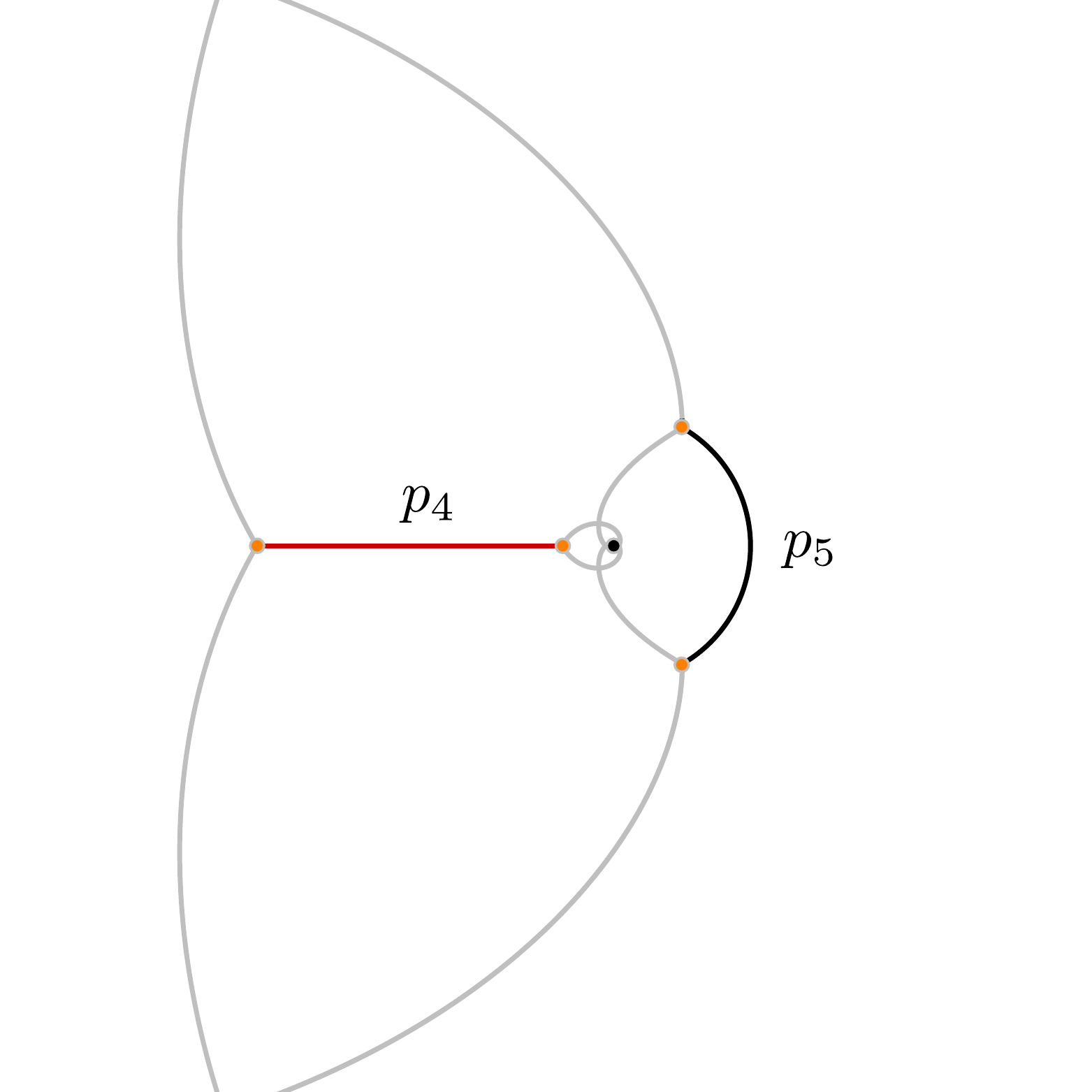}
\caption{Saddles of the exponential network at $-Q_b=Q_f=1$. On the left  $\vartheta=0$, on the right $\vartheta=\pi/2$. Only \emph{primary} walls (generated directly at branch points) are shown.}
\label{fig:exp-net-sym-point}
\end{center}
\end{figure}

At the point $\mathbf{Q}_0$ we are interested in, one finds five especially simple saddles: three appearing at $\vartheta=0$ and two at $\vartheta=\pi$, see Figure \ref{fig:exp-net-sym-point}.
Each saddle admits a unique lift to a closed cycle on $\Sigma$,\footnote{More precisely, there is an infinite tower of lifts to the logarithmic-covering of $\Sigma$, branched at $\fp_i$, and denoted $\tilde \Sigma$ in \cite{Banerjee:2018syt}. We suppress this detail here, until it will become necessary to deal with it.} whose homology class corresponds to the charge of the BPS state. Each of these is the mirror of a certain D4-D2-D0 boundstate in the mirror Calabi-Yau. 
From the identification between vanishing cycles and exceptional collections described in Section \ref{sec:exc-coll-mirror}, we deduce the following correspondence between these saddle and D-brane boundstates\footnote{Here a sum (or difference) of saddles is understood as a sum (or difference) of the homology cycles obtained by lifting the saddle to $\Sigma$. Then arrows $\to$ denote the map from $H_1(\Sigma)$ to $H^{even}( K_{\IF_0})$.}
\be\label{eq:D-brane-charge-dictionary}
\begin{split}
	p_2 + p_3 - p_4 + p_5 &\to D0 \\
	p_3 + p_5 &\to  D2_f \\
	p_3 + p_4 &\to D2_b \\
	p_3 & \to D4 
\end{split}
\qquad
\begin{split}
	p_1&\to D0\-\overline{D2}_b  \- D2_f \- \overline{D4} \\
	p_2&\to D0 \- D2_b \- \overline{D2}_f \- \overline{D4}\\
	p_3&\to D4\\
	p_4&\to D2_b \- \overline{D4}\\
	p_5&\to D2_f \- \overline{D4}\\
\end{split}
\ee
where $D2_{b,f}$ denotes a D2 brane wrapping, respectively, the base or fiber $\IP^1$.
Notice that this collection of basic saddles is invariant under fiber-base duality $D2_b\leftrightarrow D2_f$ which exchanges $p_1\leftrightarrow p_2$ and $p_4\leftrightarrow p_5$ leaving $p_3$ fixed.

The fact that multiple saddles appear at the same phase may be requires some explanation, since it may signal the presence of a wall of marginal stability.
In fact, by direct inspection one may verify that all cycles appearing at $\vartheta=0$ are mutually local, and the same holds for all those appearing at $\vartheta=\pi/2$.
Therefore we are not on a wall of marginal stability and the BPS spectrum is well-defined.

The charge lattice has rank four, we choose the following basis
\be\label{eq:basic-charges}
	\gamma_1 : [\pi^{-1}(p_3)]
	\qquad
	\gamma_2 : [\pi^{-1}(p_5)]
	\qquad
	\gamma_3 : [\pi^{-1}(p_2)]
	\qquad
	\gamma_4 : -[\pi^{-1}(p_4)]	
\ee
or, in terms of D-brane charges
\be\nonumber
	\gamma_1 : D4
	\qquad
	\gamma_2 : D2_f \- \overline{D4}
	\qquad
	\gamma_3 : D0 \- D2_b \- \overline{D2}_f \- \overline{D4}
	\qquad
	\gamma_4 : \overline{D2}_b \- {D4}	\,.
\ee
The matrix of intersection pairings reads
\be\label{eq:quiver-pairing-matrix-F0}
	\langle\gamma_i,\gamma_j\rangle
	= \left(\begin{array}{cccc}
	0 & -2 & 0 & 2 \\
	2  & 0 & -2 & 0 \\
	0 & 2 & 0 & -2\\
	-2 & 0  & 2 & 0
	\end{array}
	\right)
\ee
which is compatible with the pairing among D-brane charges
\be
	\langle D2_{b,f}, D4\rangle = 2\,,
	\qquad 0 \  \ \text{otherwise} \,.
\ee

\begin{figure}[h!]
\begin{center}
\raisebox{40pt}{\includegraphics[width=0.35\textwidth]{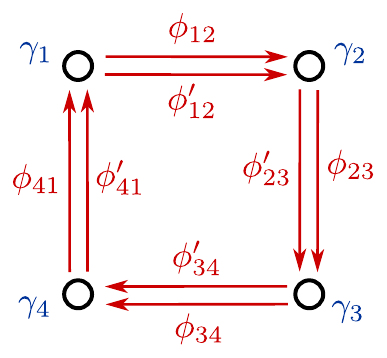}}
\hspace*{0.05\textwidth}
\includegraphics[width=0.20\textwidth]{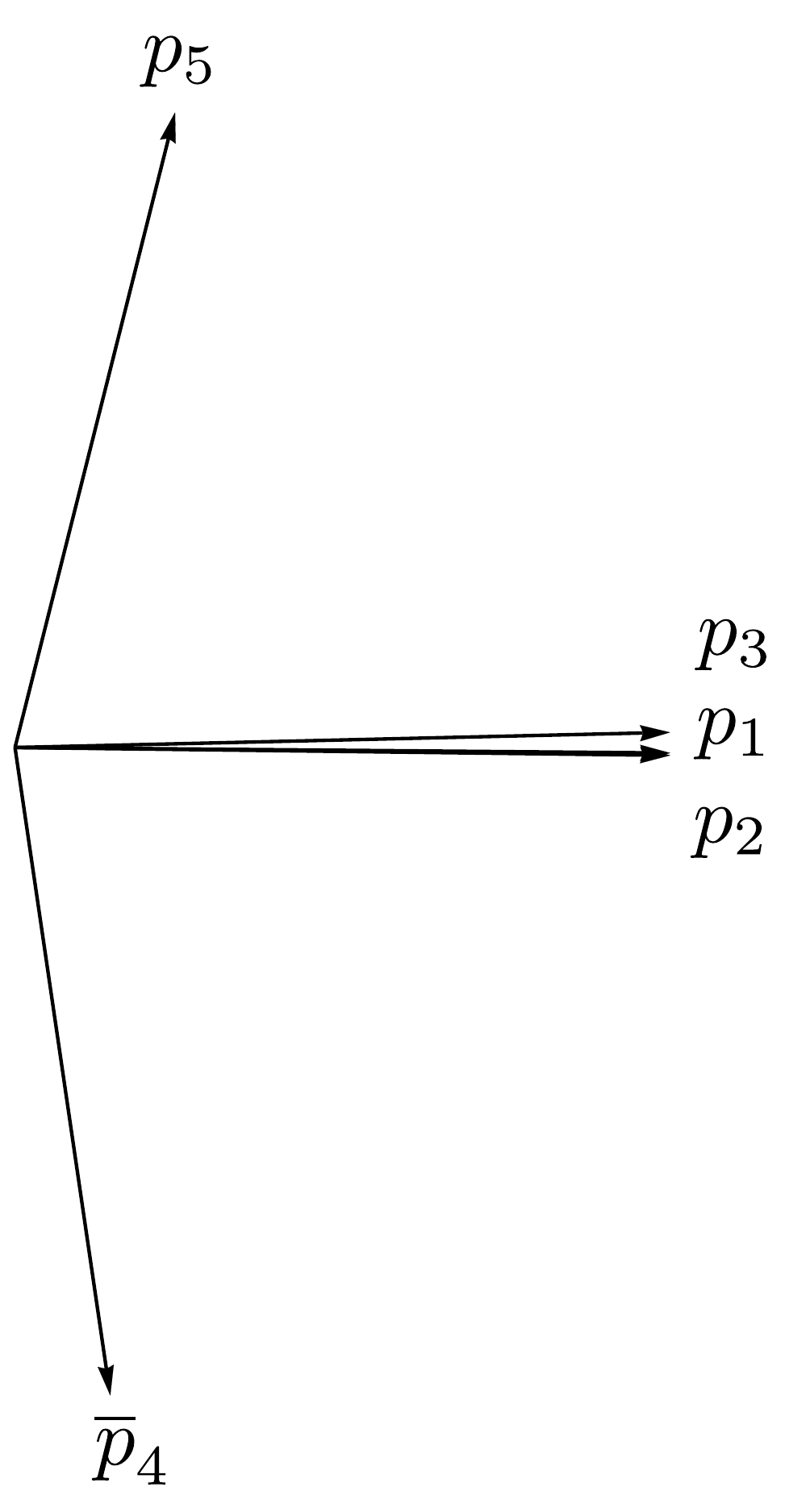}
\caption{Left: quiver. Right: central charges of basic BPS states corresponding to the five basic saddles, at the perturbed point $(Q_b, Q_f)=(-1, 1+0.1i)$. }
\label{fig:quiver}
\end{center}
\end{figure}

The BPS quiver can be obtained by identifying the four distinguished saddles with nodes, and deducing the arrows from the intersection pairing matrix. The result, shown in Figure \ref{fig:quiver}, coincides with descriptions found in the literature such as \cite[(4.14)]{Hanany:2001py}.
One way to motivate the choice of saddles corresponding to quiver nodes, is via the discussion of Section \ref{sec:exc-coll-mirror},
that is, we consider distinguished paths in moduli space, which start from the point $\mathbf{Q}_0 = (Q_b, Q_f) = (-1,1)$ and end at each of the four divisors $\CD_i$ without crossing any of the branch cuts.\footnote{Branch cuts in moduli space, that trivialize the charge lattice, must be chosen in a way that is compatible with the (pullback) of the Lefshetz thimbles. We gave the thimbles for $Q_b=i, Q_f=1$ in Section \ref{sec:exc-coll-mirror}. Here we work at $Q_f=1, Q_b=-1$, so one should deform the thimbles and the branch cuts accordingly.} 
The vanishing cycles at the divisors are precisely the (lifts of) saddles corresponding to nodes of the quiver.
It is important to note that the periods of the saddles at $(Q_b, Q_f)=(-1,1)$ still lie within the same half-plane as the periods of the same saddles at $(Q_b,Q_f)=(i,1)$, cf. Figures \ref{fig:quiver} and \ref{fig:pinching-cycles-Zs}. Therefore, the quiver description is exactly the same, no mutations are involved.

\begin{figure}[h!]
\begin{center}
\includegraphics[width=0.65\textwidth]{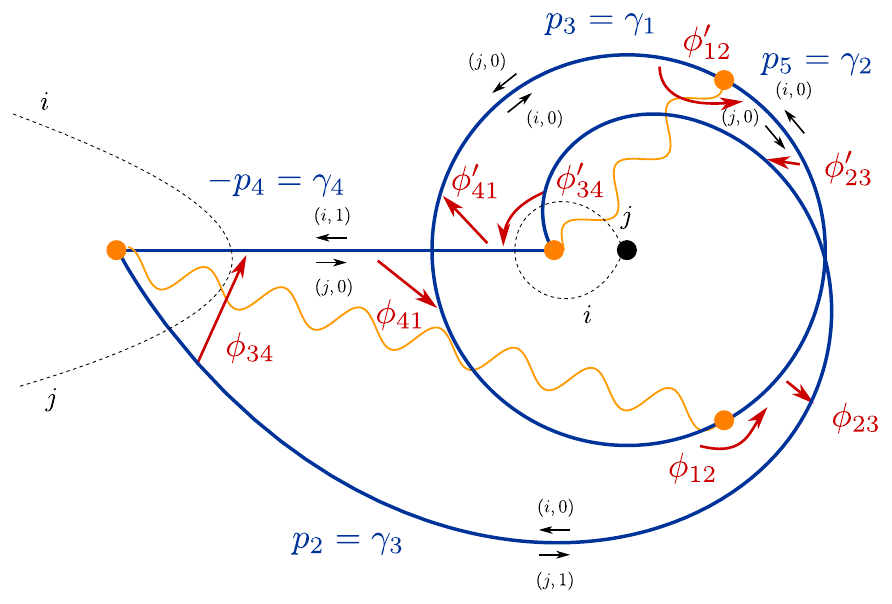}
\caption{Superposition of basic saddles. A choice of trivialization is also shown, with square-root cuts represented by wavy lines and logarithmic cuts (on $\Sigma$) by dashed lines. Arrows of the quiver, corresponding to intersection points of basic saddles with charges $\gamma_i,\gamma_j$ are labeled $\phi_{ij},\phi_{ij}'$. Here, $\phi_{41}$ and $\phi_{32}$ label interesections occurring on sheet $i$, while $\phi_{41}',\phi_{23}'$ label intersections on sheet $j$.
Labels of saddles, including logarithmic ones, are indicated by $(i,0)$, $(j,1)$ etc. Their orientation determines that of holomorphic disks.}
\label{fig:quiver-saddles}
\end{center}
\end{figure}

\begin{figure}[h!]
\begin{center}
\includegraphics[width=0.48\textwidth]{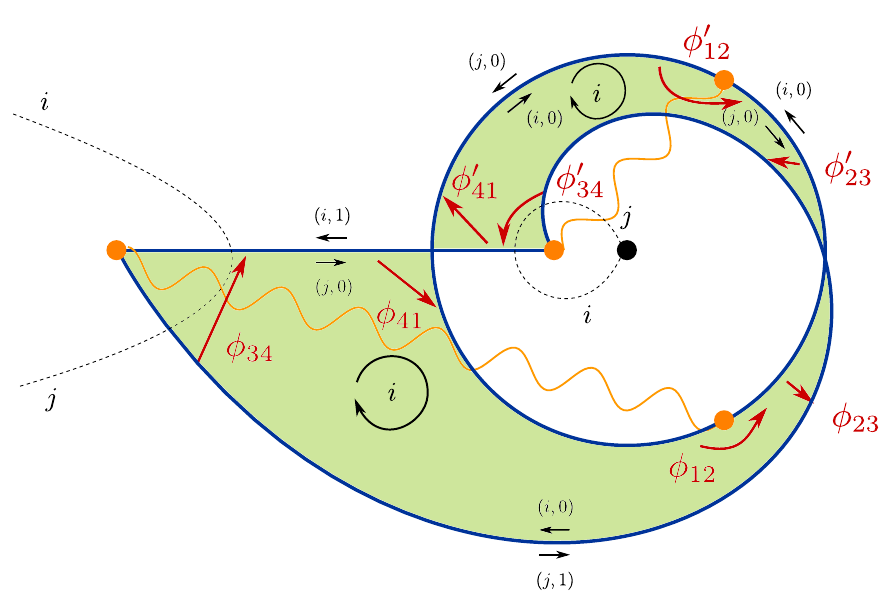}
\includegraphics[width=0.48\textwidth]{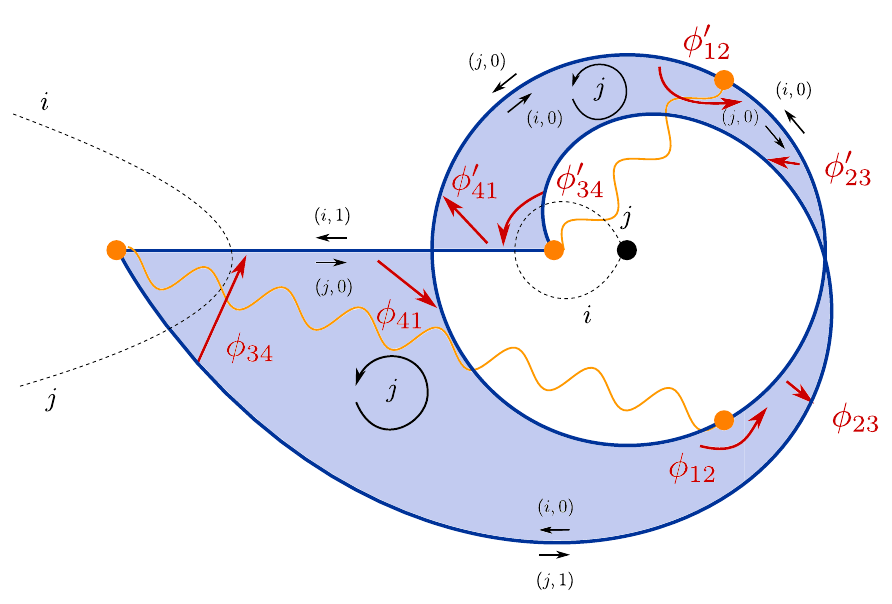}
\caption{Four holomorphic disks generating the quiver potential.}
\label{fig:quiver-potential-disks}
\end{center}
\end{figure}

To complete the quiver description we need to compute its potential. 
In \cite{Banerjee:2019apt} we explained how to do this, by going to a point in moduli space where central charges BPS states are maximally aligned, following the relation between BPS graphs and BPS quivers \cite{Gabella:2017hpz} (also see \cite{2013arXiv1302.7030B}).
Here we choose a different approach, following \cite{Eager:2016yxd} we super-impose diagrams of the four basic saddles as in Figure \ref{fig:quiver-saddles}.
The potential is generated by finite-area holomorphic disks bounded by the saddles, which corresponding to special Lagrangians in the mirror geometry \cite{Eager:2016yxd, Banerjee:2019apt}. There are four disks, shown in Figure \ref{fig:quiver-potential-disks}. The potential is thus the sum of four terms
\be\label{eq:quiver-potential}
	W = 
	\phi_{12}\phi_{23}\phi_{34}\phi'_{41} + \phi'_{12}\phi'_{23}\phi'_{34}\phi_{41}
	- \phi_{12}\phi'_{23}\phi_{34}\phi_{41} - \phi'_{12}\phi_{23}\phi'_{34}\phi'_{41} \,.
\ee
Let us describe how the first term is derived, the other three can be obtained in a similar way.
The term $\phi_{12}\phi_{23}\phi_{34}\phi'_{41} $ comes the bottom green  disk: starting from the bottom-right branch point ($\phi_{12}$) its boundary proceeds along saddle $\gamma_2$ on sheet $i$, until the intersection with $\gamma_3$, where it turns $(\phi_{23})$ and proceeds along $\gamma_3$ still on sheet $i$; the disk boundary eventually reaches the leftmost branch point and turns onto $\gamma_4$ (contribution $\phi_{34}$) after crossing  the branch cut and therefore passing to sheet $j$, it then meets $\gamma_1$ at the intersection $\phi_{41}'$ (since we are now on sheet $j$), then proceeds along $\gamma_1$ crosing the branch cut again, and closes off.
The sign of each disk is determined by its orientation. The potential (\ref{eq:quiver-potential}) agrees with results from the literature \cite[eq (6.30)]{Morrison:1998cs}.

\subsection{Saddle topologies}

Having identified the basic saddles, we begin to move on to the rest of the BPS spectrum.
At the  point $\mathbf{Q}_0$ the spectrum is infinite, and BPS saddles come in a variety of shapes and topologies.
In this subsection we collect a few of of the relevant topological types of saddles that appear, and explain how to compute their BPS indices.
Later on we will identify these with actual saddles from exponential networks.

\subsubsection*{Type-0 saddle}
The simplest type of saddle consists of a pair of $\CE$-wall of opposite types $(ij,n)$ and $(ji,-n)$ running anti-parallel to each other between two branch points, which source respectively each  of the two $\CE$-walls.  Each of the saddles already encountered in Figure \ref{fig:exp-net-sym-point} is of this type, other examples can be found in Appendix \ref{app:conifold-point}.
The analysis of soliton data, which can be found in the companion paper \cite{Banerjee:2019apt}, is quite straightforward and leads to the answer
\be
	\Omega(\gamma) = 1 \,.
\ee

\subsubsection*{Type-1 saddle}
\begin{figure}[h!]
\begin{center}
\includegraphics[width=0.7\textwidth]{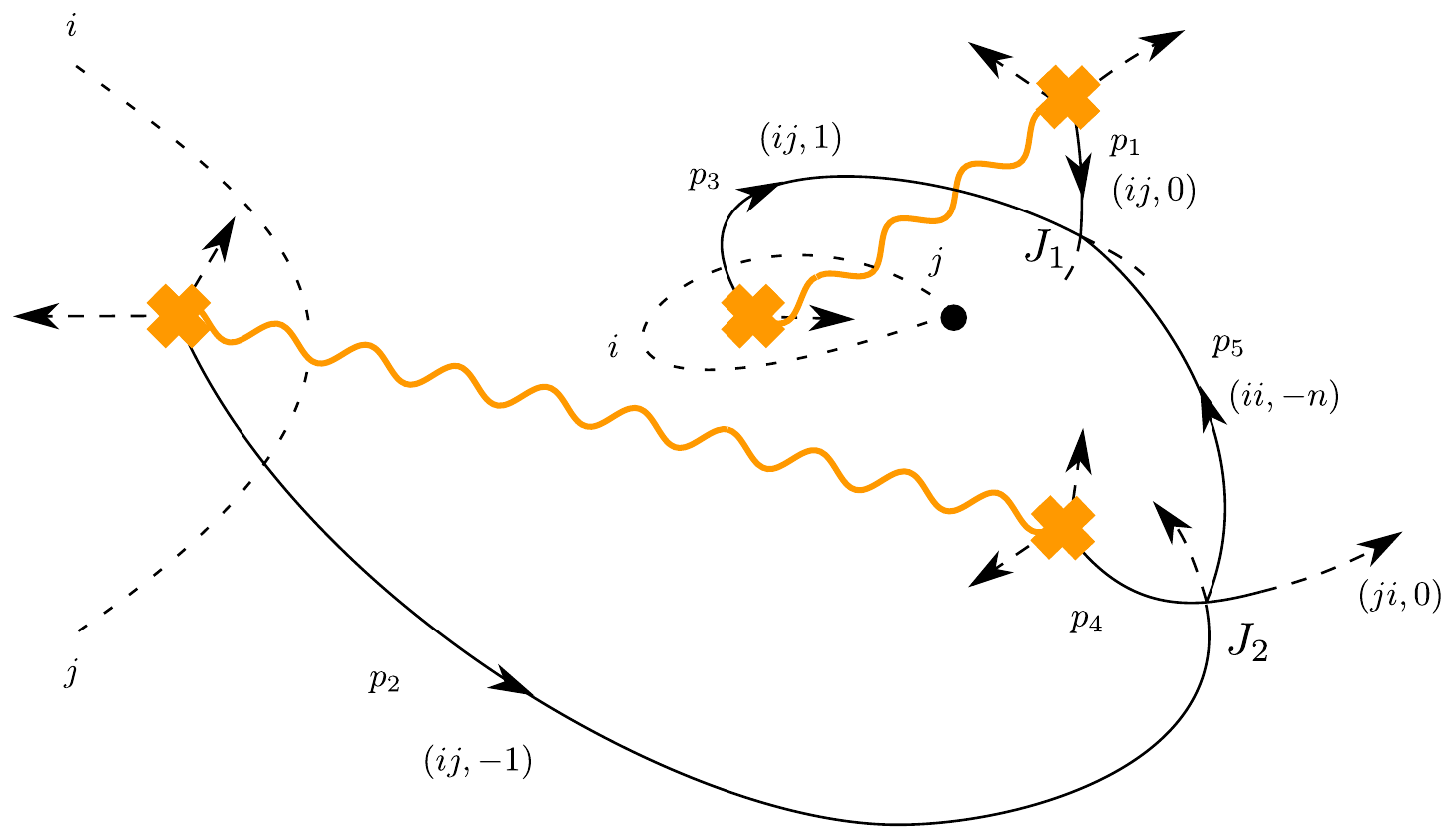}
\caption{A saddle of Type-1.}
\label{fig:type-1-saddle}
\end{center}
\end{figure}

An interesting novelty is the type-1 saddle, shown in Figure \ref{fig:type-1-saddle}. Real-world examples of this type of saddle can be seen in Figures \ref{fig:0110} and \ref{fig:0112}.
To study it, let us start with a detailed description.
This type of saddle involves the presence of four branch points, as well as logarithmic cuts, running around two of the branch points, and depicted as dashed lines. 
Recall that logarithmic cuts are cuts for the differential $\lambda$, and therefore are defined on $\Sigma$, as opposed to $\IC^*_x$: for this reason we indicate on which sheet the logarithmic cut lies, namely sheet $i$ in the top of the picture, and sheet $j$ in the bottom.

There are four distinguished $\CE$-walls, originating from the branch points, and denoted $p_1, p_2, p_3, p_4$. As indicated, they are of types $(ij,0)$, $(ij,-1)$, $(ji,1)$ (at junction $J_1$, after crossing the branch cut) and $(ji,0)$ respectively. 
Saddles with this type of topology are frequently encountered in spectral networks, see e.g. \cite[Figure 4]{Galakhov:2013oja} where they would have BPS index equal to $1$, corresponding to hypermultiplets. As we will see, this is \emph{not} the case for the saddle we are considering here: A special feature of exponential networks, as opposed to spectral networks, is the existence of an $ij-ji$ junction, extensively studied in our previous work \cite[Section 3.3]{Banerjee:2018syt}. 
The Type-1 saddle features \emph{two} such junctions: one between $p_1$ and $p_3$  denoted $J_1$, and one between $p_2$ and $p_4$ denoted $J_2$. There is a symmetry between the two, so let us focus on the former one. 
When  $p_3$, which is of type $(ji,1)$, meets $p_1$, which is of type $(ij,0)$, they generate infinite families of new $\CE$-walls.
The presence of these walls is determined by wall-crossing of the 3d-5d framed BPS spectrum, see \cite{Banerjee:2018syt} for a derivation.
Among them, there are two towers of $\CE$-walls of types $(ii,n)$ and $(jj,n)$ for $n\geq 1$.
A similar story applies to $p_2, p_4$ which generate $(ii,-n)$ and $(jj,-n)$ trajectories. Together these imply that $p_5$ is a double-wall, i.e. part of the saddle.
The $(ii/jj,-n)$ walls intersecting $p_1$ generate a descendant of type $(ji,0)$ at $J_1$ (see Appendix \ref{app:ii-ij-junction}), which runs anti-parallel to $p_1$ making it a double-wall as well.
Similarly $p_3$ is made into a double wall by the interaction of $p_1$ and $(ii/jj,-n)$, and likewise for $p_2, p_4$.

Identifying which walls are degenerate (\emph{a.k.a.} ``two-way streets'') is only the first step for computing the BPS index. 
The second step involves computing soliton data for each of $p_1, p_2, p_3, p_4, p_5$.
For this purpose we need to determine the outgoing soliton data in terms of incoming one at each of the two junctions.

A full analysis of the junction is provided in Appendix \ref{app:type-1-saddle}, here we sketch the main result. 
For this purpose denote by $\Upsilon_i$ the generating function of solitons supported on $p_i$, running from the branch point towards the junction, and let $\Delta_i$ be the generating function of solitons oriented in the opposite way.\footnote{These were generically denoted $\Xi_{ij,n}$ in our previous work \cite{Banerjee:2018syt}.}
Each of these generating functions can be determined by considering flatness constraints for the nonabelianization map at junctions and branch points.
At the branch points we simply have
\be
\begin{split}
	\Upsilon_i & = X_{a_i} \qquad i=1,\dots, 4
\end{split}
\ee
where $a_{i}$ denote `simpletons' paths obtained by lifting the wall $p_i$, and connecting the two strands at the branch point that sources the wall.\footnote{\label{foot:simpletons}Working on the logarithmic covering $\tilde\Sigma\to\Sigma$, a wall of type $(ij,0)$ can be lifted to sheets labeled by $(i,N)$ (with the opposite orientation) and to $(j,N)$ (with the same orientation). The ``simpleton'' path runs on these two lifts with the corresponding orientations, passing through the ramification point \cite{Banerjee:2018syt}.}

Before proceeding let us comment on a technical point, which can be safely skipped on a first reading.
The above equation is actually slightly over-simplified, since we are suppressing all information about logarithmic branching of $\lambda$ over $\Sigma$. 
Due to a shift symmetry, the dependence on the branch of the logarithm turns out to be somewhat trivial \cite{Banerjee:2018syt}, and can be effectively ignored to keep notation lighter. 
In actual computations one should however keep track of this dependence.
This can be unambiguously recovered from the equations presented here by reintroducing dependence on $N$, for example $X_{a_i}\to \sum_{N\in \IZ} X_{a_{i,N}}$ etc.

The flatness constraints at junction $J_1$, connecting $p_1, p_3, p_5$ are 
\be\label{eq:junction-J1-type1-saddle}
\begin{split}
	\Delta_1 & =   \frac{2 + \Upsilon_5^{(jj,-1)}\Upsilon_3\Upsilon_1}{(1 +  \Upsilon_5^{(jj,-1)}\Upsilon_3\Upsilon_1)^2}\, \Upsilon_5^{(jj,-1)}\Upsilon_3  
	\\
	\Delta_3 & =  \Upsilon_1\Upsilon_5^{(jj,-1)}  \, \frac{2 + \Upsilon_5^{(jj,-1)}\Upsilon_3\Upsilon_1}{(1 +  \Upsilon_5^{(jj,-1)}\Upsilon_3\Upsilon_1)^2}
	\\
	\Delta_{5}^{(ii,k)} & = \frac{1}{k} (-\Upsilon_1\Upsilon_3)^k 
	\qquad
	\Delta_{5}^{(jj,k)} =  -\frac{1}{k} (-\Upsilon_3\Upsilon_1)^k \,,
\end{split}
\ee
where $\Upsilon_{5}^{jj,n}, \Delta_{5}^{jj,n}$ are generating functions of $(jj,n)$ solitons supported on $p_5$, see Appendix~\ref{app:type-1-saddle} for the relation between this generating function and Stokes matrices of the underlying $\CE$-walls.
It is useful to note that $\Upsilon_5^{(jj,-1)} \Xi_{ji} = - \Xi_{ji} \Upsilon_5^{(ii,-1)}$ if $\Xi_{ji}$ is any shift-symmetric soliton generating function \cite{Banerjee:2018syt}.
The equations describing junction $J_2$, which connects $p_2,p_4,p_5$, have a  similar form
\be\label{eq:type-1-saddle-J2-sol}
\begin{split}
	\Delta_4 & = \Delta_5^{(ii,1)}\Upsilon_2\,  \frac{2 + \Delta_5^{(ii,1)}\Upsilon_2\Upsilon_4}{(1+\Delta_5^{(ii,1)}\Upsilon_2\Upsilon_4)^2} \\
	\Delta_2 & =  \frac{2 + \Delta_5^{(ii,1)}\Upsilon_2\Upsilon_4}{(1+\Delta_5^{(ii,1)}\Upsilon_2\Upsilon_4)^2} \,  \Upsilon_4 \Delta_5^{(ii,1)}\\
	\Upsilon_{5}^{(ii,-k)} & = -\frac{1}{k} (-\Upsilon_2\Upsilon_4)^k 
	\qquad
	\Upsilon_{5}^{(jj,-k)} =  \frac{1}{k} (-\Upsilon_4\Upsilon_2)^k \,.
\end{split}	
\ee

Let  $\gamma$ denote the closure of the concatenation of $a_{1}\cdot a_{3}\cdot a_{2}\cdot a_{3}$, where all endpoints are understood to be transported to a common position and joined (e.g. along $p_5$). 
It then follows directly from the above expressions that
\be
	Q(p_1) = 1-\Upsilon_1 \cdot \Delta_1 = 1+X_{\gamma}\frac{2  - X_{\gamma}}{(1-X_{\gamma})^{2}} = (1-X_{\gamma})^{-2}\,,
\ee
and similarly $Q(p_2)=Q(p_3) = Q(p_4) = Q(p_1)$.

The factorization $Q(p) = \prod_{n\geq 1}(1-X_{n\gamma})^{\alpha_{n\gamma}(p)}$ leads to $\alpha_{\gamma}(p_i)=-2$ for $i=1,2,3,4$. 
The BPS index is computed by the ratio $\Omega(n\gamma) = [L(n\gamma)] / n\gamma$, where $L(n\gamma) = \bigcup_{i}\alpha_{n\gamma}(p_i)\cdot \pi^{-1}(p_i)$.
Notice that $[\bigcup_{i=1}^{5}\pi^{-1}(p_i)]=\gamma$, therefore the saddle of Type-1 gives an overall contribution to the BPS index \footnote{We did not determine $\alpha(p_5)$ because this cannot be determined for $ii/jj$ streets, see \cite{Banerjee:2018syt} for a discussion.}
\be
	\Delta\Omega(\gamma) = -2\,.
\ee
Whenever this saddle appears simultaneously with other saddles, the total index is the sum of each contribution.

\subsubsection*{Type-2 saddle}
\begin{figure}[h!]
\begin{center}
\includegraphics[width=0.7\textwidth]{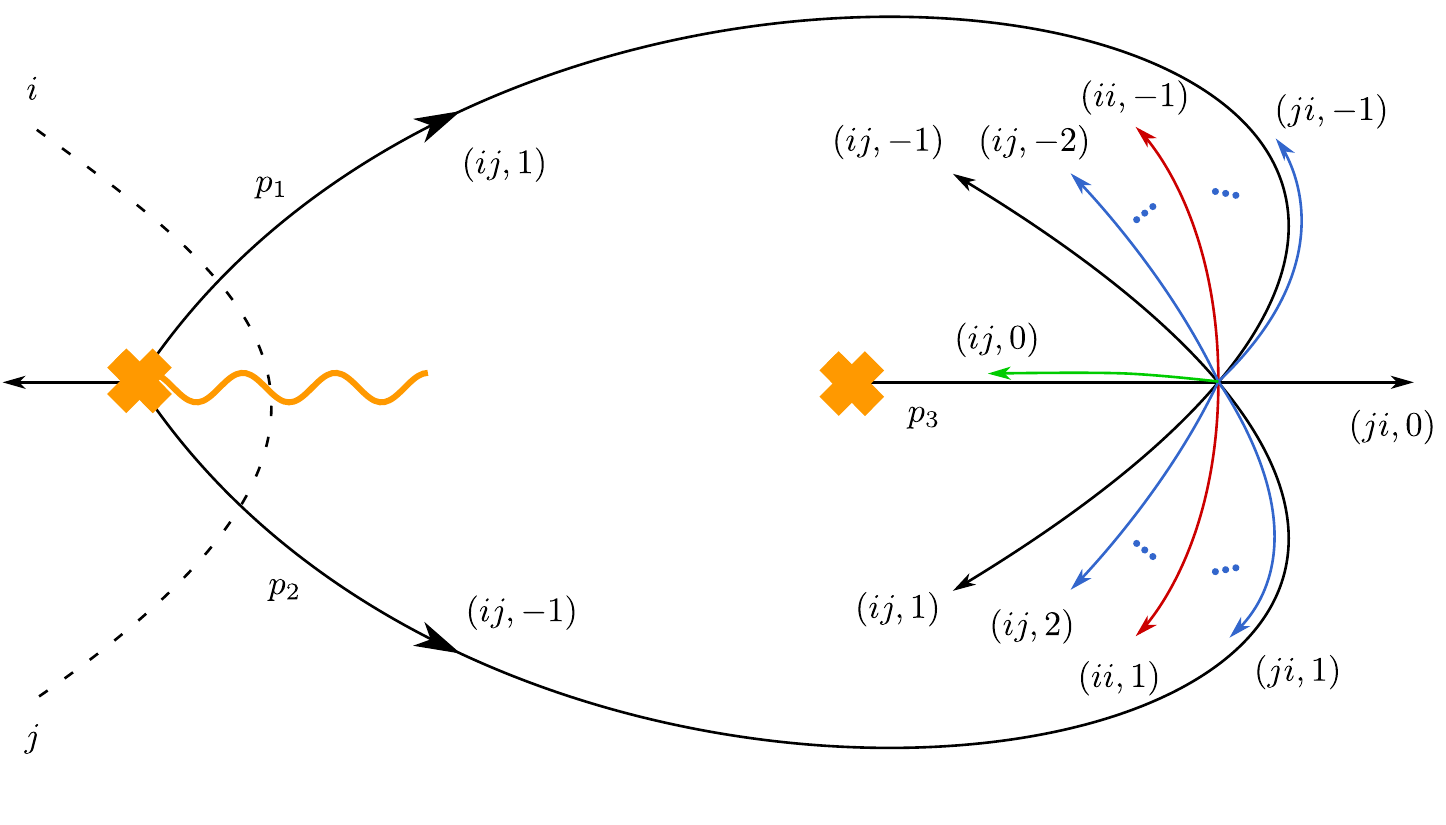}
\caption{A saddle of Type-2.}
\label{fig:type-2-saddle}
\end{center}
\end{figure}

Another interesting novelty is the type-2 saddle, shown in Figure \ref{fig:type-2-saddle}. Real-world examples of this type of saddle can be seen in Figures \ref{fig:0011} and \ref{fig:0211}.
To understand it, let us start with a detailed description.
This type of saddle involves the presence of two branch points, as well as a logarithmic cut, running around one of the branch points, and depicted as a dashed line on the left. Recall that logarithmic cuts are cuts for the differential $\lambda$, and therefore are defined on $\Sigma$, as opposed to $\IC^*_x$: for this reason we indicate on which sheet the logarithmic cut lies, namely sheet $i$ in the top of the picture, and sheet $j$ in the bottom.

There are three distinguished $\CE$-walls, originating from the branch points, and denoted $p_1, p_2, p_3$. As indicated, they are of types $(ij,\pm1)$ and $(ji,0)$. A special feature of exponential networks, as opposed to spectral networks, is the existence of an $ij-ji$ junction, extensively studied in our previous work \cite[Section 3.3]{Banerjee:2018syt}.
The Type-2 saddle features \emph{two} such junctions: one between $p_1$ and $p_3$, and one between $p_2$ and $p_4$. There is a symmetry between the two, so let us focus on the former one. 
When  $p_3$, which is of type $(ji,0)$, $p_1$, which is of type $(ij,1)$, they generate infinite families of new $\CE$-walls.
Shown in red,  is a tower of walls of types $(ii,k)$ and $(jj,k)$ for $k>0$ running from below the junction.\footnote{We only depict the wall of type $ii,1$, as walls for other values of $k$ are exactly overlapping.}
Shown in blue is a tower of walls of types $(ij,k)$ for $k>1$ located between $p_1$ and the red $\CE$-walls. Also in blue, is a tower of walls of types $(ji,k)$ for $k>0$ located between $p_3$ and the red $\CE$-walls.
The presence of these walls is determined by wall-crossing of the 3d-5d framed BPS spectrum, see \cite{Banerjee:2018syt} for a derivation.

One should note that the walls generated by $p_1$ and $p_3$ have a chance to intersect $p_2$, generating further descendants. We only plot a green $\CE$-wall of type $(ij,0)$, arising from a junction between $p_1$ and the $(ii,-1)$ red wall, as well as the $(jj,-1)$ wall, if one works in \emph{American resolution}.
The presence of this green wall makes $p_3$ a two-way street, hence part of the saddle. 
Likewise, $p_1$ is also a 2-way street due to the presence of the anti-parallel $(ji,-1)$ wall generated by $p_2$ and $p_3$, and similarly $p_2$ is also a two-way street. 

Identifying which walls are degenerate (\emph{a.k.a.} ``two-way streets'') is only the first step for computing the BPS index. 
The second step involves computing soliton data for each of $p_1, p_2, p_3$.
This involves two stages: first, solving the outgoing soliton data in terms of incoming one at the junction among $p_1,p_2,p_3$; second, taking into account identities for soliton data coming from other endpoints of the saddle, namely branch points.

A full analysis of the junction is provided in Appendix \ref{app:saddle-type-2-eqs}, here we sketch the main result.
For this purpose denote by $\Upsilon_i$ the generating function of solitons supported on $p_i$, running from the branch point towards the junction, and let $\Delta_i$ be the generating function of solitons oriented in the opposite way.\footnote{These were generically denoted $\Xi_{ij,n}$ in our previous work \cite{Banerjee:2018syt}.}
These are related as follows  
\be
\begin{split}
	\Upsilon_1 & = X_{a_1} (1 -  X_{a_2} \Delta_2) \qquad
	\Upsilon_2  = X_{a_{2}}  \qquad
	\Upsilon_3  = X_{a_{3}} \,,\\
	\Delta_1 & = - \Upsilon_3\cdot \Upsilon_2\cdot \Upsilon_3\qquad
	\Delta_2  = - \Upsilon_3\cdot \Upsilon_1\cdot \Upsilon_3 \qquad
	Q(p_3)    = Q(p_1)^2 = Q(p_2)^2 
\end{split}
\ee
Here, the first equation follows from homotopy across a branch point, see (\ref{eq:branch-homotopy-equation}), and $a_{i}$ denote `simpletons' paths obtained by lifting the wall $p_i$, and connecting the two strands at the branch point that sources the wall, see footnote \ref{foot:simpletons}.
We also denote $Q(p_i) = 1 - \Delta_i\Upsilon_i$.

For the sake of keeping notation light we will, as above, suppress information about logarithmic branching of soliton data, keeping in mind that this can be restored in each of the following expressions in a non-ambiguous way.

Returning to the computation of the BPS index, let  $\gamma$ denote the closure of the concatenation of $a_{1}\cdot a_{3}\cdot a_{2}\cdot a_{3}$. 
Combining equations for $\Upsilon_1$ and $\Delta_2$, one immediately deduces that $\Upsilon_1 = X_{a_1}(1-X_{\gamma})^{-1}$. Therefore
\be
	Q(p_1) = 1-\Upsilon_1 \cdot \Delta_1 = 1+X_{\gamma}(1-X_{\gamma})^{-1} = (1-X_{\gamma})^{-1}\,,
\ee
which also implies
\be
	Q(p_2) = (1-X_{\gamma})^{-1}\,,\qquad
	Q(p_3) = (1-X_{\gamma})^{-2}\,.
\ee 
The factorization $Q(p) = \prod_{n\geq 1}(1-X_{n\gamma})^{\alpha_{n\gamma}(p)}$ leads to $\alpha_{\gamma}(p_1)=\alpha_{\gamma}(p_2)=-1$ and $\alpha_{\gamma}(p_3) = -2$. The BPS index is computed by the ratio $\Omega(n\gamma) = [L(n\gamma)] / n\gamma$, where $L(n\gamma) = \bigcup_{i}\alpha_{n\gamma}(p_i)\cdot \pi^{-1}(p_i)$.
Notice that $[\pi^{-1}(p_1) \cup \pi^{-1}(p_2) \cup (2\times \pi^{-1}(p_3))]=\gamma$, therefore the saddle of Type-2 gives an overall contribution to the BPS index 
\be
	\Delta\Omega(\gamma) = -1\,.
\ee
Often, this saddle appears simultaneously with other saddles of Type-2. When this is the case, the total index is the sum of each contribution.

\subsubsection*{Type-3 saddle}
\begin{figure}[h!]
\begin{center}
\includegraphics[width=0.99\textwidth]{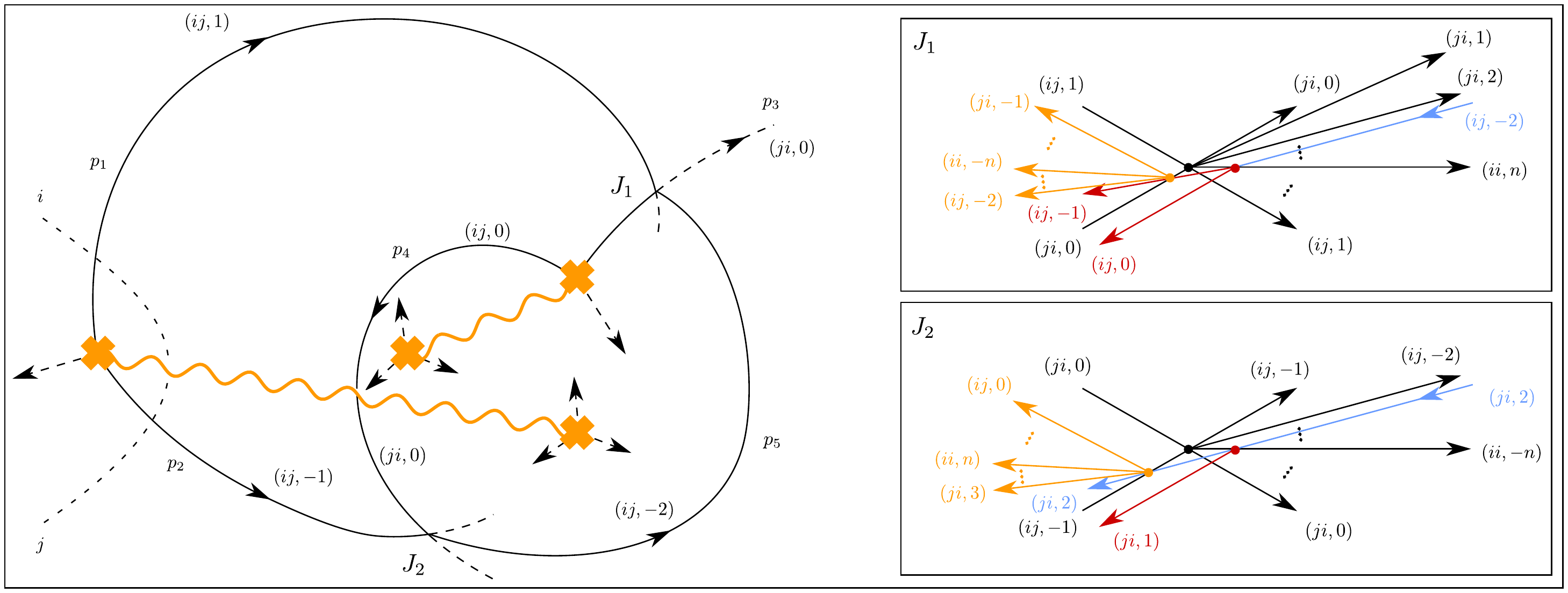}
\caption{A saddle of Type-3.}
\label{fig:type-3-saddle}
\end{center}
\end{figure}

Another interesting example is the Type-3 saddle, shown in Figure~\ref{fig:type-3-saddle}. A real-world example of this type of saddle can be seen in Figure \ref{fig:1122}.
As for Type-2 saddles, there are two branch points supporting the saddle, as well as a logarithmic cut running around one of them, depicted as a dashed line on the left.

There are five distinguished $\CE$-walls, originating from the points and junctions, and denoted $p_1, \dots, p_5$. 
There are two junctions, denoted $J_1$ and $J_2$ in the left frame.
Upon closer inspection, and resolving the phase from the critical one $\vartheta=\vartheta_c$ to British resolution $\vartheta=\vartheta_c + \epsilon$ we see that each of $J_i$ actually features a lot of sub-structure. This is depicted in the right-top and right-bottom frames.

Let us describe $J_1$ in detail, hopefully readers can infer the structure of $J_2$ form the picture, along similar lines.
Junction $J_1$ is generated by the intersection of $p_1$ (of type $(ij,1)$) and $p_3$ (of type $(ji,0)$). These form an $ij-ji$ junction which spawns, among other walls, an $\CE$-wall of type $(ji,2)$.
Due to an incoming $(ij,-2)$ wall (depicted in light blue) sourced from $J_2$, this will be a two-way street and therefore will be part of the saddle, it will be labeled $p_5$.
Furthermore, the incoming blue wall intersects the family of $(ii,n)$ walls generated at the $p_1$-$ p_3$ junction.
Interacting with the $n=2$ wall, they give rise to a $(ij,0)$-wall (depicted in red), which runs anti-parallel to $p_3$: this makes $p_3$ a double-wall, and therefore part of the saddle.
On the other hand, when the blue wall interacts with the $(ii,1)$ wall, they generate an $(ij,-1)$ wall (depicted in red), which goes on to intersect with $p_3$ to generate an $ij-ji$ junction marked by an orange dot.
Among the descendants generated from this junction, is an $(ji,-1)$ wall (depicted in orange), which runs anti-parallel to $p_1$. This makes $p_1$ a double wall, and therefore part of the saddle.
We are only considering a few of the intersections among one-way walls that occur near $J_1$, omitting those which are irrelevant for the study of the saddle.
An exhaustive analysis, taking into account the whole sub-structure, requires use of the flatness equations and is carried out in Appendix \ref{app:saddle-type-3-eqs}. 
The analysis of $J_2$ is qualitatively similar, although there are subtle differences in how walls intersect one another. This is also studied in appendix.

To compute the BPS index, our task is to study the soliton data on the two-way streets $p_1, \dots,  p_5$.
Let us collect our conventions for soliton generating functions
\be
\begin{array}{c|c|c}
 & \text{type} & \text{description} \\
 \hline
p_1 \ \   
\begin{array}{c}
\Upsilon_1 \\ \Delta_1
 \end{array} 
 &
 \begin{array}{c}
(ij,1)  \\ (ji,-1)
 \end{array} 
 &
 \begin{array}{c}
\text{from branch point to $J_1$} \\ \text{opposite}
 \end{array} 
 \\
 \hline
 p_2 \ \   
\begin{array}{c}
\Upsilon_2 \\ \Delta_2
 \end{array} 
 &
 \begin{array}{c}
(ij,-1)  \\ (ji,1)
 \end{array} 
 &
 \begin{array}{c}
\text{from branch point to $J_2$} \\ \text{opposite}
 \end{array} 
 \\
 \hline
 p_3 \ \   
\begin{array}{c}
\Upsilon_3 \\ \Delta_3
 \end{array} 
 &
 \begin{array}{c}
(ji,0)  \\ (ij,0)
 \end{array} 
 &
 \begin{array}{c}
\text{from branch point to $J_1$} \\ \text{opposite}
 \end{array} 
 \\
 \hline
 p_4 \ \   
\begin{array}{c}
\Upsilon_4 \\ \Delta_4
 \end{array} 
 &
 \begin{array}{c}
(ji,0)  \\ (ij,0)
 \end{array} 
 &
 \begin{array}{c}
\text{from branch point to $J_2$} \\ \text{opposite}
 \end{array} 
 \\
 \hline
 p_5 \ \   
\begin{array}{c}
\Upsilon_5 \\ \Delta_5
 \end{array} 
 &
 \begin{array}{c}
(ij,-2)  \\ (ji,2)
 \end{array} 
 &
 \begin{array}{c}
\text{from $J_2$ to $J_1$} \\ \text{opposite}
 \end{array} 
\end{array}
\ee

We can determine these generating functions through the consistency conditions for the nonabelianization map, see Appendix \ref{app:saddle-type-3-eqs}.
The consistency conditions at the left-most branch point are
\be
\begin{split}
	\Upsilon_1 & = X_{a_1}\,,
	\qquad
	\Upsilon_2 = X_{a_2}(1 - X_{a_1} \Delta_1 )\,,
\end{split}
\ee
by a direct application of (\ref{eq:branch-homotopy-equation}). Similarly, at the top-right branch point one has
\be
\begin{split}
	\Upsilon_3 & = X_{a_3} (1 - X_{a_4} \Delta_4)\,,
	\qquad
	\Upsilon_4 = X_{a_4}\,,
\end{split}
\ee
where again we suppress information about logarithmic branching to keep notation light, and   $a_i$ denote `simpletons', see footnote \ref{foot:simpletons}.
At junction $J_1$ we find the following relations  
\be\label{eq:junction-J1-type3-saddle}
\begin{split}
	\Delta_5 & = \Upsilon_3 \Upsilon_1 \Upsilon_3  \Upsilon_1  \Upsilon_3  \\
	\Delta_3 & =
	Q(p_5)^2 \Upsilon_1\Upsilon_3\Upsilon_1\Upsilon_3\Upsilon_5  + 
	Q(p_5) \Upsilon_1\Upsilon_3\Upsilon_5\Upsilon_3\Upsilon_1 
	+ \Upsilon_5 \Upsilon_3 \Upsilon_1\Upsilon_3 \Upsilon_1\\ 
	\Delta_1 & = \Upsilon_3 \Upsilon_1 \Upsilon_3 \Upsilon_5 \Upsilon_3\(1 -  \Upsilon_5 \Upsilon_3 \Upsilon_1 \Upsilon_3 \Upsilon_1 \Upsilon_3\) +  \Upsilon_3 \Upsilon_5 \Upsilon_3 \Upsilon_1 \Upsilon_3   \\
\end{split}
\ee
where $Q(p_i) = 1 - \Delta_i\Upsilon_i$, and 
\be
	Q(p_1) = Q(p_5)^2\,,
	\qquad Q(p_3) = Q(p_5)^3 \,.
\ee

A similar computation, also spelled out in Appendix, yields  junction $J_2$, which turn out to imply
\be
\begin{split}
	\Upsilon_5 & = -\Upsilon_2 \Upsilon_4 \Upsilon_2   \\
	\Delta_2 & = - \Delta_5 \Upsilon_2  \Upsilon_4   - \(1+\Delta_5\Upsilon_2\Upsilon_4\Upsilon_2\) \Upsilon_4 \Upsilon_2 \Delta_5   \\ 
	\Delta_4 & = - \Upsilon_2\Delta_5\Upsilon_2   \\
\end{split}	
\ee
as well as
\be
	Q(p_2) = Q(p_5)^2 \,,\qquad Q(p_4) = Q(p_5) \,.
\ee

Now let us turn to the computation of $\Omega$. 
Recall from \cite{Banerjee:2018syt} that $\Omega(\gamma) = [L(\gamma)] / \gamma$ where $L(\gamma) = \sum_{i=1}^{5} \alpha_\gamma(p_i) \pi^{-1}(p_i)$ is obtained by concatenation of lifts of walls $p_i$ with coefficients $\alpha_\gamma(p_i)$ determined by factorization $Q(p_i) = \prod_{n\geq 1} (1 \pm X_{n\gamma})^{\alpha_{n\gamma}(p_i)} $.
Given the above relations, we have
\be
	\alpha_{n\gamma}(p_1) = \alpha_{n\gamma}(p_2) = 2 \alpha_{n\gamma}(p_5)
	\qquad
	\alpha_{n\gamma}(p_3) = 3\alpha_{n\gamma}(p_5)
	\qquad
	\alpha_{n\gamma}(p_4) = \alpha_{n\gamma}(p_5)
\ee
and therefore
\be
	\gamma = \pi^{-1} [2p_1 +2p_2+3p_3+p_4+p_5]\,.
\ee
This is consistent with formulae. Indeed, let us focus on $Q(p_5)$, since all other $Q(p_i)$ are obtained from this in a straightforward way. By perturbative substitution, we find that
\be
	Q(p_5) = 1 - \Upsilon_5\Delta_5 = 1 + X_{a_1a_3a_1a_3a_1a_2a_4a_2} + \dots = 1 + X_\gamma + O(X_\gamma^2)
\ee
reflecting the fact that $\gamma = {\rm cl}(a_1a_3a_1a_3a_1a_2a_4a_2)$ is indeed the lift of 2 copies of $p_1$ and $p_2$,  3 copies of $p_3$, and one copy of $p_4$ and $p_5$.

To solve for the full generating function $Q(p_5)$ we proceed as follows. 
Let 
\be
	\Upsilon_\gamma = - \Upsilon_3\Upsilon_1\Upsilon_3\Upsilon_1\Upsilon_3\Upsilon_2\Upsilon_4\Upsilon_2\,,
	\qquad
	X_\gamma = X_{a_1a_3a_1a_3a_1a_2a_4a_2}
\ee
up to cyclic reorderings of factors in each expression. Note that $Q(p_5) = 1-\Upsilon_\gamma$.
It is now straightforward to solve the equations by direct substitution: 
one readily obtains $\Upsilon_2 = X_{a_2} (1-\Upsilon_\gamma)^2$ and $\Upsilon_3 = X_{a_3} (1-\Upsilon_\gamma)$, leading to $\Upsilon_\gamma =- X_\gamma \, (1-\Upsilon_\gamma)^7$. In terms of $Q(p_5)$ this implies 
\be\label{eq:Qp5-alg-eq}
	{ Q(p_5) = 1 + X_\gamma Q(p_5)^7 }\,.
\ee
From here we can extract $\Omega(n\gamma)$. For example
\be
	Q(p_5) = 1 + X_\gamma + 7 X_\gamma^2 + 70 X_\gamma^3 + 819 X_\gamma^4 + 10472 X_\gamma^5 + 141778 X_\gamma^6 + O(X_\gamma^7)
\ee
implying that\footnote{Recall that $L[n\gamma]$ contains $\pi^{-1}(p_5)$ precisely $n \Omega(n\gamma)$ times, this explains the exponents in this factorization. The choice of sign in front of $X_\gamma$ turns out to be the only one that gives integer $\Omega(n\gamma)$ for all~$n$.}
\be
	Q(p_5) = \prod_{n\geq 1}(1-X_{n\gamma})^{n\Omega(n\gamma)}
\ee
\be\label{eq:type-3-saddle-BPS-indices}
\begin{split}
	&
	\Omega(\gamma) = -1
	\qquad 
	\Omega(2\gamma) = -3
	\qquad 
	\Omega(3\gamma) = -21
	\\ 
	&
	\Omega(4\gamma) = -182
	\qquad 
	\Omega(5\gamma) = -1855
	\qquad 
	\Omega(6\gamma) = -20811
	\\
	& \dots
\end{split}
\ee
Note that equations like (\ref{eq:Qp5-alg-eq}) appeared in \cite[eq. (3.2)]{Galakhov:2013oja}. This equation is similar, but not of the same type, due to the fact that the exponent (7 in this case) isn't an exact square. Nevertheless, this indicates the presence of wild BPS states. From this equation we may easily obtain arbitrarily many BPS indices for states with charges $n\gamma$.

\subsubsection*{Type-4 saddle}
\begin{figure}[h!]
\begin{center}
\includegraphics[width=0.65\textwidth]{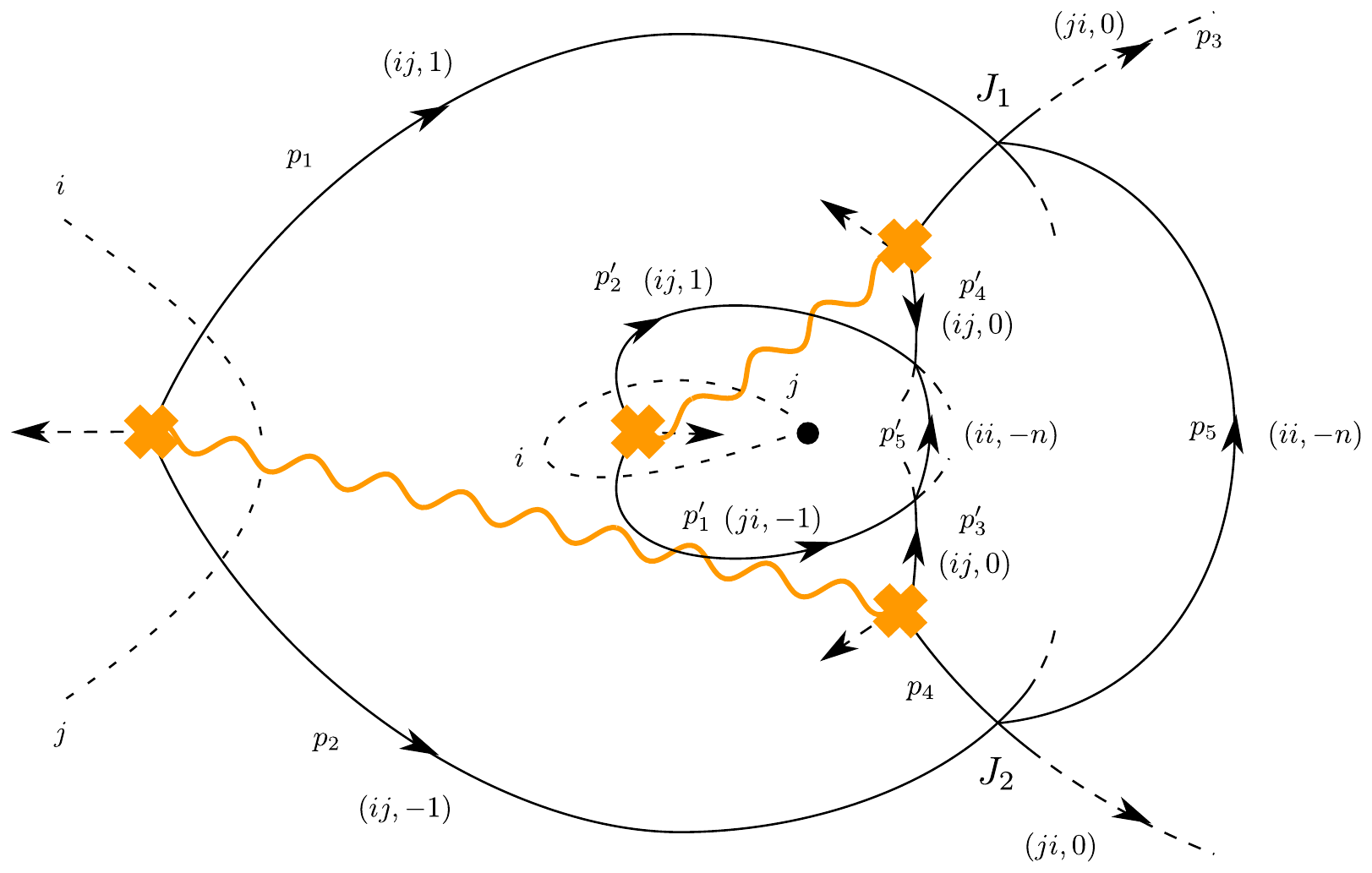}
\caption{A saddle of Type-4.}
\label{fig:type-4-saddle}
\end{center}
\end{figure}

The fourth and last type of saddle we will consider is shown in Figure~\ref{fig:type-4-saddle}. A real-world example of this type of saddle can be found in Figure \ref{fig:0111}. 
Here only two-way streets are shown (solid lines labeled $p_1,\dots, p_5$ and $p_1',\dots, p_5'$), while the picture omits infinitely many $\CE$-walls which are not part of the saddle. 

The saddle involves four junctions, two of these are labeled  $J_1$ and $J_2$.
Resolving the phase slightly away from the critical one $\vartheta=\vartheta_c+\epsilon$, one finds a lot of sub-structure at each junction. An exhaustive analysis is provided in Appendix \ref{app:type-4-saddle}.

To compute the indices of BPS states associated to this saddle, the first task is to study the soliton data of all two-way streets.
For clarity, we spell out the topological types of all soliton generating functions for $p_1\dots p_5$
\be
\begin{array}{c|c|c}
 & \text{type} & \text{description} \\
 \hline
p_1 \ \   
\begin{array}{c}
\Upsilon_1 \\ \Delta_1
 \end{array} 
 &
 \begin{array}{c}
(ij,1)  \\ (ji,-1)
 \end{array} 
 &
 \begin{array}{c}
\text{from branch point to $J_1$} \\ \text{opposite}
 \end{array} 
 \\
 \hline
 p_2 \ \   
\begin{array}{c}
\Upsilon_2 \\ \Delta_2
 \end{array} 
 &
 \begin{array}{c}
(ij,-1)  \\ (ji,1)
 \end{array} 
 &
 \begin{array}{c}
\text{from branch point to $J_2$} \\ \text{opposite}
 \end{array} 
 \\
 \hline
 p_3 \ \   
\begin{array}{c}
\Upsilon_3 \\ \Delta_3
 \end{array} 
 &
 \begin{array}{c}
(ji,0)  \\ (ij,0)
 \end{array} 
 &
 \begin{array}{c}
\text{from branch point to $J_1$} \\ \text{opposite}
 \end{array} 
 \\
 \hline
 p_4 \ \   
\begin{array}{c}
\Upsilon_4 \\ \Delta_4
 \end{array} 
 &
 \begin{array}{c}
(ji,0)  \\ (ij,0)
 \end{array} 
 &
 \begin{array}{c}
\text{from branch point to $J_2$} \\ \text{opposite}
 \end{array} 
\end{array}
\ee
The counterparts for $p_1'\dots p_5'$ can be deduced from Figure \ref{fig:type-4-saddle} in the same way.

Each of these generating functions can be determined by considering flatness constraints for the nonabelianization map at junctions and branch points.
For the sake of keeping notation light we will, as above, suppress information about logarithmic branching of soliton data, keeping in mind that this can be restored in each of the following expressions in a non-ambiguous way.

Flatness constraints at the left-most branch point are
\be
\begin{split}
	\Upsilon_1 & = X_{a_1}
	\qquad
	\Upsilon_2 = X_{a_2}(1- X_{a_1} \Delta_1 )
\end{split}
\ee
by  a direct application of (\ref{eq:branch-homotopy-equation}). 
Likewise, at the top-right branch point one finds
\be
\begin{split}
	\Upsilon_3 & = X_{a_3} 
	\qquad
	\Upsilon_4' = X_{a_4'}(1 - X_{a_3} \Delta_3)
\end{split}
\ee
Similarly, at the two other branch points, one obtains
\be
\begin{split}
	\Upsilon_1' & = X_{a_1'}
	\qquad
	\Upsilon_2' = X_{a_2'} (1 - X_{a_1'} \Delta_1' )\\
	\Upsilon_3' & = X_{a_3'} 
	\qquad
	\Upsilon_4 = X_{a_4} (1 - X_{a'_3} \Delta_3')
\end{split}
\ee
As before,  $a_i$ denote `simpleton' paths, see footnote \ref{foot:simpletons}.
The flatness constraints at junction $J_1$, connecting $p_1, p_3, p_5$ are 
\be\label{eq:junction-J1-type4-saddle}
\begin{split}
	\Delta_1 & =   \frac{2 + \Upsilon_5^{(jj,-1)}\Upsilon_3\Upsilon_1}{(1 +  \Upsilon_5^{(jj,-1)}\Upsilon_3\Upsilon_1)^2}\, \Upsilon_5^{(jj,-1)}\Upsilon_3  
	\\
	\Delta_3 & =  \Upsilon_1\Upsilon_5^{(jj,-1)}  \, \frac{2 + \Upsilon_5^{(jj,-1)}\Upsilon_3\Upsilon_1}{(1 +  \Upsilon_5^{(jj,-1)}\Upsilon_3\Upsilon_1)^2}
	\\
	\Delta_{5}^{(ii,k)} & = \frac{1}{k} (-\Upsilon_1\Upsilon_3)^k 
	\qquad
	\Delta_{5}^{(jj,k)} =  -\frac{1}{k} (-\Upsilon_3\Upsilon_1)^k \,,
\end{split}
\ee
where $\Upsilon_{5}^{jj,n}, \Delta_{5}^{jj,n}$ are generating functions of $(jj,n)$ solitons supported on $p_5$, see Appendix~\ref{app:type-4-saddle} for the relation between this generating function and Stokes matrices of the underlying $\CE$-walls.
It is useful to note that $\Upsilon_5^{(jj,-1)} \Xi_{ji} = - \Xi_{ji} \Upsilon_5^{(ii,-1)}$ if $\Xi_{ji}$ is any shift-symmetric soliton generating function \cite{Banerjee:2018syt}.
The equations describing junction $J_2$, which connects $p_2,p_4,p_5$, have a  similar form
\be\label{eq:type-4-saddle-J2-sol}
\begin{split}
	\Delta_4 & = \Delta_5^{(ii,1)}\Upsilon_2\,  \frac{2 + \Delta_5^{(ii,1)}\Upsilon_2\Upsilon_4}{(1+\Delta_5^{(ii,1)}\Upsilon_2\Upsilon_4)^2} \\
	\Delta_2 & =  \frac{2 + \Delta_5^{(ii,1)}\Upsilon_2\Upsilon_4}{(1+\Delta_5^{(ii,1)}\Upsilon_2\Upsilon_4)^2} \,  \Upsilon_4 \Delta_5^{(ii,1)}\\
	\Upsilon_{5}^{(ii,-k)} & = -\frac{1}{k} (-\Upsilon_2\Upsilon_4)^k 
	\qquad
	\Upsilon_{5}^{(jj,-k)} =  \frac{1}{k} (-\Upsilon_4\Upsilon_2)^k \,.
\end{split}	
\ee
Thanks to the $\IZ_2$ symmetry relating $x\to x^{-1}$, the equations for the two additional junctions, can be inferred from the ones for $J_1, J_2$
\be
\begin{split}
	\Delta_1' & =   
	{\Delta'}_5^{(ii,1)}\Upsilon'_3 \frac{2 + {\Delta'}_5^{(ii,1)}\Upsilon'_3 \Upsilon'_1}{(1+ {\Delta'}_5^{(ii,1)}\Upsilon'_3 \Upsilon'_1)^2} 
	\\
	\Delta_3' & =  
	\frac{2 + {\Delta'}_5^{(ii,1)}\Upsilon'_3 \Upsilon'_1}{(1+ {\Delta'}_5^{(ii,1)}\Upsilon'_3 \Upsilon'_1)^2}
	\Upsilon'_1 {\Delta'}_5^{(ii,1)}
	\\
	{\Upsilon'}_{5}^{(ii,-k)} & = - \frac{1}{k} (-\Upsilon'_3\Upsilon'_1)^k 
	\qquad
	{\Upsilon'}_{5}^{(jj,-k)} =  \frac{1}{k} (-\Upsilon'_1\Upsilon'_3)^k \,.
\end{split}
\ee
\be
\begin{split}
		\Delta'_4 & =  
		{\Upsilon'}_5^{(jj,-1)}\Upsilon'_2\,
		\frac{2 + {\Upsilon'}_5^{(jj,-1)}\Upsilon'_2\Upsilon_4' }{(1+ {\Upsilon'}_5^{(jj,-1)}\Upsilon'_2\Upsilon_4' )^2}
		\\
	\Delta'_2 & = 
	\frac{2 + {\Upsilon'}_5^{(jj,-1)}\Upsilon'_2\Upsilon_4' }{(1+ {\Upsilon'}_5^{(jj,-1)}\Upsilon'_2\Upsilon_4' )^2} \,
	\Upsilon'_4 {\Upsilon'}_5^{(jj,-1)} 
	\\
	{\Delta'}_{5}^{(ii,k)} & = \frac{1}{k} (-\Upsilon'_4\Upsilon'_2)^k 
	\qquad
	{\Delta'}_{5}^{(jj,k)} =  -\frac{1}{k} (-\Upsilon'_2\Upsilon'_4)^k \,.
\end{split}
\ee

It follows from the above equations that  
\be\label{eq:type-3-saddle-Qpi}
\begin{split}
	Q(p_i) = Q(p_i') &= Q(X_\gamma) = (1+\Upsilon_\gamma)^{-2}
\end{split}	
\ee
where $Q(p_i) = 1 - \Delta_i\Upsilon_i$ (similarly for $p'_i$) and we introduced $\gamma =[ {\rm cl}(a_3 a_1  a_4 a_2) ]=  [{\rm cl}(a_3' a_1'  a_4' a_2')]$\footnote{Physical charges are identified by a quotient of a homology (sub-)lattice of $\Sigma$, which identifies cycles with the same $\lambda$-period  \cite{Banerjee:2018syt}.} 
as well as
\be
	\Upsilon_\gamma 
	= \Upsilon_5^{(jj,-1)}\Upsilon_3\Upsilon_1 
	= \Delta_5^{(ii,1)}\Upsilon_2\Upsilon_4 
	= {\Delta'_5}^{(ii,1)}\Upsilon'_3\Upsilon'_1 
	= {\Upsilon'}_5^{(ii,-1)}\Upsilon'_4\Upsilon'_2 \,.
\ee
From the equations it is straightforward to obtain an expression for $\Upsilon_\gamma$
\be
\begin{split}
	\Upsilon_\gamma 
	& = \Upsilon_5^{(jj,-1)}\Upsilon_3\Upsilon_1 \\
	& = -\Upsilon_2\Upsilon_4  \Upsilon_3\Upsilon_1\\
	& = -X_{a_2} (1- X_{a_1}\Delta_1) \, X_{a_4}(1- X_{a_3'} \Delta_3') X_{a_3} X_{a_1} \\
	& = -X_{a_2} Q(p_1) \, X_{a_4} Q(p_3') X_{a_3} X_{a_1} \\
	& = -X_\gamma (1+\Upsilon_\gamma)^{-4}
\end{split}
\ee
In terms of $Q(p)$, we arrive at the following equation
\be
	Q = (1 - X_\gamma \, Q^2)^{-2} \,.
\ee
From here we may easily obtain a series expansion for $Q(X_\gamma)$ to arbitrary order
\be
\begin{split}
	Q(X_\gamma) 
	& = 
	1 
	+2 X_\gamma 
	+11 X_\gamma^2
	+80 X_\gamma^3
	+665 X_\gamma^4
	+5980 X_\gamma^5\\
	&
	+56637 X_\gamma^6
	+556512 X_\gamma^7
	+5620485  X_\gamma^8
	+O(X_\gamma^9)
\end{split}
\ee

To compute BPS indices, we must factorize this generating function. Choosing 
\be
	Q(X_\gamma)  = \prod_{n\geq 1} (1-(-1)^nX_{n\gamma})^{n\, \beta_{n\gamma}}
\ee
with integer coefficients 
\be
\begin{split}
	& 
	\beta_1= 2,
	\qquad
	\beta_2= -5
	\qquad
	\beta_3= 20
	\qquad
	\beta_4= -120
	\\
	& \beta_5= 850
	\qquad
	\beta_6= -6602
	\qquad
	\beta_7= 54894
	\qquad
	\beta_8= -480624
	\\
	& \dots
\end{split}
\ee
The contribution of the saddle to the BPS index is $\Delta\Omega(n\gamma) = [L(n\gamma)] / (n\gamma)$, where  $L(n\gamma) = \bigcup_{p} \alpha_{n\gamma}(p) \pi^{-1}(p)$ where $\alpha_{n\gamma}(p) = n\beta_{n\gamma}$ are the exponents of the factorization.
Taking into account both streets $p_i$ and $p_i'$ this yields
\be\label{eq:type-4-saddle-index}
	\Delta\Omega(n\gamma)  = 2\beta_n \,.
\ee

\subsection{Description of the spectrum}\label{eq:spectrum-description}

Having completed our preliminary analysis of a selection of saddle types, we are now in the position to explore the geometry of exponential networks at the point $\mathbf{Q}_0$. 

We will adopt a mixed approach, blending the use of exponential networks with quiver representation theory and wall-crossing.
In principle, exponential networks are expected to capture all BPS states. 
However, for practical reasons, it is more convenient to rely on all available techniques. 
There are two main technical difficulties involved with working with exponential networks: the fact that the spectrum  is infinite at $\mathbf{Q}_0$, and the fact that the saddles of certain BPS states can be somewhat complicated to study.
On the other hand, there are also important advantages: it is not easy to obtain information about BPS states with charges $\gamma = \sum_{i=1}^4 d_i\gamma_i$ where all $d_i>0$ using quivers, while from the viewpoint of networks these states are just like all others. Even better, for charges $\gamma$ such that $\Omega(n\gamma)\neq 0$ for several (possibly all) $n>0$, networks often encode the full generating function of all BPS indices in a neat algebraic equation. We will come across an illustration of both these points shortly.

Without further ado, let us list the BPS states at $\mathbf{Q}_0$, where we make a note of which techniques were used to compute them.
First of all, there are the basic saddles $p_1, \dots, p_5$ encountered in Figure \ref{fig:exp-net-sym-point} 
\be\label{eq:basic-saddles-spectrum}
\begin{array}{c|c|c|c}
\gamma & \text{D-brane charge}  & \Omega(\gamma) & \text{Figure}\\
\hline
(1,0,0,0) & {D4} & 1 & \ref{fig:exp-net-sym-point}\\
(0,1,0,0) & {D2_f\-\overline{D4}} & 1& \ref{fig:exp-net-sym-point}\\ 
(0,0,1,0) & D0\-D2_b\-\overline{D2}_f\-\overline{D4} & 1& \ref{fig:exp-net-sym-point}\\
(0,0,0,1) & \overline{D2}_b\-D4 & 1& \ref{fig:exp-net-sym-point}\\
\hline
(0,2,1,2) & D0\-\overline{D2}_b\-{D2}_f\-\overline{D4} & 1& \ref{fig:exp-net-sym-point}\\
\end{array}
\ee

Next, we consider states supported on pairs of nodes $\gamma_a, \gamma_b$ of the quiver. 
Besides plotting their saddles, it is worth noting that any pair of nodes forms a sub-quiver that is either trivial (no arrows) and hence without boundstates, or a Kronecker quiver with two arrows. The representation theory of the Kronecker quiver is well known, and is fully determined by the relative ordering of $\arg Z_{\gamma_a} , \arg Z_{\gamma_b}$, see Figure \ref{fig:quiver} for comparison.
\be\label{eq:2-node-subquiver-spectrum}
\begin{array}{c|c|c|c}
\gamma & \text{D-brane charge}  & \Omega(\gamma) & \text{Figure}\\
\hline
(0,0,1,1) & {D0\-\overline{D2}_f} & -2 & \ref{fig:0011}\\ 
(0,0,n,n+1) & {nD0\-\overline{D2}_b\-n\overline{D2}_f\-{D4}} & 1 & \ref{fig:0012}, \ref{fig:0023}\\ 
(0,0,n+1,n) & {(n+1)D0\-D2_b\-(n+1)\overline{D2}_f\-\overline{D4}} & 1 & \ref{fig:0021}, \ref{fig:0032}\\
\hline
(0,1,1,0) & {D0\-D2_b\-2\overline{D4}} & -2 & \ref{fig:0110}\\ 
(0,n,n+1,0) & {(n+1)D0\-(n+1)D2_b\-\overline{D2}_f\-(2n+1)\overline{D4}} & 1 & 
\\ 
(0,n+1,n,0) & {nD0\-nD2_b\-{D2}_f\-(2n+1)\overline{D4}} & 1 & \ref{fig:0210}, \ref{fig:0320}
\end{array}
\ee
This table illustrates the advantage of using quiver representation theory over networks, for those BPS states supported on a sub-quiver whose representations are well understood. 

Next we consider the class of BPS states whose charges are supported on three nodes of the quiver, in other words $\gamma = \sum_{i=1}^{4} d_i\gamma_i$ and exactly one of the $d_i=0$.
We will henceforth suppress D-brane charges, these can always be recovered by the dictionary (\ref{eq:D-brane-charge-dictionary}).
For charges with only $d_1=0$ we find, up to $d_2+d_3+d_4\leq 9$,
\be\label{eq:3-node-subquiver-spectrum-A}
\begin{array}{c|c|c}
\gamma & \Omega(\gamma) & \text{Figure}\\
\hline
(0, 1,1,1) & \Omega=4& \ref{fig:0111}
\\(0, 2,1,1) & \Omega=-2 & \ref{fig:0211}
\\(0, 1,2,1) & \Omega=-4
\\(0, 1,1,2) & \Omega=-2 & \ref{fig:0112}
\\(0, 2,2,1) & \Omega=10
\\(0, 1,2,2) & \Omega=10
\\(0, 2,1,2) & \Omega=1
\\(0, 3,2,1) & \Omega=-4
\\(0, 2,3,1) & \Omega=-14
\\(0, 2,2,2) & \Omega=-12& \ref{fig:0111}
\\(0, 1,3,2) & \Omega=-14
\\(0, 1,2,3) & \Omega=-4
\\(0, 4,2,1) & \Omega=1
\\(0, 3,3,1) & \Omega=20
\\(0, 3,2,2) & \Omega=10
\\(0, 2,4,1) & \Omega=6
\end{array}
\qquad
\begin{array}{c|c}
\gamma & \Omega(\gamma) \\
\hline
(0, 2,3,2) & \Omega=27
\\(0, 1,4,2) & \Omega=6
\\(0, 1,3,3) & \Omega=20
\\(0, 2,2,3) & \Omega=10
\\(0, 1,2,4) & \Omega=1
\\(0, 4,3,1) & \Omega=-6
\\(0, 3,4,1) & \Omega=-24
\\(0, 3,3,2) & \Omega=-72
\\(0, 2,4,2) & \Omega=-16
\\(0, 1,4,3) & \Omega=-24
\\(0, 2,3,3) & \Omega=-72
\\(0, 3,2,3) & \Omega=-4
\\(0, 1,4,3) & \Omega=-6
\\(0, 1,3,4) & \Omega=-6
\\(0, 5,3,1) & \Omega=4
\\(0, 5,4,0) & \Omega=1
\end{array}
\qquad
\begin{array}{c|c|c}
\gamma & \Omega(\gamma) & \text{Figure}\\
\hline
(0, 4,5,0) & \Omega=1
\\(0, 4,4,1) & \Omega=35
\\(0, 4,3,2) & \Omega=49
\\(0, 4,2,3) & \Omega=1  & \ref{fig:0423}
\\(0, 3,5,1) & \Omega=16
\\(0, 3,4,2) & \Omega=172
\\(0, 3,3,3) & \Omega=60 & \ref{fig:0111}
\\(0, 2,4,3) & \Omega=172
\\(0, 1,5,3) & \Omega=16
\\(0, 3,2,4) & \Omega=1
\\(0, 2,3,4) & \Omega=49
\\(0, 1,4,4) & \Omega=35
\\(0, 0,5,4) & \Omega=1
\\(0, 0,4,5) & \Omega=1
\\(0, 1,3,5) & \Omega=4
\\ \vdots & \vdots
\end{array}
\ee
This part of the spectrum is obtained as follows. The three-node sub-quiver obtained by dropping node $\gamma_1$ admits a choice of moduli, corresponding to $\arg Z_{\gamma_4}>\arg Z_{\gamma_3}>\arg Z_{\gamma_2}$, for which the only stable representations correspond to $\Omega(\gamma_2)=\Omega(\gamma_3)=\Omega(\gamma_4)=1$.
This enables to write down the Kontsevich-Soibelman invariant \cite{Kontsevich:2008fj}, or \emph{spectrum generator} \cite{Gaiotto:2009hg}, for this sub-quiver 
\be\label{eq:minimal-chamber-3-node-subquiver}
	\mathbb{S} = \CK_{\gamma_4}\CK_{\gamma_3}\CK_{\gamma_2} \,.
\ee
BPS states in (\ref{eq:3-node-subquiver-spectrum-A}), in fact any of those with charges $(0,d_2,d_3,d_4)$, can be obtained by factorizing $\mathbb{S}$ at $\mathbf{Q}_0$. This requires knowing the exact values of $Z_{\gamma_i}$, we provide numerical estimates in (\ref{eq:sym-point-periods}) and (\ref{eq:D-brane-Z-sym-pt}).
Given the simplicity of (\ref{eq:minimal-chamber-3-node-subquiver}), it would be just as straightforward to write down the \emph{motivic} spectrum generator, whose factorization would in turn yield \emph{protected spin characters} (PSCs).
\footnote{
A similar strategy may also be applied to subquivers defined by suppressing one of the nodes $\gamma_2$, $\gamma_3$ or $\gamma_4$. 
In this way, the BPS indices (or their motivic deformations to PSC) 
can be directly computed. However we obtain no new information, due to how central charges are arranged.
For $\Omega(d_1 \gamma_1 + d_2 \gamma_2 +d_3 \gamma_3 )$ we find that the only non-vanishing boundstates are the ones we already found, namely $\Omega(\gamma_2 + \gamma_3 )= - 2$ and $\Omega(\gamma_2 + n(\gamma_2 + \gamma_3) )=\Omega(\gamma_3 + n(\gamma_2 + \gamma_3) )= 1$. 
For $\Omega(d_1 \gamma_1 + d_1 \gamma_2 +d_1 \gamma_4 )$ we find that this always vanishes unless $d_1+d_2+d_4=1$.
For $\Omega(d_1 \gamma_1 + d_3 \gamma_3 +d_4 \gamma_4 )$ we find that the only non-vanishing boundstates are the ones we already found, namely $\Omega(\gamma_3 + \gamma_4 )= - 2$ and $\Omega(\gamma_3 + n(\gamma_3 + \gamma_4) )=\Omega(\gamma_4 + n(\gamma_3 + \gamma_4) )= 1$. 
Also worth noting, is that these results may, at least in part, be obtained using Reineke's formula \cite{2003InMat.152..349R}.}

Let us comment on how the results obtained by factorization of the spectrum generator compare with results obtained from exponential networks.
We just focus on states with charges $(0,n,n,n)$. These states are represented by a Type-4 saddle like the one of Figure \ref{fig:type-4-saddle}, see Figure \ref{fig:0111}.
For $n=1$ we obtain $\Omega=4$ from the analysis of the saddle, see (\ref{eq:type-4-saddle-index}), which matches the expectation from the spectrum generator. For $n>1$ however the result obtained from analyzing the saddle systematically undercounts the prediction from the spectrum generator (comparing $|\Omega|$). A possible explanation is that the Type-4 saddle is only a \emph{part} of the BPS state, which may consists of additional 2-way streets that we haven't detected in Figure \ref{fig:0111}. This issue illustrates how, sometimes, computing BPS indices with exponential networks can be rather subtle due to the intricacy of the network.

Finally, we consider states whose charges are supported on all four nodes. This is where exponential networks are most useful, since we cannot invoke simple results for the quiver representation theory.
For example, in Figure \ref{fig:1122} we observe a pair of saddles of Type-3. We deduce the following spectrum from (\ref{eq:type-3-saddle-BPS-indices})
\be\label{eq:4-node-subquiver-spectrum-A}
\begin{array}{c|c|c}
\gamma & \Omega(\gamma) & \text{Figure}\\
\hline
(1,1,2,2) & \Omega=-2& \ref{fig:1122}
\\(2,2,4,4) & \Omega=-6 & \ref{fig:1122}
\\(3,3,6,6) & \Omega=-42 & \ref{fig:1122}
\\(4,4,8,8) & \Omega=-364 & \ref{fig:1122}
\\ \vdots & \vdots &
\end{array}
\ee
This is a case in which an infinite tower of BPS states is encoded by a single saddle (or two disjoint copies thereof). The generating function of BPS indices is elegantly  encoded by the algebraic equation (\ref{eq:Qp5-alg-eq}).
The same phenomenon was observed in spectral networks, in the context of SU(3) super Yang-Mills theory \cite{Galakhov:2013oja}. In \cite{Galakhov:2014xba} it was shown how this sort of equations can be deformed to a \emph{functional equation} for the generating function of protected spin characters.

Notice that the first state in this series, namely $(1,1,2,2)$ corresponds to the vectormultiplet $(0,0,1,1) = D0\-\overline{D2}_f$ plus a $D0$, and has the same BPS index as $(0,0,1,1)$. It thus seems natural to ask whether there is a whole KK tower of such states, namely $\Omega(n,n,n+1,n+1) = -2$ for all $n\geq 0$.

Bearing in mind the subtleties encountered previously with the BPS index of $\gamma = (0,n,n,n)$, a word of caution is in order: it is possible that, once again, we may be missing parts of the saddle in plotting Figure \ref{fig:1122}. If this were the case, the (absolute values of) BPS indices (\ref{eq:4-node-subquiver-spectrum-A}) would presumably underestimate of the actual answer.  It would be interesting to verify these results independently.
Another natural question that arises, is to what extent the above description of the spectrum is exhaustive. The full spectrum is certainly infinite, and it certainly includes infinitely many states that are missing in the description above.
For charges $\gamma=(d_1,d_2,d_3,d_4)$ with $\sum_i d_i\leq 6$, we believe our spectrum is exhaustive, with the exception of three states: $\Omega(1,1,1,1)=-4$ corresponding to a D0 brane, $\Omega(1,1,2,1)=4$ and $\Omega(1,2,2,1)=-2$.\footnote{We obtained these using the `Coulomb branch formula' with stability parameters calculated on central charges (\ref{eq:sym-point-periods}), using the mathematica code \cite{CoulombHiggs}.} In principle, these states should be visible in the networks tuned to the appropriate phase $\vartheta = \arg Z_\gamma$.

\subsection{Fiber-base symmetry}\label{eq:fiber-base-duality}

As noted below (\ref{eq:D-brane-charge-dictionary}), the set of basic saddles in Figure \ref{fig:exp-net-sym-point} is manifestly invariant under the exchange of fiber and base $\IP^1$s.
In the quiver picture, this duality is hidden: this is because the quiver description is based on a choice of `basis cycles' (\ref{eq:basic-charges}) which is not invariant under fiber-base exchange.
Nevertheless, since we are studying BPS states at a fiber-base symmetric point, we expect the spectrum  to be invariant under exchange of $D2_b\leftrightarrow D2_f$.

Using the charge dictionary from (\ref{eq:basic-charges}) we can verify that the spectrum is indeed symmetric under exchange of fiber and base.
For example consider the infinite towers of states in (\ref{eq:2-node-subquiver-spectrum}). The first tower of states corresponds to the cohort of the vectormultiplet $(0,0,1,1)$ which corresponds to $p_2\cup \overline {p_4}$ (where $\overline p_i$ denotes the orientation-reversal of $p_i$). Recall that, under fiber-base exchange, $p_2$ maps to $p_1$ and $p_4$ maps to $p_5$, then we expect another tower of states obtained by mapping
\be\label{eq:fb-duality}
\begin{split}
	& p_2 : (0,0,1,0) \quad \leftrightarrow \quad p_1 : (0,2,1,2)\\
	& p_4 : (0,0,0,-1) \quad \leftrightarrow \quad  p_5 : (0,1,0,0)
\end{split}	
\ee
While fiber-base exchange acts in a very simple way on saddles $p_i$ from Figure~\ref{fig:exp-net-sym-point}, its action on quiver charges appears to be quite involved: in particular it mixes states with positive dimension vector to ones with negative dimension. This is an artifact of the quiver description, and the BPS spectrum is invariant under fiber-base exchange.

The fiber-base map (\ref{eq:fb-duality}) maps the states in (\ref{eq:2-node-subquiver-spectrum}) to the following ones 
\be\label{eq:2-node-subquiver-spectrum-FB-dual}
\begin{array}{c|c|c|c}
\gamma & \text{D-brane charge}  & \Omega(\gamma) & \text{Figure}\\
\hline
(0,1,1,2) & {D0\-\overline{D2}_b} & -2 & \ref{fig:0112}\\ 
(0,n-1,n,2n) & {nD0\-\overline{D2}_f\-n\overline{D2}_b\-{D4}} & 1 & \ref{fig:0012}\\
(0,n+2,n+1,2n+2) & {(n+1)D0\-D2_f\-(n+1)\overline{D2}_b\-\overline{D4}} & 1 & 
\\
\hline
(0,2,1,1) & {D0\-D2_f\-2\overline{D4}} & -2 & \ref{fig:0211}\\ 
(0,2n,n,n-1) & {(n+1)D0\-(n+1)D2_f\-\overline{D2}_b\-(2n+1)\overline{D4}} & 1 & \ref{fig:0210}
\\ 
(0,2n+2,n+1,n+2) & {nD0\-nD2_f\-{D2}_b\-(2n+1)\overline{D4}} & 1 & \ref{fig:0423}, \ref{fig:0634}
\end{array}
\ee
Unlike the argument leading to (\ref{eq:2-node-subquiver-spectrum}), based on Kronecker-subquivers visible in Figure \ref{fig:quiver}, the infinite towers of states predicted by the fiber-base exchange map are highly nontrivial to see from the viewpoint of the quiver.
This prediction can nonetheless be checked directly. We plot the saddles of both vectormultiplets and of some of the hypermultiplets in Appendix \ref{app:conifold-point} (see links to figures in the table above). We also find the states $(0,1,2,4)$, $(0,3,2,4)$, $(0,4,2,1)$ already present in (\ref{eq:3-node-subquiver-spectrum-A}) with the correct BPS indices, as part of the prediction from the wall-crossing formula.
This partial evidence for (\ref{eq:2-node-subquiver-spectrum-FB-dual}) supports the prediction these new towers of states that would be otherwise challenging to obtain.

Both (\ref{eq:2-node-subquiver-spectrum}) and (\ref{eq:2-node-subquiver-spectrum-FB-dual}) have an especially simple description in terms of saddles. Let us introduce the notion of `cohort' following \cite{Galakhov:2013oja}: given $\gamma,\gamma'$ their 2-cohort  is the set of BPS states
\be
	\CC_2(\gamma,\gamma') = \{\Omega(\gamma+\gamma')=-2, \ \Omega(\gamma+n(\gamma+\gamma'))=\Omega(\gamma'+n(\gamma+\gamma'))=1\,, \ n\geq 1\}
\ee
then (\ref{eq:2-node-subquiver-spectrum}) and (\ref{eq:2-node-subquiver-spectrum-FB-dual}) are simply the cohorts formed by saddles
\be
	\CC_2(p_2,\overline p_4)\,,\ 
	\CC_2(p_2, p_5)\,,\ 
	\CC_2(p_1,\overline p_5)\,,\ 
	\CC_2(p_1, p_4)\,.
\ee
The simplicity of this description, as opposed to the organization by quiver charges or even D-brane charges, suggests that saddles of Figure \ref{fig:exp-net-sym-point} provide an especially natural basis for the BPS spectrum at the point $\mathbf{Q}_0$.

Fiber-base symmetry may be also studied on the BPS states of table (\ref{eq:3-node-subquiver-spectrum-A}). 
The map relates some BPS states with the same BPS index, as follows
\be
\begin{array}{l}
(0, 0,1,0) \leftrightarrow (0, 2,1,2) \\
(0, 1,1,0) \leftrightarrow (0, 2,1,1) \\
(0, 0,1,1) \leftrightarrow (0, 1,1,2) \\
(0, 1,2,0) \leftrightarrow (0, 4,2,3) \\
(0, 0,2,1) \leftrightarrow (0, 3,2,4) 
\end{array}
\qquad
\begin{array}{l}
(0, 1,2,1) \leftrightarrow (0, 3,2,3) \\
(0, 3,2,0) \leftrightarrow (0, 4,2,1) \\
(0, 2,2,1) \leftrightarrow (0, 3,2,2) \\
(0, 1,2,2) \leftrightarrow (0, 2,2,3) \\
(0, 0,2,3) \leftrightarrow (0, 1,2,4) \\
\end{array}
\ee
There are also charges which are invariant under fiber-base symmetry, these include
\be\nonumber
\begin{split}
	& (0, 0,1,2), (0, 2,1,0) , (0, n,n,n), (0, 3,2,1), (0, 1,2,3),\\
	&  (0, 5,3,1), (0, 4,3,2), (0, 2,3,4), (0, 1,3,5) \,.
\end{split}
\ee
Finally, the remaining charges are dual to states which we have not encountered so far. This gives predictions for new BPS states, based on the assumption that the spectrum enjoys fiber-base symmetry:
\be
\begin{array}{c|c}
\gamma & \Omega(\gamma) \\
\hline
 (0, 2,3,5) & \Omega = -6 \\
 (0, 2,3,6) & \Omega = 1 \\
 (0, 3,3,4) & \Omega = -72 \\
 (0, 3,3,5) & \Omega = 20 \\
 (0, 3,4,8) & \Omega = 1 \\
 (0, 4,3,3) & \Omega = -72 \\
 (0, 4,3,4) & \Omega = 27 \\
 (0, 4,3,5) & \Omega = -14 \\
 (0, 4,3,6) & \Omega = 1 \\
 (0, 4,4,7) & \Omega = 35 \\
 (0, 5,3,2) & \Omega = -6 \\
 (0, 5,3,3) & \Omega = 20 \\
 (0, 5,3,4) & \Omega = -14 \\
 (0, 5,4,6) & \Omega = 172 \\
 (0, 5,4,7) & \Omega = -24 \\
\end{array}
\qquad
\begin{array}{c|c}
\gamma & \Omega(\gamma) \\
\hline
 (0, 5,4,8) & \Omega = 1 \\
 (0, 6,3,2) & \Omega = 1 \\
 (0, 6,3,4) & \Omega = 1 \\
 (0, 6,4,5) & \Omega = 172 \\
 (0, 6,4,6) & \Omega = -16 \\
 (0, 6,4,7) & \Omega = 6 \\
 (0, 6,5,10) & \Omega = 1 \\
 (0, 7,4,4) & \Omega = 35 \\
 (0, 7,4,5) & \Omega = -24 \\
 (0, 7,4,6) & \Omega = 6 \\
 (0, 7,5,9) & \Omega = 16 \\
 (0, 8,4,3) & \Omega = 1 \\
 (0, 8,4,5) & \Omega = 1 \\
 (0, 9,5,7) & \Omega = 16 \\
 (0, 10,5,6) & \Omega = 1 \\
\end{array}
\ee

Likewise, applying fiber-base symmetry to the BPS states in (\ref{eq:4-node-subquiver-spectrum-A}) yields the prediction
\be\label{eq:4-node-subquiver-spectrum-A-FB}
\begin{array}{c|c}
\gamma & \Omega(\gamma) \\
\hline
(1,2,2,3) & \Omega=-2
\\(2,4,4,6) & \Omega=-6 
\\(3,6,6,9) & \Omega=-42 
\\(4,8,8,12) & \Omega=-364
\\ \vdots & \vdots
\end{array}
\ee
One should note that the first state can be viewed as a boundstate of the vectormultiplet $(0,1,1,2) = D0-\overline{D2}_b$ and a $D0$ brane, so it is natural to ask whether there is a whole KK tower of states with charges $(n,n+1,n+1,n+2)$. This would of course be the fiber-base symmetric image of the tower mentioned below (\ref{eq:4-node-subquiver-spectrum-A}).

\section{BPS spectrum in various limits}\label{eq:limits-of-spectrum}

This section studies how exponential networks, and saddles representing BPS states, behave under two different limits of the mirror geometry.
We will first consider a four-dimensional limit obtained by shrinking the $S^1$ radius to zero. 
Later we will also consider a factorization limit into a half-geometry obtained by sending either of the complex moduli to zero or to infinity.
Since BPS states of the models appearing in either limit are well-understood, the study of these limits will provide further checks on brane-charge dictionary (\ref{eq:D-brane-charge-dictionary}) for saddles of $K_{\IF_0}$.

\subsection{Spectral networks in 4d $\CN=2$ $SU(2)$ super Yang-Mills theory}\label{sec:4d-theory}

\begin{figure}[h!]
\begin{center}
\includegraphics[width=0.4\textwidth]{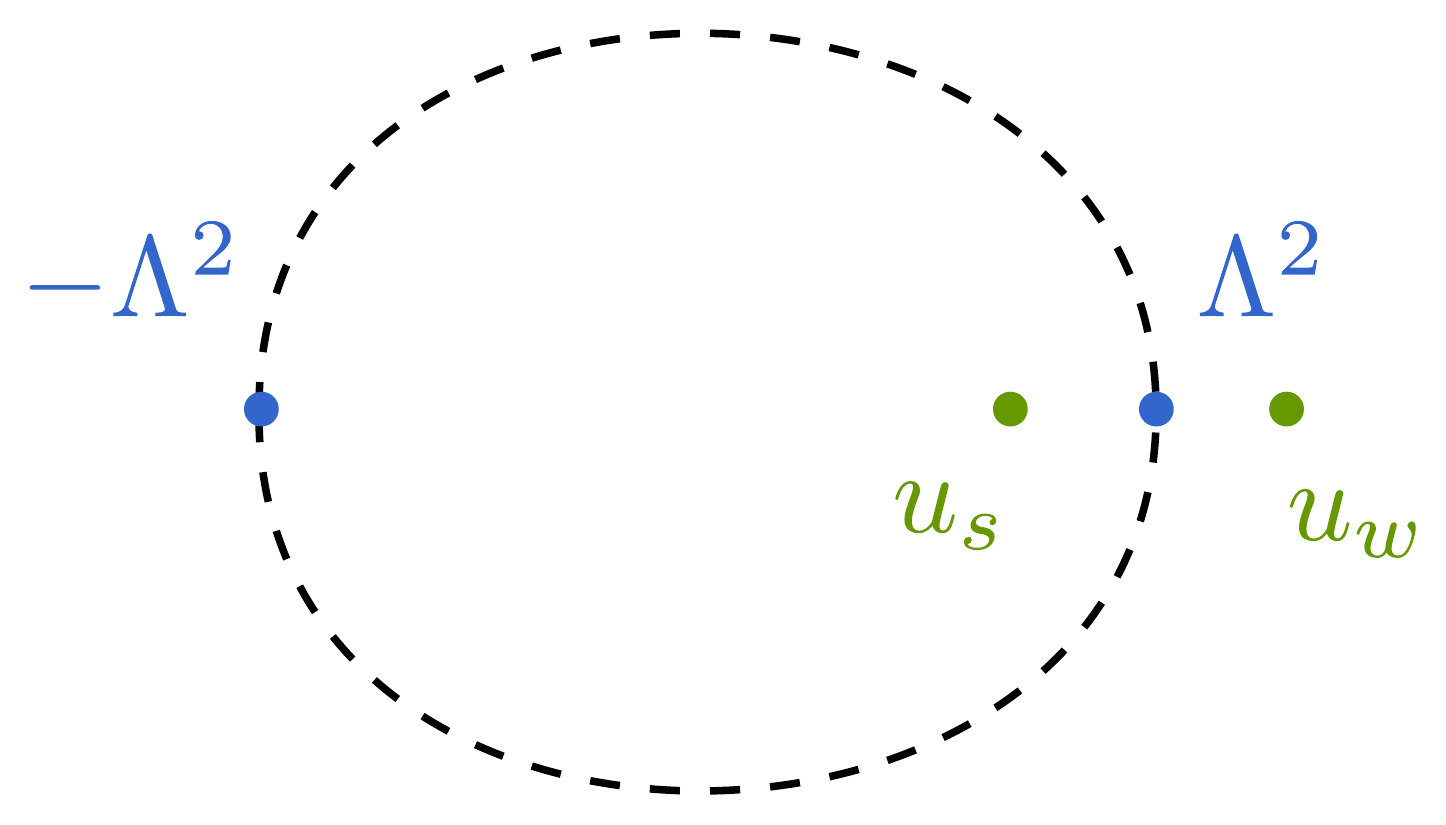}
\caption{Coulomb branch of 4d $\CN=2$ SU(2) Yang-Mills theory.}
\label{fig:SW-coulomb}
\end{center}
\end{figure}

Before we consider the four-dimensional limit of exponential networks, let us briefly recollect basic results about spectral networks for 4d $\CN=2$ super Yang-Mills with gauge group $SU(2)$.
The class $\CS$ presentation of Seiberg-Witten theory can be written in the following form \cite{Gaiotto:2009hg}
\be\label{eq:4d-curve}
	\lambda^2 = \(-\frac{\Lambda^2}{x^3} + \frac{2u}{x^2} - \frac{\Lambda^2}{x} \) \, dx^2
\ee
where $x$ is a local coordinate on the UV curve, $u$ is the Coulomb branch modulus and $\Lambda$ is the dynamical sQCD scale.
This curve has a $\IZ_2$-symmetry sending $x\to 1/x$, as well as a $\IZ_{2}$ symmetry acting as $\lambda\to-\lambda$.
The Coulomb branch also has a $\IZ_2$ symmetry, sending $u\to -u$.\footnote{Combining this with a change of coordinates $(\lambda, x)\to (i\lambda,-x)$ leaves the curve invariant.}
This curve has two branch points, where $\lambda = 0$. The branch points coincide if $u = \Lambda^2$ (the dyon point) or if $u=-\Lambda^2$ (the monopole point).\footnote{Assuming these conventions are compatible with those of \cite{Seiberg:1994rs}, where this identification is made on p.28.}
The chamber structure of the Coulomb branch consists of two chambers, separated by walls of marginal stability connecting the monopole and dyon points, see Figure \ref{fig:SW-coulomb}.

\begin{figure}[h!]
\begin{center}
\includegraphics[width=0.35\textwidth]{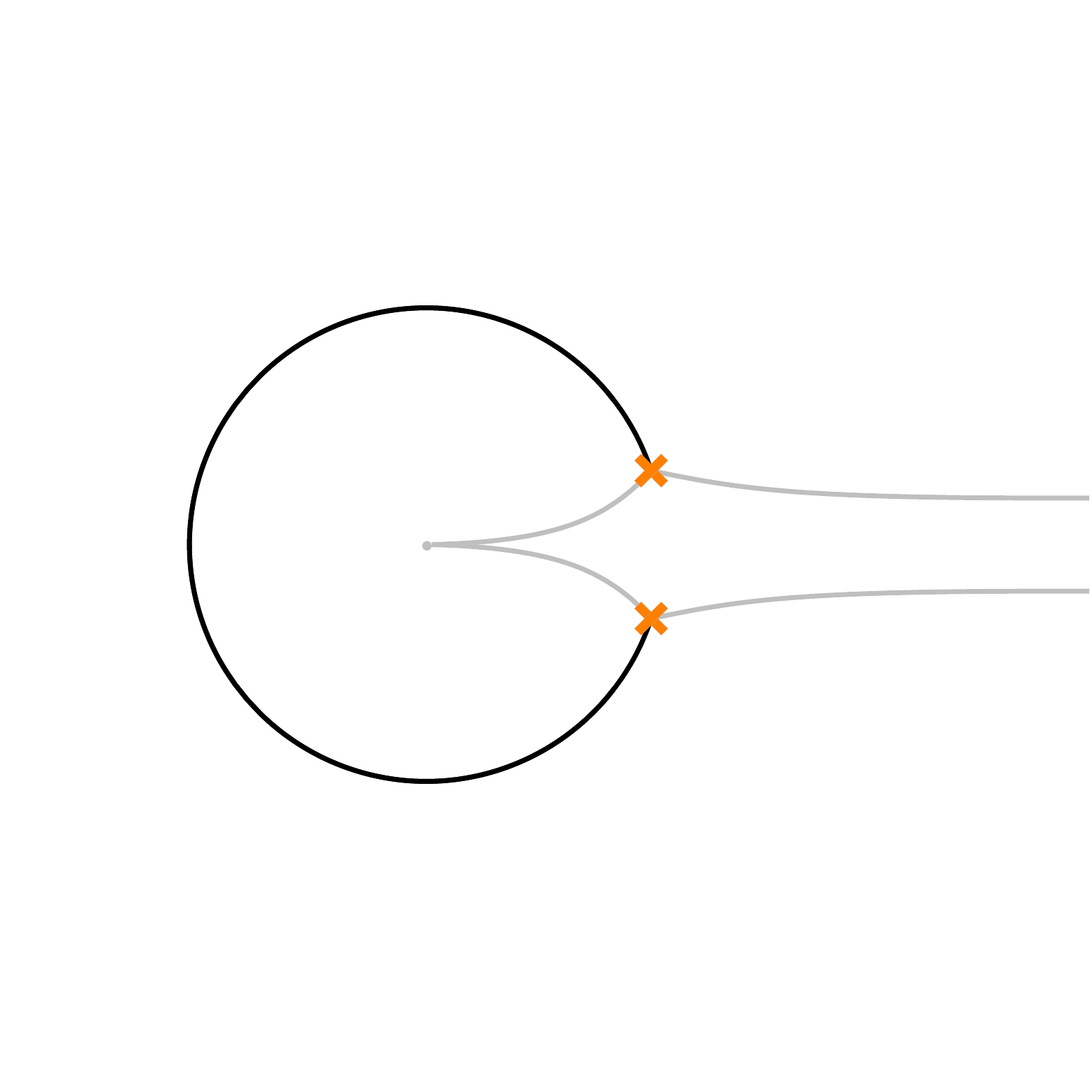}
\hspace*{30pt}
\includegraphics[width=0.35\textwidth]{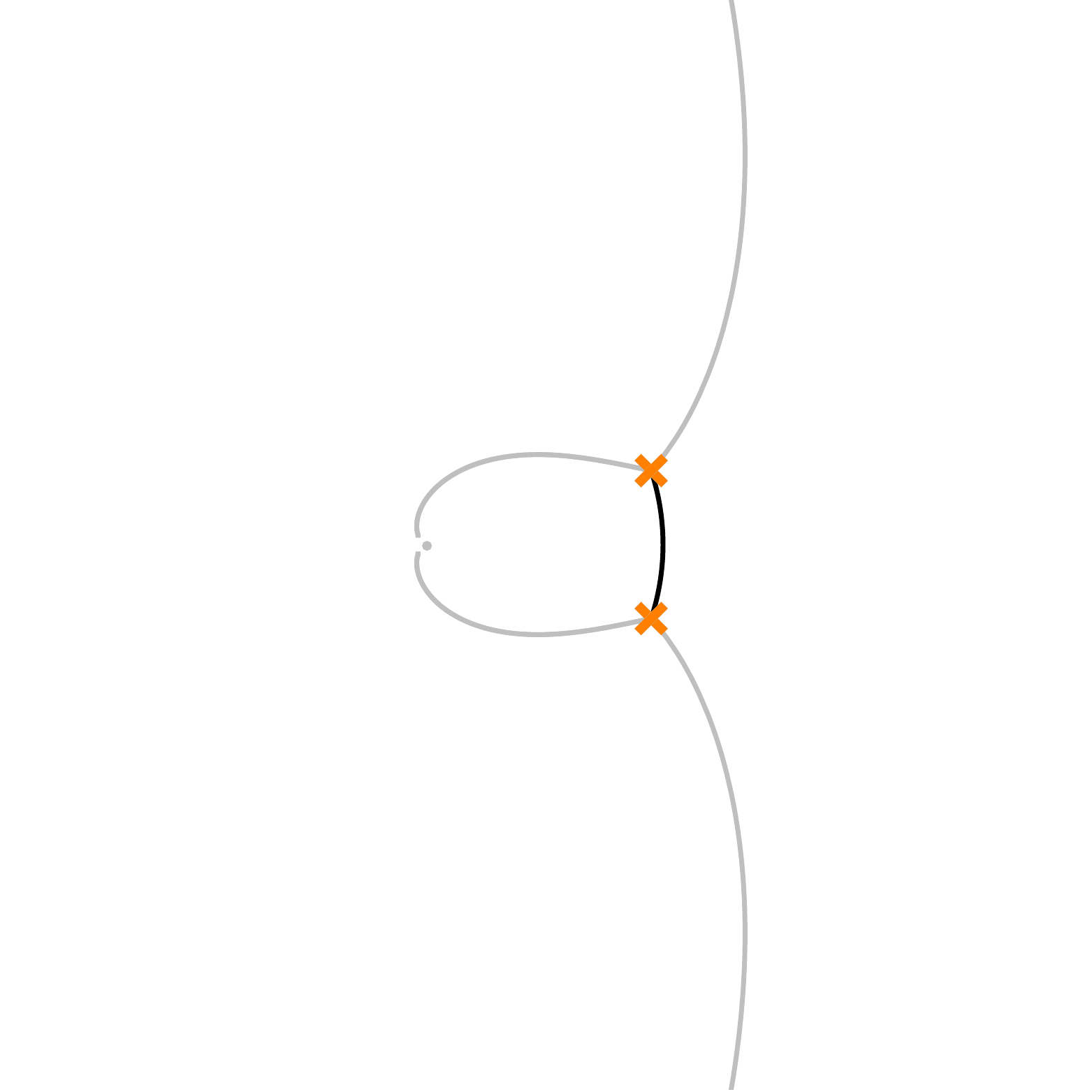}
\caption{Left: monopole appearing for $\vartheta=0$. Right: dyon appearing for $\vartheta=\pi/2$.}
\label{fig:4d-sc}
\end{center}
\end{figure}

\begin{figure}[h!]
\begin{center}
\begin{subfigure}{.175\textwidth}
\includegraphics[width=\textwidth]{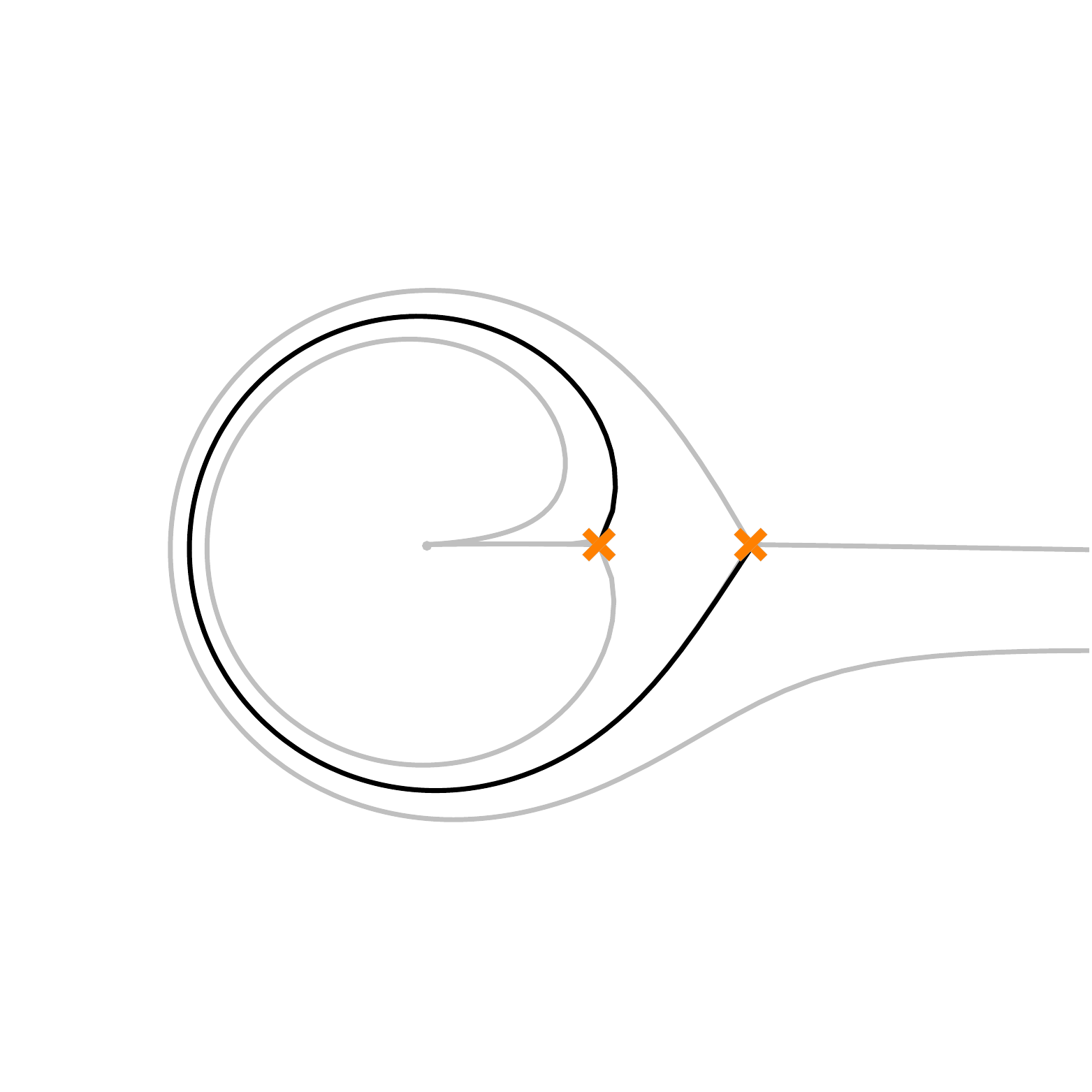}
\caption{Monopole}
\end{subfigure}
\begin{subfigure}{.175\textwidth}
\includegraphics[width=\textwidth]{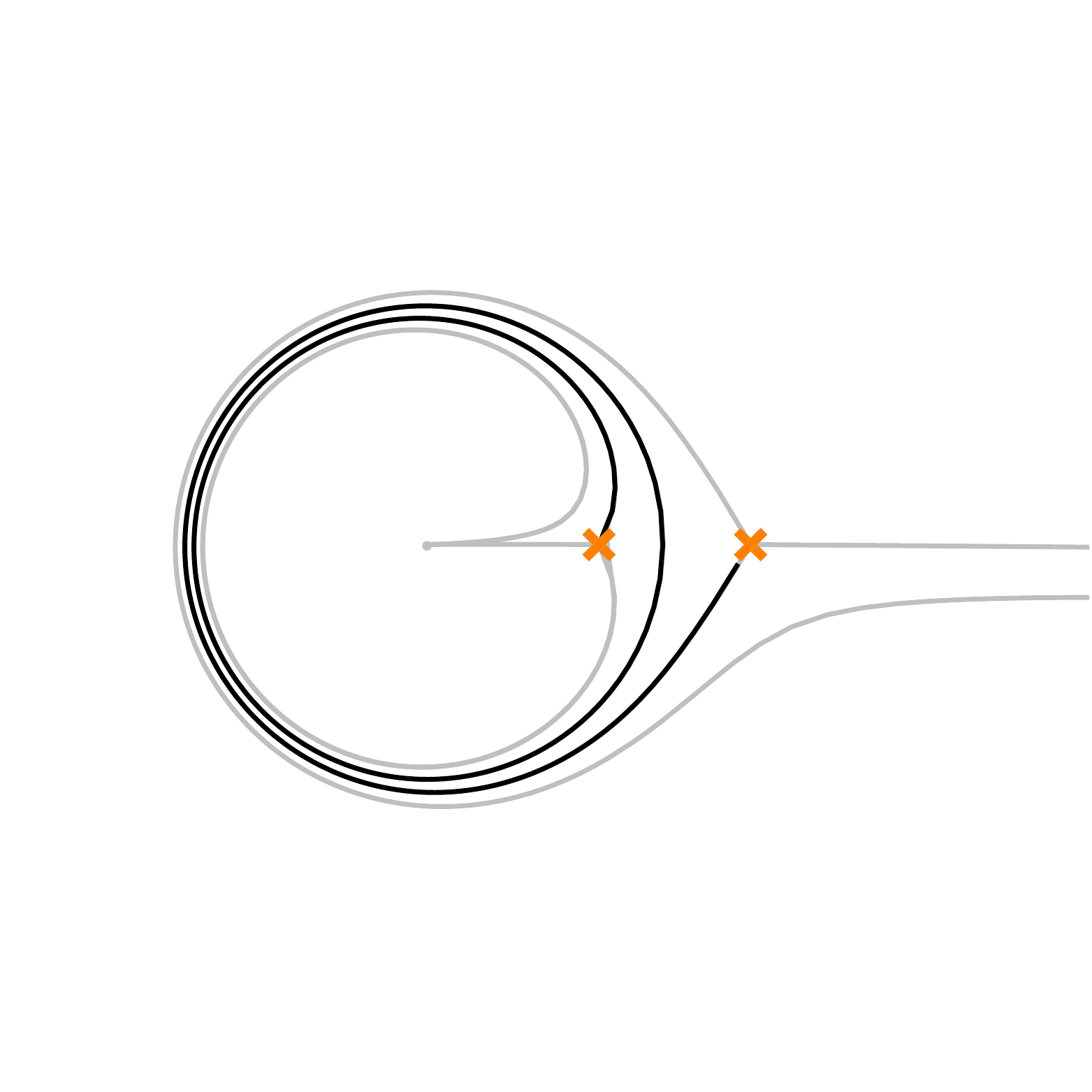}
\caption{Mono.+v.m.}
\end{subfigure}
\begin{subfigure}{.175\textwidth}
\includegraphics[width=\textwidth]{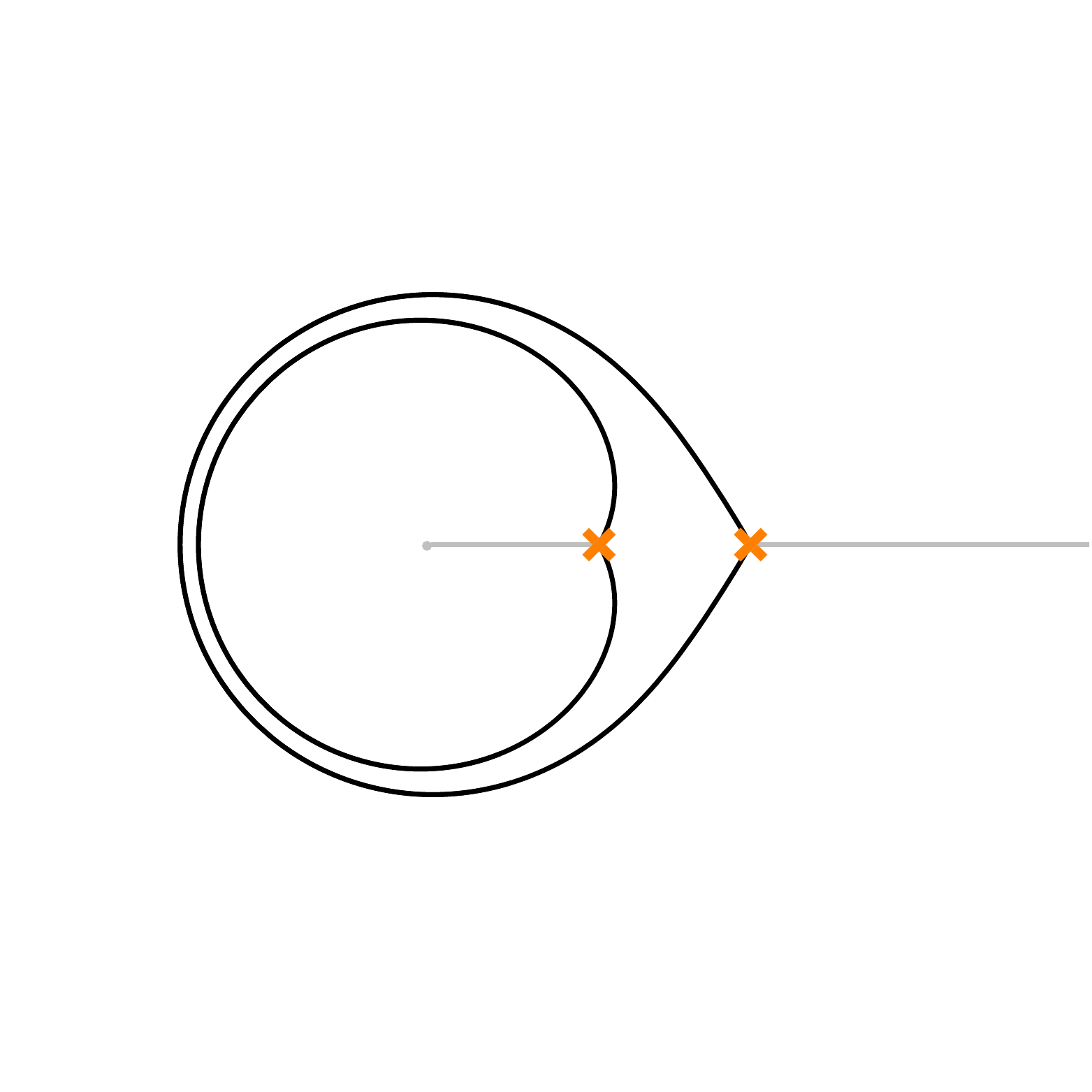}
\caption{V.m.}
\end{subfigure}
\begin{subfigure}{.175\textwidth}
\includegraphics[width=\textwidth]{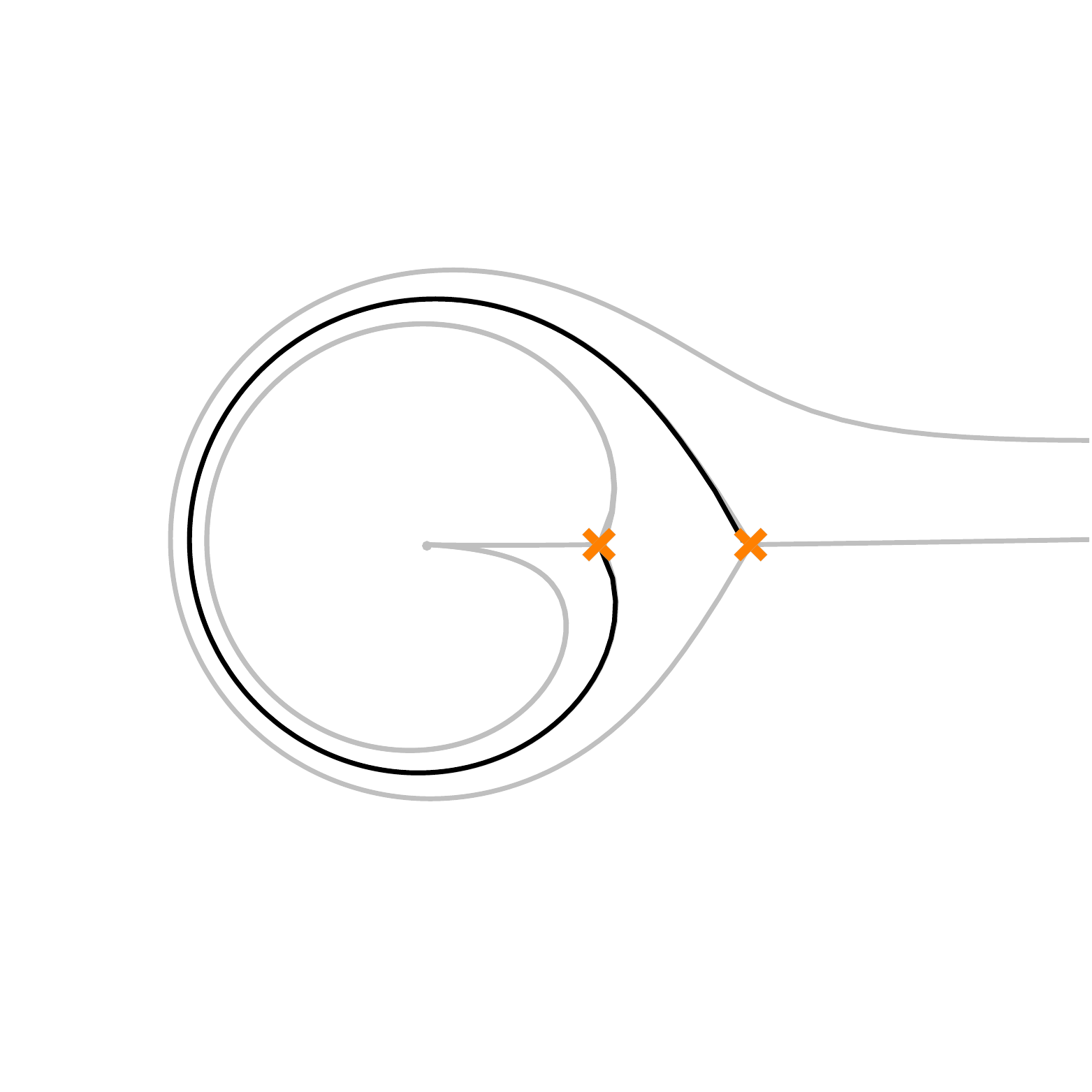}
\caption{Dyon+v.m.}
\end{subfigure}
\begin{subfigure}{.175\textwidth}
\includegraphics[width=\textwidth]{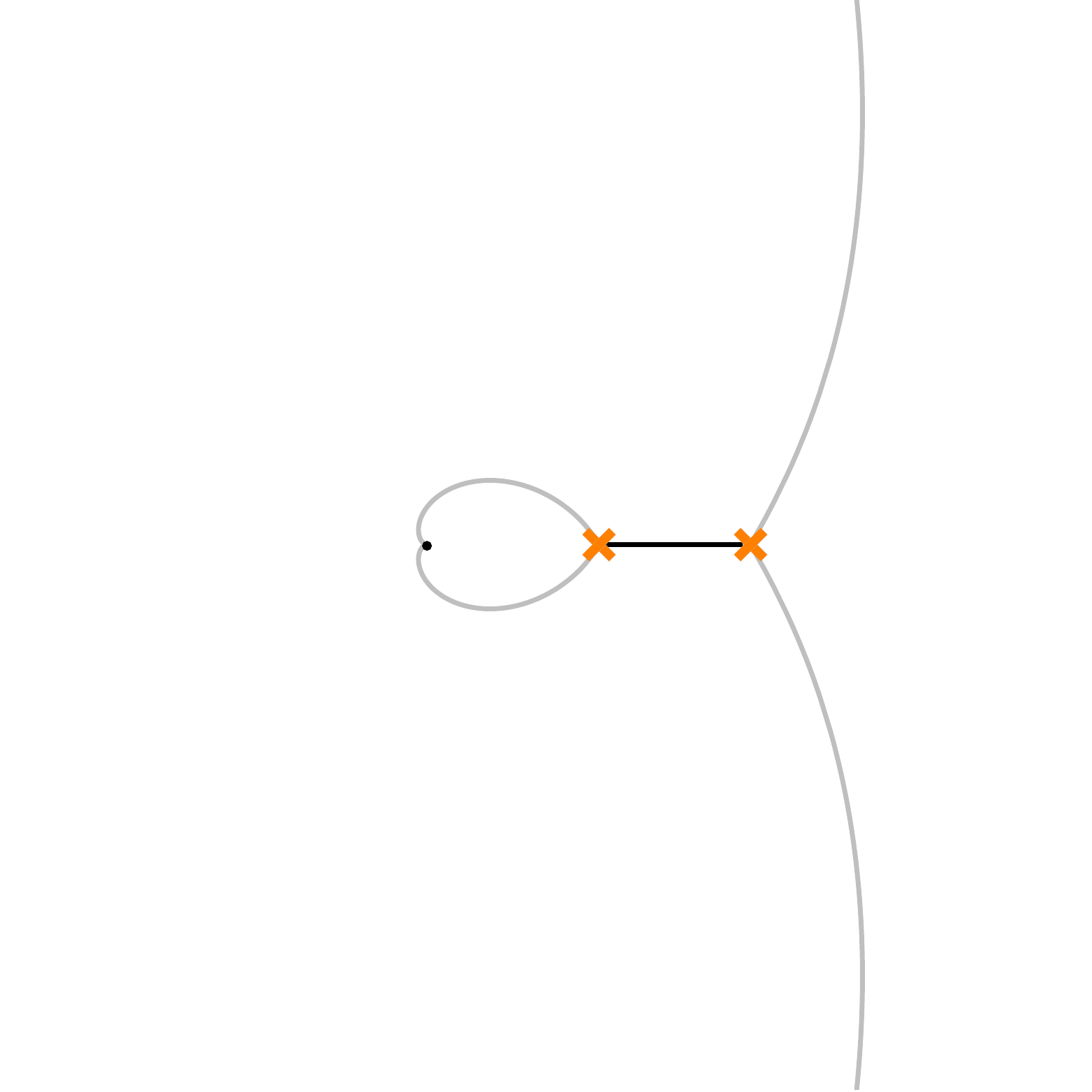}
\caption{Dyon}
\end{subfigure}
\caption{Weak coupling BPS states.}
\label{fig:4d-wc}
\end{center}
\end{figure}

The BPS spectrum in the strong coupling chamber consists of two BPS states: a monopole and a dyon, both are BPS hypermultiplets with $\Omega=1$. To illustrate this we plot the spectral network at a point $u_s$ in the Coulomb branch, and find two saddles.
Choosing $u_s$ on the positive real axis and close to the dyon point\footnote{We fix $u_s = \frac{95}{100} \Lambda^2$} we find the dyon and monopole as shown in Figure \ref{fig:4d-sc}.
Moving to the weak coupling chamber, the BPS spectrum jumps and includes infinitely many BPS states. In addition to the monopole and the dyon there are several boundstates, including a vectormultiplet with $\Omega = -2$. 
To illustrate this we plot the spectral network at a point $u_w$ on the positive real axis and close to the dyon point\footnote{We fix $u_w = \frac{105}{100} \Lambda^2$} we find the BPS spectrum partially shown in Figure \ref{fig:4d-wc}.

In the limit $u\gg \Lambda^2$, e.g. along the positive real axis,  the separation between  two branch points increases arbitrarily. 
The system eventually decouples into two separate parts, where only one BPS state remains of finite mass: it is a \emph{half} of the vectormultiplet, consisting of a circular saddle supported on a single branch point. In fact, what remains is the spectral network of the $\mathbb{CP}^1$ 2d sigma model, reflecting the fact that after decoupling the 4d theory only the 2d $(2,2)$ defect theory remains dynamical \cite{Gaiotto:2011tf, Gaiotto:2013sma}.

\subsection{Four-dimensional limits to Seiberg-Witten theory} \label{sec:4d-limit-spectrum}

In section \ref{sec:4d-limit-geometry} we reviewed how dependence on the radius of compactification may be introduced in the geometry of the mirror curve, showing that the latter approaches the Seiberg-Witten curve of 4d $\CN=2$ super Yang-Mills as $R\to 0$. 
In this section we explore how exponential networks behave in this liimit.

\subsubsection*{Limit to strong coupling}

The Coulomb branch of the 4d theory has a strong coupling region, corresponding to  $u/\Lambda^2 \lesssim 1$. Going all the way to the center of the strong coupling region corresponds to sending this ratio to zero, which in terms of 5d moduli implies
\be
	\frac{Q_f}{Q_b} = -\frac{1}{2} \(\frac{1}{\Lambda^2 R^2} + 2\frac{u}{\Lambda^2} + \dots\)\,.
\ee
We will henceforth specialize $Q_f \sim -{Q_b}/ ({2\Lambda^2 R^2})$, and study the limit $R\to 0$ keeping $\Lambda$ fixed.
Let us focus on a small patch of the moduli space near one of the singularities, namely near the divisor $\CD_2$  in (\ref{eq:discriminant-locus}), where $Q_f+Q_b =1/2$. 
As is clear from (\ref{eq:5d4dsubs}), this corresponds in the 4d limit to the dyon point $u=\Lambda^2$.\footnote{Due to symmetries of the curve, corresponding to $(x,Q_b)\to (-x,-Q_b)$ and $(y,Q_f)\to (-y,-Q_f)$, the analysis of divisor $\CD_3$ will be essentially identical. Likewise divisors $\CD_1, \CD_4$ should be equivalent.}

\begin{figure}[h!]
\begin{center}
\includegraphics[width=0.45\textwidth]{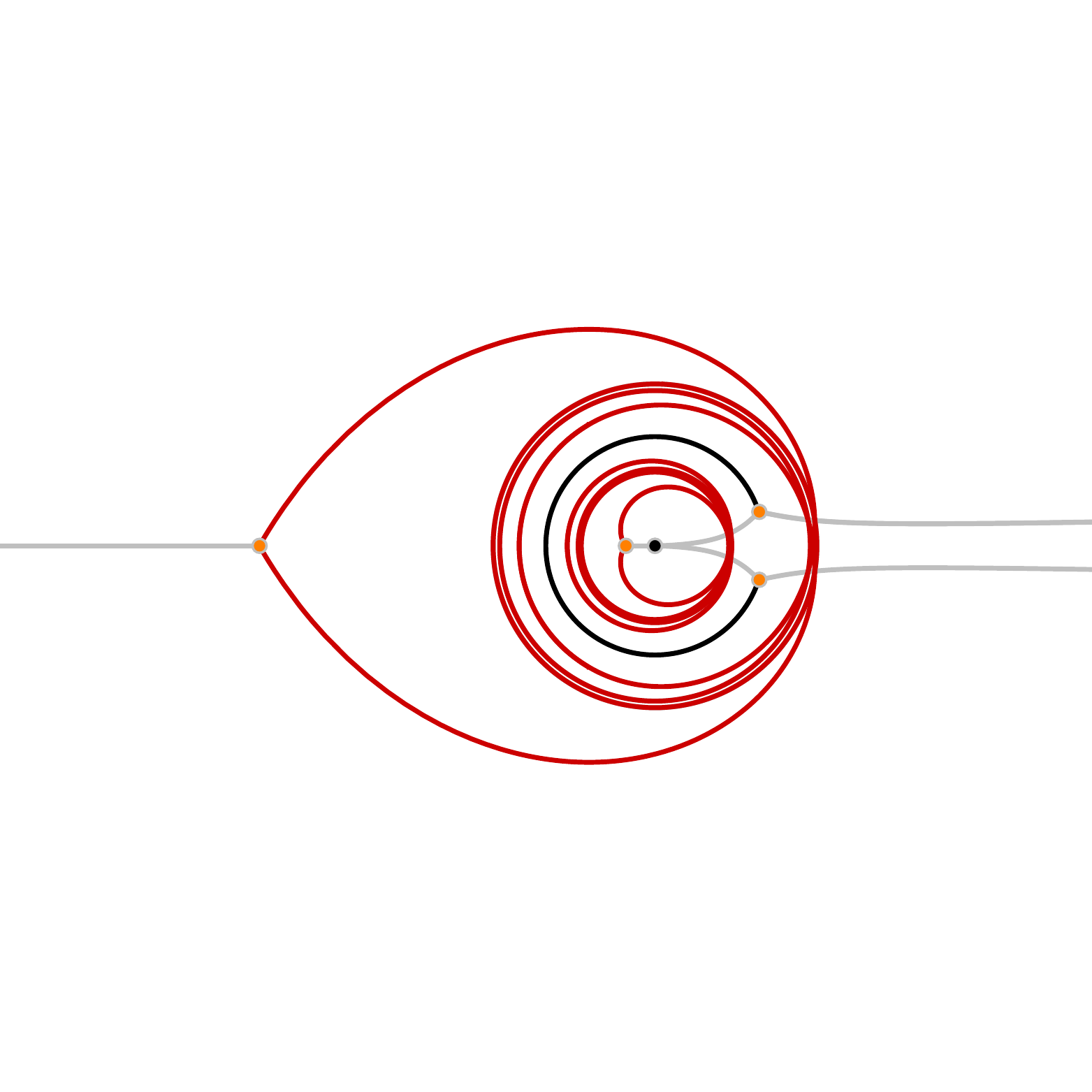}
\includegraphics[width=0.45\textwidth]{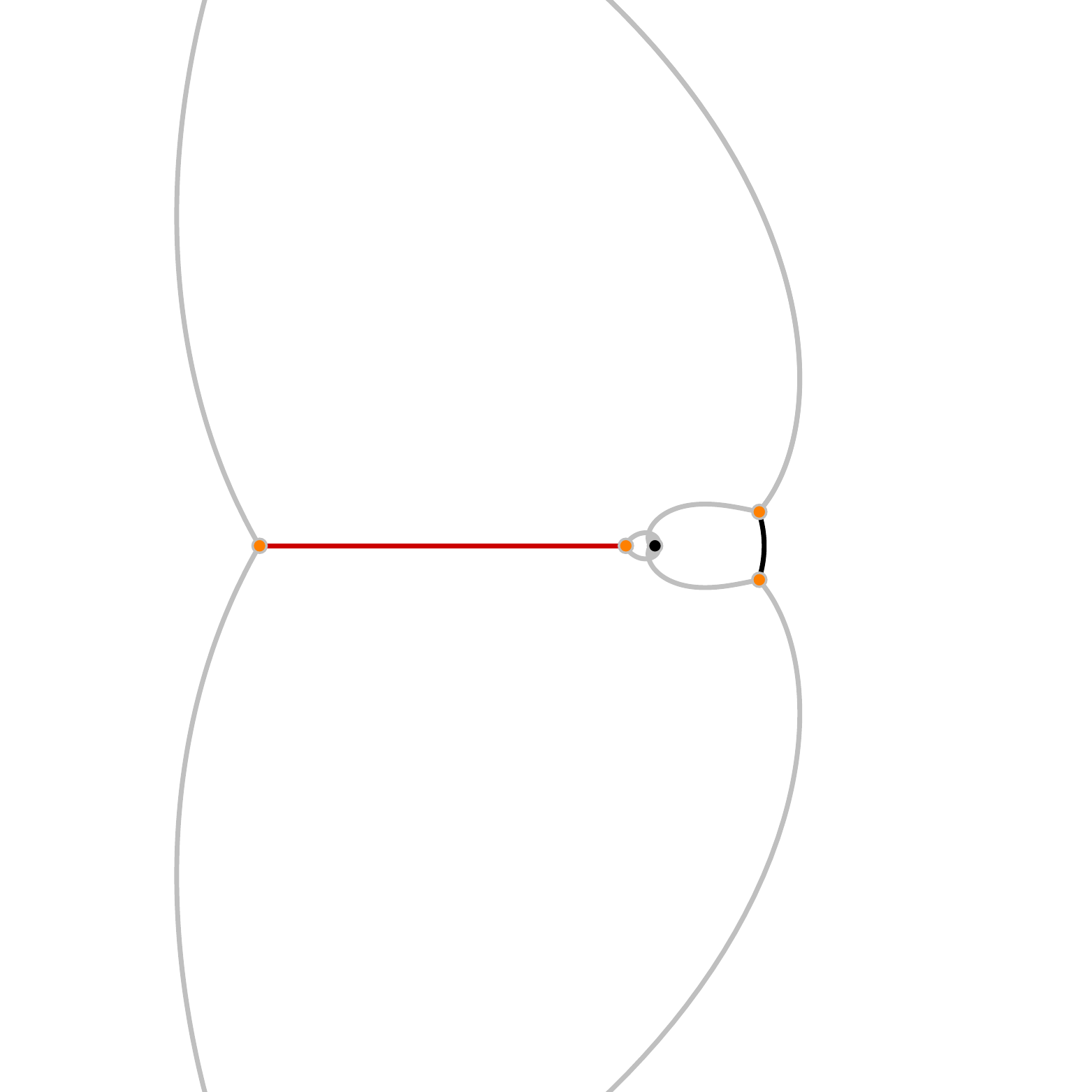}
\caption{Left: $\vartheta=0$. Right: $\vartheta=\pi/2$}
\label{fig:radius-1-basic-states-sc}
\end{center}
\end{figure}

We shall fix 4d moduli to be
\be\label{eq:5d-moduli-strong-cplg}
	\Lambda = 1,\qquad
	u = 0.95 \Lambda^2, 
\ee
which, at radius $R=1$ correspond approximately to $Q_b = -1$, $Q_f = 1/2+u = 1.45$.
Focusing on the simplest BPS saddles encountered in Section \ref{sec:spectrum-conifold-point}, we consider exponential networks at $\vartheta=0,\pi/2$ shown in Figure \ref{fig:radius-1-basic-states-sc}. 

These should be compared with the saddles of Figure \ref{fig:exp-net-sym-point}. 
Indeed, highlighted in black we recognize $p_3$ at $\vartheta=0$ and $p_5$ at $\vartheta=\pi/2$.
As we argued above, the two branch points supporting $p_3, p_5$ survive in the 4d limit, while the two other branch points will end up disappearing into punctures.
In terms of BPS states, whose charges can be read from   (\ref{eq:D-brane-charge-dictionary}), this means the following
\begin{itemize}
\item $D4$, which corresponds to $p_3$,  remains of finite mass. This should descend to a purely magnetic state, the BPS monopole of Seiberg-Witten theory.
\item likewise $D2_f\-\overline{D4}$, which corresponds to $p_5$, remains of finite mass. This should descend to a dyon with $(e,m) = (1,-1)$, the BPS dyon of Seiberg-Witten theory
\item instead $D0$ branes, which correspond to red towers of saddles at $\vartheta=0$ will become infinitely massive and eventually disappear into the punctures 
\item likewise $D2_b$, corresponding to the red saddle at $\vartheta=\pi/2$ (or $p_4$ in Figure \ref{fig:exp-net-sym-point}) becomes infinitely massive due to the fact that the two branch points that support it become infinitely separated and disappear
\end{itemize}

\begin{figure}[h!]
\begin{center}
\fbox{\includegraphics[width=0.175\textwidth]{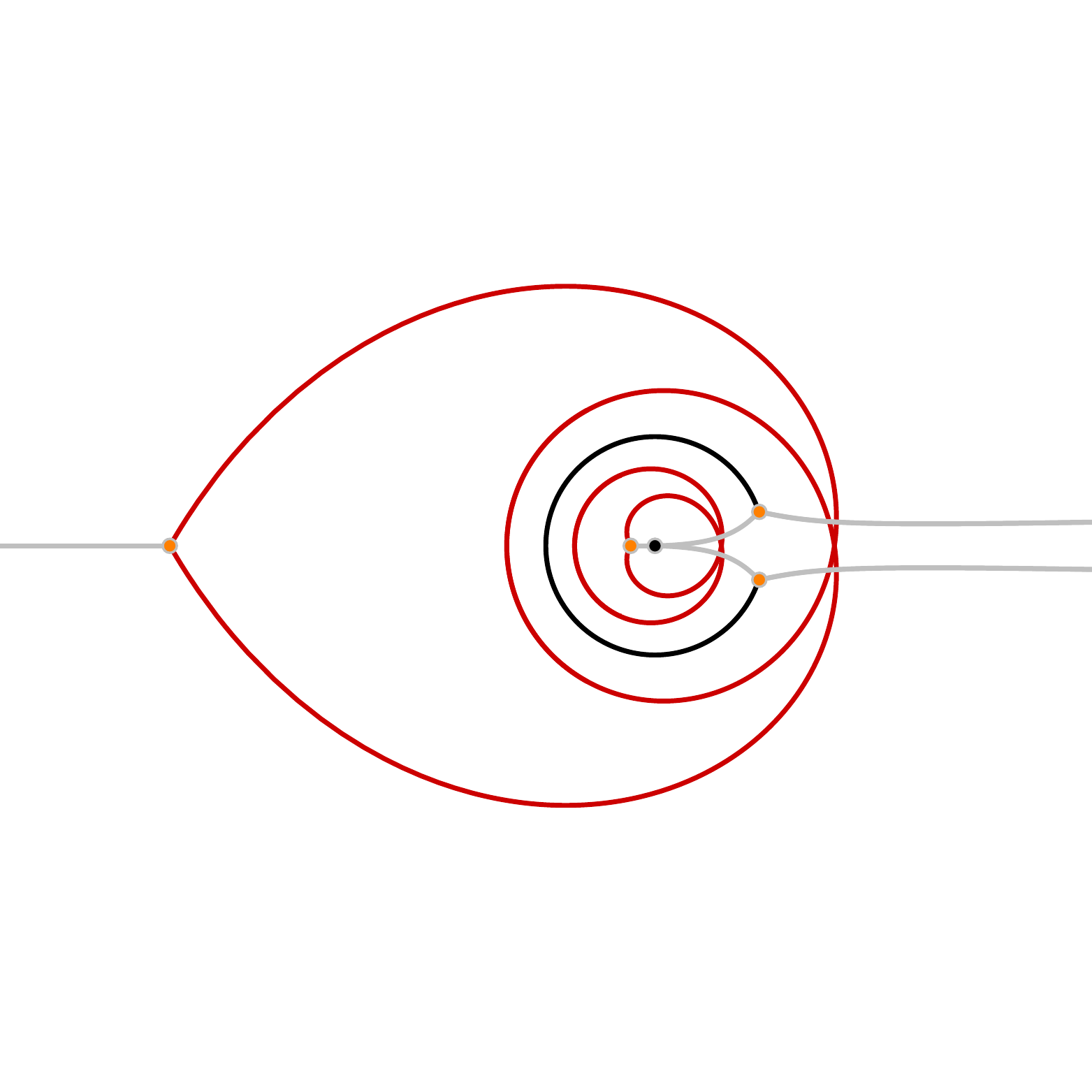}}
\fbox{\includegraphics[width=0.175\textwidth]{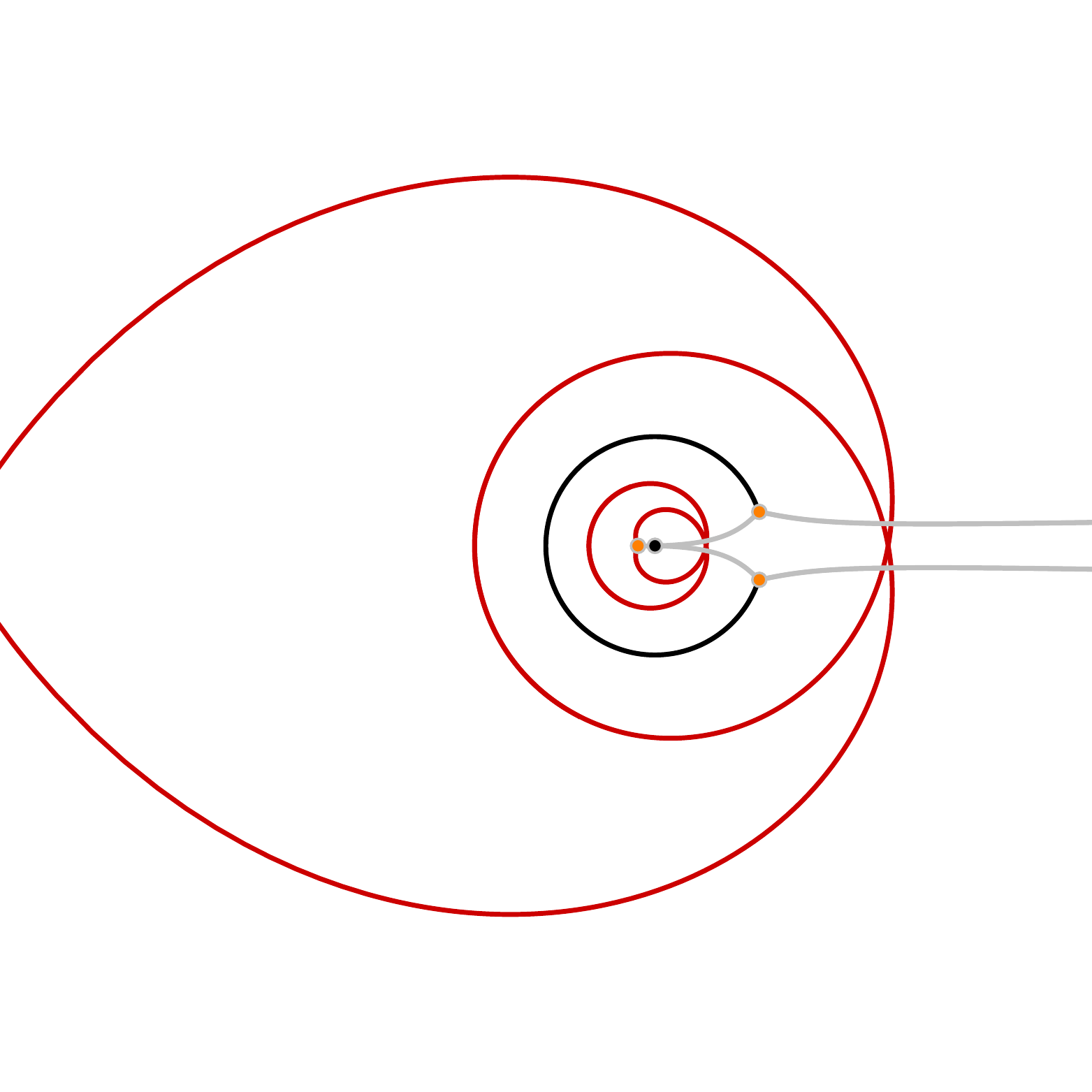}}
\fbox{\includegraphics[width=0.175\textwidth]{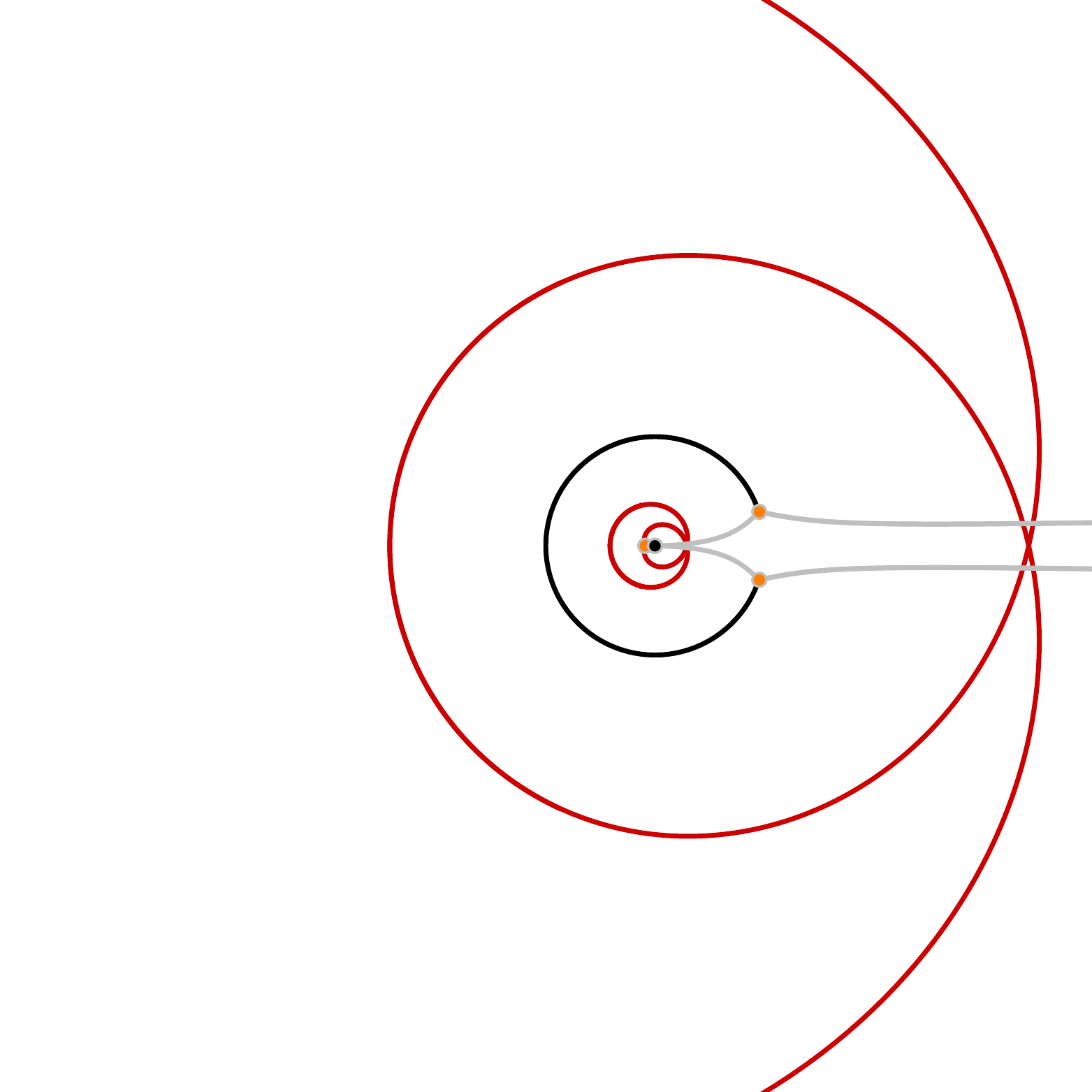}}
\fbox{\includegraphics[width=0.175\textwidth]{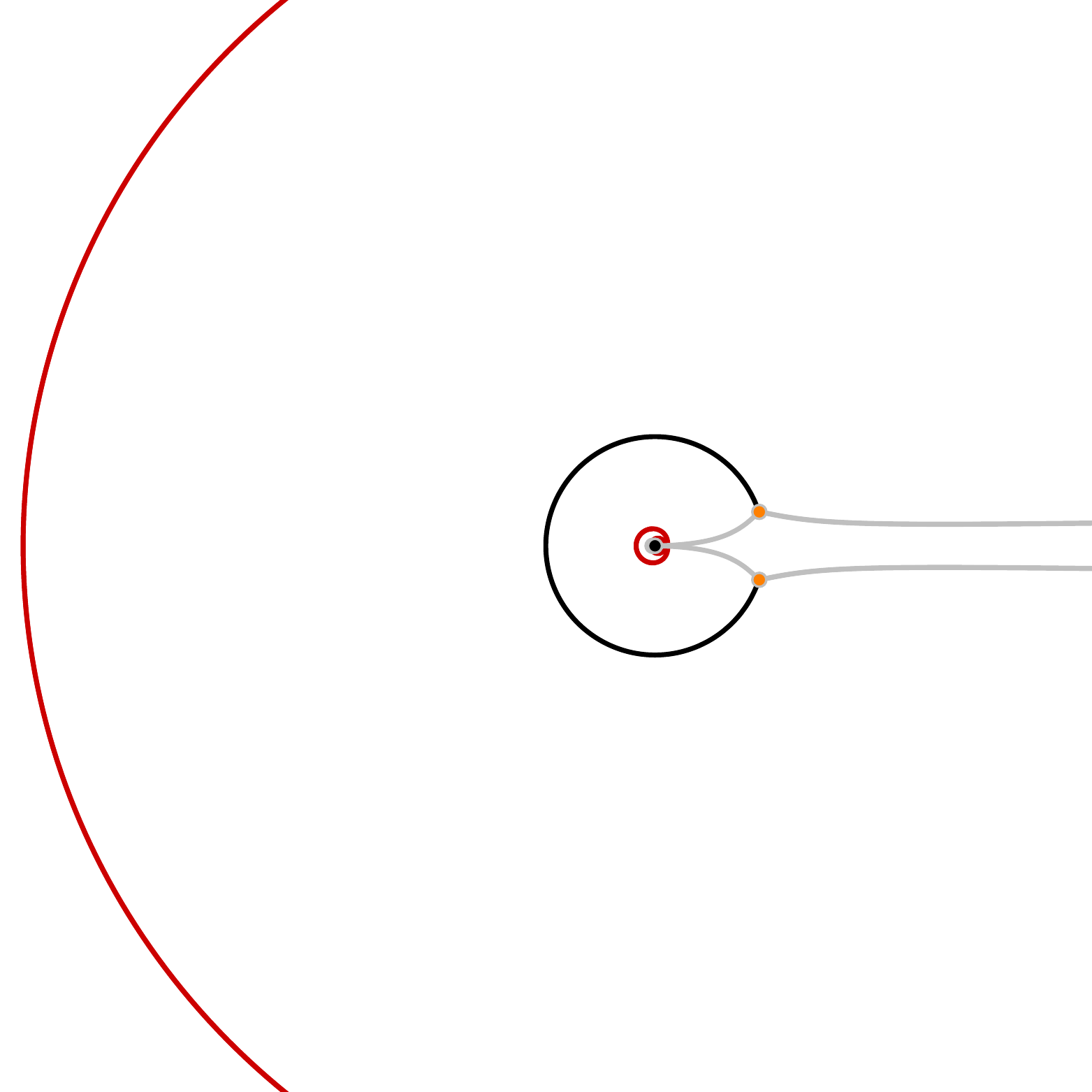}}
\fbox{\includegraphics[width=0.175\textwidth]{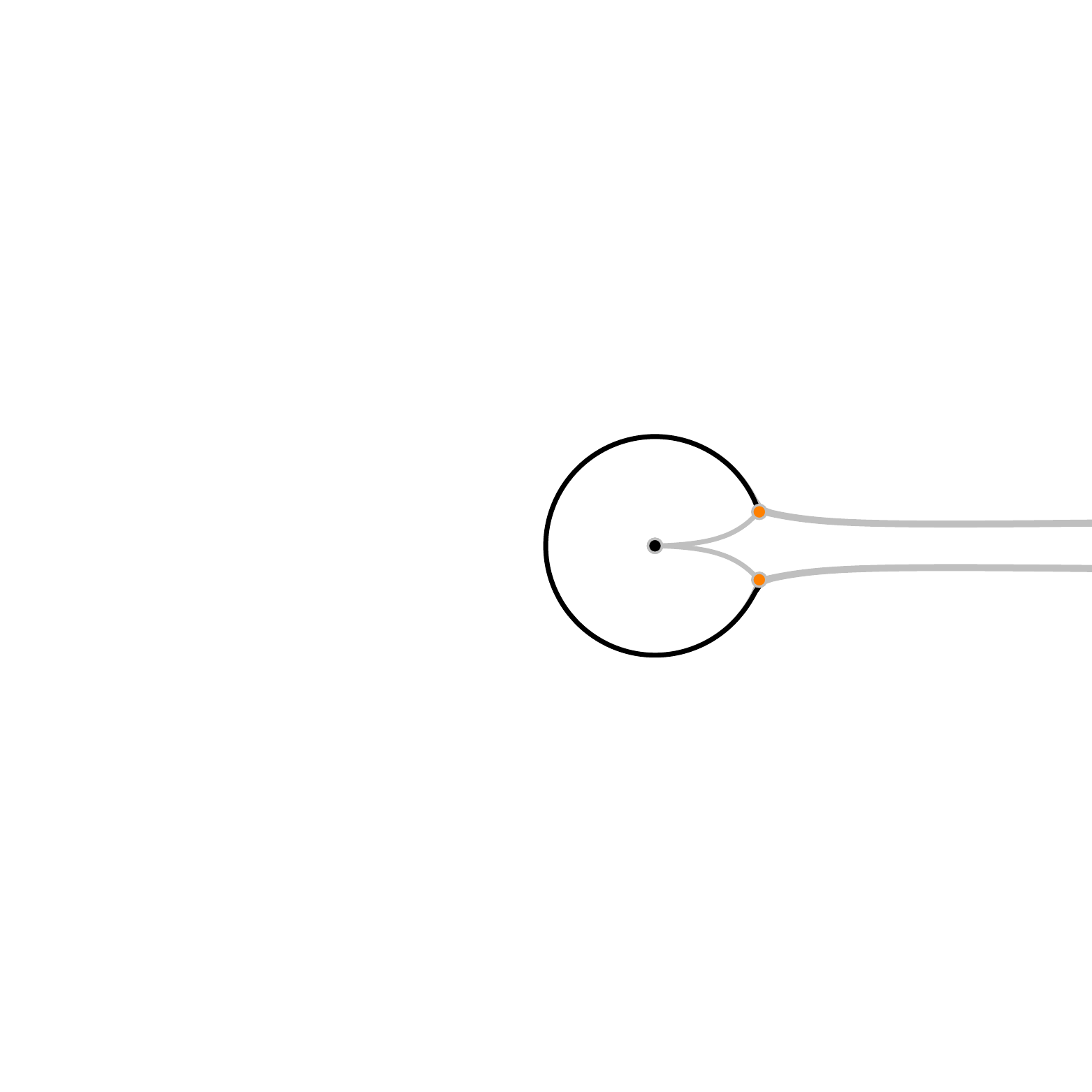}}
\caption{Exponential networks at $\vartheta=0$, for $R=$ \{0.85, 0.65, 0.45, 0.25, 0.05\}. Only primary walls are shown. This interpolates between Figures \ref{fig:radius-1-basic-states-sc} and \ref{fig:4d-sc}.}
\label{fig:strong-coupling-phase-zero}
\end{center}
\end{figure}

\begin{figure}[h!]
\begin{center}
\fbox{\includegraphics[width=0.175\textwidth]{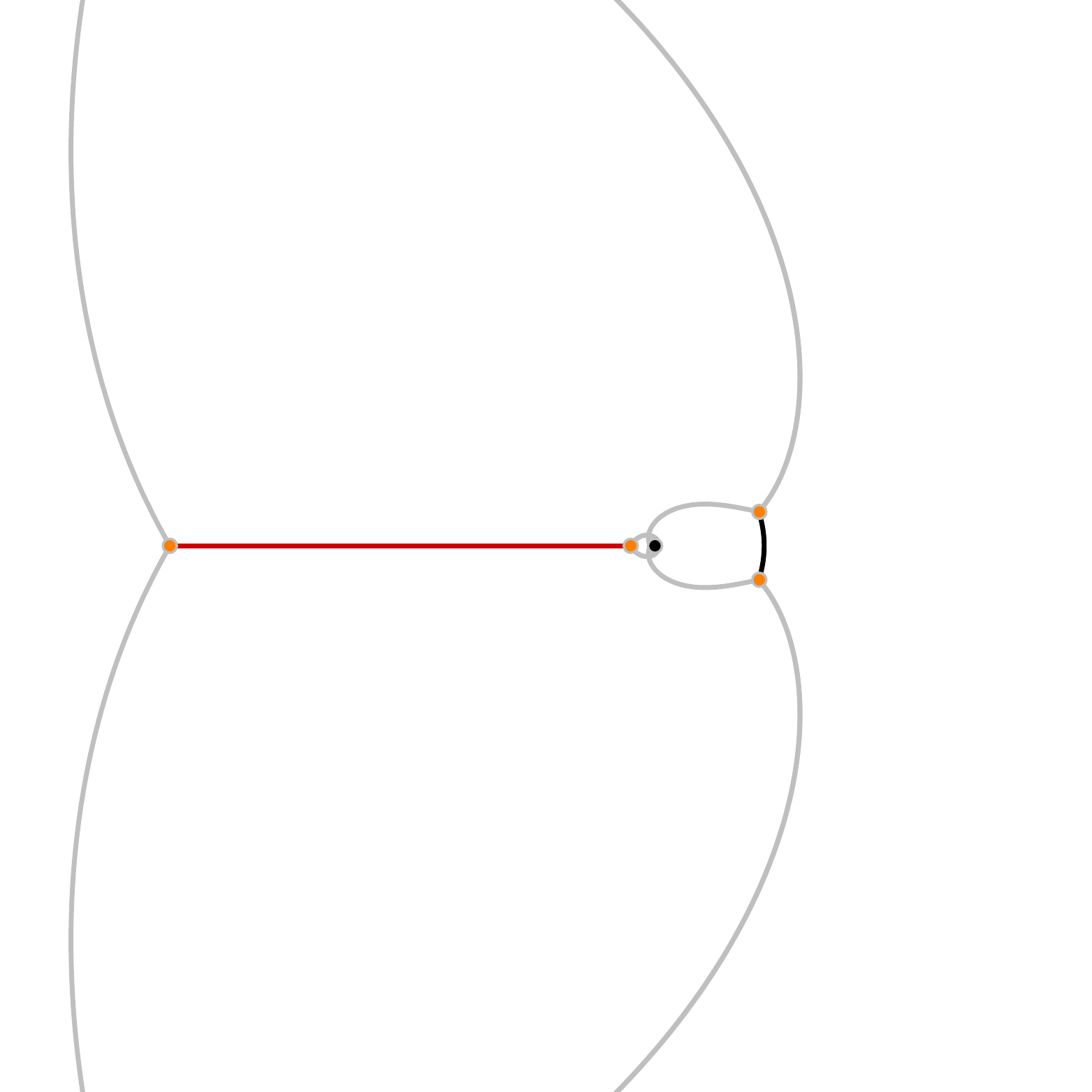}}
\fbox{\includegraphics[width=0.175\textwidth]{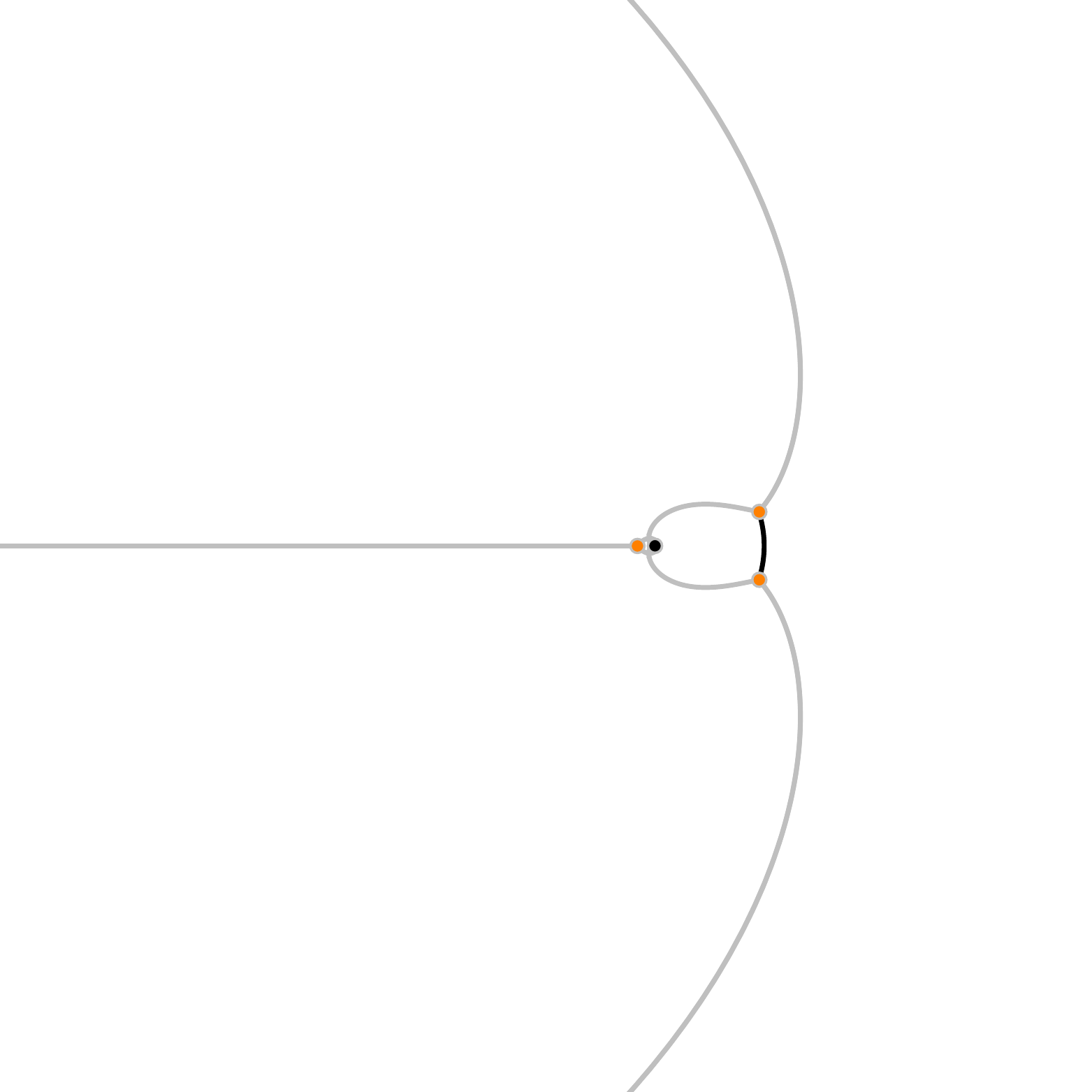}}
\fbox{\includegraphics[width=0.175\textwidth]{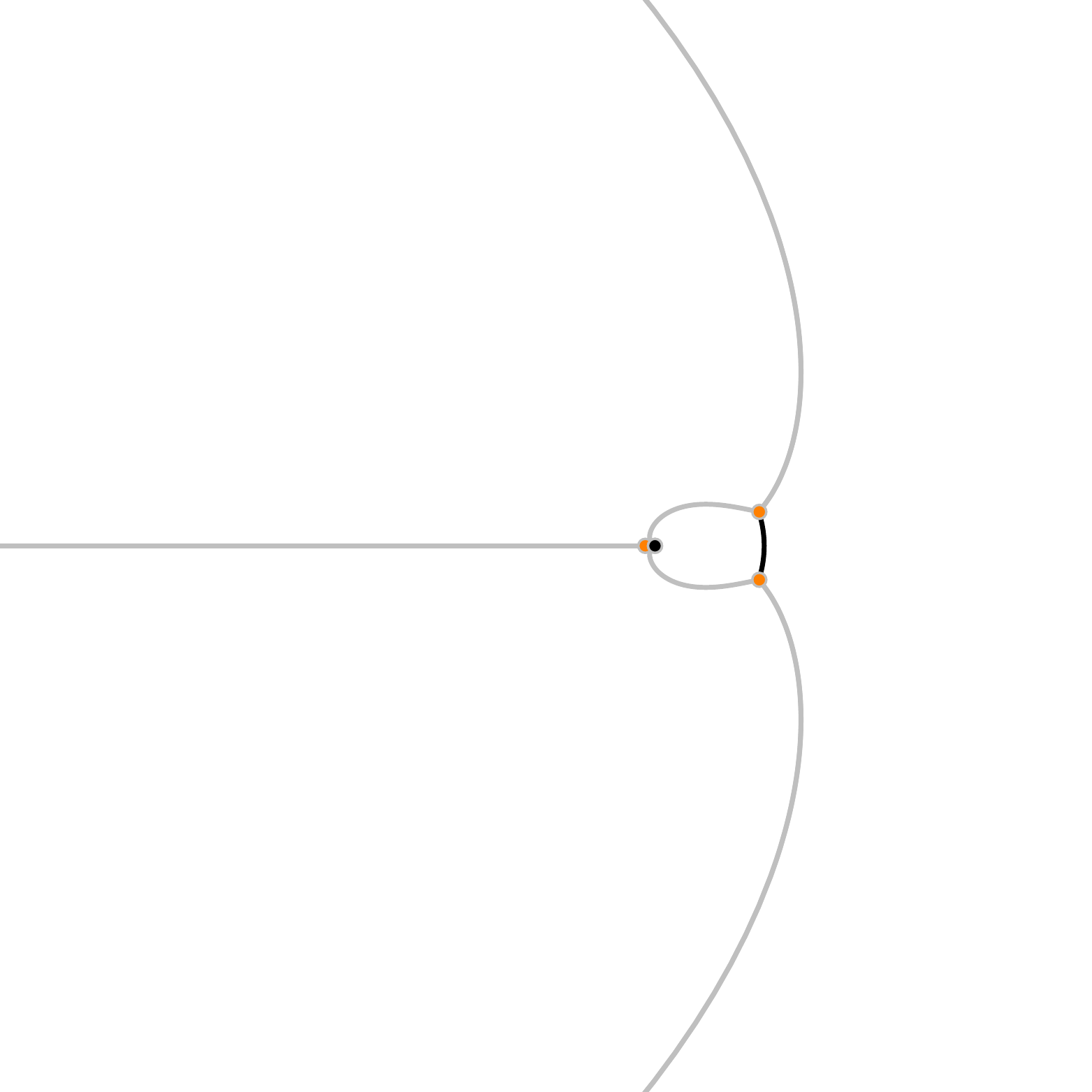}}
\fbox{\includegraphics[width=0.175\textwidth]{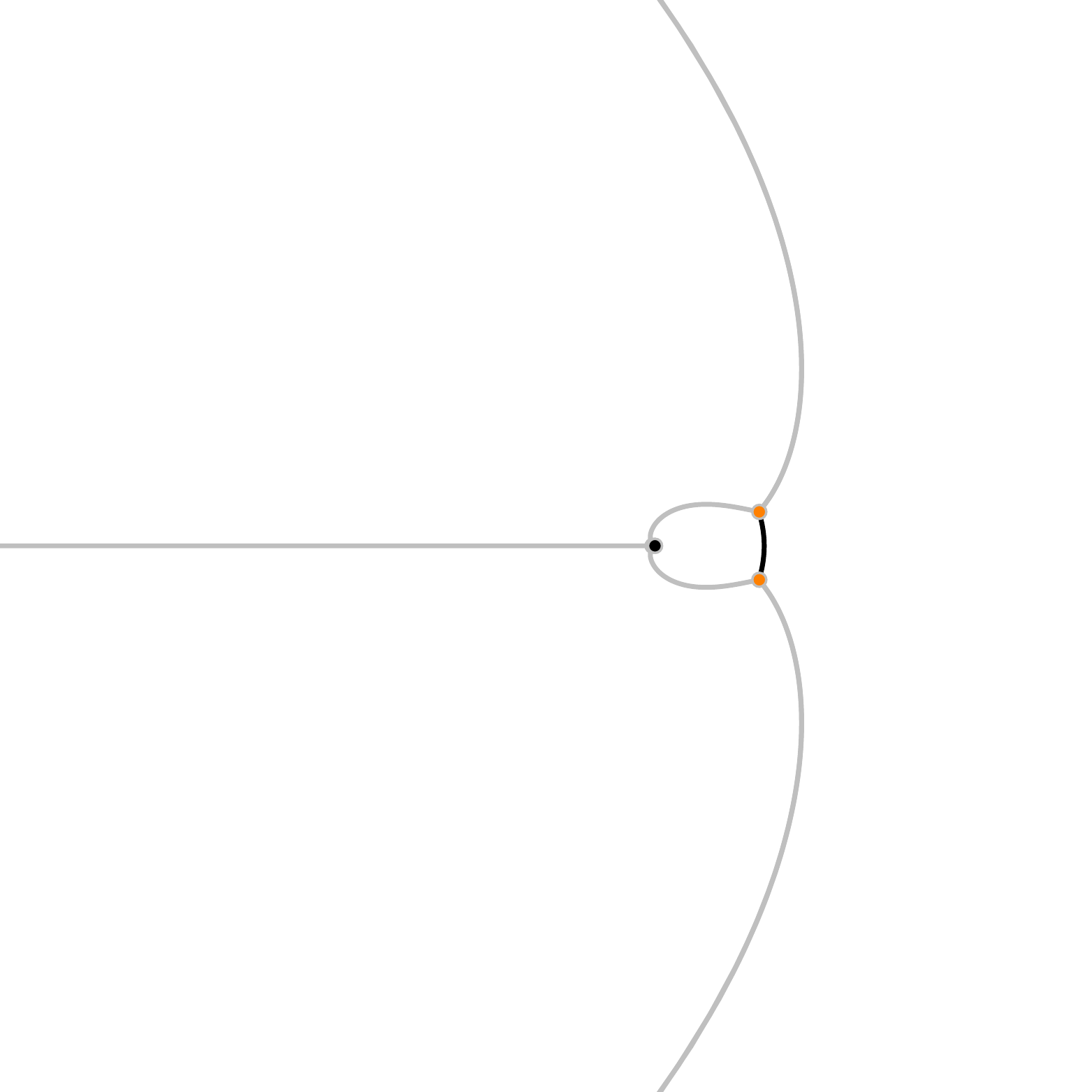}}
\fbox{\includegraphics[width=0.175\textwidth]{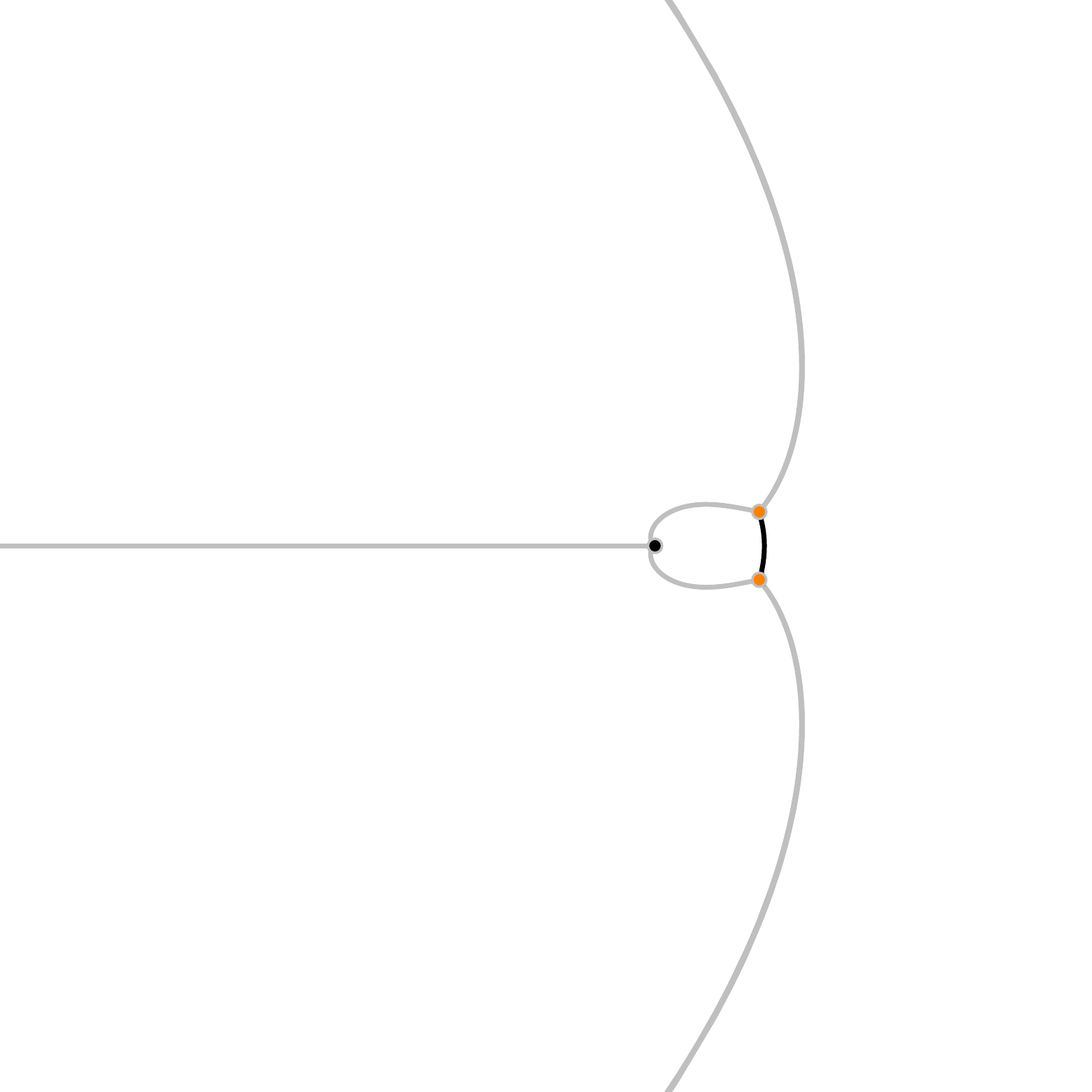}}
\caption{Exponential networks at $\vartheta=\pi/2$, for  $R=$ \{0.85, 0.65, 0.45, 0.25, 0.05\}. Only primary walls are shown. This interpolates between Figures \ref{fig:radius-1-basic-states-sc} and \ref{fig:4d-sc}.}
\label{fig:strong-coupling-phase-piover2}
\end{center}
\end{figure}

These expectations can be verified directly, as shown in Figures \ref{fig:strong-coupling-phase-zero} and \ref{fig:strong-coupling-phase-piover2}. 
As the radius decreases, the exponential network approaches precisely the spectral network of SU(2) Seiberg-Witten theory, corresponding to the last frame, which coincide with saddles shown in Figure \ref{fig:4d-sc}.
Comparing with spectral networks offers another check of our identification of the $p_3$ saddle with the monopole (hence a D4 brane) and of the $p_5$ saddle with a dyon (hence a $D2_f\-\overline{D4}$ boundstate). These are the only states present in the strong coupling chamber, providing an unambiguous check of (\ref{eq:D-brane-charge-dictionary}).


\subsubsection*{Limit to weak coupling}

\begin{figure}[h!]
\begin{center}
\includegraphics[width=0.45\textwidth]{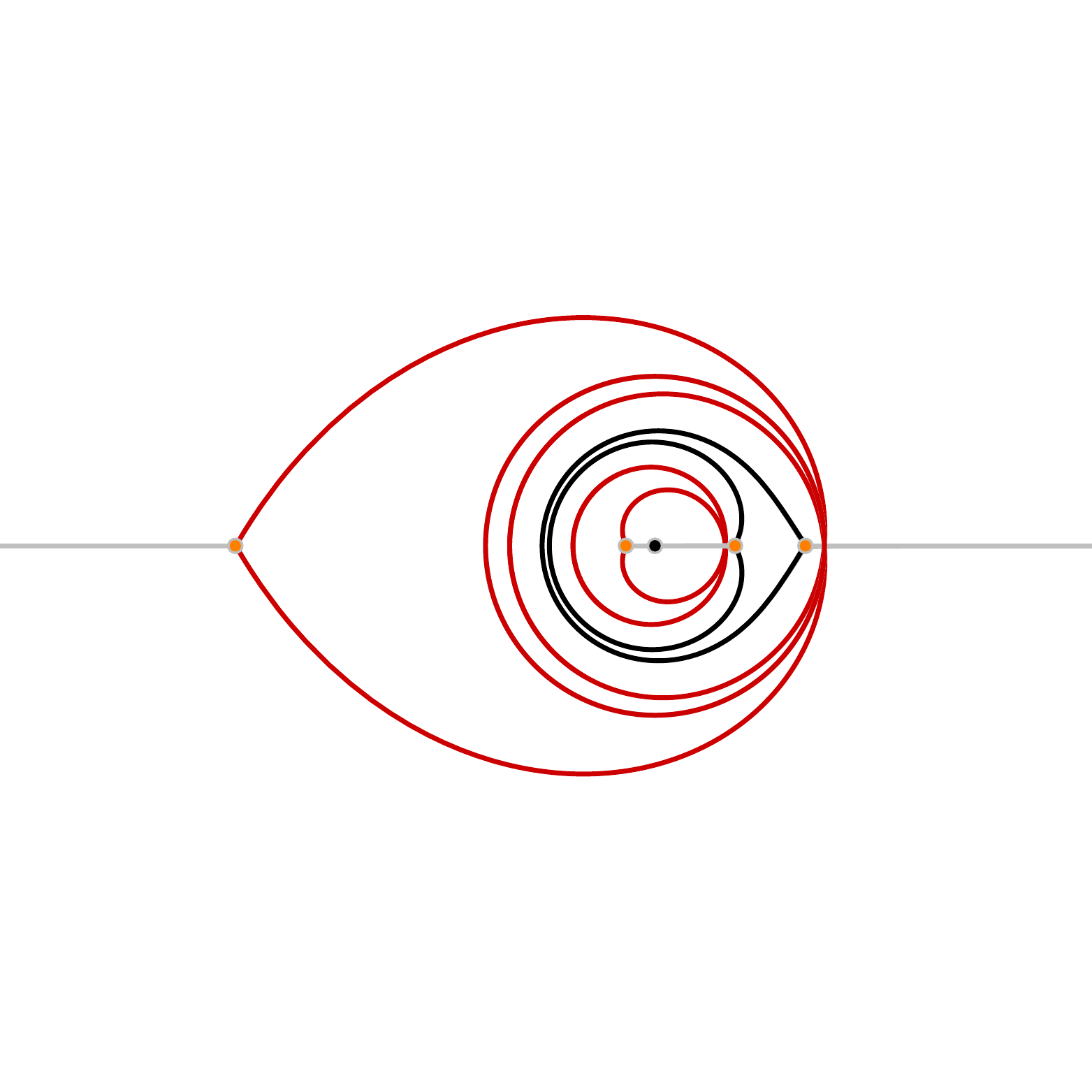}
\includegraphics[width=0.45\textwidth]{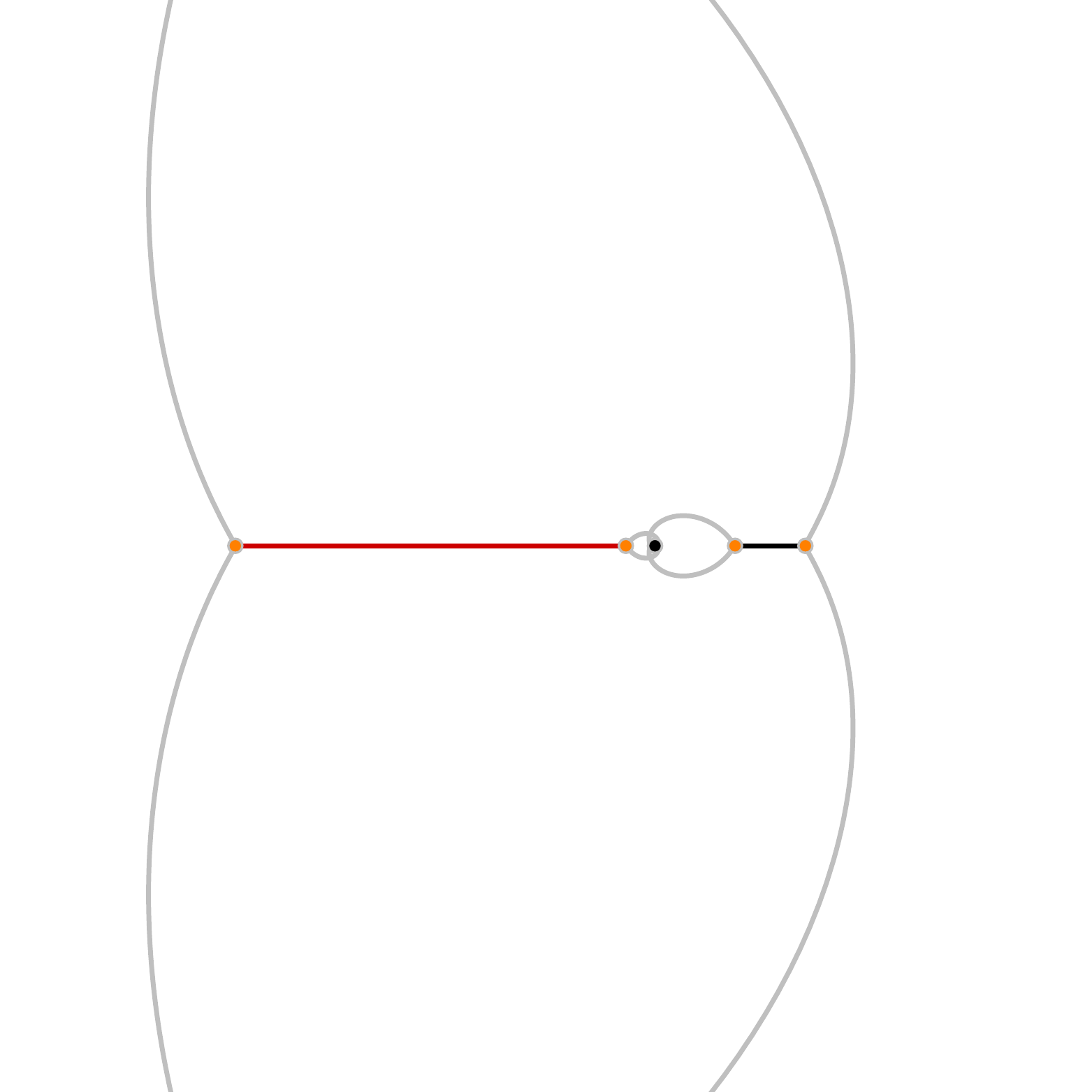}
\caption{Left: $\vartheta=0$. Right: $\vartheta=\pi/2$. Some of the states surviving in the 4d limit are visible already at $R=1$, compare with \ref{fig:4d-wc}.}
\label{fig:radius-1-basic-states-wc}
\end{center}
\end{figure}

We next take a 4d limit that approaches the weak coupling chamber of SU(2) Seiberg-Witten theory.
For this purpose, we switch from moduli chosen as in (\ref{eq:5d-moduli-strong-cplg}) to
\be\label{eq:5d-moduli-weak-cplg}
	\Lambda = 1,\qquad
	u = 1.05 \Lambda^2\,,
\ee
which, at radius $R=1$, correspond approximately to $Q_b = -1$, $Q_f = 1/2+u = 1.55$.
Comparing to the strong coupling point analyzed earlier, this point is on the other side of the dyon singularity $\CD_2$ at $u=\Lambda^2$, where $p_5$ shrinks to zero length.
Again focusing on the simplest BPS saddles, we consider exponential networks at $\vartheta=0,\pi/2$ shown in Figure \ref{fig:radius-1-basic-states-wc}.

These saddles consist of just a tiny sample of the full BPS spectrum, but we may again recognize some familiar states.
\begin{itemize}
\item $D2_f$, which corresponds to the black saddle at $\vartheta=0$, arises as a boundstate of $p_3\cup p_5$, and remain of finite mass. This should descend to a purely electric state, the W-bosons in the BPS vectormultiplet of Seiberg-Witten theory.
\item likewise $D2_f\-\overline{D4}$, which corresponds to $p5$ (after a flop across the dyon singularity) remains of finite mass. This again descends to a dyon with $(e,m) = (-1,1)$, the BPS dyon of Seiberg-Witten theory
\item as in the case of strong coupling, $D0$ branes, which correspond to red towers of saddles at $\vartheta=0$ will become infinitely massive and eventually disappear into the punctures 
\item likewise $D2_b$, corresponding to the red saddle at $\vartheta=\pi/2$ (or $p_4$ in Figure \ref{fig:exp-net-sym-point}) becomes infinitely massive due to the fact that the two branch points that support it become infinitely separated and disappear
\end{itemize}
Comparing with the strong coupling spectrum, we observe a change in a part of the spectrum corresponding to the well-known wall-crossing phanomenon in Seiberg-Witten theory. The two regions are separated by a wall of marginal stability for  $D4$ and $D2_f\-\overline{D4}$ states, the vectormultiplet $D2_f$ is one of the boundstates formed in this process.
The match with spectral networks at weak coupling offers yet another check of our identification of D-brane charges: the vectormultiplet should indeed carry pure electric charge under the $U(1)$ gauge symmetry associated with the fiber $\IP^1$.

While we omit the details, it is worth noting that these expectations can be verified directly, along the same lines of Figures \ref{fig:strong-coupling-phase-zero} and \ref{fig:strong-coupling-phase-piover2}. Plotting exponential networks for decreasing values of $R$, one observes that red saddles grow infinitely large, while black ones remain finite.  Black saddles can be identified directly with the vectormultiplet and dyon from Figure \ref{fig:4d-wc}.

\subsection{The half-geometry}

The mirror curve admits two types of factorization limits, corresponding to $|\log Q_b|\to \infty$ or $|\log Q_f|\to \infty$. 
In each case it reduces to the mirror curve of $\CO(0)\oplus\CO(-2)\to \IP^1$. 
Let us start by briefly recalling salient features of the exponential networks for this half-geometry, a more comprehensive analysis can be found in \cite{Banerjee:2019apt}.
The mirror curve\footnote{This corresponds to the curve studied in \cite{Banerjee:2019apt} after a change of framing.}
\be\label{eq:C3-mod-Z2-curve}
	1-y-x y + \frac{Q}{(1+Q)^2} y^2 = 0
\ee
enjoys a $\IZ_2$ symmetry, which becomes manifest after rescaling $y'= \sqrt{Q}(1+Q)^{-1} y$, exchanging $y' \leftrightarrow 1/y'$
There are two logarithmic punctures above $x=\infty$, as well as two regular punctures at $x=0$ where $y_\pm \to  c_\pm$ for constants $c_\pm$. 

\begin{figure}[h!]
\begin{center}
\includegraphics[width=0.45\textwidth]{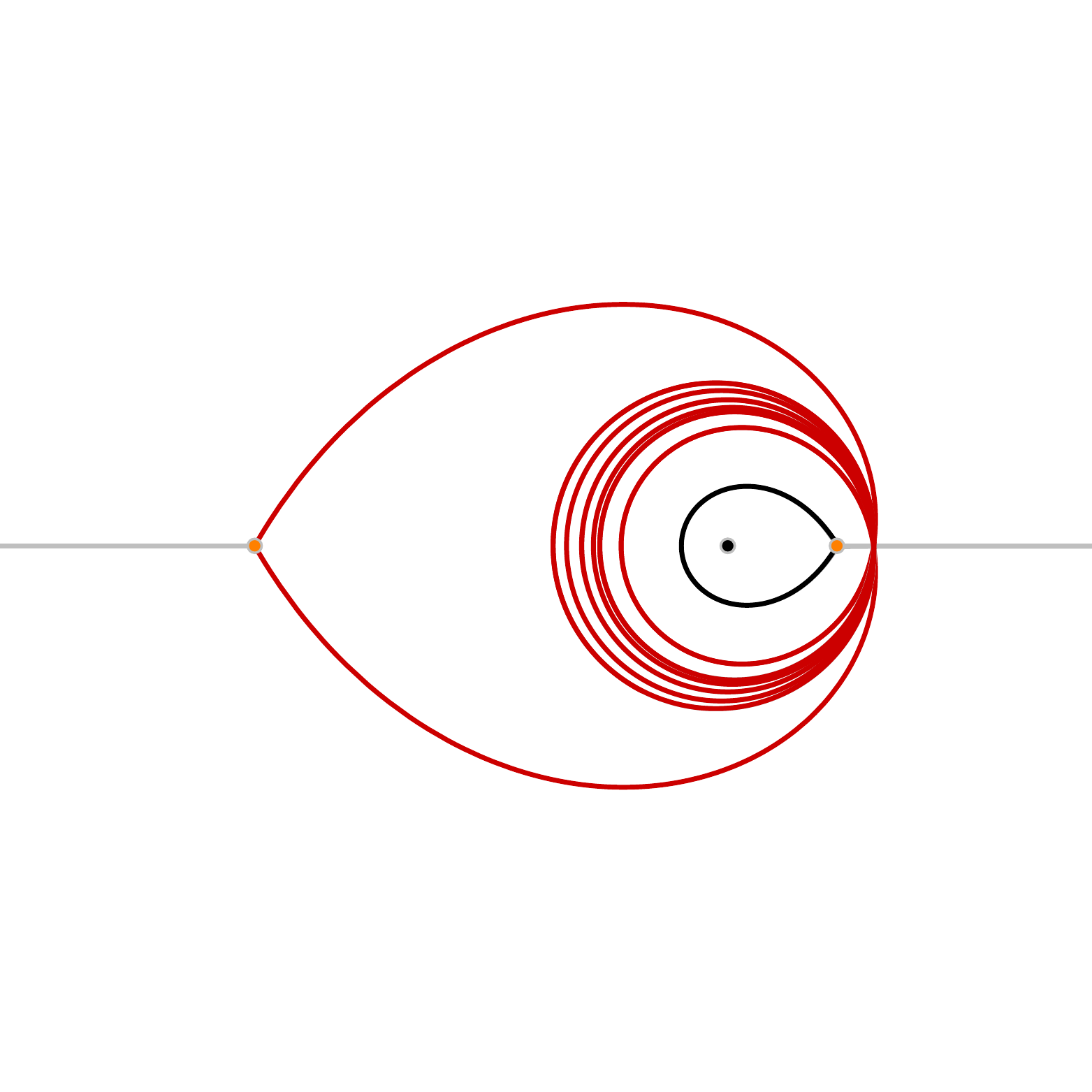}
\caption{Exponential network of $\CO(0)\oplus\CO(-2)\to\mathbb{CP}^1$ at $\vartheta=0$.}
\label{fig:C3modZ2-network}
\end{center}
\end{figure}

The BPS spectrum consists of boundstates of D2 and D0 branes. All states are mutually local, this allows to capture the whole set of saddles by going to a point in the moduli space where $\arg Z_{D2} = \arg Z_{D0}$. 
Introducing $Q_f$ defined by $Q_f^2 = Q(1+Q)^{-2}$, we fix  $Q_f = 0.8$.  
Then all BPS states have real positive central charges, corresponding to $Z_{D0} = 2\pi$ and $Z_{D2} =  i\, \log Q$ (at $R=1$).
All saddles appear therefore at $\vartheta=0$, see Figure \ref{fig:C3modZ2-network}.
Highlighted in red is an infinite tower of double walls corresponding to pure D0 boundstates, while highlighted in black is a simpler saddle corresponding to a D2 brane wrapping the $\IP^1$.

\subsubsection*{Limit $Q_b\to 0$}

\begin{figure}[h!]
\begin{center}
\fbox{\includegraphics[width=0.28\textwidth]{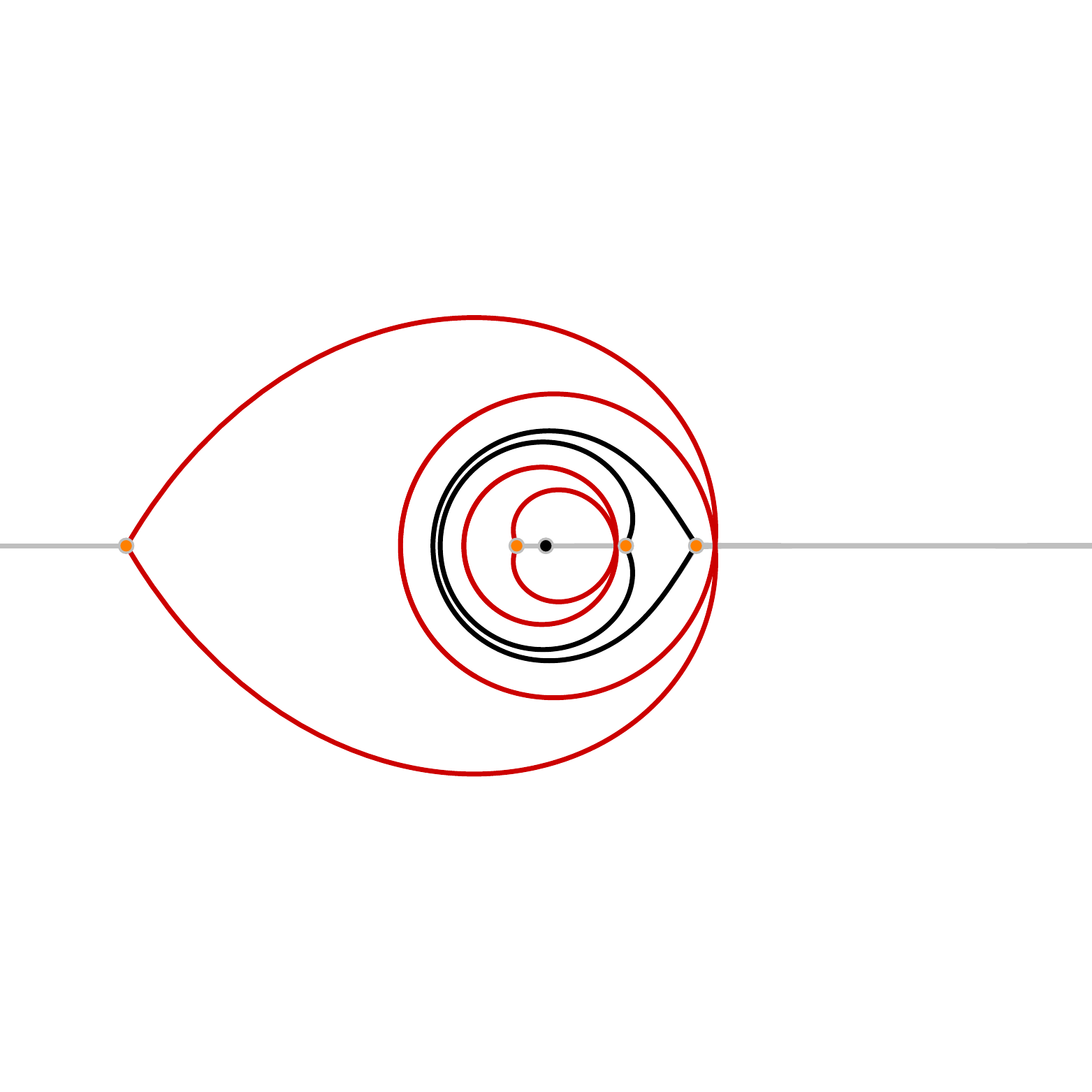}}
\fbox{\includegraphics[width=0.28\textwidth]{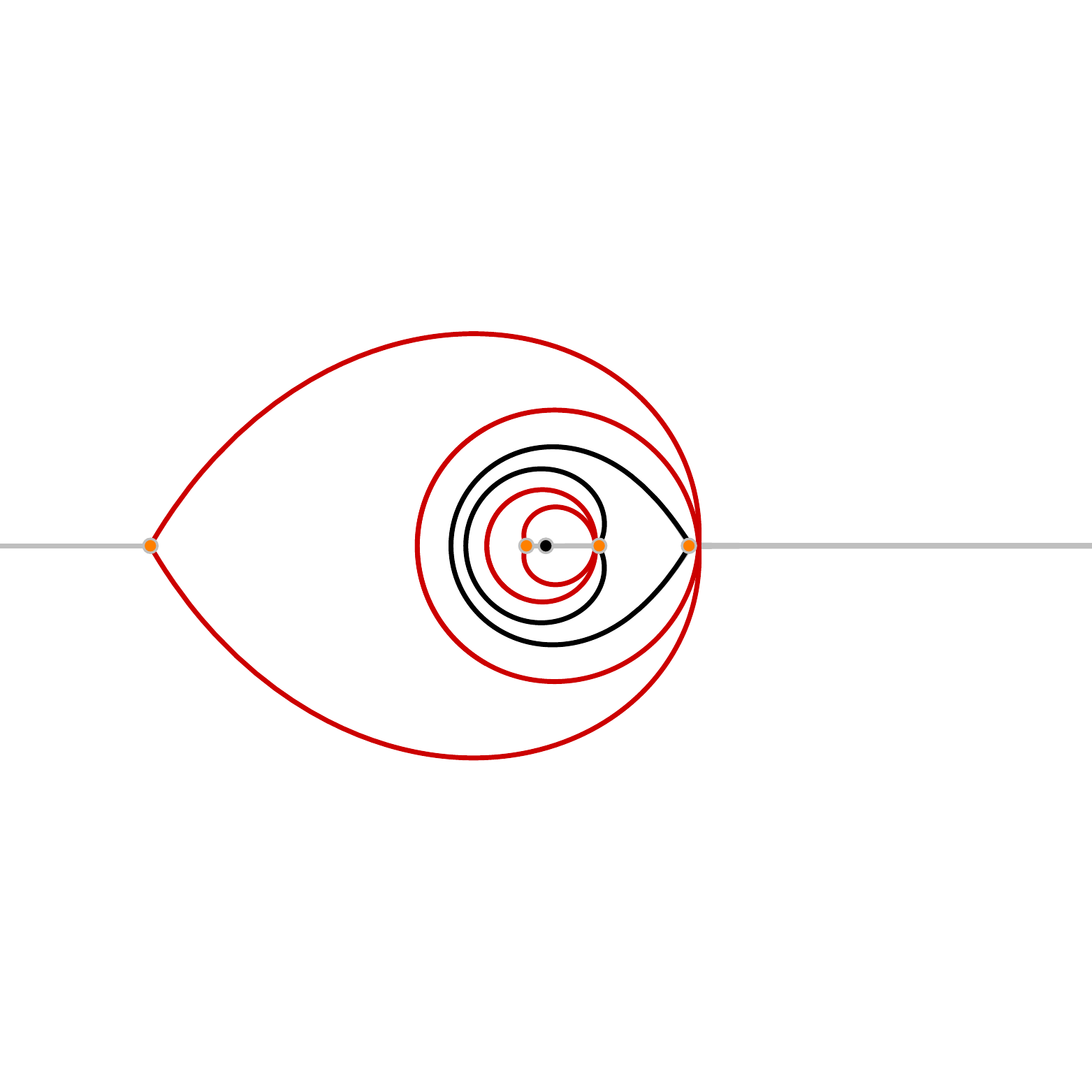}}
\fbox{\includegraphics[width=0.28\textwidth]{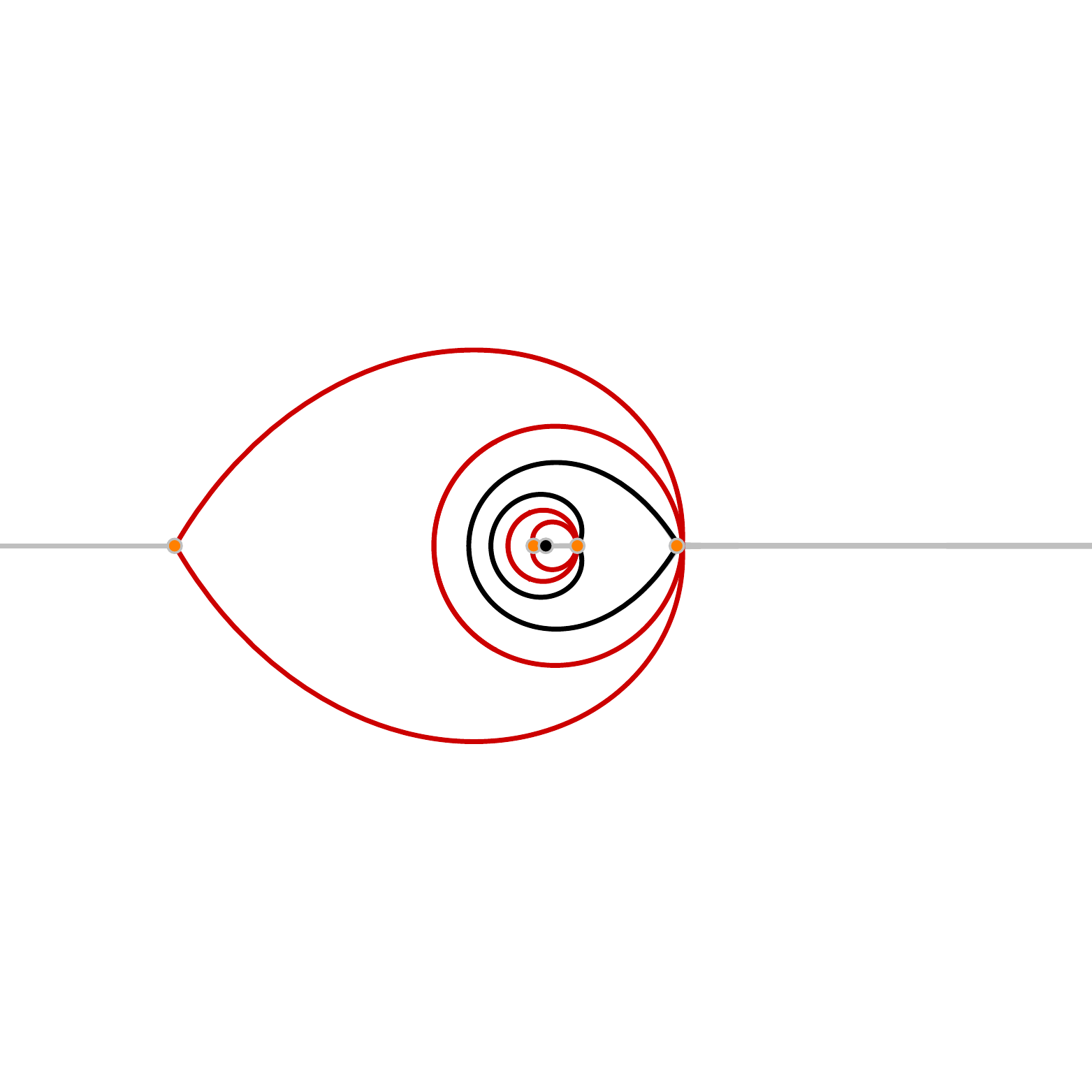}}\\
\fbox{\includegraphics[width=0.28\textwidth]{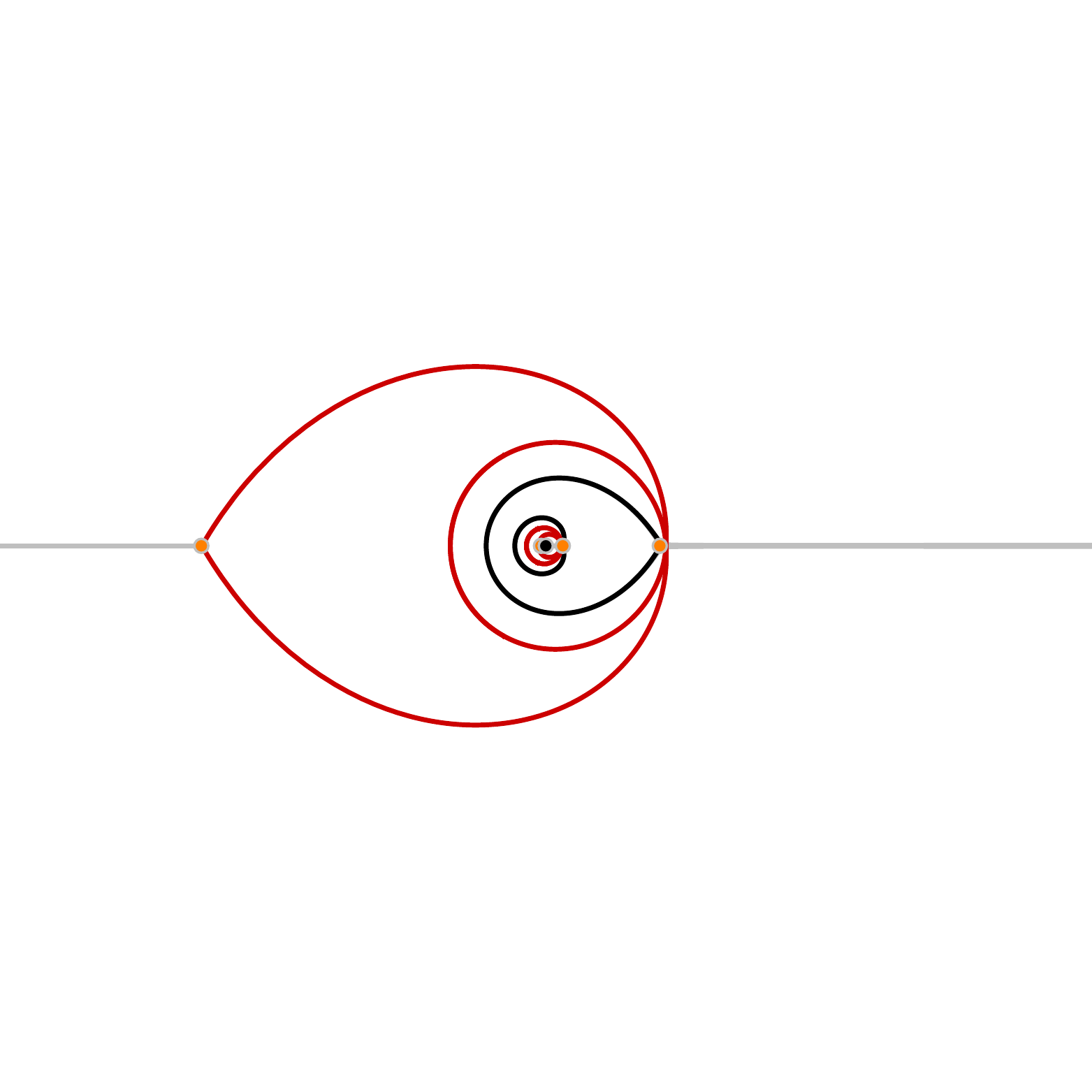}}
\fbox{\includegraphics[width=0.28\textwidth]{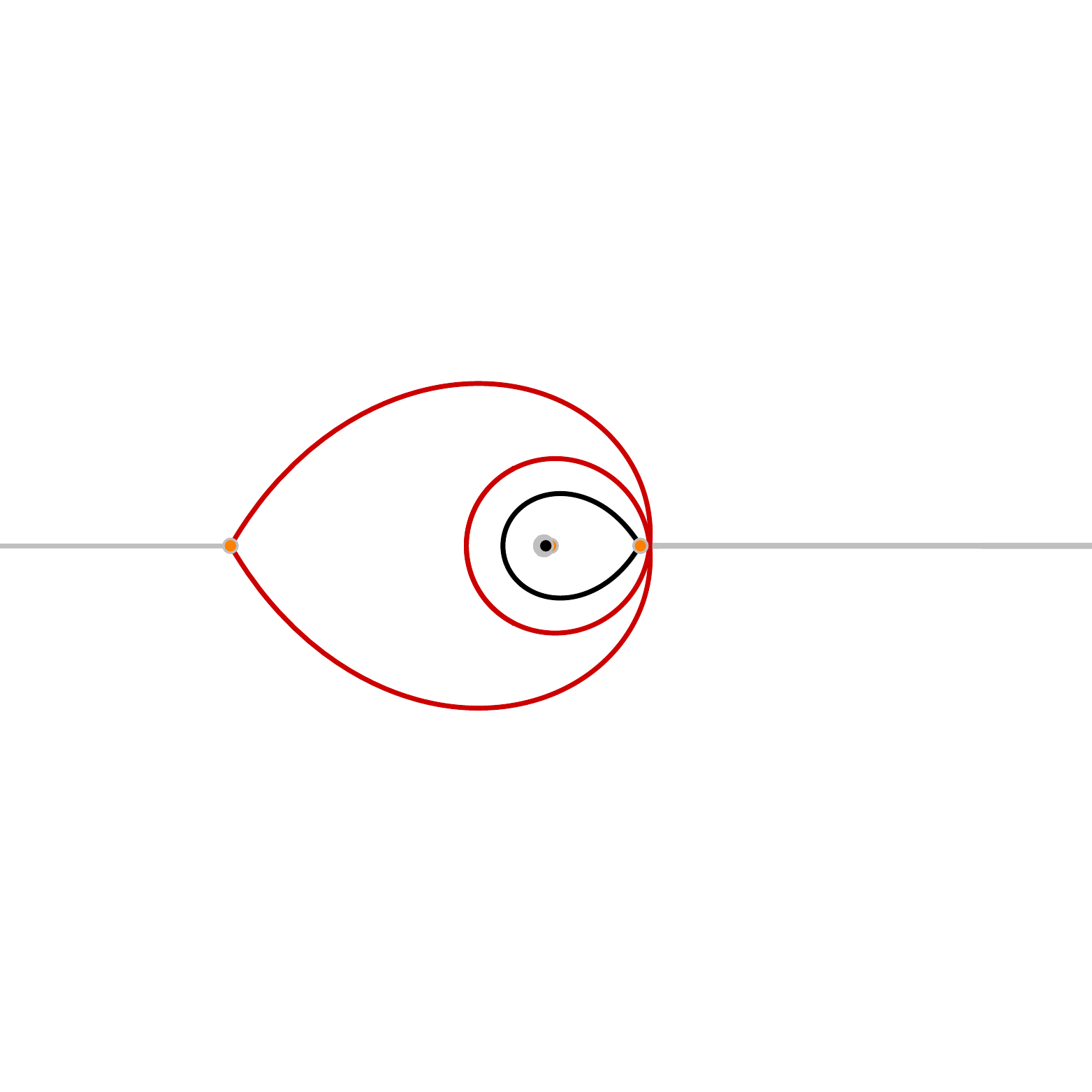}}
\fbox{\includegraphics[width=0.28\textwidth]{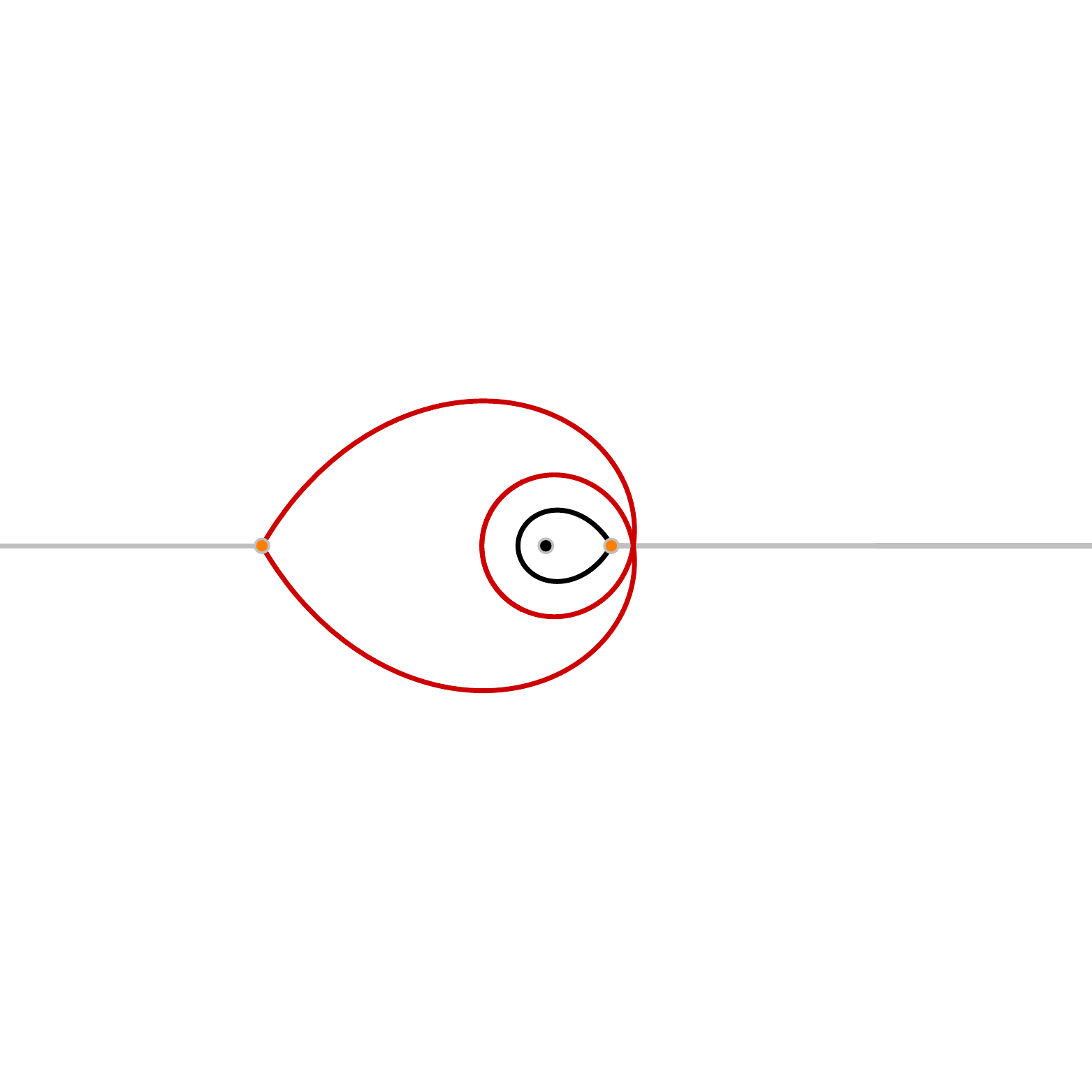}}
\caption{Exponential networks at $\vartheta=0$ for $(Q_b, Q_f) = (-1+ 0.2\cdot k, 1.55-0.15 \cdot k)$ and $k = 0,1,2,3,4,5$. Only primary walls are shown. This interpolates between Figures \ref{fig:4d-wc} and \ref{fig:C3modZ2-network}.}
\label{fig:factorization-limit}
\end{center}
\end{figure}

Going back to the full geometry $K_{\IF_0}$, we now study how it degenerates to  $\CO(0)\oplus\CO(-2)\to \IP^1$ and what happens to BPS states using exponential networks.
Consider the `weak coupling' regime defined by moduli (\ref{eq:5d-moduli-weak-cplg}). Starting from here we will take $Q_b\to 0$ along the negative real axis, while $Q_f$ is kept finite.
In this limit, two of the branch points fall into $x=0$ with positions $x = \frac{Q_b}{1\pm 2 Q_f}$, while (by $\IZ_2$-symmetry $x\to x^{-1}$) the other two tend to infinity as $x= \frac{1\pm 2 Q_f}{Q_b}$.
Changing variables to $x=x'/Q_b$ to focus near the region $x=\infty$ where $x'$ is finite, we observe the limiting behavior of the exponential network shown in Figure \ref{fig:factorization-limit}. 
In the limit,  the exponential network of the half-geometry emerges (cf. Figure \ref{fig:C3modZ2-network}). By $\IZ_{2}$ symmetry a similar picture emerges near $x=0$.

\begin{figure}[h!]
\begin{center}
\fbox{\includegraphics[width=0.28\textwidth]{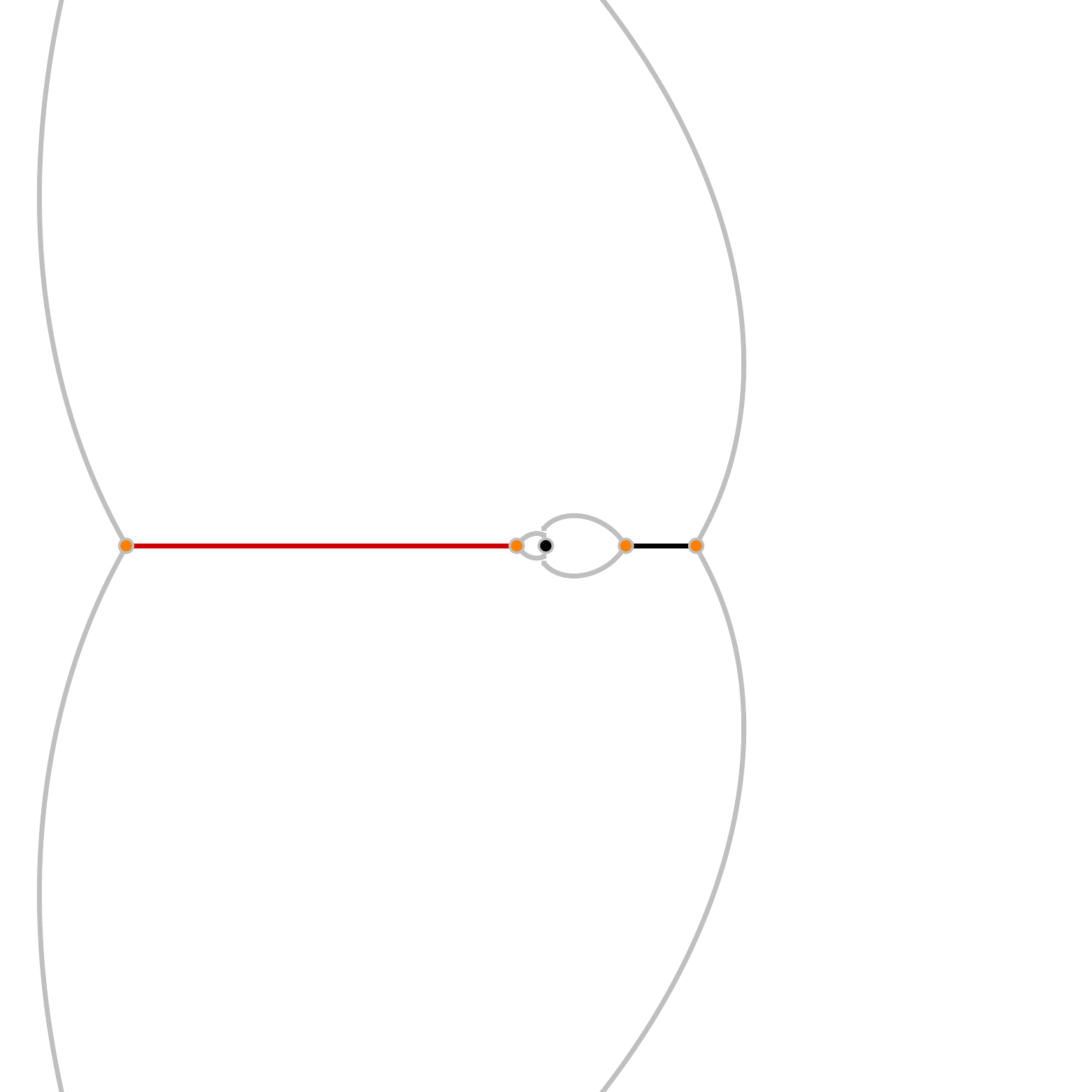}}
\fbox{\includegraphics[width=0.28\textwidth]{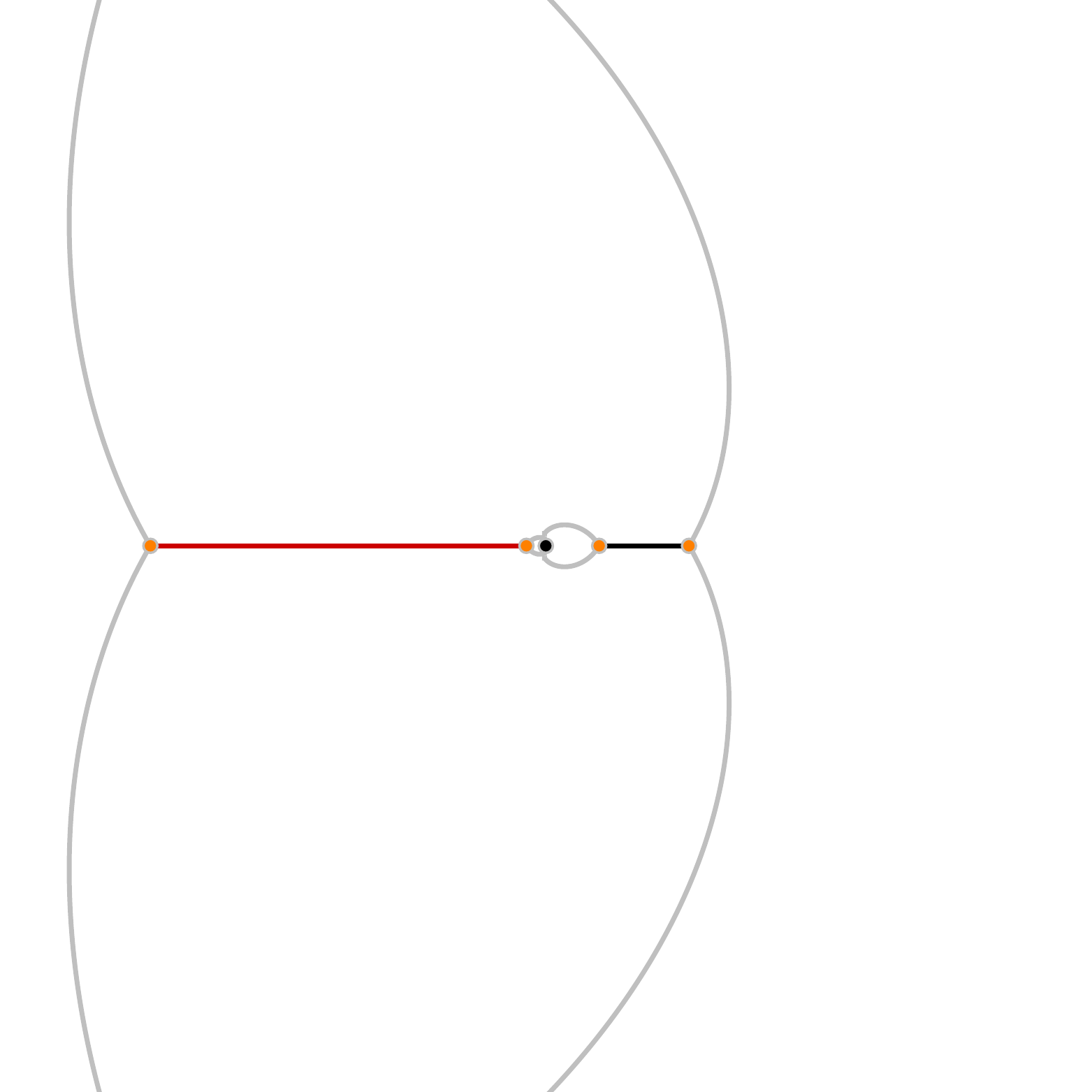}}
\fbox{\includegraphics[width=0.28\textwidth]{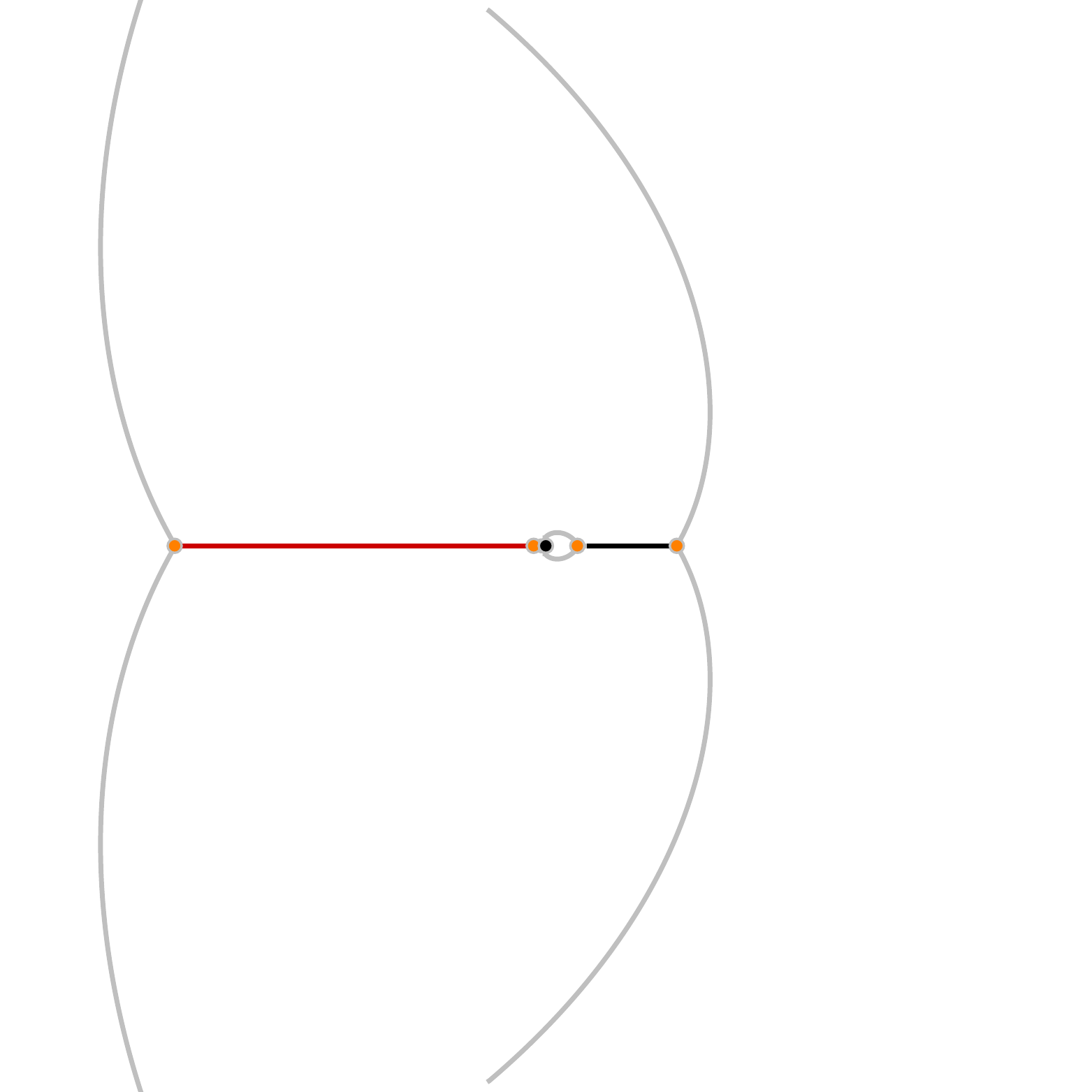}}\\
\fbox{\includegraphics[width=0.28\textwidth]{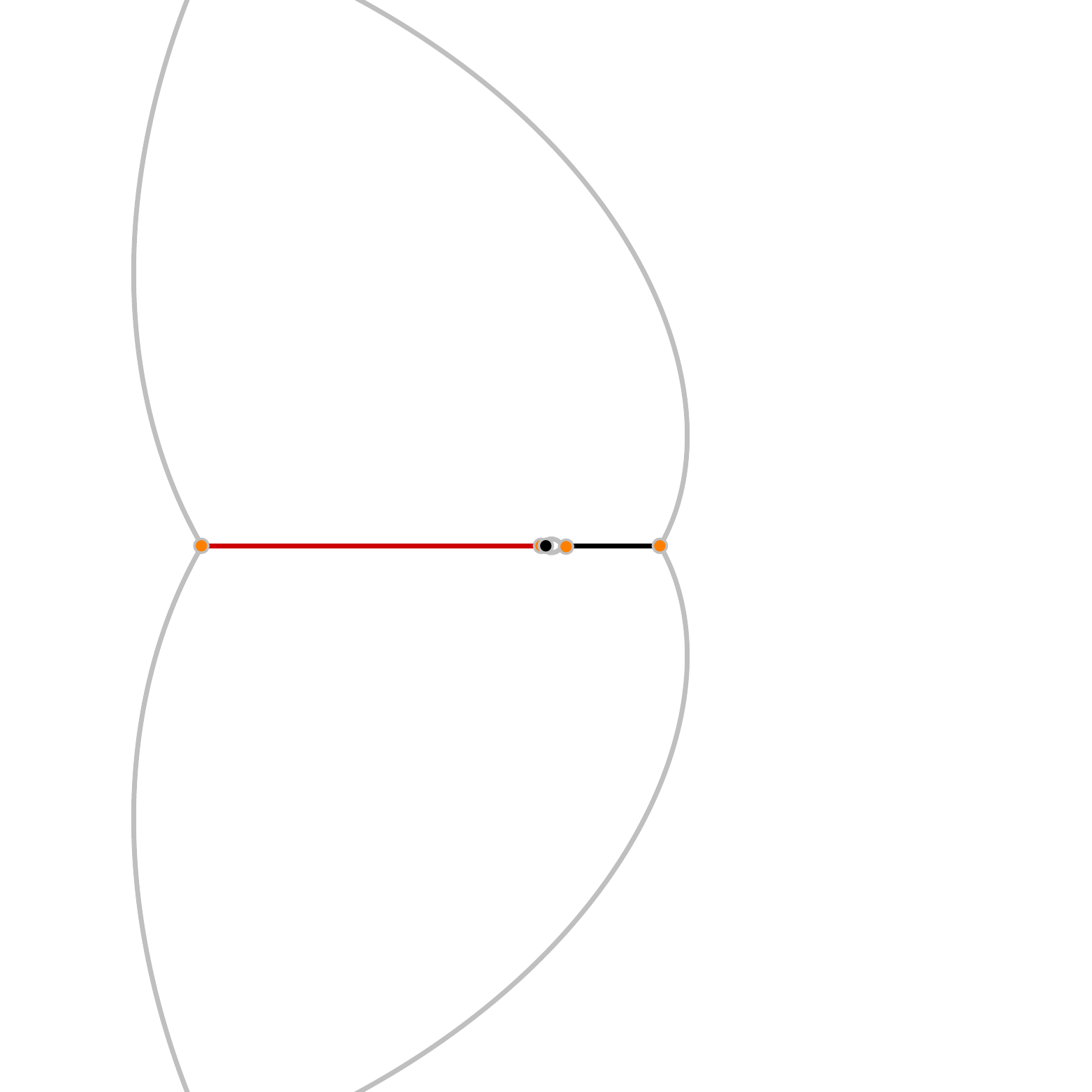}}
\fbox{\includegraphics[width=0.28\textwidth]{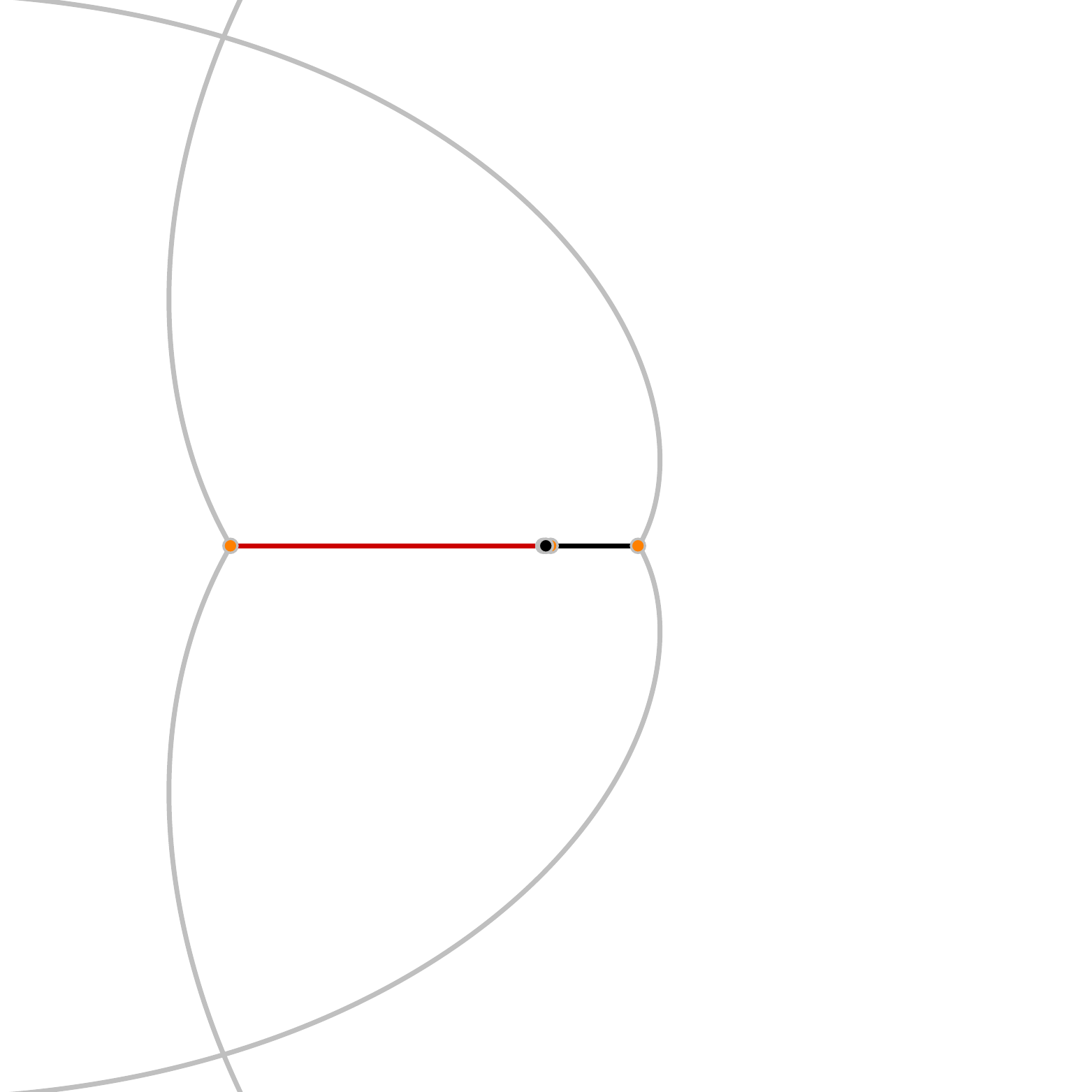}}
\fbox{\includegraphics[width=0.28\textwidth]{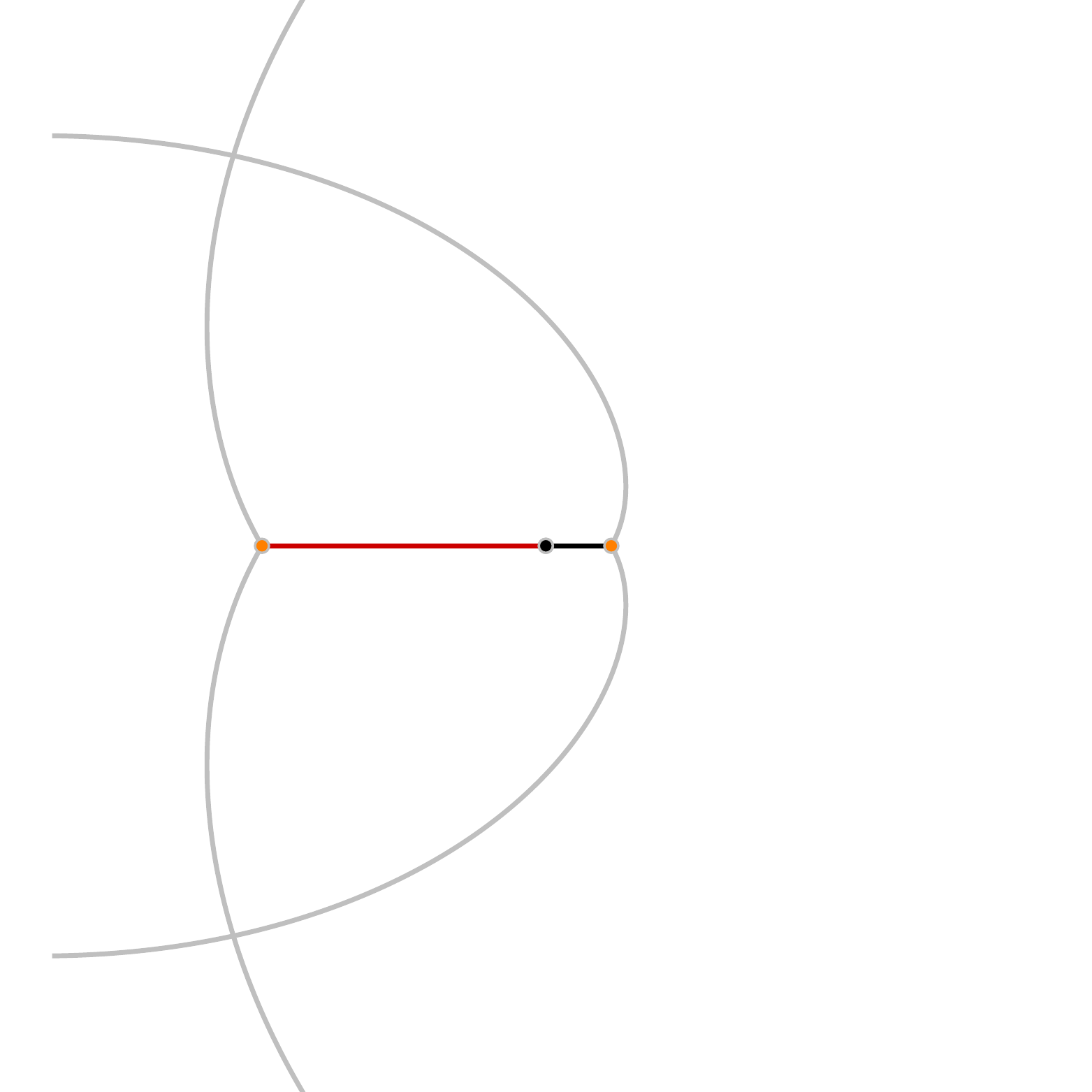}}
\caption{Exponential networks at $\vartheta=\pi/2$ for $(Q_b, Q_f) = (-1+ 0.2\cdot k, 1.55-0.15 \cdot k)$ and $k = 0,1,2,3,4,5$. Only primary walls are shown}
\label{fig:factorization-limit-heavy-states}
\end{center}
\end{figure}

The vectormultiplet corresponding to a D2 wrapping $\IP^1_f$ (the black saddle) starts out with BPS index $\Omega=-2$, but in the limit only half of it remains, ending up with $\Omega=-1$ as expected \cite{Banerjee:2019apt}. Comparing with the known spectrum for the half-geometry, this yields another check that our identification of $D2_f$ is correct.
Likewise, the spectrum of D0-branes is halved in the limit, as half of the red saddles disappears into the puncture.
The way the BPS spectrum changes in this degeneration limit is quite different from what we observed in the 4d limit. While in the 4d limit certain BPS states became very heavy and disappeared altogether (as in the case of D0), here they remain of finite mass but their BPS index changes.
It still happens that certain BPS states become infinitely heavy. This is the fate of both saddles previously at $\vartheta=\pi/2$ in Figure \ref{fig:radius-1-basic-states-wc}, as shown in  Figure \ref{fig:factorization-limit-heavy-states}. The heavy states correspond to the $D4$ (black saddle) and to $D2_b\-\overline{D4}$, in line with expectations from the mirror side where $\IP^1_b$ and the whole $\IF_0$ grow to infinite size.

Also worth of notice is the behavior of punctures and branch points: in the 4d limit we observed one branch point falling into $x=0$ and one into $x=\infty$. Here (after a rescaling of $x$) there are two branch points falling into $x=0$, resulting into a \emph{simple} puncture $\log y \frac{dx}{x}\sim \frac{dx}{x}$ instead of an \emph{irregular} one (as in the 4d limit). At the same time, the puncture at $x=\infty$ remains logarithmic. This behavior is consistent with expectations about the degeneration of the mirror curve from the corresponding factorization of the toric diagram.

\vfill
\pagebreak
\cleardoublepage

\section{Conclusions}\label{sec:conclusions}

In this paper we investigated the BPS spectrum of M-theory on $X\times S^1\times \IR^4$ when $X= K_{\IF_0}$. More precisely, we focused on the sector of BPS states obtained by wrapping M2 and M5 branes on compact holomorphic two- and four-cycles on $X$.
Via geometric engineering, this is directly related to the BPS spectrum of instanton-particles and monopole-strings in 5d $\CN=1$ SU(2) Yang-Mills theory on $S^1\times \IR^4$.
Due to wall-crossing, the spectrum depends on a choice of moduli, and we focused on the special point $\mathbf{Q}_0$ (corresponding to $Q_b = -1, Q_f=1$) in the moduli space of the mirror curve (\ref{eq:mirror-curve}). 

Our approach to this problem is based on exponential networks. For practical reasons we used a combination of techniques to obtain extensive information about BPS states, although in principle the same information (and more) could be obtained just by studying networks.
The spectrum contains infinitely many BPS states, including `wild BPS states', making it challenging to provide a closed-form description. We computed part of the spectrum , corresponding to the lowest-energy (more precisely, lowest-charge) states.
A detailed description of results can be found in Section \ref{eq:spectrum-description}, here we recall the a few salient features.

We found a system of `basic' BPS saddles at $\vartheta=0$ and $\vartheta=\pi/2$, corresponding to specific D-brane boundstates $D4, D2_f\-\overline{D4}, D0\-D2_b\-\overline{D2}_f\-\overline{D4}$ and the fiber-base duals obtained by switching $D2_b\leftrightarrow D2_f$. Here $D2_f$ and $D2_b$ denote $D2$ branes wrapping the fiber and base $\IP^1$ in $\IF_0$.
We found evidence that the BPS spectrum enjoys fiber-base symmetry. In fact, the structure of the spectrum appears to simplify when  expressed in terms of the five basic BPS saddles, presumably because both the spectrum and the setof basic saddles are invariant under fiber-base symmetry.

We also derived the BPS quiver and its potential from exponential networks. The quiver has four nodes, corresponding to four of the five basic saddles. This choice however breaks fiber-base symmetry, which is no longer manifest when the spectrum is described in terms of quiver representations.

Having obtained the spectrum at the point $\mathbf{Q}_0$, we studied how BPS states behave in different limits. 
First, we considered a 4d limit obtained by shrinking the M-theory circle. One can see how certain BPS states grow to infinite size in this limit. For example the $D0$ brane central charge  is a unit of Kaluza-Klein momentum $Z_{D0} = 2\pi/R$, causing all boundstates including one or more $D0$ branes to disappear from the spectrum. In the strict 4d limit we recovered \emph{spectral networks} for 4d $\CN=2$ SU(2) Yang-Mills theory, both at strong and at weak coupling. For each value of the radius one can follow the fate of individual states, as some grow to infinite size, others may disappear (or appear) because of wall-crossing and so on.\footnote{Similar questions were posed, and explored  with different techniques, in \cite{Kachru:2018nck}.} 
We also consider the limit to a half-geometry, by sending $Q_b\to 0$, and recovering exponential networks for $\CO(0)\oplus\CO(-2)\to \IP^1$.  Once again one can follow the behavior of each BPS saddle in the spectrum, tracking which saddles grow infinitely large and which ones disappear because of wall-crossing.

Besides obtaining the spectrum itself, one of the motivations for this work was the goal of validating and exploring the framework of exponential networks developed in \cite{Eager:2016yxd, Banerjee:2018syt}. In particular, after studying geometries with compact two-cycles in \cite{Banerjee:2019apt}, in this work we explore for the first time a geometry with compact four-cycles.
From the viewpoint of enumerative geometry, exponential networks provide a systematic way to compute rank-zero numerical Donaldson-Thomas invariants of $X$. When the geometry features compact four-cycles, these include Vafa-Witten invariants.  
It would be interesting to refine BPS counting with exponential networks to compute \emph{protected spin characters} (or \emph{motivic} Donalson-Thomas invariants).
A natural attempt in this direction would be to promote approaches developed in the context of spectral networks \cite{Galakhov:2014xba, Neitzke:2020jik}.
A further step forward would be to switch from the study of single BPS states to the computation of the motivic Kontsevich-Soibelman invariant. 
As shown in  \cite{Gaiotto:2009hg, Longhi:2016wtv} this may be achieved directly from spectral networks at certain points in moduli space. 
This construction should admit a direct lift to exponential networks, and would be certainly interesting to explore.
More ambitiously, one may ask for a full-fledged categorification, to really make contact with the description of D-branes in the context of homological mirror symmetry.
We expect this to be possible, and believe that the recent developments of \cite{Khan:2020hir}, based on \cite{Gaiotto:2015aoa}, would provide a good starting point towards this goal. 
Another interesting direction is the study of framed BPS states. It is unclear to us how to include non-compact $D2$, $D4$ or $D6$ branes on the mirror side, although we expect that it would involve some version of a wrapped Fukaya category. Direct computations of framed BPS states in presence of non-compact D-branes have been performed by various methods, see e.g. \cite{Denef:2007vg, Jafferis:2008uf, Aganagic:2010qr, Cirafici:2010bd, Chuang:2013wt, Nishinaka:2013mba}, and it would be interesting to reproduce them with exponential networks. This also points to interesting applications to the study of black hole entropies in connection to \cite{Maldacena:1996gb, Maldacena:1997de, deBoer:2006vg}. Indeed, one of our results is the fact that the growth of BPS degeneracies with mass, along a fixed direction in the charge lattice, can be described in terms of algebraic equations like (\ref{eq:algebraic-eq-7}). As pointed out in \cite{Galakhov:2013oja}, where similar equations were found in the context of 4d $\CN=2$ Yang-Mills theories, it is possible to deduce information about asymptotic growth of BPS degeneracies.

\subsection*{Relation to other work}

Over the past year, several works studying related questions have appeared. 

The authors of \cite{Closset:2019juk} considered the same geometry studied in this paper using BPS quivers \cite{Douglas:1996sw, Douglas:2000qw, Fiol:2000wx, Denef:2002ru,  Alim:2011ae, Alim:2011kw, Cecotti:2012se}. 
The computation of BPS states relies on the assumption of existence of stability conditions corresponding to a `tame' chamber, where the mutation method can be applied to great effect. It would be interesting to compare the results with ours using wall-crossing formulae. For this one would need to know the exact central charges corresponding to the putative tame chamber. Interesting extensions of the mutation method were considered in \cite{Cecotti:2014zga, Bonelli:2020dcp}, based on the discrete-time evolution of the integrable system associated to the quiver.\footnote{Another very interesting approach connected to quivers is based on scattering diagrams \cite{2016arXiv160300416B}, this has been recently applied to great effect to the study of sheaves on $\IP^2$ \cite{Bousseau:2019ift}.}

Another approach based on quivers was taken in \cite{Beaujard:2020sgs}, using the `Coulomb branch' and  the `flow tree' formula of \cite{Manschot:2010qz, Manschot:2011xc, Manschot:2012rx, Manschot:2013sya} and \cite{Alexandrov:2018iao}. The workings of this formula are rather involved, with complexity growing quickly with the dimension of the quiver representation. 
Using the mathematica package \cite{CoulombHiggs} provided with the paper, we were able to check their predictions against our results, for charges $\gamma=(d_1,d_2,d_3, d_4)$ with $\sum_{i}d_i \leq 6$, and found an exact match.

The first main difference with these approaches, is that exponential networks only rely on information of the mirror geometry. 
Connecting with enumerative invariants on the toric side, such as Vafa-Witten invariants, then requires establishing a map between D-branes between the toric side and the mirror side. The details of such a map are spelled out in Section \ref{sec:exc-coll-mirror}.
On the other hand, one of the advantages is that it is straightforward to obtain information about central charges (hence stability parameters for the quiver) from the mirror geometry, where they are encoded by periods of $\lambda = \log y\, d\log x$ on $\Sigma$. This sidesteps any assumptions on the existence of suitable chambers, or the necessity to study very special limits of the spectrum, since networks are well-defined over the entire moduli space (with the exception of singular divisors (\ref{eq:discriminant-locus})).
Another nice feature is the fact that, in some cases, a single BPS saddle encodes information about whole towers of BPS states: the corresponding BPS indices (or rank-zero numerical Donaldson-Thomas invariants) are encoded in algebraic equations like (\ref{eq:algebraic-eq-7}).
Perhaps the main drawback, is that whenever the BPS spectrum is complicated, saddles tend to be harder to study. 
Overall, each of these techniques has its strengths, and we found that the the most effective approach was to use a combination of exponential networks, quiver representation theory and wall-crossing identities of \cite{Kontsevich:2008fj,Joyce:2008pc}.

A third approach to compute the Vafa-Witten invariants is through exploiting modularity. The elliptic genus for the M5-branes on $\mathcal{D} \times T^2$, where 
$\mathcal{D}$ corresponds to a divisor, is expected to be a (higher depth, mock) modular form of certain weight \cite{Maldacena:1997de, Dijkgraaf:1997ce,  Gaiotto:2006wm, Gaiotto:2007cd}. On the type IIA side, this descends to the 
$D4\-D2\-D0$ branes boundstate configuration of the $\CN=2$ supersymmetric black holes. From the fact that this partition function has specific modularity
properties, one can then derive the BPS degeneracies explicitly, which in mathematical terms provides predictions for Donaldson-Thomas invariants for 
sheaves of various ranks. This is a strategy which has been implemented to great extent in several recent works \cite{Alexandrov:2016tnf, Alexandrov:2019rth}. 
In particular this was invoked in the papers \cite{Alexandrov:2020bwg,Alexandrov:2020dyy,Beaujard:2020sgs}. Several recent works of mathematicians 
also take this approach for computing the DT invariants, for example the works \cite{Manschot:2011dj, Manschot:2011ym, Mozgovoy:2013zqx, Gholampour:2012qp,Gholampour:2013jfa, Gholampour:2013hfa, 2017arXiv170108902G, Gottsche:2018meg,Sheshmani:2019pby,Bouchard:2016lfg, Tanaka:2017jom, Tanaka:2017bcw, Gottsche:2017vxs, Thomas:2018lvm}.
This remarkable connection of the DT invariants was in fact a source of inspiration for recent developments in the field of modular forms  \cite{Alexandrov:2016enp}, which provide a remarkable generalization of Zwegers' construction of mock modular forms \cite{Zwegers:2008zna}. However, this connects to our story of exponential networks in a slightly non-trivial fashion. In particular, this powerful modularity property holds in the chamber 
corresponding to the so called large volume point. From there, coming to our chamber involves intricate wall-crossing phenomena. The compatibility
of our results with that of this large volume chamber, as deduced by the match with \cite{Beaujard:2020sgs} mentioned above, corroborates the remarkable consistency of several of these methods. 

One more approach worth of mention is based on the mirror geometry like exponential networks, and revolves around the study of quantum periods of the (quantized) mirror curve. The connection to BPS counting is not entirely clear to us at this point, but it seems plausible that applying the techniques recently developed in  \cite{Garoufalidis:2020xec, Garoufalidis:2020nut} to quantum periods of mirror curves considered e.g. in  \cite{Hatsuda:2013oxa, Kallen:2013qla, Grassi:2019coc}, should lead to a computation of the same kind of BPS indices studied in this paper. These expectations are based on the better-understood relation between the two frameworks in the context of 4d $\CN=2$ theories, and quantum periods (obtained via exact  WKB analysis) for their Seiberg-Witten curves \cite{Gaiotto:2008cd, Gaiotto:2009hg}.

\subsection*{Acknowledgements}

We are grateful to Tobias Ekholm, Jan Manschot, Marcos Mari\~no, Boris Pioline, Daniel Pomerleano, Vivek Shende and Johannes Walcher for helpful discussions.
The work of SB is supported by Alexander von Humboldt foundation for researchers and he acknowledges 
the hospitality of the University of Cologne where he is based. 
MR acknowledges support from the Research Fund for International Young Scientists, NSFC grant No. 11950410500.
The work of PL was supported by NCCR SwissMAP, funded by the Swiss National Science Foundation.

\appendix

\section{Central charges and plots of BPS saddles at $\mathbf{Q}_0$}\label{app:conifold-point}

Central charges of BPS saddles can be evaluated numerically, by integrating 
\be
	\lambda_{ij} = (\log y_j - \log y_i+2\pi n) \, \frac{dx}{x}
\ee 
along a double-wall (saddle) composed of underlying $\CE$-walls $(ij,n)/(ji,-n)$.
The direction of integration corresponds to the orientation of the $(ij,n)$ wall.  

The periods of the five basis saddles, obtained by integrating $\frac{1}{2\pi}\lambda_{ij}$ along each saddle, are reported below together with the corresponding central charges of D-branes. We  evaluated these numerically, both both at the point $\mathbf{Q}_0$ and at a slightly perturbed point in moduli space. The normalization corresponds to $Z_{D0} = 2\pi / R$ independent of moduli, where $R=1$.
\be\label{eq:sym-point-periods}
\begin{array}{|c|c|}
	\hline
	(Q_b, Q_f) & (-1, 1) \\
	\hline\hline
	Z_1 & 4.34600\\\hline
	Z_2 & 4.34600 \\\hline
	Z_3 & 1.93718\\\hline
	Z_4 & 0.502661 i\\\hline
	Z_5 & 0.50266 i \\
	\hline
\end{array}
\qquad
\begin{array}{|c|c|}
	\hline
	(Q_b, Q_f) & (-1, 1+0.1 i) \\
	\hline\hline
	Z_1 & 4.54916\, - 0.0546797 i \\\hline
	Z_2 & 4.15047\, - 0.034779 i  \\\hline
	Z_3 & 1.93337\, + 0.0447295  i\\\hline
	Z_4 & -0.0743764 \, + 0.506038 i \\\hline
	Z_5 &  0.124961 \,+ 0.496088 I i\\
	\hline
\end{array}
\ee

\be\label{eq:D-brane-Z-sym-pt}
\begin{array}{|c|c|}
	\hline
	(Q_b, Q_f) & (-1, 1) \\
	\hline\hline
	Z_{D0} &6.28318   \\\hline
	Z_{D2_f} & 1.93718 + 0.502661 i  \\\hline
	Z_{D2_b} &1.93718 + 0.502661 i \\\hline
	Z_{D4}& 1.93718  \\\hline
\end{array}
\qquad
\begin{array}{|c|c|}
	\hline
	(Q_b, Q_f) & (-1, 1+0.1 i) \\
	\hline\hline
	Z_{D0} &6.28318 \\\hline
	Z_{D2_f} &2.05833 + 0.540817 i  \\\hline
	Z_{D2_b} & 1.85899 + 0.550767 i\\\hline
	Z_{D4}&1.93337 + 0.0447295 i\\\hline
\end{array}
\ee

This information is essential for studying BPS states with exponential networks, since to see the corresponding saddles one needs to tune  the phase $\vartheta$ to the one of the central charge $\arg Z_\gamma$. Below we collect plots of saddles for some of the BPS states found at $\mathbf{Q}_0$.

\begin{figure}[h!]
\begin{center}
\includegraphics[width=0.85\textwidth]{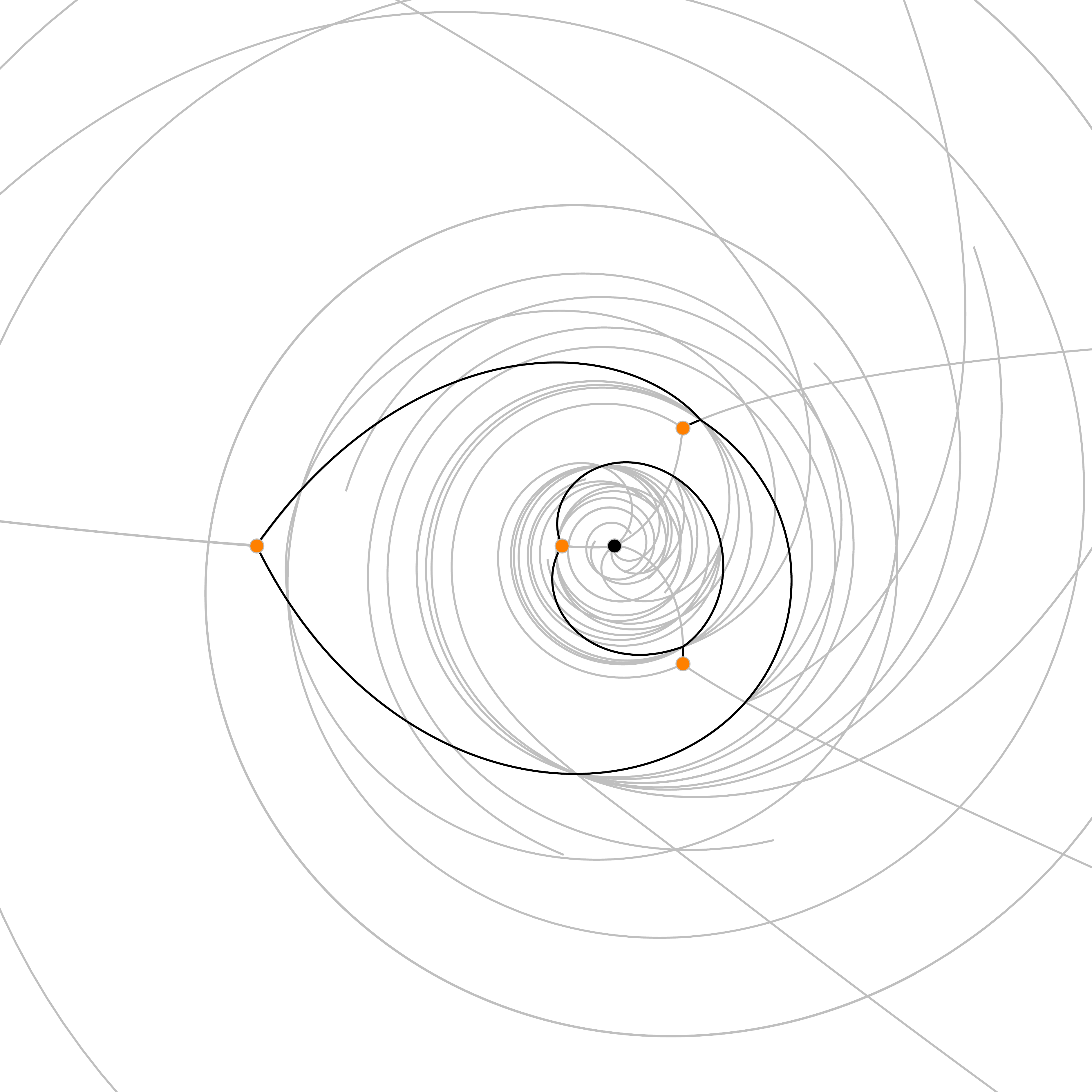}
\caption{Saddle of BPS state with charge $\gamma=(0,0,1,1)$, corresponding to $D0$-$\overline{D2}_f$.
There are two saddles of Type-2, giving overall $\Omega(\gamma) = -2$, corresponding to a BPS vectormultiplet. }
\label{fig:0011}
\end{center}
\end{figure}

\begin{figure}[h!]
\begin{center}
\includegraphics[width=0.85\textwidth]{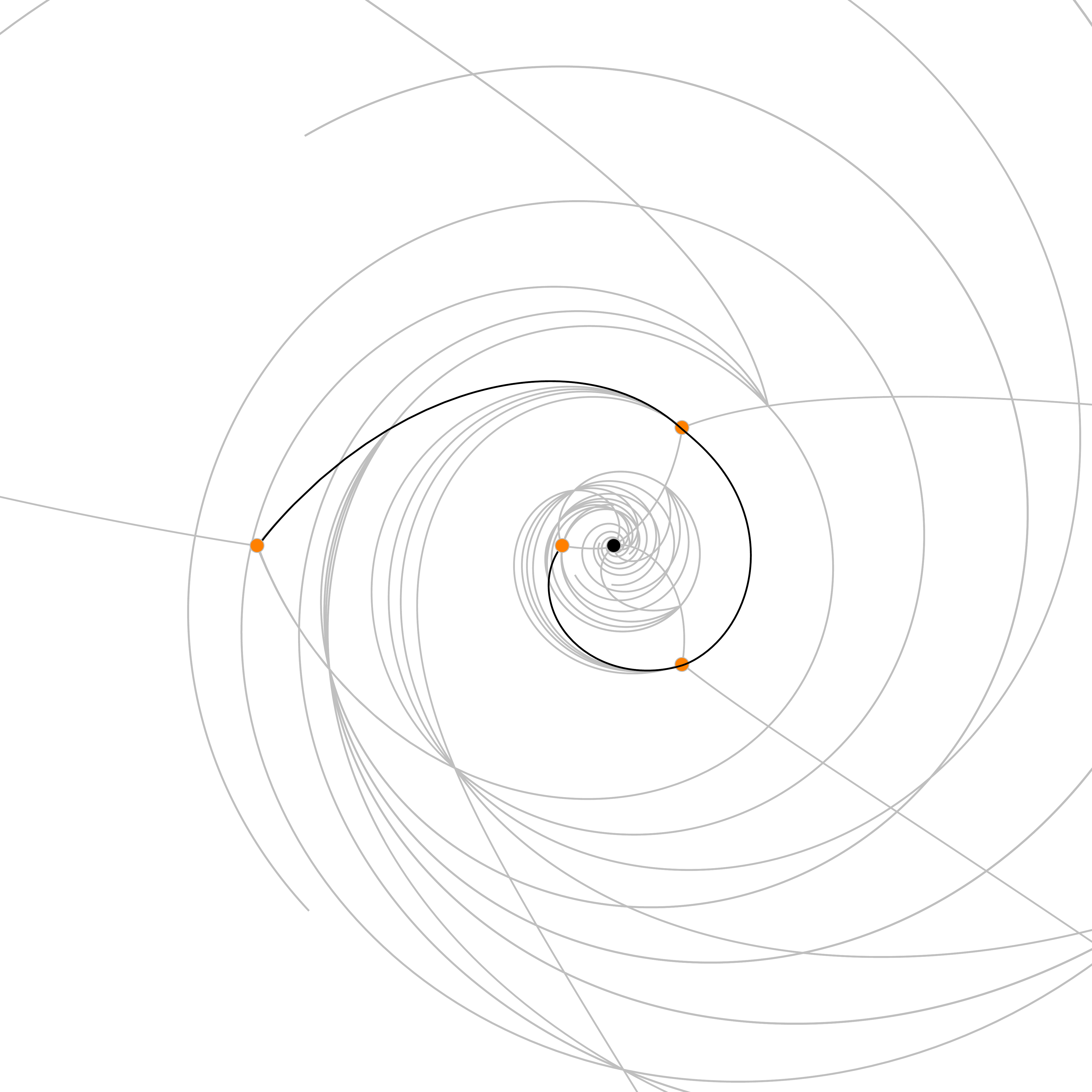}
\caption{Saddle of BPS state with charge $\gamma = (0,0,1,2)$, corresponding to $D0$-$\overline{D2}_b$-$\overline{D2}_f$-${D4}$. This is a saddle of Type-0, so $\Omega(\gamma)=1$ corresponding to a hypermultiplet.}
\label{fig:0012}
\end{center}
\end{figure}

\begin{figure}[h!]
\begin{center}
\includegraphics[width=0.85\textwidth]{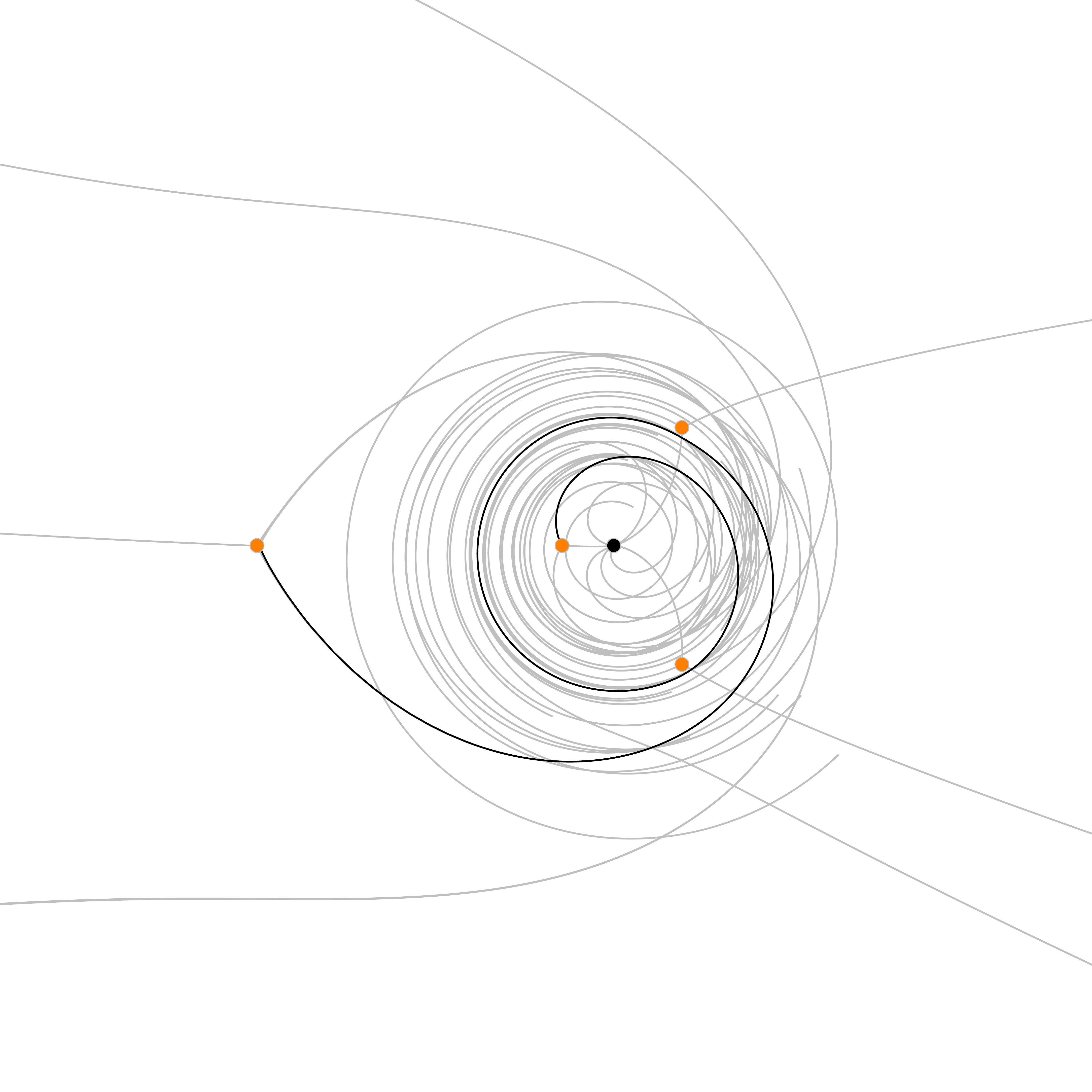}
\caption{Saddle of BPS state with charge $\gamma = (0,0,2,1)$, corresponding to $2D0$-${D2}_b$-$2\overline{D2}_f$-$\overline{D4}$. This is a saddle of Type-0, so $\Omega(\gamma)=1$ corresponding to a hypermultiplet.}
\label{fig:0021}
\end{center}
\end{figure}

\begin{figure}[h!]
\begin{center}
\includegraphics[width=0.85\textwidth]{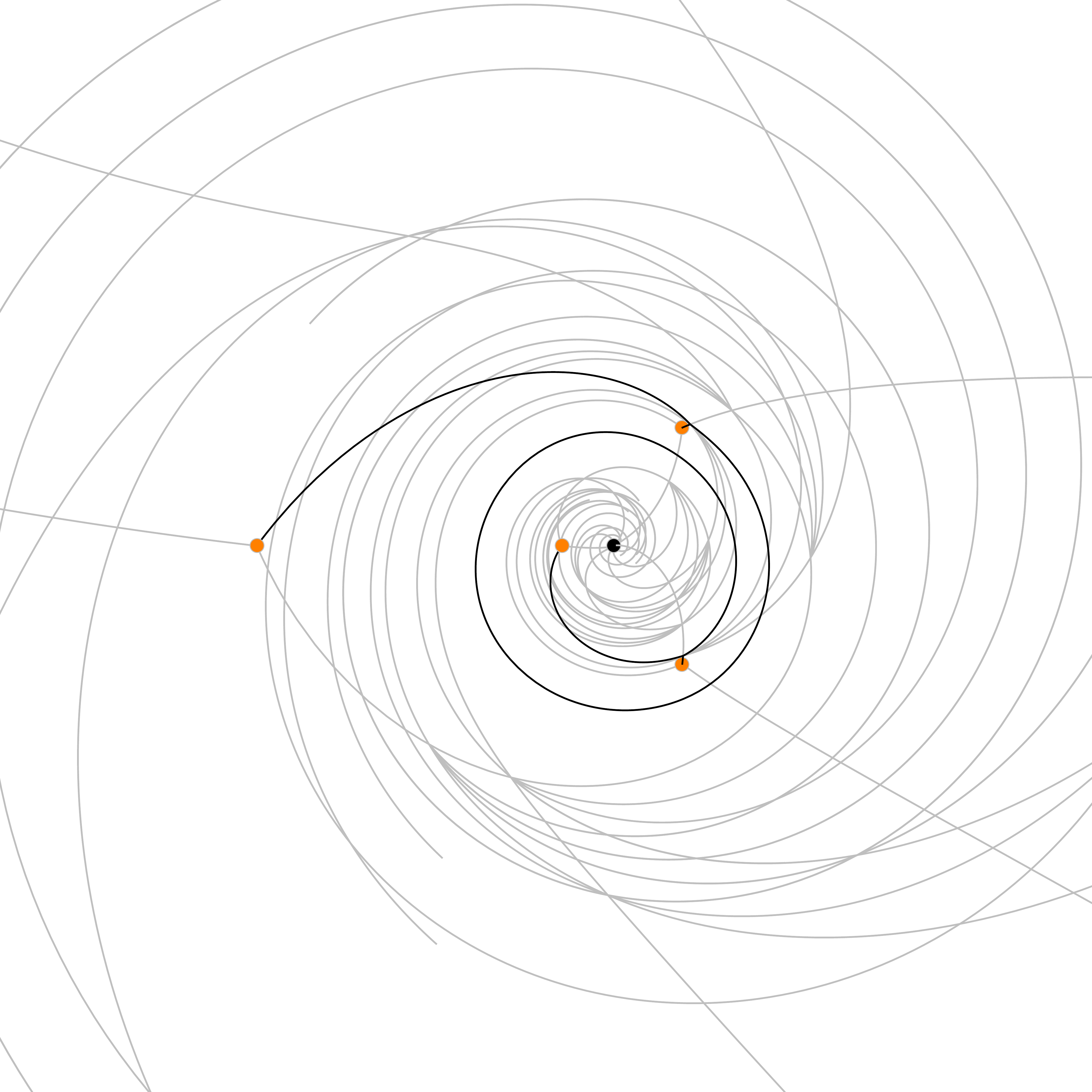}
\caption{Saddle of BPS state with charge $\gamma = (0,0,2,3)$, corresponding to $2D0$-$\overline{D2}_b$-$2\overline{D2}_f$-${D4}$. This is a slight variant of a saddle of Type-0 (see e.g. \cite[Figure 35]{Gaiotto:2012rg}), so $\Omega(\gamma)=1$ corresponding to a hypermultiplet.}
\label{fig:0023}
\end{center}
\end{figure}

\begin{figure}[h!]
\begin{center}
\includegraphics[width=0.85\textwidth]{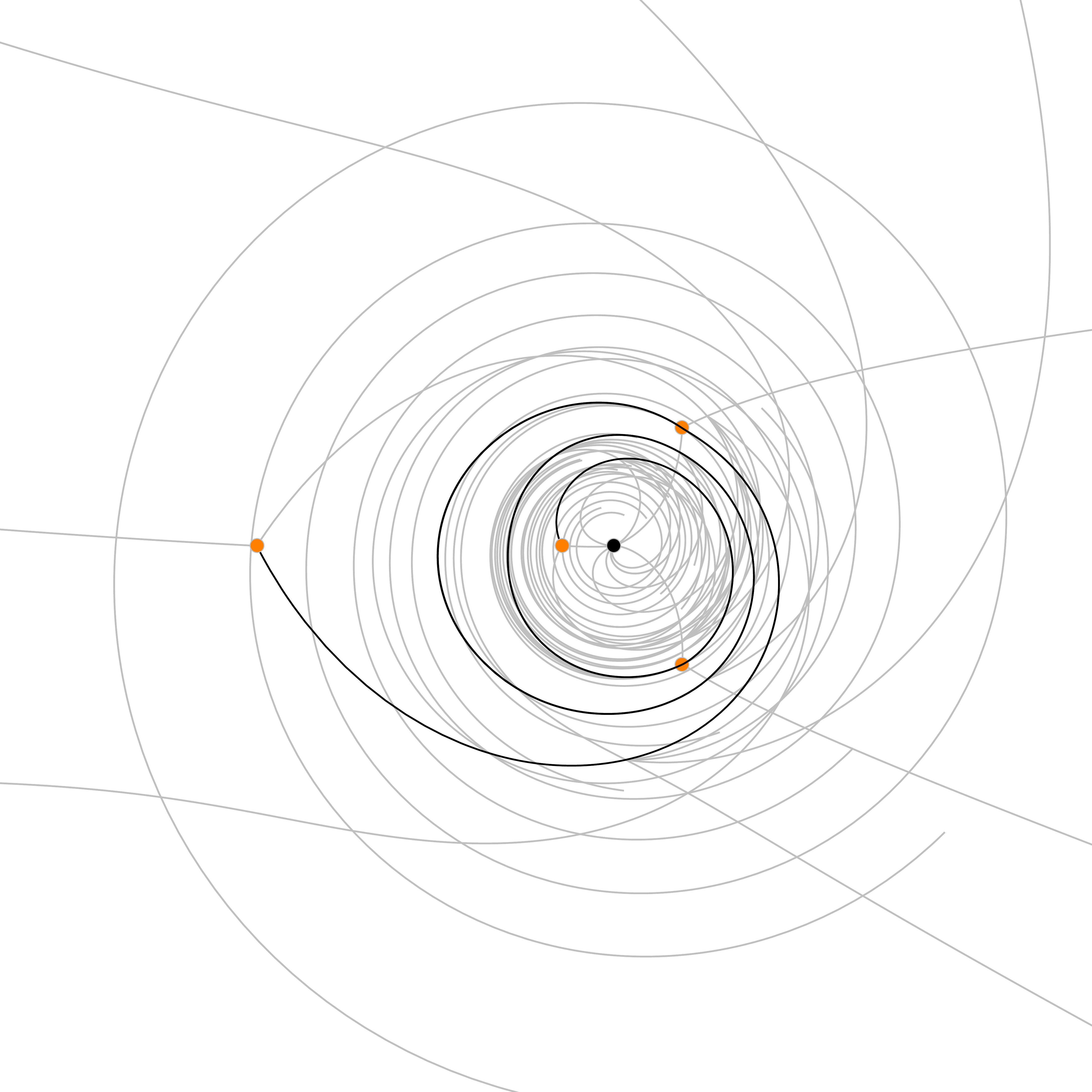}
\caption{Saddle of BPS state with charge $\gamma = (0,0,3,2)$, corresponding to $3D0$-${D2}_b$-$3\overline{D2}_f$-$\overline{D4}$. This is a saddle of Type-0, so $\Omega(\gamma)=1$ corresponding to a hypermultiplet.}
\label{fig:0032}
\end{center}
\end{figure}

\begin{figure}[h!]
\begin{center}
\includegraphics[width=0.85\textwidth]{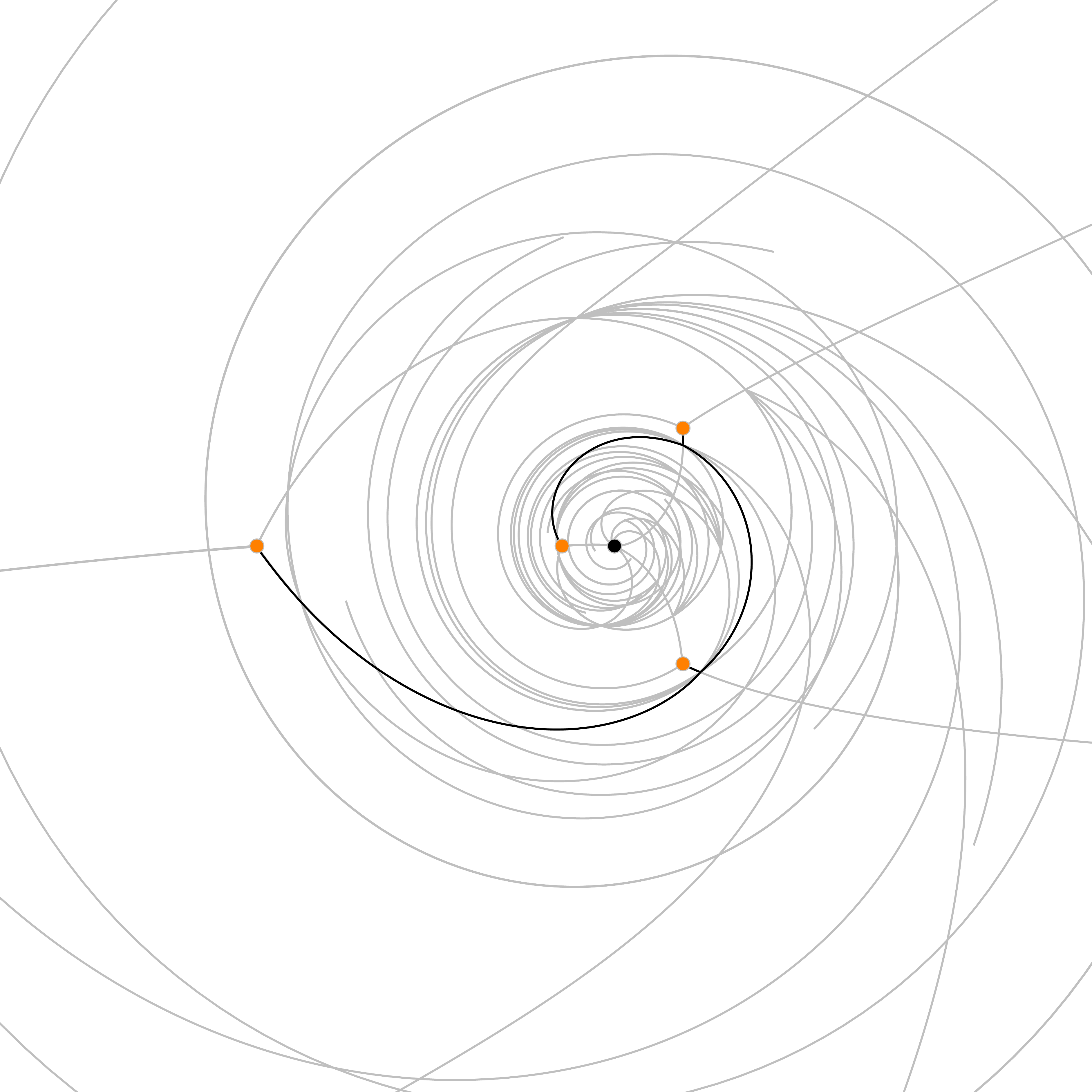}
\caption{Saddle of a BPS state with charge $\gamma=(0,1,1,0)$, corresponding to $D0$-$D2_b$-$2\overline{D4}$.
This is a saddle of Type-1, giving overall $\Omega(\gamma) = -2$, corresponding to a BPS vectormultiplet.}
\label{fig:0110}
\end{center}
\end{figure}

\begin{figure}[h!]
\begin{center}
\includegraphics[width=0.85\textwidth]{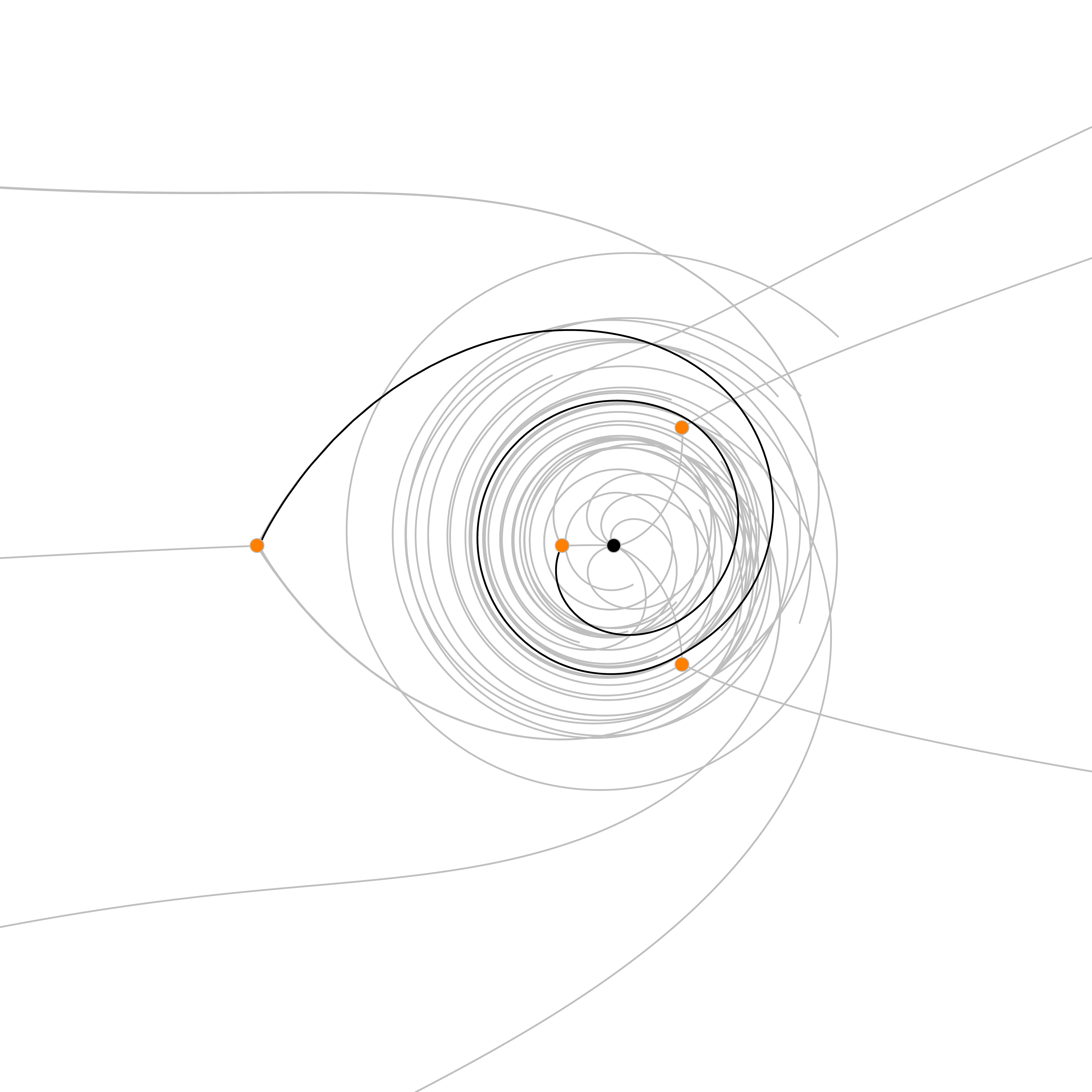}
\caption{Saddle of BPS state with charge $\gamma = (0,4,2,3)$, corresponding to $2D0$-$\overline{D2}_b$-$2{D2}_f$-$3\overline{D4}$. This is a saddle of Type-0, so $\Omega(\gamma)=1$ corresponding to a hypermultiplet.}
\label{fig:0423}
\end{center}
\end{figure}

\begin{figure}[h!]
\begin{center}
\includegraphics[width=0.85\textwidth]{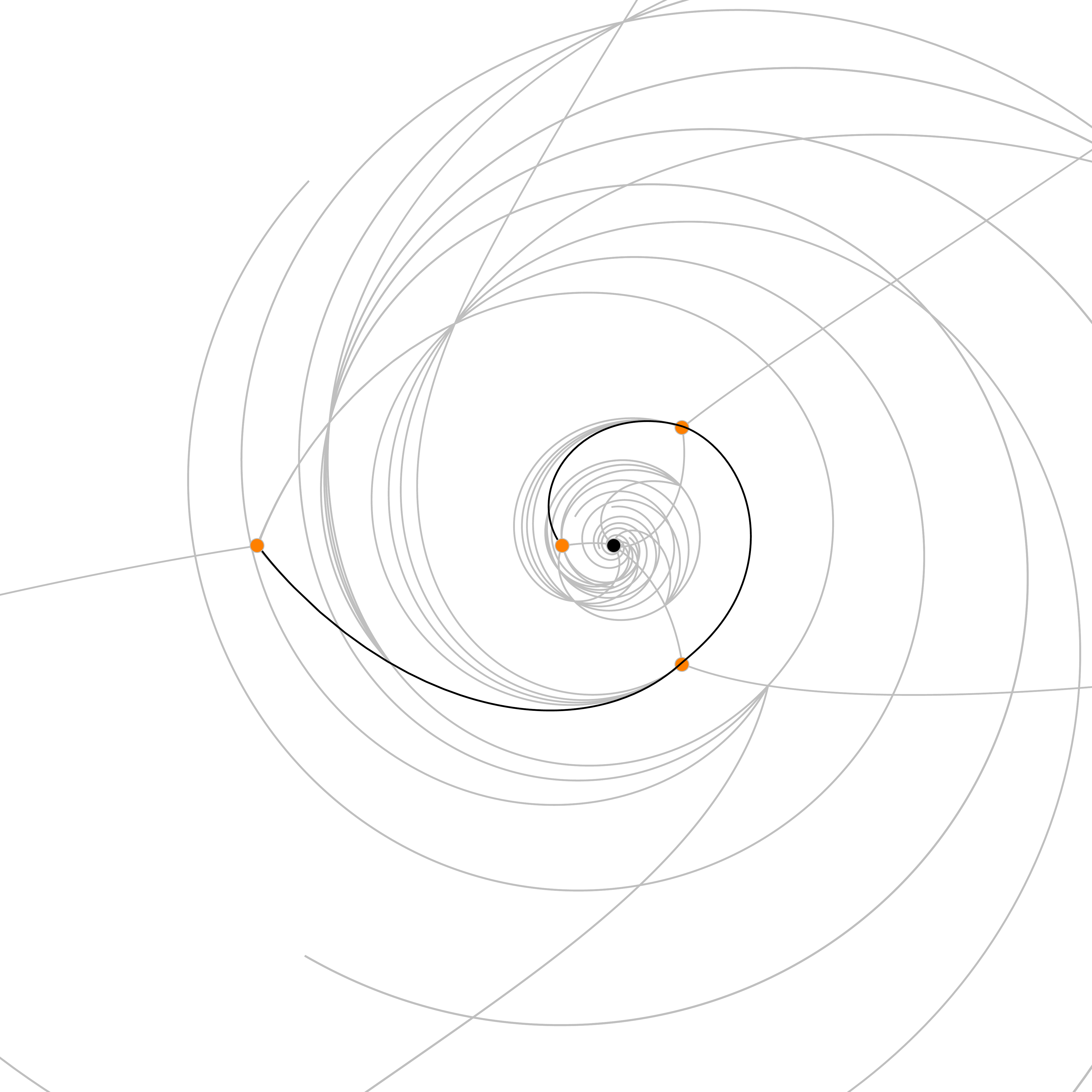}
\caption{Saddle of BPS state with charge $\gamma = (0,2,1,0)$, corresponding to $D0$-$D2_b$-${D2}_f$-$3\overline{D4}$. This is a saddle of Type-0, so $\Omega(\gamma)=1$ corresponding to a hypermultiplet.}
\label{fig:0210}
\end{center}
\end{figure}

\begin{figure}[h!]
\begin{center}
\includegraphics[width=0.85\textwidth]{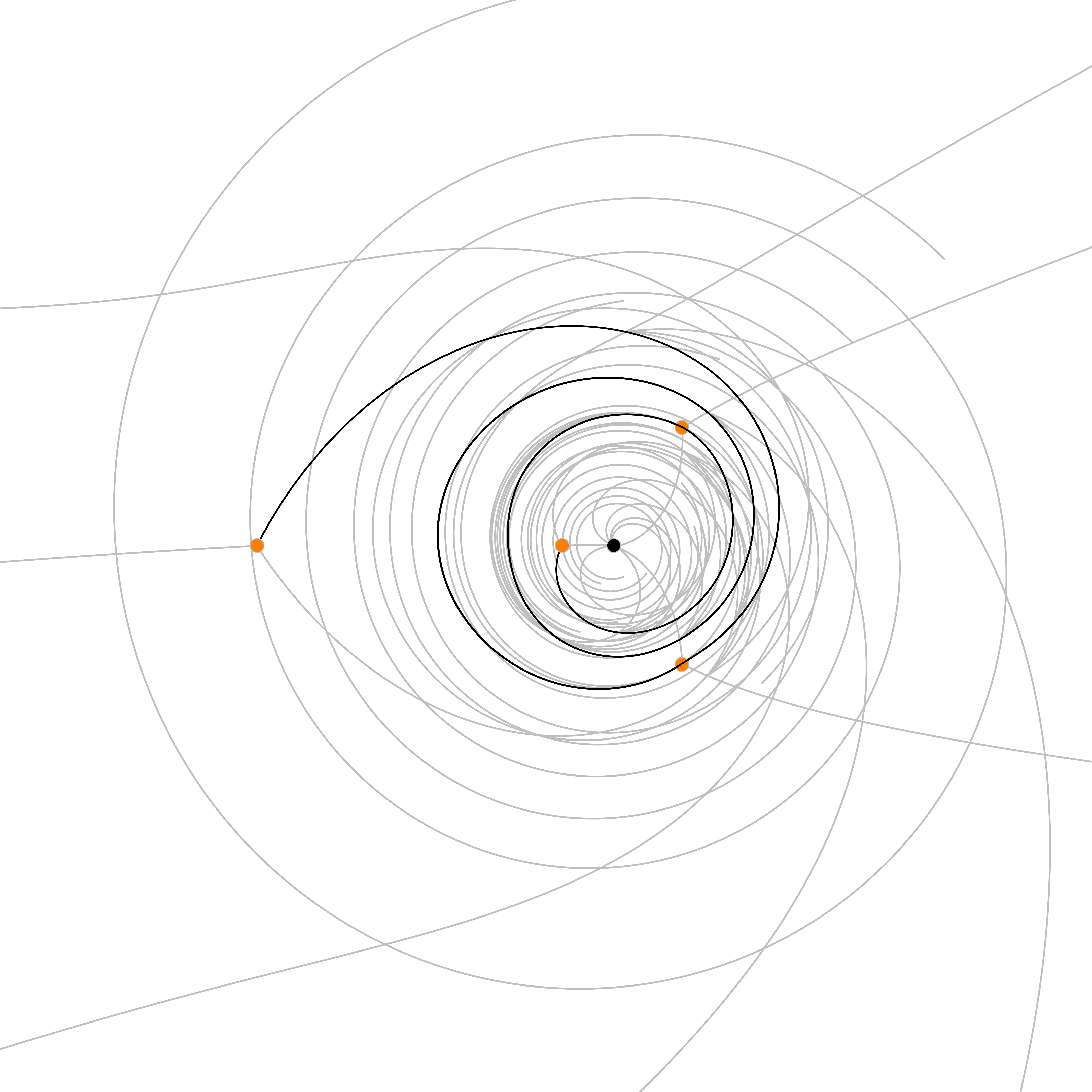}
\caption{Saddle of BPS state with charge $\gamma = (0,6,3,4)$, corresponding to $3D0$-$\overline{D2}_b$-$3{D2}_f$-$5\overline{D4}$. This is a saddle of Type-0, so $\Omega(\gamma)=1$ corresponding to a hypermultiplet.}
\label{fig:0634}
\end{center}
\end{figure}

\begin{figure}[h!]
\begin{center}
\includegraphics[width=0.85\textwidth]{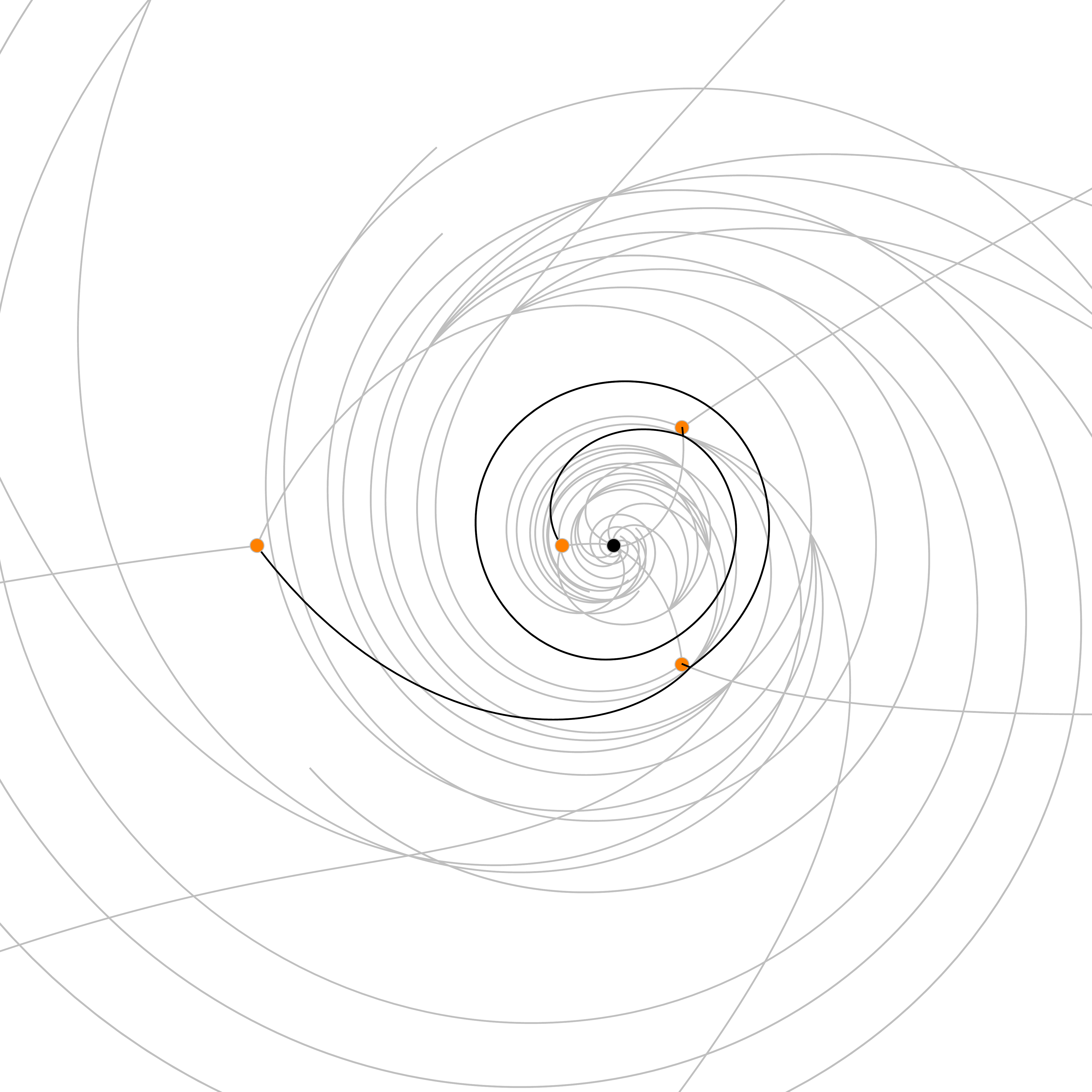}
\caption{Saddle of BPS state with charge $\gamma = (0,3,2,0)$, corresponding to $2D0$-$2D2_b$-${D2}_f$-$5\overline{D4}$.  This is a slight variant of a saddle of Type-0 (see e.g. \cite[Figure 35]{Gaiotto:2012rg}), so $\Omega(\gamma)=1$ corresponding to a hypermultiplet.}
\label{fig:0320}
\end{center}
\end{figure}

\begin{figure}[h!]
\begin{center}
\includegraphics[width=0.85\textwidth]{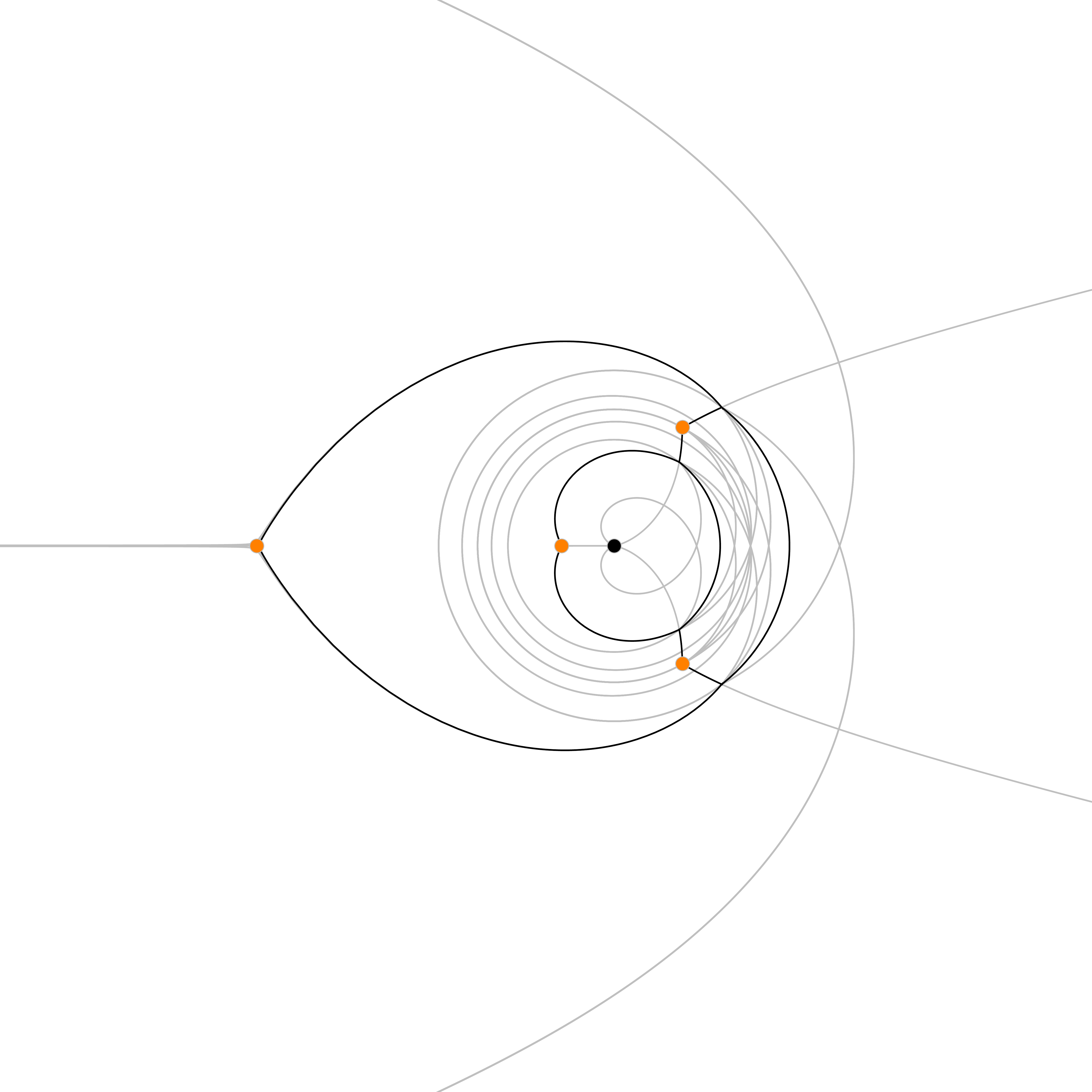}
\caption{
Saddle for the BPS states of charge $n\gamma = (0,n,n,n)$ with $n\geq 1$, corresponding to $n\times (D0$-$\overline{D4})$. This is a saddle of Type-4, therefore $\Omega(\gamma) = 4$. Due to the special choice of moduli $(Q_b=-1, Q_f=1)$, central charges satisfy the relation  $Z_{\gamma_2}+Z_{\gamma_4}=0$, and for this reason this saddle appears at the same exact phase of $\gamma_1,\gamma_2,\gamma_3$ (and many other saddles), making it hard to define from this picture alone To properly identify this saddle, we worked at the perturbed point $Q_b=-0.9-0.1 i, Q_f = 1.1 + 0.1 i$ where this becomes well-distinguished from other saddles.}
\label{fig:0111}
\end{center}
\end{figure}

\begin{figure}[h!]
\begin{center}
\includegraphics[width=0.85\textwidth]{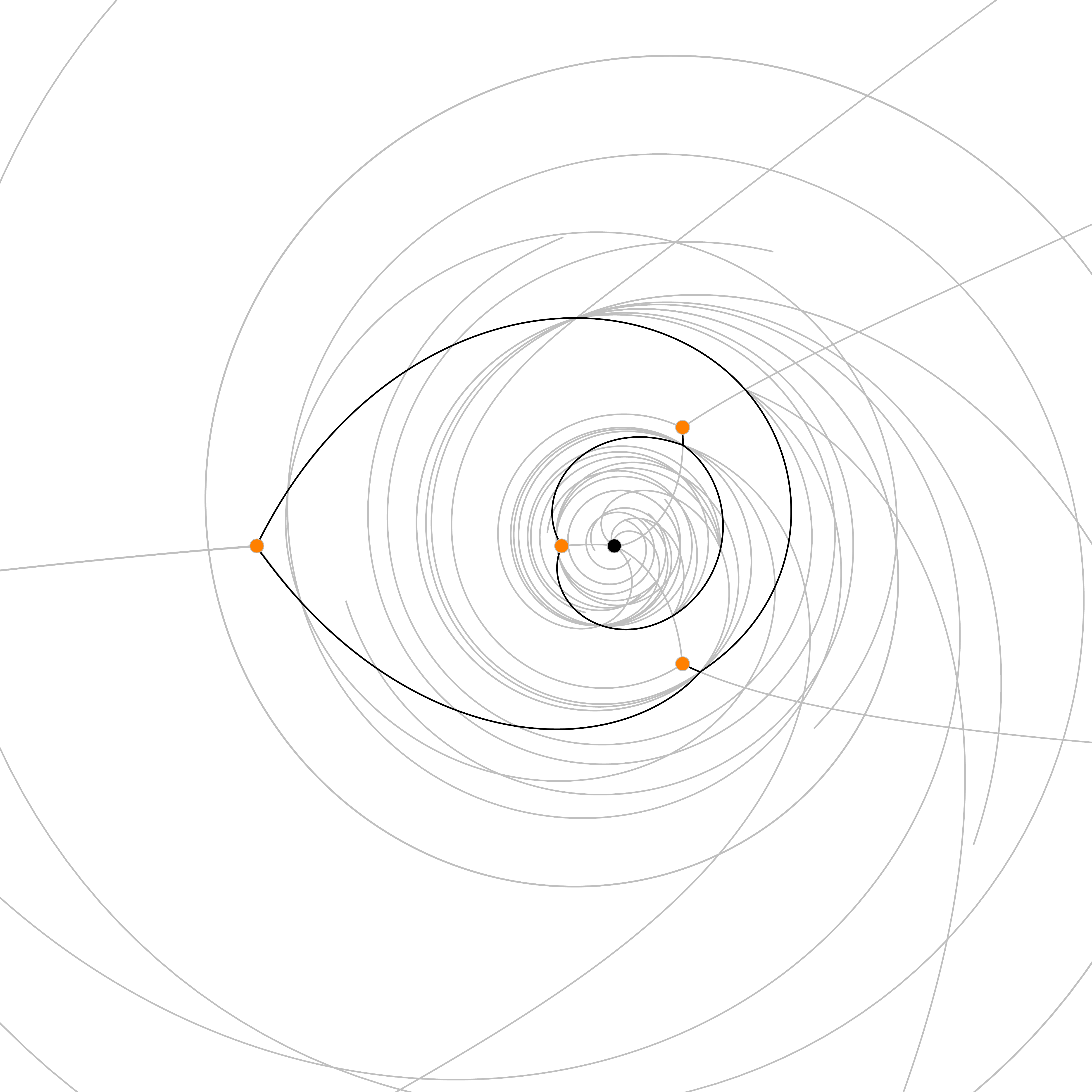}
\caption{Saddle of a BPS state with charge $\gamma=(0,2,1,1)$, corresponding to $D0$-$D2_f$-$2\overline{D4}$.
There are two saddles of Type-2, giving overall $\Omega(\gamma) = -2$, corresponding to a BPS vectormultiplet.}
\label{fig:0211}
\end{center}
\end{figure}

\begin{figure}[h!]
\begin{center}
\includegraphics[width=0.85\textwidth]{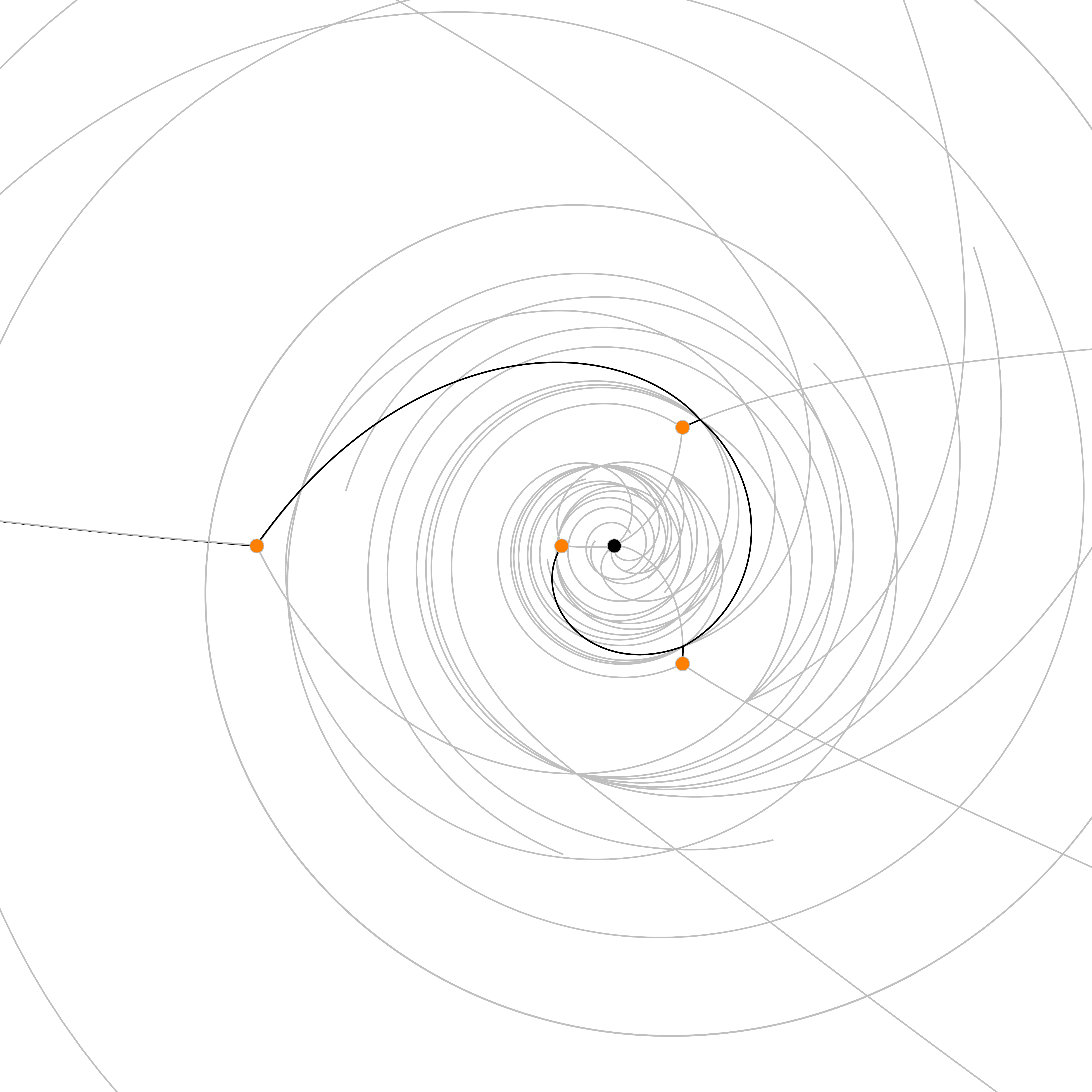}
\caption{Saddle of a BPS state with charge $\gamma=(0,1,1,2)$, corresponding to $D0$-$\overline{D2}_b$.
This is a saddle of Type-1, giving overall $\Omega(\gamma) = -2$, corresponding to a BPS vectormultiplet.}
\label{fig:0112}
\end{center}
\end{figure}

\begin{figure}[h!]
\begin{center}
\includegraphics[width=0.85\textwidth]{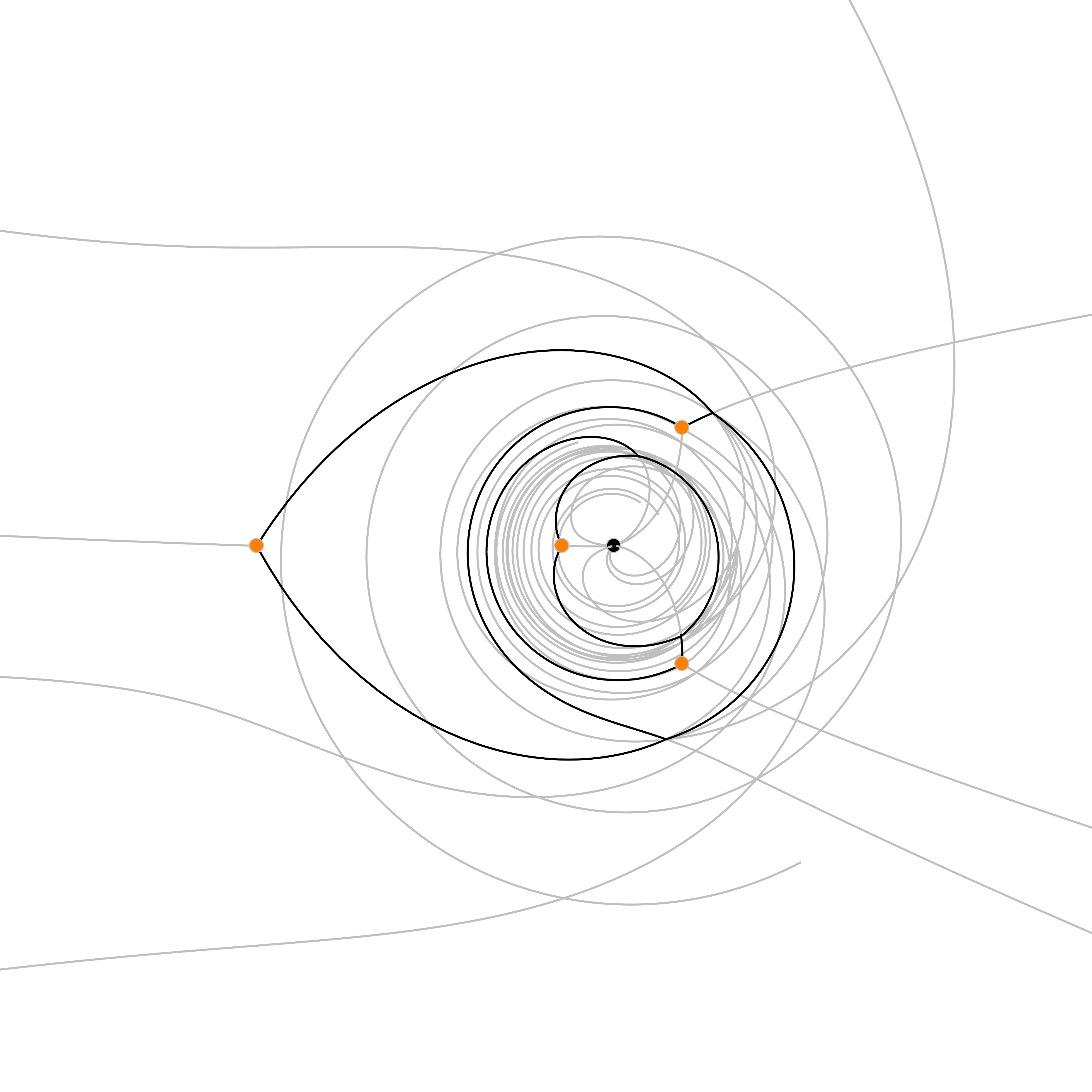}
\caption{Saddle of a BPS state with charge $n\gamma=(n,n,2n,2n)$ for $n\geq 1$, corresponding to $n\times (2 D0$-$\overline{D2}_f)$.
There are two saddles of Type-3, giving overall $\Omega(\gamma) = -2$, corresponding to a BPS vectormultiplet.}
\label{fig:1122}
\end{center}
\end{figure}

\cleardoublepage

\section{Soliton equations for selected BPS saddles}

\subsection{Type-1}\label{app:type-1-saddle}

\begin{figure}[h!]
\begin{center}
\includegraphics[width=0.50\textwidth]{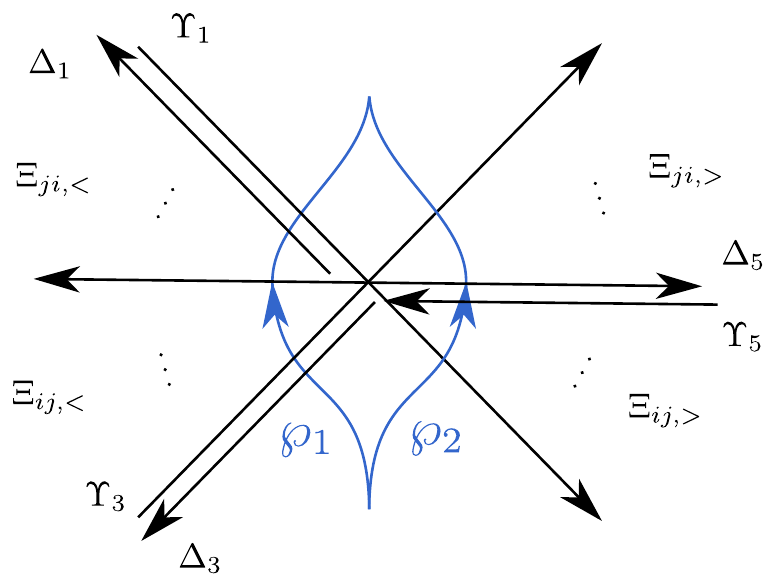}
\caption{Junction $J_1$ in the saddle of Type-1.}
\label{fig:type-1-saddle-junction-J1}
\end{center}
\end{figure}

Here we study the junctions appearing in the Type-1 saddle.
Junction $J_1$ is depicted in Figure \ref{fig:type-1-saddle-junction-J1}. There are three double-walls, and infinitely many one-way walls.

The soliton labels for each wall are as follows
\be
\begin{array}{c|c|c|c|c|c}
\Upsilon_1 & \Delta_1 & \Upsilon_3 & \Delta_3 & \Upsilon_5^{(ii,-n)} /\Upsilon_5^{(jj,-n)}
& \Delta_5^{(ii,n)} /\Delta_5^{(jj,n)}%
\\
\hline
(ij,0) & (ji,0) & (ji,1) & (ij,-1) & (ii,-n) /  (jj,-n) & (ii,n)/(jj,n)
\end{array}
\ee
We consider paths $\wp_1, \wp_2$ across the junction, as shown in Figure \ref{fig:type-4-saddle-junction-J1}.
To write down the parallel transport along $\wp_1$ let us introduce the following conventions on notation. We will work with two-dimensional matrices, with index 1 corresponding to sheet $i$, and index 2 corresponding to sheet $j$. 
We will keep track of the logarithmic shift carried by a soliton by introducing an auxiliary variable $z$. For example, the transition function associated to a wall of type $(ij,n)$ would be
\be
	e^{\Xi_{ij,n}} = 1+ \Xi_{ij,n} \to \left(\begin{array}{cc}
	1 & z^n \Xi_{ij,n} \\
	0 & 1
	\end{array}\right)
\ee
The transport $F(\wp_1)$ then has the following form
\be
\begin{split}
	F(\wp_1)
	& = e^{\Delta_3}  e^{-\Upsilon_3} 
	\left[  \prod_{k\geq 1} e^{\Xi_{ij,-(k+1)}}  \right]
	\left[  \prod_{k\geq 1} e^{\Xi_{ii,-k}} e^{\Xi_{jj,-k}}  \right]
	\left[  \prod_{k\geq 1} e^{\Xi_{ji,-k}}  \right]
	e^{\Delta_1} e^{-\Upsilon_1}
	\\
	& = 
	\left(\begin{array}{cc}
	1-\Delta_3\Upsilon_3 & z^{-1}\Delta_3 \\
	-z\Upsilon_3 & 1 
	\end{array}\right)
	\left(\begin{array}{cc}
	1 & \Xi_{ij,<}(z) \\
	0 & 1
	\end{array}\right)
	\left(\begin{array}{cc}
	S_{ii,<}(z)  & 0 \\
	0 & S_{jj,<}(z)
	\end{array}\right)
	\\
	&\times
	\left(\begin{array}{cc}
	1 & 0 \\
	\Xi_{ji,<}(z) & 1
	\end{array}\right)
	\left(\begin{array}{cc}
	1 & -\Upsilon_1\\
	\Delta_1 & 1 - \Delta_1\Upsilon_1
	\end{array}\right)
\end{split}
\ee
where we introduced
\be
\begin{split}
	&
	\Xi_{ji,<}(z) = \sum_{k\geq 1} z^{-k} \Xi_{ji,-k}\,,
	\qquad
	\Xi_{ij,<}(z) = \sum_{k\geq 1} z^{-(k+1)} \Xi_{ij,-(k+1)}\,,
	\\
	&
	S_{ii,<}(z) = \prod_{k\geq 1} e^{z^{-k}\Xi_{ii, -k}} \,,
	\qquad\qquad\quad\ 
	S_{jj,<}(z) = \prod_{k\geq 1}e^{z^{-k}\Xi_{jj, -k}}\,.
\end{split}
\ee
A bit of algebra yields the following expressions for components of the transport matrix
\be
\begin{split}
	F(\wp_1)_{ii}(z)
	& = (1-\Delta_3\Upsilon_3) 
	S_{ii,<}(z)
	\\
	& + \left[z^{-1}\Delta_3 + (1-\Delta_3\Upsilon_3) \Xi_{ij,<}(z)\right] 
	S_{jj,<}(z)
	\left[ \Delta_1  + \Xi_{ji,<} \right]
	\\
	F(\wp_1)_{ij}(z)
	& = 
	\left[z^{-1}\Delta_3 + (1-\Delta_3\Upsilon_3) \Xi_{ij,<}(z)\right] 
	S_{jj,<}(z)
	\left[
	1
	-
	\(  \Delta_1 + \Xi_{ji,<}(z) \) \, \Upsilon_1
	\right]
	\\
	&
	- (1-\Delta_3\Upsilon_3) \, 
	S_{ii,<}(z)
	\, \Upsilon_1
	\\
	F(\wp_1)_{ji}(z)
	& =  
	-z\Upsilon_3 
	S_{ii,<}(z)
	\\
	& + \(1-z\Upsilon_3 \Xi_{ij,<}(z)\) \, 
	S_{jj,<}(z)
	\(\Delta_1 + \Xi_{ji,<}(z)\) 
	\\
	F(\wp_1)_{jj}(z)
	& = 
	z\Upsilon_3 
	S_{ii,<}(z)
	\, \Upsilon_1 
	\\
	&+ \(1-z\Upsilon_3 \Xi_{ij,<}(z)\)  
	S_{jj,<}(z)
	\left[
	1-\(\Delta_1 + \Xi_{ji,>}(z)\) \Upsilon_1
	\right]
\end{split}
\ee

Similarly, the formal parallel transport along $\wp_2$ is
\be
\begin{split}
	F(\wp_2)
	& = 
	\left[  \prod_{k\geq 0} e^{-\Xi_{ij,k}}  \right]
	\left[  \prod_{k\geq 1} e^{\Upsilon_5^{(ii, -k)}} e^{\Upsilon_5^{(jj,-k)}}  \right]
	\left[  \prod_{k\geq 1} e^{-\Delta_5^{(ii, k)}} e^{-\Delta_5^{(jj,k)}}  \right]
	\left[  \prod_{k\geq 0} e^{-\Xi_{ji,k+1}}  \right]
	\\
	& = 
	\left(\begin{array}{cc}
	1 & -\Xi_{ij,>}(z)  \\
	0 & 1
	\end{array}\right)
	\left(\begin{array}{cc}
	S^{p_5}_{ii,<}(z) S^{p_5}_{ii,>}(z)^{-1} & 0 \\
	0 & S^{p_5}_{jj,<}(z) S^{p_5}_{jj,>}(z)^{-1}
	\end{array}\right)
	\left(\begin{array}{cc}
	1 & 0 \\
	-\Xi_{ji,>}(z)  & 1
	\end{array}\right)
\end{split}
\ee
where we introduced 
\be\label{eq:dict-1-type-1}
\begin{split}
	&
	\Xi_{ji,>}(z) = \sum_{k\geq 0} z^{k+1} \Xi_{ji,k+1}\,,
	\qquad\qquad\quad\ \ \ 
	\Xi_{ij,>}(z) = \sum_{k\geq 0} z^{k} \Xi_{ij,k}\,,
	\\
	&
	S^{p_5}_{ii,>}(z) = \prod_{k\geq 1} e^{z^{k}\Delta_5^{(ii,k)}} \,,
	\qquad\qquad\quad\,
	S^{p_5}_{jj,>}(z) = \prod_{k\geq 1} e^{z^{k}\Delta_5^{(jj,k)}} \,,
	\\
	&
	S^{p_5}_{ii,<}(z) = \prod_{k\geq 1} e^{z^{-k}\Upsilon_5^{(ii,-k)}} \,,
	\qquad\qquad\quad\,
	S^{p_5}_{jj,<}(z) = \prod_{k\geq 1} e^{z^{-k}\Upsilon_5^{(jj,-k)}} \,.
\end{split}
\ee
Components of the transport matrix are as follows
\be
\begin{split}
	F(\wp_2)_{ii}(z)
	& = 
	S^{p_5}_{ii,<}(z)S^{p_5}_{ii,>}(z)^{-1}
	+ 
	\Xi_{ij,>}(z) 
	S^{p_5}_{jj,<}(z)S^{p_5}_{jj,>}(z)^{-1}
	\Xi_{ji,>}(z) 
	\\
	F(\wp_2)_{ij}(z)
	& = 
	-\Xi_{ij,>}(z) 
	S^{p_5}_{jj,<}(z)S^{p_5}_{jj,>}(z)^{-1}
	\\
	F(\wp_2)_{ji}(z)
	& =  
	-S^{p_5}_{jj,<}(z)S^{p_5}_{jj,>}(z)^{-1}
	\Xi_{ji,>}(z) 
	\\
	F(\wp_2)_{jj}(z)
	& = 
	S^{p_5}_{jj,<}(z)S^{p_5}_{jj,>}(z)^{-1}
\end{split}
\ee

Now studying the equations $F(\wp_1)=F(\wp_2)$ for each matrix element, and term by term in $z$, yields the generating functions of outgoing solitons in terms of those of incoming ones.

Without loss of generality, we introduce $\Theta, \overline\Theta$ as follows\footnote{\label{eq:foot-reparam-inverse}This parametrization appears to be the inverse of the one used in \cite[Sec. 3.3]{Banerjee:2018syt}. 
The difference can be tracked to the choice of sign rule in (\ref{eq:sign-rule}): repeating the computation of \cite{Banerjee:2018syt} with this rule would give (\ref{eq:ii-jj-param}). This boils down to a change of conventions for the signs of soliton degeneracies, since $\log S_{ii/jj,<}$ is a linear combination of them. See also footnote \ref{foot:signs-S-ii}. 
}
\be\label{eq:ii-jj-param-type-1}
	S_{ii,<}^{p_5}(z) = \frac{1}{1 + z^{-1}\Theta} \,,
	\qquad
	S_{jj,<}^{p_5}(z) = {1+ z^{-1}\overline\Theta} \,.
\ee
These generating functions are formally inverses of each other, as should be expected by the fact that all transport matrices have unit determinant. There is thus no loss of generality in expressing the transport in this way. The notation chose here is inspired to reflect our earlier work on the $ij-ji$ junction in \cite[Section 3]{Banerjee:2018syt}.

After a bit of algebra\footnote{\label{foot:insuff-eqs}The equations for $\wp_1,\wp_2$ as written above seem to be insufficient. We actually used a resolution of the network, and chose several more refined choices of paths. This allowed us to focus separately on certain sub-junctions that appear in the resolution, individually. This gives far more equations than the whole big junction at once. It makes the problem solvable, at the price of introducing more equations.}, we obtain the following solution
\be\label{eq:full-solution-type-1-saddle}
\begin{split}
	\Delta_1 
	& = - (2 - \overline\Theta\Upsilon_3\Upsilon_1) Q(p)  \, \overline\Theta\Upsilon_3
	\\
	\Delta_3
	& = - \Upsilon_1\overline\Theta\,  (2 - \overline\Theta\Upsilon_3\Upsilon_1) Q(p)  
	\\
	\Xi_{ij,>}(z)
	&=  \Upsilon_1 \frac{1}{1+z \Upsilon_3\Upsilon_1}  Q(p) \,,\\
	\Xi_{ji,>}(z) 
	& = \frac{1}{1+z\Upsilon_3\Upsilon_1}\,z\Upsilon_3\,,\\
	S^{p_5}_{ii,>} 
	& = \frac{1}{1+ z\Upsilon_1\Upsilon_3}\,,
	\\
	S^{p_5}_{jj,>} 
	& = 1 + z\Upsilon_3\Upsilon_1\,,\\
	\Xi_{ij,<}(z)
	&=  
	 \, z^{-2}\Upsilon_1 \overline\Theta ^2 \frac{
			3  
	  		+ 2 z^{-1} \overline\Theta  (1- z \Upsilon_3 \Upsilon_1)
			-  z^{-1} \,\overline\Theta ^2  \Upsilon_3 \Upsilon_1
		}{
			1
			+ 2 z^{-1} \,\overline\Theta  (1-z\, \Upsilon_3 \Upsilon_1)
			+ z^{-2} \, \overline\Theta ^2 \left(1  -z \, \Upsilon_3 \Upsilon_1+ z^{2} (\Upsilon_3 \Upsilon_1)^2\right)
		}
	\\
	\Xi_{ji,<}(z)
	& =
	 \frac{
			3  
	  		+ 2 z^{-1} \overline\Theta  (1- z \Upsilon_3 \Upsilon_1)
			-  z^{-1} \,\overline\Theta ^2  \Upsilon_3 \Upsilon_1
		}{
			1
			+ 2 z^{-1} \,\overline\Theta  (1-z\, \Upsilon_3 \Upsilon_1)
			+ z^{-2} \, \overline\Theta ^2 \left(1  -z \, \Upsilon_3 \Upsilon_1+ z^{2} (\Upsilon_3 \Upsilon_1)^2\right)
		}
		  \, z^{-1}\, \overline\Theta ^2 \Upsilon_3 \cdot Q(p)
	\\
	S_{ii,<}(z)
	& = 
	\frac{
		1 + z^{-1}\Theta 
	}{
		1
		+ 2 z^{-1} \,\Theta  (1-z\, \Upsilon_3 \Upsilon_1)
		+ z^{-2} \, \Theta ^2 \left(1  -z \, \Upsilon_3 \Upsilon_1+ z^{2} (\Upsilon_3 \Upsilon_1)^2\right)
	}\, Q(p)^{-1}	
	\\
	S_{jj,<}(z)
	& = 
	\frac{
		1 
		+ 2z^{-1} \overline\Theta  (1- z\,\Upsilon_1 \Upsilon_3)
		+ z^{-2}\overline\Theta ^2 \left(1  - z\Upsilon_1\Upsilon_3 + z(\Upsilon_1 \Upsilon_3)^2\right)
		}{
			1 + z^{-1}\overline\Theta 
		}\, Q(p)
	\\
\end{split}
\ee
where
\be
	Q(p) = Q(p_1) =Q(p_3) = (1-\Delta_1\Upsilon_1) = (1-\Delta_3\Upsilon_3) = (1-\overline\Theta\Upsilon_3\Upsilon_1)^{-2} \,.
\ee

Let us comment on the limiting behavior of this solution. When $\Upsilon_1$ is set to zero, this becomes
\be\label{eq:full-solution-type-1-saddle-specialized-Upsilon1}
\begin{split}
	\Delta_1 
	& = - 2  \, \overline\Theta\Upsilon_3
	\\
	\Delta_3
	& = 0 
	\\
	\Xi_{ij,>}(z)
	&=  0 \,,\\
	\Xi_{ji,>}(z) 
	& = z\Upsilon_3\,,\\
	S^{p_5}_{ii,<} & = 	S^{p_5}_{jj,<}  = 1\\
	\Xi_{ij,<}(z) &=  0
	\\
	\Xi_{ji,<}(z)
	& =
	 \frac{
			3  
	  		+ 2 z^{-1} \overline\Theta  
		}{
			(1 +  z^{-1} \,\overline\Theta )^2
		}
		  \, z^{-1}\, \overline\Theta ^2 \Upsilon_3 \, 
	\\
	S_{ii,<}(z)
	& = (1 + z^{-1} \Theta)^{-1}
	\\
	S_{jj,<}(z)
	& = 1 + z^{-1} \overline\Theta
	\\
\end{split}
\ee
Notice that this agrees with the computation in Appendix \ref{app:ii-ij-junction}.
A similar limit can be checked to hold when $\Upsilon_3$ is set to zero.

One may also set $\Theta=\overline\Theta=0$, this yields
\be\label{eq:full-solution-type-1-saddle-specialized-Theta}
\begin{split}
	\Delta_1 
	& = 0
	\\
	\Delta_3
	& = 0
	\\
	\Xi_{ij,>}(z)
	&=  \Upsilon_1 \frac{1}{1+z \Upsilon_3\Upsilon_1}   \,,\\
	\Xi_{ji,<}(z) 
	& = \frac{1}{1+z\Upsilon_3\Upsilon_1}\, z\Upsilon_3\,,\\
	S^{p_5}_{ii,>} 
	& = \frac{1}{1+ z \Upsilon_1\Upsilon_3}\,,
	\\
	S^{p_5}_{jj,>} 
	& = 1 + z\Upsilon_3\Upsilon_1\,,\\
	\Xi_{ij,>}(z)
	&=  
	0
	\\
	\Xi_{ji,>}(z)
	& =
	0
	\\
	S_{ii,>}(z)
	& = 
	1
	\\
	S_{jj,>}(z)
	& = 
	1
	\\
\end{split}
\ee
recovering exactly the descendant wall structure of the $ij-ji$ junction \cite[Section 3]{Banerjee:2018syt}.

\subsection{Type-2} \label{app:saddle-type-2-eqs}
\begin{figure}[h!]
\begin{center}
\includegraphics[width=0.45\textwidth]{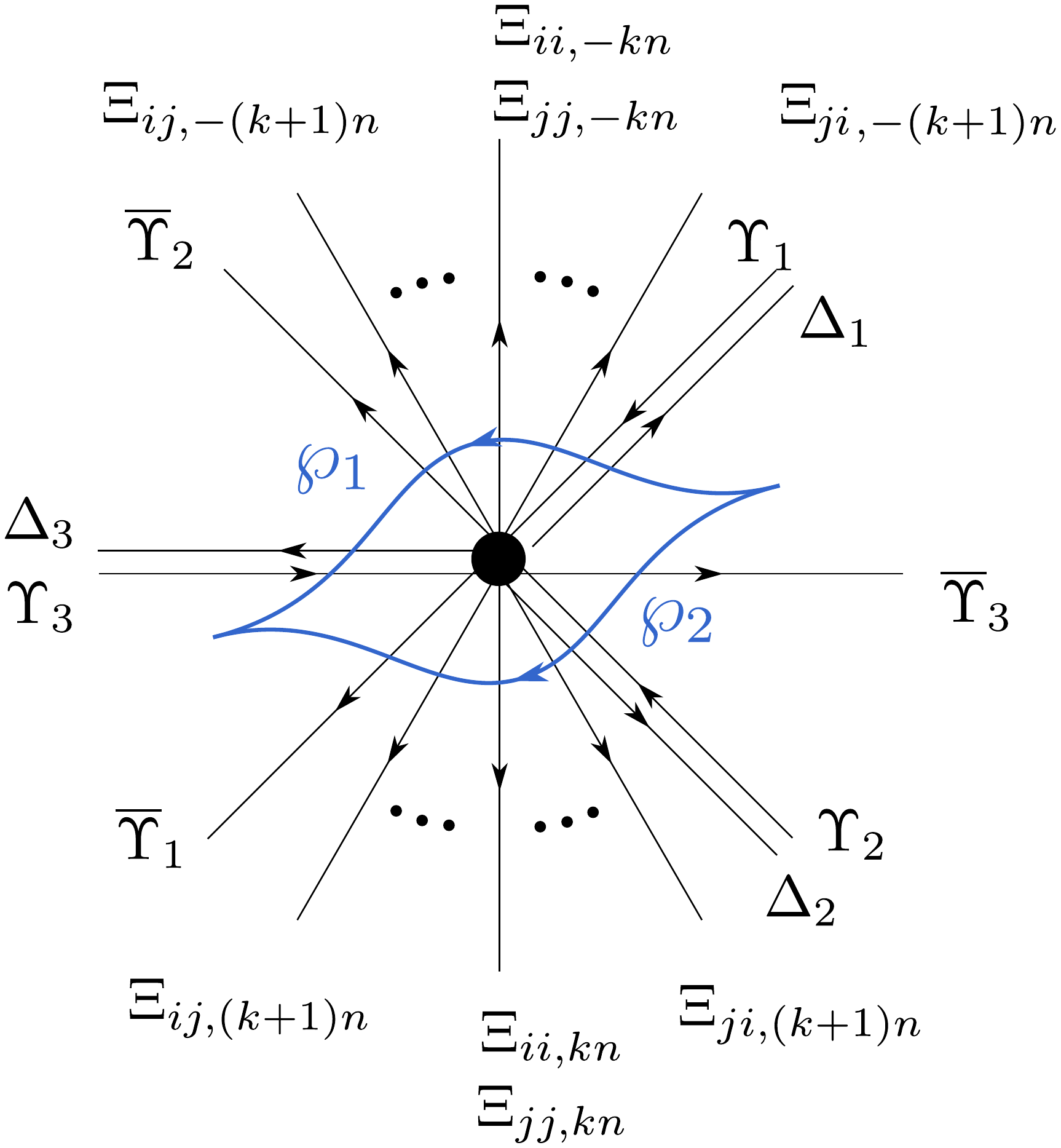}
\caption{}
\label{fig:type-2-saddle-junction}
\end{center}
\end{figure}

Here we work out the soliton equations for the junction appearing in the Type-2 saddle, depicted in Figure \ref{fig:type-2-saddle-junction}. There are three double-walls, and infinitely many one-way walls.

The soliton labels for each wall are as follows
\be
\begin{array}{c|c|c|c|c|c}
\Upsilon_1 & \Delta_1 & \Upsilon_2 & \Delta_2 & \Upsilon_3 & \Delta_3 \\
\hline
(ij,n) & (ji,-n) & (ij,-n) & (ji,n) & (ji,0) & (ij,0)
\end{array}
\ee
We consider paths $\wp_1, \wp_2$ across the junction, as shown in Figure \ref{fig:type-2-saddle-junction}.
As above, we adopt matrix notation to keep track of finite indices $i,j$ and of a formal variable $z$ to keep track of logarithmic indices.
We can then write $F(\wp_1)$ as follows
\be
\begin{split}
	F(\wp_1)
	& = e^{-\Delta_1}  e^{\Upsilon_1} 
	\left[  \prod_{k\geq 1} e^{-\Xi_{ji,-n(k+1)}}  \right]
	\left[  \prod_{k\geq 1} e^{-\Xi_{ii,-n k}} e^{-\Xi_{jj,-n k}}  \right]
	\left[  \prod_{k\geq 1} e^{-\Xi_{ij,-n(k+1)}}  \right]
	e^{-\overline\Upsilon_2}
	e^{-\Delta_3} e^{\Upsilon_3}
	\\
	& = 
	\left(\begin{array}{cc}
	1 & z^n \Upsilon_1 \\
	- z^{-n}\Delta_1 & 1 - \Delta_1\Upsilon_1
	\end{array}\right)
	\left(\begin{array}{cc}
	1 & 0 \\
	-\Xi_{ji,<}(z) & 1
	\end{array}\right)
	\left(\begin{array}{cc}
	S_{ii,<}(z)^{-1} & 0 \\
	0 & S_{jj,<}(z)^{-1}
	\end{array}\right)
	\\
	&\times
	\left(\begin{array}{cc}
	1 & -\Xi_{ij,<}(z) \\
	0 & 1
	\end{array}\right)
	\left(\begin{array}{cc}
	1 & - z^{-n} \overline\Upsilon_2 \\
	0 & 1
	\end{array}\right)
	\left(\begin{array}{cc}
	1-\Delta_3\Upsilon_3 & -\Delta_3\\
	\Upsilon_3 & 1
	\end{array}\right)
\end{split}
\ee
where we introduced
\be
\begin{split}
	&
	\Xi_{ji,<}(z) = \sum_{k\geq 1} z^{-n(k+1)} \Xi_{ji,-n(k+1)}\,,
	\qquad
	\Xi_{ij,<}(z) = \sum_{k\geq 1} z^{-n(k+1)} \Xi_{ij,-n(k+1)}\,,
	\\
	&
	S_{ii,<}(z) = \prod_{k\geq 1} e^{z^{-n k}\Xi_{ii,-n k}} \,,
	\qquad\qquad\qquad\quad\,
	S_{jj,<}(z) = \prod_{k\geq 1}e^{z^{-nk}\Xi_{jj,-n k}}\,.
\end{split}
\ee
A bit of algebra yields the following expressions for components of the transport matrix
\be
\begin{split}
	F(\wp_1)_{ii}(z)
	& = \left[1- z^n \Upsilon_1 \Xi_{ji,<}(z)\right] \, S_{ii,<}(z)^{-1} \, \left[1-\Delta_3\Upsilon_3 - \(z^{-n} \overline{\Upsilon}_2+ \Xi_{ij,<}(z) \)\Upsilon_3\right] \\
	&+ z^n \Upsilon_1\, S_{jj,<}(z)^{-1}\, \Upsilon_3
	\\
	F(\wp_1)_{ij}(z)
	& = \left[1- z^n \Upsilon_1 \Xi_{ji,<}(z)\right] \, S_{ii,<}(z)^{-1} \, \left[-\Delta_3-  
	\(z^{-n} \overline{\Upsilon}_2+\Xi_{ij,<}(z) \)\right] \\
	&+ z^n \Upsilon_1\, S_{jj,<}(z)^{-1}
	\\
	F(\wp_1)_{ji}(z)
	& =  - \left[z^{-n}\Delta_1 + (1-\Delta_1\Upsilon_1) \Xi_{ji,<}(z)\right] S_{ii,<}(z)^{-1} \left[1 -\Delta_3\Upsilon_3- \(z^{-n}\overline{\Upsilon}_2+ \Xi_{ij,<}(z) \)\Upsilon_3 \right] \\
	&+ \left[1-\Delta_1\Upsilon_1\right]\, S_{jj,<}(z)^{-1} \, \Upsilon_3
	\\
	F(\wp_1)_{jj}(z)
	& =  \left[z^{-n}\Delta_1 + (1-\Delta_1\Upsilon_1) \Xi_{ji,<}(z)\right] S_{ii,<}(z)^{-1} \, \left[\Delta_3 + \(z^{-n} \overline{\Upsilon}_2+ \Xi_{ij,<}(z)\)\right]\\
	& + \left[1-\Delta_1\Upsilon_1\right] S_{jj,<}(z)^{-1}
\end{split}
\ee

Similarly, the formal parallel transport along $\wp_2$ is
\be
\begin{split}
	F(\wp_2)
	& = e^{\overline{\Upsilon}_3} e^{-\Upsilon_2}  e^{\Delta_2} 
	\left[  \prod_{k\geq 1} e^{\Xi_{ji,n(k+1)}}  \right]
	\left[  \prod_{k\geq 1} e^{\Xi_{ii,n k}} e^{\Xi_{jj,n k}}  \right]
	\left[  \prod_{k\geq 1} e^{\Xi_{ij,n(k+1)}}  \right]
	e^{\overline\Upsilon_1}
	\\
	& = 
	\left(\begin{array}{cc}
	1 & 0 \\
	\overline{\Upsilon}_3 & 1
	\end{array}\right)
	\left(\begin{array}{cc}
	1 - \Upsilon_2\Delta_2 & -z^{-n} \Upsilon_2 \\
	z^{n}\Delta_2 & 1 
	\end{array}\right)
	\left(\begin{array}{cc}
	1 & 0 \\
	\Xi_{ji,>}(z) & 1
	\end{array}\right)
	\\
	&\times
	\left(\begin{array}{cc}
	S_{ii,>}(z) & 0 \\
	0 & S_{jj,>}(z)
	\end{array}\right)
	\left(\begin{array}{cc}
	1 & \Xi_{ij,>}(z) \\
	0 & 1
	\end{array}\right)
	\left(\begin{array}{cc}
	1 & z^{n} \overline{\Upsilon}_1 \\
	0 & 1
	\end{array}\right)
\end{split}
\ee
where we introduced 
\be
\begin{split}
	&
	\Xi_{ji,>}(z) = \sum_{k\geq 1} z^{n(k+1)} \Xi_{ji,n(k+1)}\,,
	\qquad
	\Xi_{ij,>}(z) = \sum_{k\geq 1} z^{n(k+1)} \Xi_{ij,n(k+1)}\,,
	\\
	&
	S_{ii,>}(z) = \prod_{k\geq 1} e^{z^{nk}\Xi_{ii,n k}} \,,
	\qquad\qquad\qquad\quad\,
	S_{jj,>}(z) = \prod_{k\geq 1}e^{z^{nk} \Xi_{jj,n k}}\,.
\end{split}
\ee
A bit of algebra yields the following expressions for components of the transport matrix
\be
\begin{split}
	F(\wp_2)_{ii}(z)
	& = \(1-\Upsilon_2\Delta_2\) \, S_{ii,>}(z) - z^{-n}\Upsilon_2\, \Xi_{ji,>}(z) \, S_{ii,>}(z)
	\\
	F(\wp_2)_{ij}(z)
	& = \(1-\Upsilon_2\Delta_2\) \, S_{ii,>}(z)\, \(\Xi_{ij,>}(z) + z^n \overline{\Upsilon}_1\) - z^{-n}\Upsilon_2 \, S_{jj,>}(z)
	\\
	F(\wp_2)_{ji}(z)
	& =  \left[\overline{\Upsilon}_3 \, \(1-\Upsilon_2\Delta_2\) - z^n \Delta_2\right]\,  S_{ii,>}(z) 
	+ \(1-z^{-n} \overline{\Upsilon}_3 \Upsilon_2\)\Xi_{ji,>}(z) \, S_{ii,>}(z)
	\\
	F(\wp_2)_{jj}(z)
	& =  \left[-\overline{\Upsilon}_3 \, \(1-\Upsilon_2\Delta_2\) + z^n \Delta_2\right]\,  S_{ii,>}(z) \, \left[-\Xi_{ij,>}(z) - z^n \overline{\Upsilon}_1\right]
	+ \(1-z^{-n} \overline{\Upsilon}_3 \Upsilon_2\) \, S_{jj,>}(z)
\end{split}
\ee

Now studying the equations $F(\wp_1)=F(\wp_2)$ for each matrix element, and term by term in $z$, yields the generating functions of outgoing solitons in terms of those of incoming ones.
Let us adopt the following definitions
\be
\begin{split}
	S_{ii,<} & = 1 + \sum_{k\geq 1} z^{-kn} S_{ii,-kn}
	\qquad
	S_{ii,>} = 1 + \sum_{k\geq 1} z^{kn} S_{ii,kn}
	\\
	\Xi_{ij,<} & = \sum_{k\geq 2} z^{-kn} \Xi_{ij,-kn}
	\qquad
	\Xi_{ij,>} = \sum_{k\geq 2} z^{kn} \Xi_{ij,kn}
\end{split}
\ee
and similarly with $i \leftrightarrow j$.

Our goal is now to derive expression for $\Delta_1,\Delta_2$ in terms of $\Upsilon_1,\Upsilon_2,\Upsilon_3$. 
To compute $\Delta_1$, let us consider the $(ji,-n)$ equation,  that is the equation $F(\wp_1)_{ji} = F(\wp_2)_{ji}$ restricted to the coefficient of $z^{-n}$:
\be
	\Delta_1 \(1-\Delta_3\Upsilon_3\) - (1-\Delta_1\Upsilon_1) \, S^{-1}_{jj,-n} \, \Upsilon_3 = 0 \,.
\ee
Next let us note that the $(jj,-n)$ equation, which reads
\be
	\Delta_1\Delta_3 +(1-\Delta_1\Upsilon_1) \, S^{-1}_{jj,-n}  =- \overline{\Upsilon}_3\Upsilon_2
\ee
can be multiplied from the right by $\Upsilon_3$, then used to substitute into the $(ji,-n)$ equation to obtain
\be
	\Delta_1  = -\overline{\Upsilon}_3\Upsilon_2\Upsilon_3  \,.
\ee
Now using the $(ji,0)$ equation, which reads
\be
	(1-\Delta_1\Upsilon_1) \Upsilon_3 = \overline{\Upsilon}_3 (1-\Upsilon_2\Delta_2)
\ee
we arrive at
\be
	\Delta_1  = -\Upsilon_3\Upsilon_2\Upsilon_3 \, \frac{1-\Delta_1\Upsilon_1}{1-\Upsilon_2\Delta_2}
\ee
where we used the fact that $Q(p_i) = 1-\Delta_i\Upsilon_i$ can be formally traded for generating series of closed paths, which commute with other soliton generating functions.

Similarly, to obtain $\Delta_2$ we consider the $(ji,n)$ equation 
\be
	-\Delta_2 - \overline{\Upsilon}_3 (1-\Upsilon_2\Delta_2) S_{ii,n} + \overline{\Upsilon}_3\Upsilon_2 \Xi_{ji,2n} = 0\,.
\ee
Combining this with the $(ii,n)$ equation, which reads
\be
	(1-\Upsilon_2\Delta_2) S_{ii,n} - \Upsilon_2 \, \Xi_{ji,2n} = \Upsilon_1\Upsilon_3
\ee
yields
\be
	\Delta_2 = - \overline{\Upsilon}_3 \Upsilon_1 \Upsilon_3 = -\Upsilon_3\Upsilon_1\Upsilon_3 \, \frac{1-\Delta_1\Upsilon_1}{1-\Upsilon_2\Delta_2}
\ee
where we again used the $(ji,0)$ equation already invoked above.
Now from these expressions for $\Delta_1, \Delta_2$ one can immediately see that $\Delta_1\Upsilon_1 = \Upsilon_2\Delta_2$ (upon taking closure of paths), leading to 
\be\label{eq:saddle-type-2-delta-1-delta-2}
	\Delta_1  = -\Upsilon_3\Upsilon_2\Upsilon_3\,,
	\qquad
	\Delta_2  = -\Upsilon_3\Upsilon_1\Upsilon_3\,.
\ee
For later convenience, let us note that this also implies $\overline{\Upsilon}_3 = \Upsilon_3$.
Expressions (\ref{eq:saddle-type-2-delta-1-delta-2}) imply that
\be
	Q(p_1) = 1- \Upsilon_1\Delta_1 =  1- \Upsilon_2\Delta_2 = Q(p_2) \,.
\ee

Let us move on to $\Delta_3$ by considering the $(ij,0)$ equation
\be
	\Delta_3 - \Upsilon_1 \, S_{jj,-n} = \Upsilon_2 \, S_{jj,n}
\ee
Consider the $(jj,-n)$ equation
\be
	\Delta_1\Delta_3 +(1-\Delta_1\Upsilon_1) \, S_{jj,-n}  = -\overline{\Upsilon}_3\Upsilon_2\,,
\ee
solving this for $S_{jj,-n}$ and plugging into the $(ij,0)$ equation above, we obtain
\be\label{eq:Delta3-type2-partial}
	\Delta_3 + \Upsilon_1\overline{\Upsilon}_3\Upsilon_2 = (1-\Upsilon_1\Delta_1) \Upsilon_2 S_{jj,n}\,.
\ee
Now using the $(jj,0)$ equation
\be
		\Delta_1\Upsilon_1 = \overline{\Upsilon}_3\Upsilon_2 S_{jj,n}
\ee
and recalling that $\overline{\Upsilon}_3 = \Upsilon_3$, we can multiply (\ref{eq:Delta3-type2-partial}) from the left by $\Upsilon_3$ to obtain
\be
\begin{split}
	\Upsilon_3\Delta_3 & = - \Upsilon_3 \Upsilon_1\overline{\Upsilon}_3\Upsilon_2 + (1-\Upsilon_1\Delta_1) \Upsilon_3\Upsilon_2 S_{jj,n}
	\\
	& = - \Upsilon_3 \Upsilon_1{\Upsilon}_3\Upsilon_2 + (1-\Upsilon_1\Delta_1) \Delta_1\Upsilon_1
\end{split}
\ee
which can be recast as follows
\be
	1-Q(p_3)  = (1- Q(p_1)) + Q(p_1) (1-Q(p_1)) = 1-Q(p_1)^2 \,,
\ee
implying that 
\be
	Q(p_3) = Q(p_1)^2 = Q(p_2)^2\,.
\ee

\subsection{Type-3}\label{app:saddle-type-3-eqs}

Here we work out the soliton equations for the junctions appearing in the Type-3 saddle.

\subsubsection*{Junction $J_1$} 
\begin{figure}[h!]
\begin{center}
\includegraphics[width=0.75\textwidth]{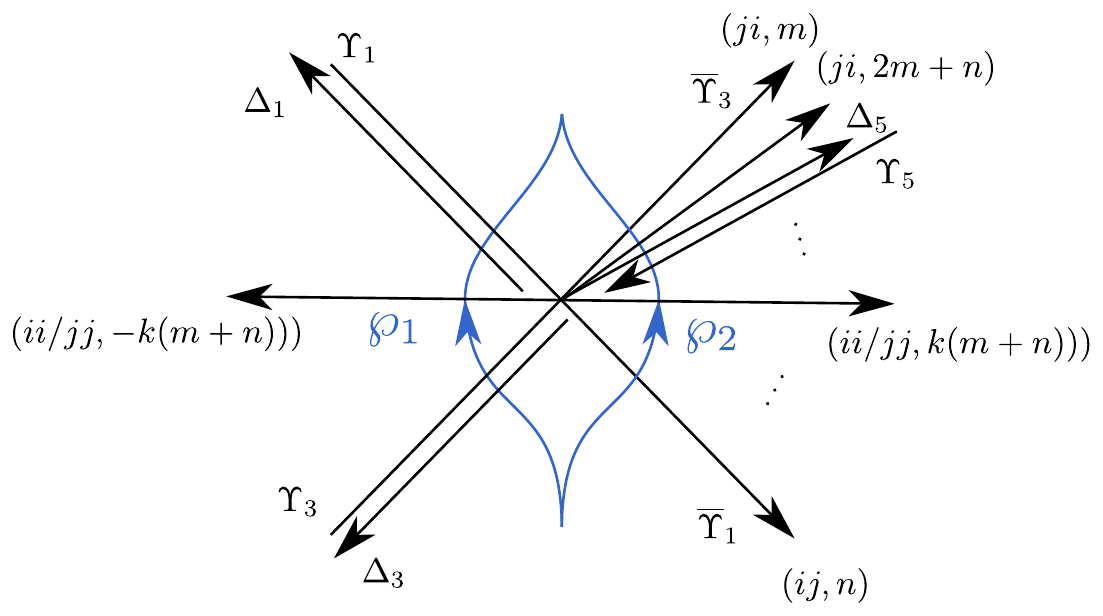}
\caption{Junction $J_1$ in the saddle of Type-3. Here $m=0,n=1$.}
\label{fig:type-3-saddle-junction-J1}
\end{center}
\end{figure}

We begin with junction $J_1$, depicted in Figure \ref{fig:type-3-saddle-junction-J1}. There are three double-walls, and infinitely many one-way walls.

The soliton labels for each wall are as follows
\be
\begin{array}{c|c|c|c|c|c}
\Upsilon_1 & \Delta_1 & \Upsilon_3 & \Delta_3 & \Upsilon_5 & \Delta_5 \\
\hline
(ij,1) & (ji,-1) & (ji,0) & (ij,0) & (ij,-2) & (ji,2)
\end{array}
\ee
We consider paths $\wp_1, \wp_2$ across the junction, as shown in Figure \ref{fig:type-3-saddle-junction-J1}.
Adopting the formal variable $z$ as above to keep track of logarithmic indices, we can write $F(\wp_1)$ as follows
\be
\begin{split}
	F(\wp_1)
	& = e^{\Delta_3}  e^{-\Upsilon_3} 
	\left[  \prod_{k\geq 1} e^{\Xi_{ij,-k}}  \right]
	\left[  \prod_{k\geq 1} e^{\Xi_{ii,-k}} e^{\Xi_{jj,-k}}  \right]
	\left[  \prod_{k\geq 1} e^{\Xi_{ji,-n(k+1)}}  \right]
	e^{\Delta_1} e^{-\Upsilon_1}
	\\
	& = 
	\left(\begin{array}{cc}
	1-\Delta_3\Upsilon_3 & \Delta_3 \\
	-\Upsilon_3 & 1 
	\end{array}\right)
	\left(\begin{array}{cc}
	1 & \Xi_{ij,<}(z) \\
	0 & 1
	\end{array}\right)
	\left(\begin{array}{cc}
	S_{ii,<}(z) & 0 \\
	0 & S_{jj,<}(z)
	\end{array}\right)
	\\
	&\times
	\left(\begin{array}{cc}
	1 & 0 \\
	\Xi_{ji,<}(z) & 1
	\end{array}\right)
	\left(\begin{array}{cc}
	1 & -z\Upsilon_1\\
	z^{-1}\Delta_1 & 1 - \Delta_1\Upsilon_1
	\end{array}\right)
\end{split}
\ee
where we introduced
\be
\begin{split}
	&
	\Xi_{ji,<}(z) = \sum_{k\geq 1} z^{-(k+1)} \Xi_{ji,-(k+1)}\,,
	\qquad
	\Xi_{ij,<}(z) = \sum_{k\geq 1} z^{-k} \Xi_{ij,-k}\,,
	\\
	&
	S_{ii,<}(z) = \prod_{k\geq 1} e^{z^{-k}\Xi_{ii,- k}} \,,
	\qquad\qquad\quad\ 
	S_{jj,<}(z) = \prod_{k\geq 1}e^{z^{-k}\Xi_{jj,- k}}\,.
\end{split}
\ee
A bit of algebra yields the following expressions for components of the transport matrix
\be
\begin{split}
	F(\wp_1)_{ii}(z)
	& = (1-\Delta_3\Upsilon_3) S_{ii,<}(z)
	+ z^{-1} \left[\Delta_3 + (1-\Delta_3\Upsilon_3) \Xi_{ij,<}(z)\right] S_{jj,<}(z) \Delta_1
	\\
	F(\wp_1)_{ij}(z)
	& =  \left[\Delta_3 + (1-\Delta_3\Upsilon_3) \Xi_{ij,<}(z)\right] 
	S_{jj,<}(z)
	\left[
	1
	-
	\( z^{-1} \Delta_1 + \Xi_{ji,<}(z) \) \, z \Upsilon_1
	\right]
	\\
	&
	 (1-\Delta_3\Upsilon_3) \, S_{ii,<}(z)  \, z \Upsilon_1
	\\
	F(\wp_1)_{ji}(z)
	& = - \Upsilon_3 S_{ii,<}(z) + \(1-\Upsilon_3 \Xi_{ij,<}(z)\) \, S_{jj,<}(z) \, 
	\(z^{-1} \Delta_1 + \Xi_{ji,<}(z)\) 
	\\
	F(\wp_1)_{jj}(z)
	& = 
	\Upsilon_3 S_{ii,<}(z) \, z\Upsilon_1 
	+ \(1-\Upsilon_3 \Xi_{ij,<}(z)\)  S_{jj,<}(z)
	\left[
	1-\(z^{-1}\Delta_1 + \Xi_{ji,<}(z)\) z\Upsilon_1
	\right]
\end{split}
\ee

Similarly, the formal parallel transport along $\wp_2$ is
\be
\begin{split}
	F(\wp_2)
	& = e^{-\overline{\Upsilon}_1}  
	\left[  \prod_{k\geq 1} e^{-\Xi_{ij,k+1}}  \right]
	\left[  \prod_{k\geq 1} e^{-\Xi_{ii, k}} e^{-\Xi_{jj,k}}  \right]
	\left[  \prod_{k\geq 3} e^{-\Xi_{ji,k}}  \right]
	e^{\Upsilon_5} e^{-\Delta_5} e^{-\Xi_{ji,1}} e^{-\overline{\Upsilon}_3}
	\\
	& = 
	\left(\begin{array}{cc}
	1 & -z\, \overline{\Upsilon}_1 \\
	0 & 1
	\end{array}\right)
	\left(\begin{array}{cc}
	1 & -\Xi_{ij,>}(z)  \\
	0 & 1
	\end{array}\right)
	\left(\begin{array}{cc}
	S_{ii,>}(z)^{-1} & 0 \\
	0 & S_{jj,>}(z)^{-1}
	\end{array}\right)
	\left(\begin{array}{cc}
	1 & 0 \\
	-\Xi_{ji,>}(z)  & 1
	\end{array}\right)
	\\
	&\times
	\left(\begin{array}{cc}
	1 - \Upsilon_5\Delta_5 & z^{-2}\Upsilon_5 \\
	-z^2\Delta_5 & 1
	\end{array}\right)
	\left(\begin{array}{cc}
	1 & 0 \\
	-z\,\Xi_{ji,1}  & 1
	\end{array}\right)
	\left(\begin{array}{cc}
	1 & 0 \\
	-\overline{\Upsilon}_3  & 1
	\end{array}\right)
\end{split}
\ee
where we introduced 
\be
\begin{split}
	&
	\Xi_{ji,>}(z) = \sum_{k\geq 3} z^{k} \Xi_{ji,k}\,,
	\qquad
	\Xi_{ij,>}(z) = \sum_{k\geq 1} z^{k+1} \Xi_{ij,k+1}\,,
	\\
	&
	S_{ii,>}(z) = \prod_{k\geq 1} e^{z^{nk}\Xi_{ii,n k}} \,,
	\qquad\qquad\qquad\quad\,
	S_{jj,>}(z) = \prod_{k\geq 1}e^{z^{nk} \Xi_{jj,n k}}\,.
\end{split}
\ee
A bit of algebra yields the following expressions for components of the transport matrix
\be
\begin{split}
	F(\wp_2)_{ii}(z)
	& = 
	S_{ii,>}(z)^{-1} \left[
	1-z^{-2} \Upsilon_5\( z^2 \Delta_5  + z\Xi_{ji,1}+\overline{\Upsilon}_3  \)
	\right]
	\\
	& +
	\left[  z\overline{\Upsilon}_1 + \Xi_{ij,>}(z) \right] S_{jj,>}(z)^{-1} 
	\Big[ 
		z^2 \Delta_5 + \overline{\Upsilon}_3 
		\\
		&
		+ \Xi_{ji,>}(z)\,  \(1 - z^{-2}\Upsilon_5\, (z^2\Delta_5 + \overline{\Upsilon}_3 ) \) 
		+ \(  1- \Xi_{ji,>}(z)\, z^{-2}\Upsilon_5  \) z \Xi_{ji,1}
	\Big]
	\\
	F(\wp_2)_{ij}(z)
	& = 
	S_{ii,>}(z)^{-1} \, z^{-2}\Upsilon_5 
	- \left[ z\overline{\Upsilon}_1 + \Xi_{ij,>}(z) \right]S_{jj,>}(z)^{-1} 
	\left[ 1 - \Xi_{ji,>}(z) \,  z^{-2}\Upsilon_5\right]
	\\
	F(\wp_2)_{ji}(z)
	& =  
	- S_{jj,>}(z)^{-1} \, \Big[
	\overline{\Upsilon}_3 + z^2\Delta_5
	+\Xi_{ji,>}(z) ( 1 - \Upsilon_5\Delta_5 - z^{-2} \Upsilon_5\overline{\Upsilon}_3) 
	\\
	&
	+ (1 - \Xi_{ji,>}(z) \, z^{-2}\Upsilon_5)\, z \Xi_{ji,1}
	\Big]
	\\
	F(\wp_2)_{jj}(z)
	& = 
	S_{jj,>}(z)^{-1} \(1 - \Xi_{ji,>}(z) \, z^{-2} \Upsilon_5 \)
\end{split}
\ee

Now studying the equations $F(\wp_1)=F(\wp_2)$ for each matrix element, and term by term in $z$, yields the generating functions of outgoing solitons in terms of those of incoming ones.
Let us adopt the following definitions 
\be
\begin{split}
	S_{ii,<} & = 1 + \sum_{k\geq 1} z^{-k} S_{ii,-k}
	\qquad
	S_{ii,>} = 1 + \sum_{k\geq 1} z^{k} S_{ii,k}
	\\
	\Xi_{ij,<} & = \sum_{k\geq 1} z^{-k} \Xi_{ij,-k}
	\qquad\ \ \ \
	\Xi_{ij,>} = \sum_{k\geq 1} z^{k+1} \Xi_{ij,k+1}
\end{split}
\ee
and similarly with $i \leftrightarrow j$.

Our goal is now to derive expressions for $\Delta_1,\Delta_3, \Delta_5$ in terms of $\Upsilon_1,\Upsilon_3,\Upsilon_5$. 
To start, let us note that the $(ji,0)$ equation reads
\be
	\overline{\Upsilon}_3 = \Upsilon_3 \,.
\ee
To compute $\Delta_1$ we consider the $(ji,-1)$ equation
\be
	\Delta_1 - \Upsilon_3 \, S_{ii,-1} = 0
\ee
To compute $\Delta_3$ we consider the $(ij,0)$ equation
\be
	-\Delta_3(1-\Delta_1\Upsilon_1)  + (1-\Delta_3  \Upsilon_3) S_{ii,-1}\Upsilon_1 = -S_{ii,2}^{-1} \, \Upsilon_5
\ee
To compute $\Delta_5$ we consider the $(ji,2)$ equation
\be\label{eq:type-3-saddle-delta-5-partial}
	\Delta_5 + S_{jj,2}^{-1} \overline\Upsilon_3 +S_{jj,1}^{-1} \, \Xi_{ji,1} - S_{jj,1}^{-1} \, \Xi_{ji,3} \, \Upsilon_5 \overline \Upsilon_3 - \Xi_{ji,3} \, \Upsilon_5 \, \Xi_{ji,1} - \Xi_{ ji,4}\Upsilon_5 \overline\Upsilon_3 = 0
\ee

From the $(ii,-1)$ equation, which reads
\be
	(1-\Delta_3\Upsilon_3) S_{ii,-1} +\Delta_3\Delta_1 =  -S_{ii,1}^{-1} \Upsilon_5\overline{\Upsilon}_3 - \Upsilon_5 \Xi_{ji,1} \,,
\ee
we can solve for $S_{ii,-1}$, and use this to plug into the $(ji,-1)$ equation above, yielding
\be
	\Delta_1 = - \Upsilon_3 S_{ii,1}^{-1} \Upsilon_5\Upsilon_3 - \Upsilon_3 \Upsilon_5 \Xi_{ji,1}
\ee
We need to evaluate both $\Xi_{ji,1}$ and $S_{ii,1}$. 
The former can be obtained by considering the  $(ji,1)$ equation
\be
	\Xi_{ji,1} =  \Xi_{ji,3} \Upsilon_5\overline{\Upsilon}_3 - S_{jj,1}^{-1}\overline{\Upsilon}_3\,,
\ee
together with the $(jj,1)$ equation
\be
	\Upsilon_3\Upsilon_1 = S_{jj,1}^{-1} - \Xi_{ji,3}\, \Upsilon_5\,,
\ee
which together imply
\be
	\Xi_{ji,1} = - \Upsilon_3\Upsilon_1\Upsilon_3\,.
\ee
To obtain $S_{ii,1}$ we consider the $(ii,1)$ equation
\be
	0 = \overline{\Upsilon}_1\overline{\Upsilon}_3 + S_{ii,1}^{-1}(1-\Upsilon_5\Delta_5) - S_{ii,3}^{-1} \Upsilon_5\overline{\Upsilon}_3 -S_{ii,2}^{-1} \Upsilon_5\Xi_{ji,1}\,.
\ee
From the $(ii,0)$ equation
\be
	\Delta_3\Upsilon_3 = \Upsilon_5\Delta_5 + S_{ii,2}^{-1} \Upsilon_5 \overline{\Upsilon}_3 + S_{ii,1}^{-1} \Upsilon_5 \Xi_{ji,1} 
\ee
we solve for $S_{ii,2}^{-1} \Upsilon_5 \overline{\Upsilon}_3$ and plug into the term $S_{ii,2}^{-1} \Upsilon_5\Xi_{ji,1} = -S_{ii,2}^{-1} \Upsilon_5\Upsilon_3\Upsilon_1\Upsilon_3$ in the above $(ii,1)$ equation.
Moreover, from the $(ij,1)$ equation
\be
	(1-\Delta_3\Upsilon_3) \Upsilon_1 = \overline{\Upsilon}_1 - S_{ii,3}^{-1} \Upsilon_5
\ee
we solve for $S_{ii,3}^{-1} \Upsilon_5$ and also use this in the above $(ii,1)$ equation.
Overall, we arrive at an explicit exact expression for $S_{ii,1}$
\be
	S_{ii,1}^{-1} = -\frac{1-\Upsilon_5\Delta_5}{1-\Upsilon_5\Delta_5 + \Upsilon_5\Upsilon_3\Upsilon_1\Upsilon_3\Upsilon_1\Upsilon_3} \, \Upsilon_1\Upsilon_3
\ee
We can finally plug this into the expression for $\Delta_1$ together with the expression found above for $\Xi_{ji,1}$ to obtain
\be
\begin{split}
	\Delta_1 & = 
	\frac{1-\Upsilon_5\Delta_5}{1-\Upsilon_5\Delta_5 + \Upsilon_5\Upsilon_3\Upsilon_1\Upsilon_3\Upsilon_1\Upsilon_3} \, \Upsilon_3\Upsilon_1\Upsilon_3\Upsilon_5\Upsilon_3 +  \Upsilon_3\Upsilon_5\Upsilon_3\Upsilon_1\Upsilon_3
\end{split}
\ee
This is not the final expression though, since it still involves $\Delta_5$, we will simplify it further below.

Next we turn to $\Delta_5$. Equation (\ref{eq:type-3-saddle-delta-5-partial}) can be simplified by employing the $(jj,2)$ equation:
\be
	S_{jj,2}^{-1} - S_{jj,1}^{-1} \Xi_{ji,3} \Upsilon_5 - \Xi_{ji,4} \Upsilon_5
\ee
which gives
\be
	\Delta_5 + (S_{jj,1}^{-1} -\Xi_{ji,3} \Upsilon_5) \Xi_{ji,1} = 0\,.
\ee
Next recall the $(ji,1)$ equation already discussed above, which implies
\be
	(S_{jj,1}^{-1} - \Xi_{ji,3}\Upsilon_5)\Upsilon_3 = - \Xi_{ji,1}\,.
\ee
Together with the explicit expression for $\Xi_{ji,1}$ obtained previously, this implies
\be
	\Delta_5 = \Upsilon_3 \Upsilon_1 \Upsilon_3\Upsilon_1\Upsilon_3 \,.
\ee
This is the desired exact expression for $\Delta_5$.

In a similar way, having obtained $\Delta_5$ we can simplify $S_{ii,1}$ as follows
\be\label{eq:type-3-saddle-J1-ii1}
	S_{ii,1}^{-1} = -({1-\Upsilon_5\Upsilon_3 \Upsilon_1 \Upsilon_3\Upsilon_1\Upsilon_3 }) \Upsilon_1\Upsilon_3
\ee

Now we can use this to further simplify $\Delta_1$
\be
\begin{split}
	\Delta_1 & = 
	\Upsilon_3\Upsilon_1\Upsilon_3\Upsilon_5\Upsilon_3 ({1-\Upsilon_5 \Upsilon_3 \Upsilon_1 \Upsilon_3\Upsilon_1\Upsilon_3}) + \Upsilon_3\Upsilon_5\Upsilon_3\Upsilon_1\Upsilon_3\,,
\end{split}
\ee
this is the desired expression for $\Delta_1$.

We can say something nice at this point: let $\Upsilon_\gamma = -\Upsilon_5 \Upsilon_3 \Upsilon_1 \Upsilon_3\Upsilon_1\Upsilon_3= -\Upsilon_1 \Upsilon_3 \Upsilon_5 \Upsilon_3\Upsilon_1\Upsilon_3$, then we deduce that
\be
	Q(p_1) = 1-\Delta_1 \Upsilon_1 = (1+\Upsilon_\gamma)^2 = Q(p_5)^2 \,. 
\ee

Next we study $\Delta_3$. Starting with the $(ij,0)$ equation written above, 
solving for $\Delta_3$ we obtain
\be
	\Delta_3 = S_{ii,2}^{-1} \Upsilon_5 + S_{ii,-1}\Upsilon_1\,.
\ee
Now consider the $(ii,2)$ equation 
\be
\begin{split}
	0 
	& = S_{ii,2}^{-1} \, Q(p_5) \\
	& + \(- S_{ii,4}^{-1} \Upsilon_5 + \overline{\Upsilon}_1 S_{jj,1}^{-1} +\Xi_{ij,2} + \overline{\Upsilon}_1 \Xi_{ji,3} \Upsilon_5\)  \overline{\Upsilon}_3 \\
	& + \(\overline{\Upsilon}_1 - S_{ii,3}^{-1} \Upsilon_5 \) \Xi_{ji,1} 
\end{split}
\ee
The term in brackets in the second line vanishes, by virtue of the $(ij,2)$ flatness equation, while the $(ij,1)$ transport equation implies that the term in brackets in the third line equals $Q(p_3) \Upsilon_1$. Overall we find
\be
	S_{ii,2}^{-1} = -\frac{Q(p_3)}{Q(p_5)} \Upsilon_1\Xi_{ji,1}\,.
\ee
Next we need to evaluate $S_{ii,-1}$. For this purpose we recall the $(ii,-1)$ equation written above, from which we deduce
\be
	S_{ii,-1} = -\frac{1}{Q(p_3)} \left[
	(1-\Upsilon_5\Upsilon_3\Upsilon_1\Upsilon_3\Upsilon_1\Upsilon_3) \Upsilon_1\Upsilon_3\Upsilon_5\Upsilon_3 +\Delta_3\Delta_1
	\right]
\ee
Plugging these into the expression for $\Delta_3$ we find
\be
\begin{split}
	\Delta_3 
	= -\frac{1}{Q(p_1)} 
	&\Bigg[
	-\frac{Q(p_3)}{Q(p_5)} \Upsilon_1\Upsilon_3\Upsilon_1\Upsilon_3\Upsilon_5  
	- (1-\Upsilon_5\Upsilon_3\Upsilon_1\Upsilon_3\Upsilon_1\Upsilon_3) \Upsilon_1\Upsilon_3\Upsilon_5\Upsilon_3\Upsilon_1 
	\\
	& 
	- \Upsilon_5 \Upsilon_3 \Upsilon_1\Upsilon_3 \Upsilon_1
	+\Delta_3\Delta_1\Upsilon_1
	\Bigg]
\end{split}
\ee
Next, let us use this to derive a formula for $Q(p_3)$. Multiplying from the right by $-\Upsilon_3$, this becomes
\be
	Q(p_3)-1 = \frac{1}{Q(p_1)} 
	\left[
	\frac{Q(p_3)}{Q(p_5)} (Q(p_5)-1)  + 
	(Q(p_5)-1) (Q(p_5)+1)
	+ (Q(p_3)-1)(Q(p_1)-1)
	\right]
\ee
Now using the fact that $Q(p_1) = Q(p_5)^2$, the above equation simplifies to
\be
	Q(p_3) = Q(p_5)^3 \,.
\ee
Plugging this into the expression for $\Delta_3$ eventually gives  
\be
	\Delta_3 =
	Q(p_5)^2 \Upsilon_1\Upsilon_3\Upsilon_1\Upsilon_3\Upsilon_5  + 
	Q(p_5) \Upsilon_1\Upsilon_3\Upsilon_5\Upsilon_3\Upsilon_1 
	+ \Upsilon_5 \Upsilon_3 \Upsilon_1\Upsilon_3 \Upsilon_1
\ee
where the expression for $Q(p_5)$ in terms of $\Upsilon_i$ was given above. This is the desired final expression for $\Delta_3$.

\subsubsection*{Junction $J_2$}

\begin{figure}[h!]
\begin{center}
\includegraphics[width=0.75\textwidth]{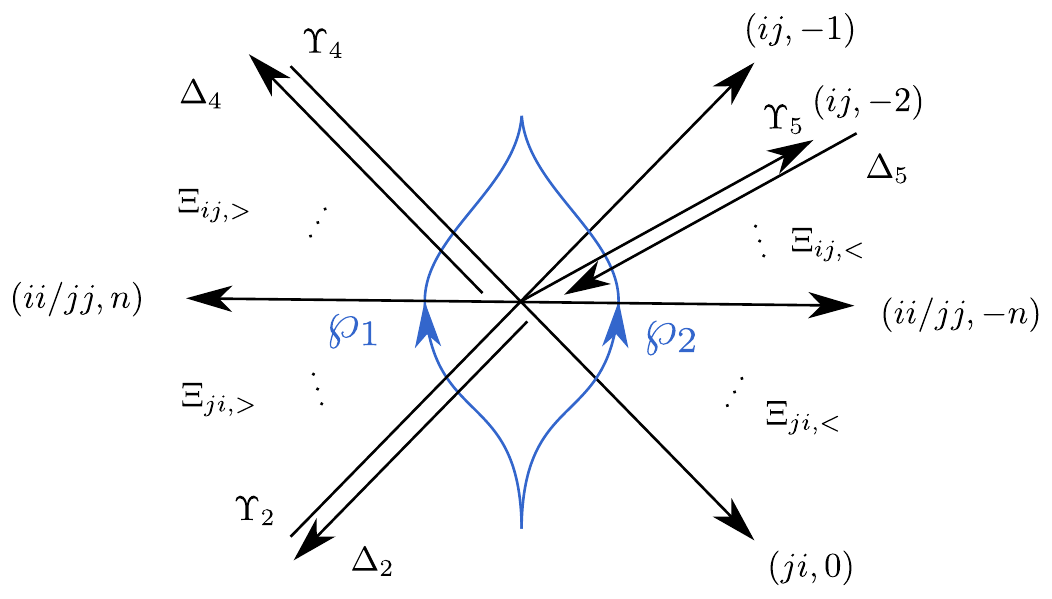}
\caption{Junction $J_2$ of the saddle of Type-3.}
\label{fig:type-3-saddle-junction-J2}
\end{center}
\end{figure}

The soliton labels for each wall are as follows
\be
\begin{array}{c|c|c|c|c|c}
\Upsilon_2 & \Delta_2 & \Upsilon_4 & \Delta_4 & \Upsilon_5 & \Delta_5 \\
\hline
(ij,-1) & (ji,1) & (ji,0) & (ij,0) & (ij,-2) & (ji,2)
\end{array}
\ee
We consider paths $\wp_1, \wp_2$ across the junction, as shown in Figure \ref{fig:type-3-saddle-junction-J2}.
Adopting the formal variable $z$ as above to keep track of logarithmic indices, we can write $F(\wp_1)$ as follows
\be
\begin{split}
	F(\wp_1)
	& = e^{\Delta_2}  e^{-\Upsilon_2} 
	\left[  \prod_{k\geq 2} e^{\Xi_{ji,k}}  \right]
	\left[  \prod_{k\geq 1} e^{\Xi_{ii,k}} e^{\Xi_{jj,k}}  \right]
	\left[  \prod_{k\geq 1} e^{\Xi_{ij,k}}  \right]
	e^{\Delta_4} e^{-\Upsilon_4}
	\\
	& = 
	\left(\begin{array}{cc}
	1 & z\Delta_2 \\
	-z^{-1}\Upsilon_2 & 1 -\Delta_2\Upsilon_2
	\end{array}\right)
	\left(\begin{array}{cc}
	1 & 0 \\
	\Xi_{ji,>}(z) & 1
	\end{array}\right)
	\left(\begin{array}{cc}
	S_{ii,>}(z) & 0 \\
	0 & S_{jj,>}(z)
	\end{array}\right)
	\\
	&\times
	\left(\begin{array}{cc}
	1 & \Xi_{ij,<}(z) \\
	0 & 1
	\end{array}\right)
	\left(\begin{array}{cc}
	1-\Delta_4\Upsilon_4 & \Delta_4\\
	-\Upsilon_4 & 1 
	\end{array}\right)
\end{split}
\ee
where we introduced generating functions $\Xi_{ij,>}(z),\Xi_{ji,>}(z), S_{ii,>}(z), S_{jj,>}(z)$ defined as usual.
A bit of algebra yields the following expressions for components of the transport matrix
\be
\begin{split}
	F(\wp_1)_{ii}(z)
	& = 
	z^{-1} \Upsilon_2 S_{jj,>}(z)\Upsilon_4
	+ 
	(1-z^{-1}\Upsilon_2 \Xi_{ji,>}(z)) S_{ii,>}(z) (1-(\Xi_{ij,>}(z) +\Delta_4) \Upsilon_4 )
	\\
	F(\wp_1)_{ij}(z)
	& = z^{-1}\Upsilon_2 S_{jj,>}(z) + (1-z^{-1}\Upsilon_2 \Xi_{ji,>}(z)) S_{ii,>}(z) (\Xi_{ij,>}(z) +\Delta_4 )
	\\
	F(\wp_1)_{ji}(z)
	& =  \left[ z\Delta_2+(1-\Delta_2\Upsilon_2)\Xi_{ji,>}(z) \right] \, S_{ii,>}(z) \left[ 1 - \(  \Xi_{ij,>}(z) + \Delta_4\) \Upsilon_4 \right]
	\\
	& - 	(1-\Delta_2\Upsilon_2)S_{jj,>}(z)  \Upsilon_4
	\\
	F(\wp_1)_{jj}(z)
	& = (1-\Delta_2\Upsilon_2)S_{jj,>}(z)
	+  \left[ z\Delta_2+(1-\Delta_2\Upsilon_2)\Xi_{ji,>}(z) \right] \, S_{ii,>}(z) \left[  \Xi_{ij,>}(z) + \Delta_4 \right]
\end{split}
\ee

Similarly, the formal parallel transport along $\wp_2$ is
\be
\begin{split}
	F(\wp_2)
	& = e^{-\overline{\Upsilon}_4}  
	\left[  \prod_{k\geq 1} e^{-\Xi_{ji,-k}}  \right]
	\left[  \prod_{k\geq 1} e^{-\Xi_{ii, -k}} e^{-\Xi_{jj,-k}}  \right]
	\left[  \prod_{k\geq 3} e^{-\Xi_{ij,-k}}  \right]
	e^{\Delta_5} e^{-\Upsilon_5} e^{-\overline{\Upsilon}_2}
	\\
	& = 
	\left(\begin{array}{cc}
	1 & 0 \\
	- \overline{\Upsilon}_4 & 1
	\end{array}\right)
	\left(\begin{array}{cc}
	1 & 0  \\
	-\Xi_{ji,<}(z) & 1
	\end{array}\right)
	\left(\begin{array}{cc}
	S_{ii,<}(z)^{-1} & 0 \\
	0 & S_{jj,<}(z)^{-1}
	\end{array}\right)
	\left(\begin{array}{cc}
	1 & -\Xi_{ij,<}(z) \\
	0  & 1
	\end{array}\right)
	\\
	&\times
	\left(\begin{array}{cc}
	1  & -z^{-2}\Upsilon_5 \\
	z^2\Delta_5 & 1 - \Delta_5\Upsilon_5
	\end{array}\right)
	\left(\begin{array}{cc}
	1 & -z^{-1}\, \overline{\Upsilon}_2 \\
	0  & 1
	\end{array}\right)
\end{split}
\ee
where we introduced generating functions $\Xi_{ji,<}(z)$ etc, defined as usual.
A bit of algebra yields the following expressions for components of the transport matrix
\be
\begin{split}
	F(\wp_2)_{ii}(z)
	& = 
	S_{ii,<}(z)^{-1} \, \(1- \Xi_{ij,<}(z) \, z^2 \Delta_5\)
	\\
	F(\wp_2)_{ij}(z)
	& = 
	- S_{ii,<}(z)^{-1} \, \left[
	z^{-1} \overline{\Upsilon}_2
	+ z^{-2} \Upsilon_5
	+ \Xi_{ij,<}(z) \( 1- \Delta_5(\Upsilon_5 + z \overline\Upsilon_2) \) 
	\right]
	\\
	F(\wp_2)_{ji}(z)
	& =  S_{jj,<}(z)^{-1} \, z^2\Delta_5 -  \( \overline\Upsilon_4 + \Xi_{ji,<}(z)\) S_{ii,<}(z)^{-1} \(1 - \Xi_{ij,<}(z) \, z^2 \Delta_5\)
	\\
	F(\wp_2)_{jj}(z)
	& = 
	\left[
	S_{jj,<}(z)^{-1} + \( \overline\Upsilon_4 + \Xi_{ji,<}(z)\) S_{ii,<}(z)^{-1}\Xi_{ij,<}(z) 
	\right]
	\( 1- \Delta_5(\Upsilon_5 + z \overline\Upsilon_2) \) 
	\\
	& +  \( \overline\Upsilon_4 + \Xi_{ji,<}(z)\) S_{ii,<}(z)^{-1} \( z^{-1} \overline\Upsilon_2 + z^{-2}\Upsilon_5\)
\end{split}
\ee

Now studying the equations $F(\wp_1)=F(\wp_2)$ for each matrix element, and term by term in $z$, yields the generating functions of outgoing solitons in terms of those of incoming ones.
Let us adopt the following definitions 
\be
\begin{split}
	S_{ii,<} & = 1 + \sum_{k\geq 1} z^{-k} S_{ii,-k}
	\qquad
	S_{ii,>} = 1 + \sum_{k\geq 1} z^{k} S_{ii,k}
	\\
	\Xi_{ij,<} & = \sum_{k\geq 1} z^{-k} \Xi_{ij,-k}
	\qquad\ \ \ \
	\Xi_{ij,>} = \sum_{k\geq 1} z^{k+1} \Xi_{ij,k+1}
\end{split}
\ee
and similarly with $i \leftrightarrow j$.

Our goal is now to derive expressions for $\Delta_2,\Delta_4, \Upsilon_5$ in terms of $\Upsilon_2,\Upsilon_4,\Delta_5$. 
To begin with, note that the $(ij,-1)$ equation implies
\be
	\Upsilon_2 = \overline\Upsilon_2\,.
\ee
To compute $\Delta_2$ we consider the $(ji,1)$ equation
\be
	-\Delta_2\(1- \Delta_4\Upsilon_4\) + (1- \Delta_2\Upsilon_2) S_{jj,1} \Upsilon_4  = -S_{jj,-1}^{-1}\Delta_5
\ee
To compute $\Delta_4$ we consider the $(ij,0)$ equation
\be
	\Delta_4 - \Upsilon_2 S_{jj,1} = 0
\ee
To compute $\Upsilon_5$ we consider the $(ij,-2)$ equation
\be\label{eq:type-3-saddle-upsilon-5-partial}
	0 = \Upsilon_5 + \(S_{ii,-1}^{-1}  - \Xi_{ij,-3} \Delta_5\)\, \overline\Upsilon_2
\ee

Let us start from $\Upsilon_5$. The $(ii,-1)$ equation 
\be
	\Upsilon_2\Upsilon_4 = S_{ii,-1}^{-1}  - \Xi_{ij,-3} \Delta_5
\ee
immediately gives the result
\be
	\Upsilon_5 = -\Upsilon_2\Upsilon_4\Upsilon_2\,.
\ee
This is the desired expression for $\Upsilon_5$.

Next consider the $(jj,1)$ equation
\be
	 \(1- \Delta_2\Upsilon_2\)S_{jj,1} + \Delta_2\Delta_4 = -\Delta_5\overline\Upsilon_2
\ee
solving for $S_{jj,1}$ and substituting into the $(ij,0)$ equation above leads to
\be
	\Delta_4 = - \Upsilon_2\Delta_5\Upsilon_2
\ee
This is the desired expression for $\Delta_4$.
Note that this implies 
\be
	Q(p_4) = Q(p_5) \,.
\ee

Substituting the $(jj,1)$ equation into the $(ji,1)$ equation above gives
\be
	\Delta_2 = -\Delta_5\Upsilon_2\Upsilon_4 + S_{jj,-1}^{-1} \Delta_5
\ee
To obtain $S^{-1}_{jj,-1}$ consider the $(jj,-1)$ equation
\be
	S_{jj,-1}^{-1}(1-\Delta_5\Upsilon_5) + \( \overline\Upsilon_4- S^{-1}_{jj,-2} \Delta_5\) \overline\Upsilon_2 = 0\,.
\ee
Using the $(ji,0)$ equation
\be
	(1-\Delta_2\Upsilon_2) \Upsilon_4 = \overline\Upsilon_4 - S^{-1}_{jj,-2} \Delta_5
\ee
we obtain 
\be
	S^{-1}_{jj,-1} = -\frac{Q(p_2)}{Q(p_5)} \Upsilon_4\Upsilon_2\,.
\ee
Plugging this into the expression for $\Delta_2$ gives
\be
	\Delta_2 =  -\Delta_5\Upsilon_2\Upsilon_4 -\frac{Q(p_2)}{Q(p_5)} \Upsilon_4\Upsilon_2 \Delta_5
\ee
Multiplying from the right by $-\Upsilon_2$ and denoting $\Upsilon_\gamma$ any cyclic permutation of $-\Delta_5\Upsilon_2\Upsilon_4\Upsilon_2$, this can be rewritten as
\be
	Q(p_2)-1 = - \(1+\frac{Q(p_2)}{Q(p_5)} \)\Upsilon_\gamma
\ee
Noting that $Q(p_5) =1-\Delta_5\Upsilon_5= 1 - \Upsilon_\gamma$ 
we obtain
\be
	Q(p_2) = Q(p_5)^2
\ee
where the right-hand side is known explicitly in terms of $\Upsilon_2,\Upsilon_4,\Delta_5$.
Finally, we have the following expression for $\Delta_2$
\be
	\Delta_2 = - \Delta_5\Upsilon_2\Upsilon_4 - (1 + \Delta_5\Upsilon_2\Upsilon_4\Upsilon_2) \Upsilon_4\Upsilon_2\Delta_5 \,.
\ee

\subsection{Type-4}\label{app:type-4-saddle}

\begin{figure}[h!]
\begin{center}
\includegraphics[width=0.50\textwidth]{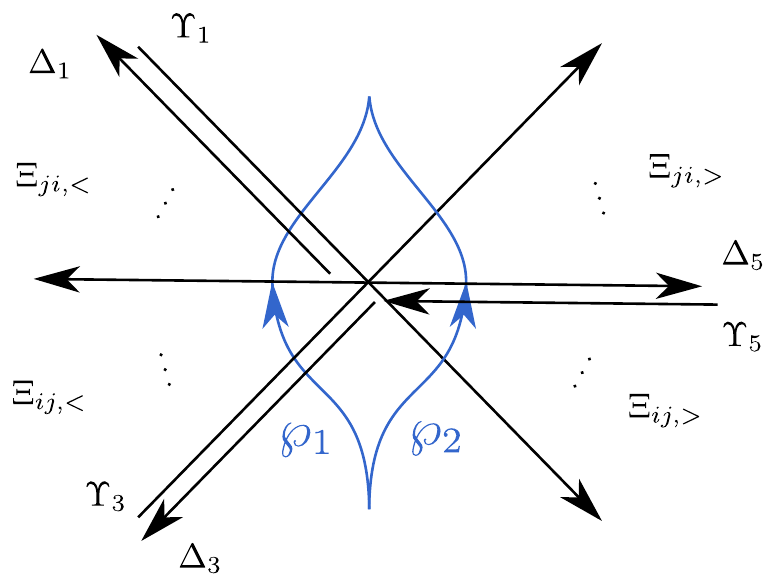}
\caption{Junction $J_1$ in the saddle of Type-4.}
\label{fig:type-4-saddle-junction-J1}
\end{center}
\end{figure}

All four junctions appearing in the Type-4 saddle are of the same type, up to changes of labels. Here we focus on junction $J_1$, depicted in Figure \ref{fig:type-4-saddle-junction-J1}. There are three double-walls, and infinitely many one-way walls. Note that this is very similar to the junction of the Type-1 saddle.

The soliton labels for each wall are as follows
\be
\begin{array}{c|c|c|c|c|c}
\Upsilon_1 & \Delta_1 & \Upsilon_3 & \Delta_3 & \Upsilon_5^{(ii,-n)} /\Upsilon_5^{(jj,-n)}
& \Delta_5^{(ii,n)} /\Delta_5^{(jj,n)}
\\
\hline
(ij,1) & (ji,-1) & (ji,0) & (ij,0) & (ii,-n) /  (jj,-n) & (ii,n)/(jj,n)
\end{array}
\ee
We consider paths $\wp_1, \wp_2$ across the junction, as shown in Figure \ref{fig:type-4-saddle-junction-J1}.
Adopting the formal variable $z$ as above to keep track of logarithmic indices, we can write $F(\wp_1)$ as follows
\be
\begin{split}
	F(\wp_1)
	& = e^{\Delta_3}  e^{-\Upsilon_3} 
	\left[  \prod_{k\geq 1} e^{\Xi_{ij,-k}}  \right]
	\left[  \prod_{k\geq 1} e^{\Xi_{ii,-k}} e^{\Xi_{jj,-k}}  \right]
	\left[  \prod_{k\geq 1} e^{\Xi_{ji,-n(k+1)}}  \right]
	e^{\Delta_1} e^{-\Upsilon_1}
	\\
	& = 
	\left(\begin{array}{cc}
	1-\Delta_3\Upsilon_3 & \Delta_3 \\
	-\Upsilon_3 & 1 
	\end{array}\right)
	\left(\begin{array}{cc}
	1 & \Xi_{ij,<}(z) \\
	0 & 1
	\end{array}\right)
	\left(\begin{array}{cc}
	S_{ii,<}(z)  & 0 \\
	0 & S_{jj,<}(z)
	\end{array}\right)
	\\
	&\times
	\left(\begin{array}{cc}
	1 & 0 \\
	\Xi_{ji,<}(z) & 1
	\end{array}\right)
	\left(\begin{array}{cc}
	1 & -z\Upsilon_1\\
	z^{-1}\Delta_1 & 1 - \Delta_1\Upsilon_1
	\end{array}\right)
\end{split}
\ee
where we introduced
\be
\begin{split}
	&
	\Xi_{ji,<}(z) = \sum_{k\geq 1} z^{-(k+1)} \Xi_{ji,-(k+1)}\,,
	\qquad
	\Xi_{ij,<}(z) = \sum_{k\geq 1} z^{-k} \Xi_{ij,-k}\,,
	\\
	&
	S_{ii,<}(z) = \prod_{k\geq 1} e^{z^{-k}\Xi_{ii, -k}} \,,
	\qquad\qquad\quad\ 
	S_{jj,<}(z) = \prod_{k\geq 1}e^{z^{-k}\Xi_{jj, -k}}\,.
\end{split}
\ee
A bit of algebra yields the following expressions for components of the transport matrix
\be
\begin{split}
	F(\wp_1)_{ii}(z)
	& = (1-\Delta_3\Upsilon_3) 
	S_{ii,<}(z)
	\\
	& + \left[\Delta_3 + (1-\Delta_3\Upsilon_3) \Xi_{ij,<}(z)\right] 
	S_{jj,<}(z)
	\left[ z^{-1} \Delta_1  + \Xi_{ji,<} \right]
	\\
	F(\wp_1)_{ij}(z)
	& = 
	\left[\Delta_3 + (1-\Delta_3\Upsilon_3) \Xi_{ij,<}(z)\right] 
	S_{jj,<}(z)
	\left[
	1
	-
	\( z^{-1} \Delta_1 + \Xi_{ji,<}(z) \) \, z \Upsilon_1
	\right]
	\\
	&
	- (1-\Delta_3\Upsilon_3) \, 
	S_{ii,<}(z)
	\, z \Upsilon_1
	\\
	F(\wp_1)_{ji}(z)
	& =  
	-\Upsilon_3 
	S_{ii,<}(z)
	\\
	& + \(1-\Upsilon_3 \Xi_{ij,<}(z)\) \, 
	S_{jj,<}(z)
	\(z^{-1} \Delta_1 + \Xi_{ji,<}(z)\) 
	\\
	F(\wp_1)_{jj}(z)
	& = 
	\Upsilon_3 
	S_{ii,<}(z)
	\, z\Upsilon_1 
	\\
	&+ \(1-\Upsilon_3 \Xi_{ij,<}(z)\)  
	S_{jj,<}(z)
	\left[
	1-\(z^{-1}\Delta_1 + \Xi_{ji,<}(z)\) z\Upsilon_1
	\right]
\end{split}
\ee

Similarly, the formal parallel transport along $\wp_2$ is
\be
\begin{split}
	F(\wp_2)
	& = 
	\left[  \prod_{k\geq 0} e^{-\Xi_{ij,k+1}}  \right]
	\left[  \prod_{k\geq 1} e^{\Upsilon_5^{(ii, -k)}} e^{\Upsilon_5^{(jj,-k)}}  \right]
	\left[  \prod_{k\geq 1} e^{-\Delta_5^{(ii, k)}} e^{-\Delta_5^{(jj,k)}}  \right]
	\left[  \prod_{k\geq 0} e^{-\Xi_{ji,k}}  \right]
	\\
	& = 
	\left(\begin{array}{cc}
	1 & -\Xi_{ij,>}(z)  \\
	0 & 1
	\end{array}\right)
	\left(\begin{array}{cc}
	S^{p_5}_{ii,<}(z) S^{p_5}_{ii,>}(z)^{-1} & 0 \\
	0 & S^{p_5}_{jj,<}(z) S^{p_5}_{jj,>}(z)^{-1}
	\end{array}\right)
	\left(\begin{array}{cc}
	1 & 0 \\
	-\Xi_{ji,>}(z)  & 1
	\end{array}\right)
\end{split}
\ee
where we introduced 
\be\label{eq:dict-1}
\begin{split}
	&
	\Xi_{ji,>}(z) = \sum_{k\geq 0} z^{k} \Xi_{ji,k}\,,
	\qquad\qquad\quad\ \ \ 
	\Xi_{ij,>}(z) = \sum_{k\geq 0} z^{k+1} \Xi_{ij,k+1}\,,
	\\
	&
	S^{p_5}_{ii,>}(z) = \prod_{k\geq 1} e^{z^{k}\Delta_5^{(ii,k)}} \,,
	\qquad\qquad\quad\,
	S^{p_5}_{jj,>}(z) = \prod_{k\geq 1} e^{z^{-k}\Delta_5^{(jj,-k)}} \,,
	\\
	&
	S^{p_5}_{ii,<}(z) = \prod_{k\geq 1} e^{z^{k}\Upsilon_5^{(ii,k)}} \,,
	\qquad\qquad\quad\,
	S^{p_5}_{jj,<}(z) = \prod_{k\geq 1} e^{z^{-k}\Upsilon_5^{(jj,-k)}} \,.
\end{split}
\ee
Components of the transport matrix are as follows
\be
\begin{split}
	F(\wp_2)_{ii}(z)
	& = 
	S^{p_5}_{ii,<}(z)S^{p_5}_{ii,>}(z)^{-1}
	+ 
	\Xi_{ij,>}(z) 
	S^{p_5}_{jj,<}(z)S^{p_5}_{jj,>}(z)^{-1}
	\Xi_{ji,>}(z) 
	\\
	F(\wp_2)_{ij}(z)
	& = 
	-\Xi_{ij,>}(z) 
	S^{p_5}_{jj,<}(z)S^{p_5}_{jj,>}(z)^{-1}
	\\
	F(\wp_2)_{ji}(z)
	& =  
	-S^{p_5}_{jj,<}(z)S^{p_5}_{jj,>}(z)^{-1}
	\Xi_{ji,>}(z) 
	\\
	F(\wp_2)_{jj}(z)
	& = 
	S^{p_5}_{jj,<}(z)S^{p_5}_{jj,>}(z)^{-1}
\end{split}
\ee

Now studying the equations $F(\wp_1)=F(\wp_2)$ for each matrix element, and term by term in $z$, yields the generating functions of outgoing solitons in terms of those of incoming ones.

Without loss of generality, we introduce $\Theta, \overline\Theta$ as follows\footnote{This parametrization appears to be the inverse of the one used in \cite[Sec. 3.3]{Banerjee:2018syt}. For a discussion see footnote \ref{eq:foot-reparam-inverse}}
\be\label{eq:ii-jj-param}
	S_{ii,<}^{p_5}(z) = \frac{1}{1 + z^{-1}\Theta} \,,
	\qquad
	S_{jj,<}^{p_5}(z) = {1+ z^{-1} \overline\Theta} \,.
\ee
These generating functions are formally inverses of each other, as should be expected by the fact that all transport matrices have unit determinant. There is thus no loss of generality in expressing the transport in this way. The notation chose here is inspired to reflect our earlier work on the $ij-ji$ junction in \cite[Section 3]{Banerjee:2018syt}.

After a bit of algebra\footnote{Similarly to what we have found for the junction of the TYpe-1 saddle, the equations for $\wp_1,\wp_2$ as written above seem to be insufficient. See footnote \ref{foot:insuff-eqs}.}, we obtain the following solution
\be\label{eq:full-solution-type-4-saddle}
\begin{split}
	\Delta_1 
	& = - (2 - \overline\Theta\Upsilon_3\Upsilon_1) Q(p)  \,  \overline\Theta\Upsilon_3
	\\
	\Delta_3
	& = - \Upsilon_1\overline\Theta\,  (2 - \overline\Theta\Upsilon_3\Upsilon_1) Q(p)  
	\\
	\Xi_{ij,>}(z)
	&=  z\Upsilon_1 \frac{1}{1+z \Upsilon_3\Upsilon_1}  Q(p) \,,\\
	\Xi_{ji,>}(z) 
	& = \frac{1}{1+z\Upsilon_3\Upsilon_1}\Upsilon_3\,,\\
	S^{p_5}_{ii,>} 
	& = \frac{1}{1+ z \Upsilon_1\Upsilon_3}\,,
	\\
	S^{p_5}_{jj,>} 
	& = 1 + z\Upsilon_3\Upsilon_1\,,\\
	\Xi_{ij,<}(z)
	&=  
	z^{-1} \, \Upsilon_1 \overline\Theta ^2 \frac{
			3  
	  		+ 2 z^{-1} \overline\Theta  (1- z \Upsilon_3 \Upsilon_1)
			-  z^{-1} \,\overline\Theta ^2  \Upsilon_3 \Upsilon_1
		}{
			1
			+ 2 z^{-1} \,\overline\Theta  (1-z\, \Upsilon_3 \Upsilon_1)
			+ z^{-2} \, \overline\Theta ^2 \left(1  -z \, \Upsilon_3 \Upsilon_1+ z^2 (\Upsilon_3 \Upsilon_1)^2\right)
		}
	\\
	\Xi_{ji,<}(z)
	& =
	 \frac{
			3  
	  		+ 2 z^{-1} \overline\Theta  (1- z \Upsilon_3 \Upsilon_1)
			-  z^{-1} \,\overline\Theta ^2  \Upsilon_3 \Upsilon_1
		}{
			1
			+ 2 z^{-1} \,\overline\Theta  (1-z\, \Upsilon_3 \Upsilon_1)
			+ z^{-2} \, \overline\Theta ^2 \left(1  -z \, \Upsilon_3 \Upsilon_1+ z^2 (\Upsilon_3 \Upsilon_1)^2\right)
		}
		  \, z^{-2}\, \overline\Theta ^2 \Upsilon_3 \cdot Q(p)
	\\
	S_{ii,<}(z)
	& = 
	\frac{
		1 + z^{-1}\Theta 
	}{
		1
		+ 2 z^{-1} \,\Theta  (1-z\, \Upsilon_3 \Upsilon_1)
		+ z^{-2} \, \Theta ^2 \left(1  -z \, \Upsilon_3 \Upsilon_1+ z^2 (\Upsilon_3 \Upsilon_1)^2\right)
	}\, Q(p)^{-1}	
	\\
	S_{jj,<}(z)
	& = 
	\frac{
		1 
		+ 2z^{-1} \overline\Theta  (1- z\,\Upsilon_1 \Upsilon_3)
		+ z^{-2}\overline\Theta ^2 \left(1  - z \Upsilon_1\Upsilon_3 + z^2(\Upsilon_1 \Upsilon_3)^2\right)
		}{
			1 + z^{-1}\overline\Theta 
		}\, Q(p)
	\\
\end{split}
\ee
where
\be
	Q(p) = Q(p_1) =Q(p_3) = (1-\Delta_1\Upsilon_1) = (1-\Delta_3\Upsilon_3) = (1-\overline\Theta\Upsilon_3\Upsilon_1)^{-2} \,.
\ee

Let us comment on the limiting behavior of this solution. When $\Upsilon_1$ is set to zero, this becomes
\be\label{eq:full-solution-type-4-saddle-specialized-Upsilon1}
\begin{split}
	\Delta_1 
	& = - 2  \overline\Theta\Upsilon_3
	\\
	\Delta_3
	& = 0 
	\\
	\Xi_{ij,>}(z)
	&=  0 \,,\\
	\Xi_{ji,>}(z) 
	& = \Upsilon_3\,,\\
	S^{p_5}_{ii,>} & = 	S^{p_5}_{jj,>}  = 1\\
	\Xi_{ij,<}(z) &=  0
	\\
	\Xi_{ji,<}(z)
	& =
	 \frac{
			3  
	  		+ 2 z^{-1} \overline\Theta  
		}{
			(1 +  z^{-1} \,\overline\Theta )^2
		}
		  \, z^{-2}\, \overline\Theta ^2 \Upsilon_3 \, 
	\\
	S_{ii,<}(z)
	& = (1 + z^{-1} \Theta)^{-1}
	\\
	S_{jj,<}(z)
	& = 1 + z^{-1} \overline\Theta
	\\
\end{split}
\ee
Notice that this agrees with the computation in Appendix \ref{app:ii-ij-junction}.
A similar limit can be checked to hold when $\Upsilon_3$ is set to zero.

One may also set $\Theta=\overline\Theta=0$, this yields
\be\label{eq:full-solution-type-4-saddle-specialized-Theta}
\begin{split}
	\Delta_1 
	& = 0
	\\
	\Delta_3
	& = 0
	\\
	\Xi_{ij,>}(z)
	&=  z\Upsilon_1 \frac{1}{1+z \Upsilon_3\Upsilon_1}   \,,\\
	\Xi_{ji,>}(z) 
	& = \frac{1}{1+z\Upsilon_3\Upsilon_1}\Upsilon_3\,,\\
	S^{p_5}_{ii,>} 
	& = \frac{1}{1+ z \Upsilon_1\Upsilon_3}\,,
	\\
	S^{p_5}_{jj,>} 
	& = 1 + z\Upsilon_3\Upsilon_1\,,\\
	\Xi_{ij,<}(z)
	&=  
	0
	\\
	\Xi_{ji,<}(z)
	& =
	0
	\\
	S_{ii,<}(z)
	& = 
	1
	\\
	S_{jj,<}(z)
	& = 
	1
	\\
\end{split}
\ee
recovering exactly the descendant wall structure of the $ij-ji$ junction \cite[Section 3]{Banerjee:2018syt}.

\section{Some technical results}

This appendix collects computations of homotopy invariance across a branch point and across a junction of one-way $\CE$-walls of types $ii$ and $ij$.
The latter was not included in our previous works, but appears in the main body of this work. 
In this paper we adopt slightly different conventions from our previous papers concerning rules for writing down parallel transport matrices, in this appendix we spell out the convention and illustrate how certain signs are fixed.

\subsection{Convention on signs for the parallel transport}

In our previous work \cite{Banerjee:2018syt, Banerjee:2019apt} we adopted the convention of working with a twisted flat connection on $\Sigma$, following \cite{Gaiotto:2012rg}.
It is however possible to keep track of signs in a more convenient manner, which is the convention we adopt in this paper. Given an $\CE$-wall $p$ of type $(ij,n)$, and a path $\wp$ intersecting the wall transversely at a point $x_0$, we define the parallel transport along $\wp$ as follows
\be\label{eq:sign-rule}
	F(\wp) = D(\wp_+) \, e^{ \sigma\,  \Xi_{ij,n} }\, D(\wp_-) \qquad \text{where }\sigma = {\rm sgn}  \(\hat \wp \wedge \hat p \)\,.
\ee
Here  $D(\wp) = \sum_{k} X_{\wp^{(k)}}$ is the diagonal transport on each sheet, and $\wp = \wp_+\circ\wp_-$ is the splitting of $\wp$ at $x_0$. The generating function $\Xi_{ij,n} = \sum_{N\in\IZ}\sum_{a\in\Gamma_{ij,N,N+n}} \mu_a X_a$  counts soliton paths supported on $p$.
The wedge product $\hat \wp \wedge \hat p$ between the two tangent (co)vectors is understood to be taken at the common fiber at $x_0$, and is dualized by the Hodge star to a scalar.\footnote{For clarity, in the $(x,y)$ plane we have $dx\wedge dy= +1$.}

The role of the sign $\sigma$ is to ensure that the inverse of the transition matrix is applied when reversing the direction in which an $\CE$-wall is crossed.
When $i\neq j$ one has  $e^{ \sigma\,  \Xi_{ij,n} } =  1 + { \sigma\,  \Xi_{ij,n} }$, but when $i=j$ al powers in the expansion fo the exponential survive \cite{Banerjee:2018syt}. 

\subsection{Homotopy at a branch point}

\begin{figure}[h!]
\begin{center}
\includegraphics[width=0.35\textwidth]{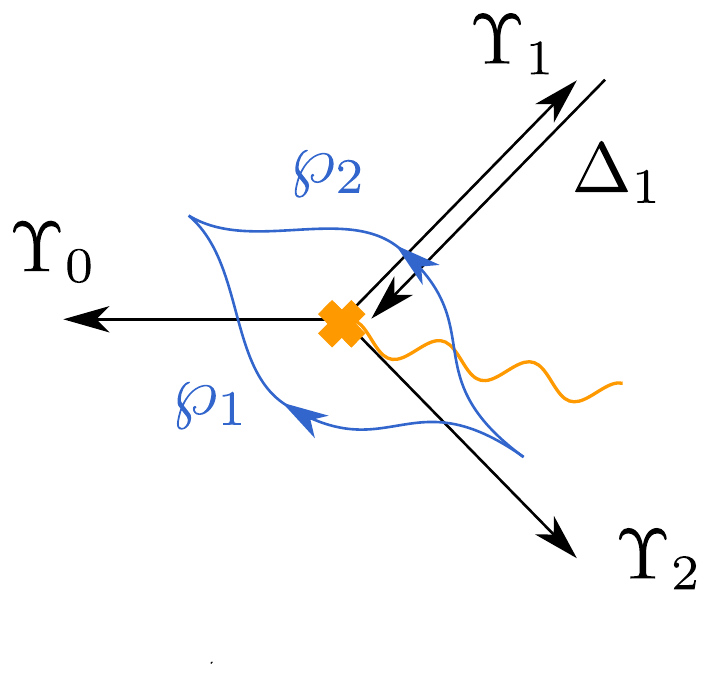}
\caption{Homotopy of transport paths $\wp_1,\wp_2$ across a branch point.}
\label{fig:hom_br1}
\end{center}
\end{figure}

Consider a branch point with one (or possibly more) double walls ending on it, as in Figure~\ref{fig:hom_br1}.
The flatness property of the nonabelianization map enables to determine the `flow of soliton data' at the branch point: all soliton generating functions of outgoing $\CE$-walls can be determined in terms of generating functions of incoming $\CE$-walls.

There are three outgoing $\CE$-walls, with soliton generating functions $\Upsilon_i$ for $i=0,1,2$. 
In this simple example we consider only one incoming wall, with soliton data encoded by $\Delta_1$ is an incoming wall giving rise to a two way street with $\Upsilon_1$ in the British resolution. 
The three outgoing walls carry types $\Upsilon_0 : (ji,0), \, \Upsilon_1,\Upsilon_2 : (ij,0)$. Correspondingly, $\Delta_1$ has type $(ji,0)$. The parallel transport along $\wp_1$ reads
\be 
\begin{split}
F(\wp_1) & = D(\wp_1^+) (1+\Upsilon_2) D(\wp_1^0) (1+\Upsilon_0) D(\wp_1^-)
\\ & 
= D(\wp_1) + X_{\wp^+_{1,i}} \Upsilon_2 X_{\wp_{1,j}^{0-}} + X_{\wp_{1,j}^{+0}}\Upsilon_0 X_{\wp_{1,i}^-}
+ X_{\wp^+_{1,i}} \Upsilon_2 X_{\wp_{1,j}^0} \Upsilon_0 X_{\wp_{1,i}^-}
\end{split}
\ee 
where we adopted the sign rule (\ref{eq:sign-rule}), and split the path at crossings as $\wp_1 = \wp_1^+\circ\wp_1^0\circ\wp_1^-$.

For $\wp_2$ one finds
\be
\begin{split} 
F(\wp_2) &= D(\wp_2^+) (1+\Delta_1)(1-\Upsilon_1) D(\wp_2^-)
\\ & 
= D(\wp_2) + X_{\wp_{2,ij}^+}\Delta_1 X_{\wp_{2,i}^-} - X_{\wp_{2,ji}^+}\Upsilon_1 X_{\wp_{2,j}^-}
- X_{\wp_{2,ij}^+} \Delta_1\Upsilon_1 X_{\wp_{2,j}^-}
\end{split} 
\ee 
where now $D(\wp_2$ is off-diagonal due to the branch cut.

Comparing the $ij$-components of the parallel transport gives
\be
	 X_{\wp^+_{1,i}} \Upsilon_2 X_{\wp_{1,j}^{0-}}
	 =
	 X_{\wp_{2,ij}}
	 - X_{\wp_{2,ij}^+} \Delta_1\Upsilon_1 X_{\wp_{2,j}^-}
\ee
now shrinking $\wp_1,\wp_2$ to infinitesimal size, with endpoints approaching the branch point, we arrive at
\be\label{eq:branch-homotopy-equation}
	\Upsilon_2 = X_{a_2} (1 - X_{a_1}\Delta_1) \,,
\ee
where $a_i$ are simpleton paths on streets $p_i$, see footnote \ref{foot:simpletons}.

Here we have performed the computation using the sign rule (\ref{eq:sign-rule}). For an alternative approach using twisted flat connections see \cite{Gaiotto:2012rg, Banerjee:2018syt}.

\subsection{The $ii-ij$ junction}\label{app:ii-ij-junction}

In this section, we focus on the interaction of the walls of type $(ii/jj,kn)$for $k\geq 1$  and $(ij,m)$. One important thing to notice for this computation is that 
the walls $ii$ and $jj$ appear together and one needs to consider the infinite set of these walls together. This follows from how they are generated in $ij-ji$ junctions \cite{Banerjee:2018syt}.

\begin{figure}[h!]
\begin{center}
\includegraphics[width=0.5\textwidth]{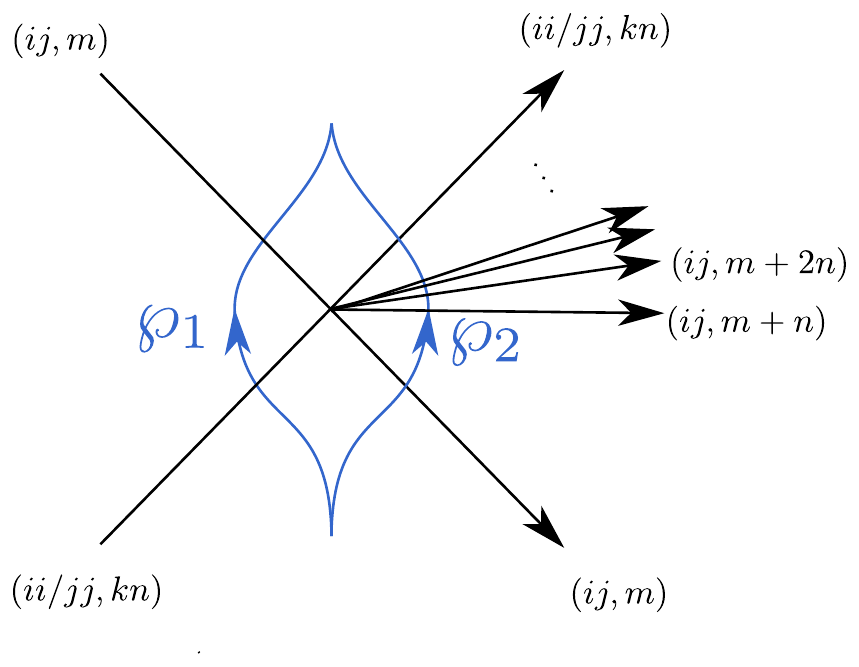}
\caption{One way junction of of type $ii-ij$. }
\label{fig:ii-ij_Jn}
\end{center}
\end{figure}

Using the auxiliary variable $z$ as a placeholder for the logarithmic label, we have 
\be
\begin{split}
F(\wp_1) &= \, \prod_{k\ge 0}\bigg[ e^{-z^{kn} \Xi_{ii,kn}}\, e^{-z^{kn} \Xi_{jj,kn}} \bigg] (1- z^m \Xi_{ij,m}) 
\\ 
F(\wp_2) &=\, \prod_{l\ge 0}\bigg(1- z^{m+ln} \bar\Xi_{ij, m+ln)}\bigg)\,\prod_{k\ge 0}\bigg[ e^{-z^k \Xi_{ii,kn}}\, e^{-z^k \Xi_{jj,kn}} \bigg]
\end{split}
\ee
Let us denote
\be \label{eq:Sigma-ii-jj}
\Sigma_{ii}(z) = \prod_{k\ge 0} e^{-z^k \Xi_{ii,kn}}-1, \quad \Sigma_{jj}(z) =  \prod_{k\ge 0} e^{-z^k \Xi_{jj,kn}}-1
\ee 
Then the flatness equations become 
\be
\begin{split}
F(\wp_1) &= \, 1+ \Sigma_{ii} + \Sigma_{jj} - (1+ \Sigma_{ii}) z^m \Xi_{ij,m}
\\ 
F(\wp_2) &=\, 1 + \Sigma_{ii} + \Sigma_{jj} - \sum_{l=0}^\infty z^{m+ln} \bar\Xi_{m+ln} (1+ \Sigma_{jj})
\end{split}
\ee

Any walls of $ii/jj$ type must be generated from an $ij-ji$ junction, from the results of \cite{Banerjee:2018syt} one therefore has\footnote{\label{foot:signs-S-ii}There is in principle an ambiguity in choosing $\Sigma_{ii}$ vs $\Sigma_{jj}$, which is unambiguously fixed by demanding that the $(ij,m)$ wall crosses the $ii/jj$ walls in the same direction as the $ij$ wall that generated the latter. More precisely, there is an extra sign in the exponent here compared to \cite{Banerjee:2018syt} arising from (\ref{eq:sign-rule}). Taking this into account would give $\prod_{k\geq 1}e^{z^n\Xi_{ii,kn}} = (1+z^n\Theta)^{-1}$ in \cite{Banerjee:2018syt}. On top of that, here in Figure \ref{fig:ii-ij_Jn} there is also an inversion of sheet labels $i\leftrightarrow j$, which switches back to $(1+z^n\Theta)$. Therefore in (\ref{eq:Sigma-ii-jj}) one gets $\Sigma_{ii} = (1+\Theta)^{-1} - 1$ leading to (\ref{eq:Sigma-ii-ij-theta}).} 
\be \label{eq:Sigma-ii-ij-theta}
\Sigma_{ii} = -\frac{z^n\Theta}{1+z^n\Theta}, \qquad \Sigma_{jj} = z^n\bar\Theta
\ee 
and as a consequence of the shift symmetry \cite{Banerjee:2018syt}, 
\be 
\Theta \Xi_{ij,p} = \Xi_{ij,p} \bar\Theta, \qquad \Xi_{ji,p} \Theta = \bar\Theta \Xi_{ji,p} \,.
\ee 
Plugging these into the flatness equation yields 
\be
\begin{split}
	\sum_{l\ge 0}z^{m+ln}\bar\Xi_{ij, m+ln} & =\, (1+z^n\Theta)^{-1} \Xi_{ij,m} (1+z^n\bar\Theta)^{-1}
\\ & 
= (1+z^n\Theta)^{-2} \Xi_{ij,m}
\end{split}
\ee 
Expanding in $z$ and matching terms with the same logarithmic shifts, we obtain 
\be	
	\bar\Xi_{ij, m+ln} \, = \, (-1)^l (l+1) \Theta^l \Xi_{ij,m},
\ee
and in particular for $l=0$, $\bar\Xi_{ij,m} = \Xi_{ij,m}$.

\bibliography{biblio}{}
\bibliographystyle{JHEP}

\end{document}